\documentclass[11pt]{article}
\usepackage{tikz}
\usepackage{endnotes}
\usepackage[hyperindex,breaklinks]{hyperref}
\newcounter{textlabels}
\usepackage{hyperendnotes}
\usepackage{float}
\usepackage{latexsym,calligra,amsmath,amssymb,stmaryrd,MnSymbol,ushort,amsthm}
\usepackage[parfill]{parskip} 
\usepackage{multicol}
\usepackage[bf, small, justification=justified,singlelinecheck=false]{caption}
\newif\ifdviwin
\usepackage{ulem}
\usepackage[all,cmtip]{xy}
\usepackage[latin1]{inputenc}
\usepackage[english]{babel}
\usepackage{indentfirst}
\usepackage[mathscr]{eucal}
\usepackage{amssymb,amsmath,amsfonts}
\usepackage{graphicx}
\usepackage{amsmath,amscd}
\usepackage{xxcolor}
\newtheorem{theorem}{Theorem}[subsection]
\newtheorem{lem}{Lemma}

\newtheorem{defi}[theorem]{Definition}

\numberwithin{equation}{subsection} 
\numberwithin{figure}{subsection} 
\numberwithin{lem}{subsection}
\DeclareMathAlphabet{\mathcalligra}{T1}{calligra}{m}{n}
\DeclareFontShape{T1}{calligra}{m}{n}{<->s*[2.2]callig15}{}

\newcommand{\mc}[1]{\mathcal{#1}}
\newcommand{\ms}[1]{\mathscr{#1}}
\newcommand{\mbf}[1]{\mathbf{#1}}
\newcommand{\mbb}[1]{\mathbb{#1}}

\newcommand{\tn}[1]{\textnormal{#1}}
\newcommand{\fsz}[1]{\footnotesize{#1}}

\def\PP{\mathbb{P}}
\def\AA{\mathbb{A}}
\def\QQ{\mathbb{Q}}
\def\ZZ{\mathbb{Z}}
\def\NN{\mathbb{N}}
\def\RR{\mathbb{R}}
\def\CC{\mathbb{C}}
\tikzset{isometricXYZ/.style={x={(-0.707cm,-0.354cm)}, y={(0.707cm,-0.354cm)}, z={(0cm,1cm)}}}
\newcommand{\comment}[1]{}
\begin{document}

\hspace*{3.1cm}{\huge On the Axioms of Causal Set Theory}

\vspace*{.3cm}

\hspace*{6.7cm}{\large Benjamin F. Dribus}

\vspace*{0cm}

\hspace*{6.27cm}{Louisiana State University}

\vspace*{-.2cm}

\hspace*{6.65cm}{\color{blue} bdribus@math.lsu.edu}

\vspace*{0cm}

\hspace*{6.8cm}{\large December 24, 2013}

\vspace*{.2cm}

\begin{abstract} 
\noindent {\it Causal set theory} is a promising attempt to model fundamental spacetime structure in a discrete order-theoretic context via sets equipped with special binary relations, called {\it causal sets.}  The elements of a causal set are taken to represent {\it spacetime events,} while its binary relation is taken to encode {\it causal relations} between pairs of events.  Causal set theory was introduced in 1987 by Bombelli, Lee, Meyer, and Sorkin, motivated by results of Hawking and Malament relating the causal, conformal, and metric structures of relativistic spacetime, together with earlier work on discrete causal theory by Finkelstein, Myrheim, and 't Hooft.   Sorkin has coined the phrase, {\it ``order plus number equals geometry,"} to summarize the causal set viewpoint regarding the roles of causal structure and discreteness in the emergence of spacetime geometry. This phrase represents a specific version of what I refer to as the {\it causal metric hypothesis,} which is the idea that the properties of the physical universe, and in particular, the metric properties of classical spacetime, arise from causal structure at the fundamental scale.\\

\vspace*{-.2cm}
\noindent Causal set theory may be expressed in terms of six axioms: the {\it binary axiom}, the {\it measure axiom}, {\it countability}, {\it transitivity}, {\it interval finiteness}, and {\it irreflexivity}.   The first three axioms, which fix the physical interpretation of a causal set, and restrict its ``size," appear in the literature either implicitly, or as part of the preliminary definition of a causal set.  The last three axioms, which encode the essential mathematical structure of a causal set, appear in the literature as the {\it irreflexive formulation of causal set theory.}  Together, these axioms represent a straightforward adaptation of familiar notions of causality to the discrete order-theoretic context.  Interval finiteness is often called {\it local finiteness} in the literature, an unfortunate misnomer.\\

\vspace*{-.2cm}
\noindent In this paper, I offer what I view as potentially critical improvements to causal set theory, including {\it changes to the axioms,} and {\it new perspectives and technical methods.}  Abandoning continuum geometry introduces new types of behavior, such as {\it irreducibility} and {\it independence} of causal relations between pairs of events, inadequately modeled by conventional order theory.  Transitive binary relations fail to resolve the subtleties of independent modes of influence between pairs of events.  Interval finiteness permits {\it locally infinite behavior} incompatible with Sorkin's version of the causal metric hypothesis, and imposes unjustified restrictions on the global structure of classical spacetime.  Intertwined with both axioms is the use of {\it causal intervals} to study local spacetime properties, despite their failure to capture or isolate local causal structure.\\

\vspace*{-.2cm}
\noindent To address these issues, I propose to replace the axioms of transitivity, interval finiteness, and irreflexivity, with {\it local finiteness,} and, under conservative assumptions, {\it acyclicity}. I refer to a binary relation satisfying these new axioms as a {\it locally finite causal preorder}; its transitive closure is the familiar {\it causal order}.  The resulting models, which I call {\it locally finite directed sets,} generalize both causal sets and Finkelstein's {\it causal nets.}  The resulting theory diverges significantly from existing approaches, particularly at the quantum level.  This is due principally to a broader interpretation of directed structure, and more generally, multidirected structure, inspired by Grothendieck's scheme-theoretic approach to algebraic geometry.  Entities more complex than spacetime events are viewed as {\it elements of higher-level multidirected sets,} in analogy to Isham's {\it topos-theoretic approach} to quantum gravity, and Sorkin's {\it quantum measure theory.}  This viewpoint leads to a new background-independent version of quantum causal theory.  The backbone of this approach is the theory of {\it co-relative histories,} an adaptation of the familiar {\it histories approach} to quantum theory.   Co-relative histories serve as the ``relations" of higher-level multidirected sets called {\it kinematic schemes,} via the principle of {\it iteration of structure.}  Systematic use of {\it relation space} circumvents the causal set problem of {\it permeability of maximal antichains,} leading to the derivation of {\it causal Schr\"{o}dinger-type equations} describing quantum spacetime dynamics in the discrete causal context. 

\end{abstract}

\newpage

\tableofcontents

\section{Introduction}\label{sectionintroduction}

Causal sets are discrete order-theoretic models of classical spacetime.  A {\it causal set} is a set $C$, assumed to be {\it countable}, whose {\it elements correspond to spacetime events,} together with a binary relation $\prec$ on $C$, satisfying the additional axioms of  {\it transitivity}, {\it irreflexivity}, and {\it interval finiteness}.  The binary relation $\prec$ defines an {\it interval finite partial order}\footnotemark\footnotetext{Section \hyperref[subsectionaxioms]{\ref{subsectionaxioms}} clarifies how the {\it irreflexive relation} $\prec$ corresponds to a {\it partial order} in the usual sense of the term.} on $C$, called the {\it causal order}, with the physical interpretation that $x\prec y$ in $C$ if and only if the event represented by $x$ exerts causal influence on the event represented by $y$. The physical interpretation of $C$ is completed by fixing a {\it discrete measure}\footnotemark\footnotetext{Here, {\it discrete measure} simply means that singletons have positive measure.} $\mu$ on $C$, assigning to each subset of $C$ a volume equal to its number of elements in fundamental units, up to {\it Poisson-type fluctuations.}\footnotemark\footnotetext{See the remarks on page 2 of Ahmed, Dodelson, Greene, and Sorkin's paper {\it Everpresent $\Lambda$} \cite{SorkinEverPresentLambda04}.} The role of $\prec$ and $\mu$ in modeling classical spacetime is summarized by the phrase, {\it ``order plus number equals geometry,"} coined by Rafael Sorkin, the foremost architect and advocate of causal set theory. This phrase represents a special case of what I refer to as the {\it causal metric hypothesis} (\hyperref[cmh]{CMH}), which is the idea that the properties of the physical universe, and in particular, the metric properties of classical spacetime, {\it arise from causal structure at the fundamental scale.}  Approaches to fundamental physics involving some form of the causal metric hypothesis may be collectively referred to as {\bf causal theory}.  Besides causal set theory, these include other discrete causal theories, such as {\it causal dynamical triangulations} and {\it causal nets,} as well as theories involving {\it interpolative} causal models of spacetime, such as {\it domain theory.} 

The purpose of this paper is to offer potentially critical improvements to the causal set program, and to causal theory in general.  This effort is based on the conviction that causal theory is among the best-motivated existing approaches to the outstanding problems of fundamental physics, and that causal set theory is perhaps the cleanest and best-balanced existing version of causal theory.   Despite important conceptual distinctions, and critical technical differences, between the existing formulation of causal set theory and the approach I offer in this paper, the two are close enough to be considered part of the same broad program.  For example, they are closer to each other than they are to the other versions of causal theory mentioned above.   

In this introductory section, I sketch a broad conceptual context for the more specific material to follow.   Section \hyperref[subsectionapproach]{\ref{subsectionapproach}} is a brief overview of causal set theory, describing its origins, outlining  its principal methods, and providing a glimpse of its current state of development.  I have included references to more thorough treatments, carefully selected for reliability and quality of exposition.  Section \hyperref[subsectionpurpose]{\ref{subsectionpurpose}} outlines in more detail the purposes of this paper, its method and approach, and its intended audience.  Section \hyperref[naturalphilosophy]{\ref{naturalphilosophy}} introduces general principles of natural philosophy used throughout the paper.  Section \hyperref[subsectionnotation]{\ref{subsectionnotation}} describes the notation and conventions of the paper in general terms; appendix \hyperref[appendixindex]{A} is much more thorough.  Section \hyperref[subsectionoutline]{\ref{subsectionoutline}} provides a hyperlinked topical outline of the succeeding material, more detailed than that provided by the table of contents above. 


\subsection{Overview of the Causal Sets Program}\label{subsectionapproach}

{\bf { Historical Antecedents.}} The study of causality long predates any formal mathematical notion of order, and it is important to resist viewing relevant early developments in a {\it teleological} sense, as mere steps along a path toward present convention.   For example, Zeno's {\it dichotomy paradox,} proposed around 450 B.C., is often described today in terms of the subdivision of continuum intervals, and ``resolved" by the convergence of geometric series.  However, this particular version of Zeno's paradox was originally stated in a {\it physical} context, in terms of {\it sequences of events}, with the physical issue at stake being whether or not fundamental causal structure is {\it interpolative.}   From this perspective, Zeno's paradox is as relevant today as it was when it was first proposed.  Finiteness axioms, such as the causal set axiom of {\it interval finiteness} (\hyperref[if]{IF}), stated in section \hyperref[subsectionaxioms]{\ref{subsectionaxioms}}, or the alternative axiom of {\it local finiteness} (\hyperref[lf]{LF}), presented in section  \hyperref[subsectionintervalfiniteness]{\ref{subsectionintervalfiniteness}}, offer {\it noninterpolative} resolutions of Zeno's paradox.  

Prior to the discover of relativity and quantum theory, the overwhelming success of continuum methods in physics relegated most alternative ideas about the basic structure of the universe to the periphery of scientific thought.  Great thinkers such as Leibniz and Riemann considered such ideas, but the scientific community as a whole followed more pragmatic lines.  Relativity brought fresh scrutiny to {\it structural issues,} elevating causality to its present central role.  For historical background on the relativistic roots of causal set theory, it is difficult to improve upon the efforts of Sorkin.  For example, in his 2005 article {\it Causal Sets: Discrete Gravity} \cite{SorkinCausalSetsDiscreteGravity05}, Sorkin details the isolation and elevation of causal structure in the relativity literature as early as the 1930's, and even quotes Einstein's misgivings about continuum geometry, and Riemann's early remarks about {\it discrete manifolds.}   David Finkelstein also provides useful historical context reaching back before 1950, in his 1988 paper {\it ``Superconducting" Causal Nets} \cite{Finkelstein88}.  Finkelstein's 1969 paper {\it Space-Time Code} \cite{Finkelstein69} is perhaps the earliest ``modern" paper on discrete causal theory, described by Sorkin \cite{SorkinCausalSetsDiscreteGravity05} as having ``adumbrated'' the causal sets program.  Results of Hawking and Malament in the late 1970's, discussed further below, provided new evidence of the quintessential role of causal structure in determining the properties of classical spacetime.  Soon after this, ideas very similar to causal set theory appeared in Myrheim's 1978 CERN preprint {\it Statistical Geometry}, \cite{Myrheim78} and 't Hooft's 1978 notes {\it Quantum Gravity: A Fundamental Problem and some Radical Ideas} \cite{tHooft78}. These efforts were spurred by the continued failure of continuum methods to provide a viable avenue toward a theory of quantum gravity, and the growing expectation that fundamental spacetime structure is non-smooth.  Finally, causal set theory was introduced by Bombelli, Lee, Meyer, and Sorkin, in their 1987 paper {\it Space-Time as a Causal Set} \cite{Sorkinetal87}.\footnotemark\footnotetext{Crediting Finkelstein, Myrheim, and 't Hooft's earlier efforts, Bombelli, Lee, Meyer, and Sorkin write that {\it ``The picture of space-time as a causal set is by no means new,"} but go on to explain that Myrheim and 't Hooft's proposals are {\it ``undeveloped,"} and that Finkelstein's {\it ``led to formulations in which the issues we would like to address here were not dealt with."}} 


{\bf Malament's Theorem; Metric Recovery.} An important foundational result directly preceding the emergence of causal set theory is a 1977 theorem of David Malament, relating the causal and conformal structures of classical spacetime in the context of general relativity \cite{Malament77}.  Hawking, King, and McCarthy \cite{Hawking76} had already established that {\it topological structure determines conformal structure} under suitable hypotheses; Malament showed that causal structure, in turn, determines topological structure.  Combining these two results, Malament concluded that the only obstruction to the recovery of pseudo-Riemannian spacetime geometry from causal structure is a {\it lack of appropriate data about scale;} i.e., a missing conformal factor.  More precisely, let $X$ and $X'$ be connected, four-dimensional, smooth manifolds without boundary, and let $g$ and $g'$ be smooth pseudo-Riemannian metrics of Lorentz signature on $X$ and $X'$, respectively.  In this context, Malament proved that {\it if $f:X\rightarrow X'$ is a bijection such that $f$ and $f^{-1}$ preserve future-directed continuous timelike curves,\footnotemark\footnotetext{In general relativity, a {\it causal curve} is a curve that is either timelike or null.  Malament's Lemma 1 (\cite{Malament77}, page 1400), building on Hawking's theorem \cite{Hawking76}, establishes that timelike curves suffice in this context.} then $f$ is a conformal isometry.}  Subsequent results in causal set theory and domain theory have clarified and broadened this picture.  For example, Luca Bombelli and David Meyer's 1989 paper {\it The origin of Lorentzian geometry} \cite{Bombelli89}, and Keye Martin and Prakash Panangaden's 2006 paper {\it A Domain of Spacetime Intervals in General Relativity} \cite{Martin06}, both contain results pertaining to the recovery of geometry from the causal structure of a {\it countable dense subset} of classical spacetime. 


{\bf Sorkin: ``Order Plus Number Equals Geometry."} Malament's theorem provides clear motivation for Sorkin's phrase, {\it ``order plus number equals geometry,"} which encapsulates Sorkin's version of the causal metric hypothesis (\hyperref[cmh]{CMH}).  Here, {\it order} is represented by the binary relation $\prec$ on a causal set $C$, which corresponds to the family of causal curves on a pseudo-Riemannian manifold in Malament's theorem, while {\it number} is endowed with volume via the discrete measure $\mu$ on $C$, which corresponds to Malament's missing conformal factor.  The idea of using the {\it counting measure} as a proxy for volume data appeared almost immediately after Malament's paper \cite{Malament77}.  In his 1978 preprint \cite{Myrheim78}, Myrheim writes, 
\begin{quotation}\noindent{\it``If spacetime is assumed to be discrete, then the counting measure is the natural measure, and the causal ordering is the only structure needed. Coordinates and metric may be derived as secondary, statistical concepts."} (page 1)
\end{quotation}
In their 1987 paper \cite{Sorkinetal87}, Bombelli, Lee, Meyer, and Sorkin write,
\begin{quotation}\noindent{\it``In this view volume is number, and macroscopic causality reflects a deeper notion of order in terms of which all the ``geometrical" structures of space-time must find their ultimate expression."} (page 522)
\end{quotation}
These statements, which implicitly {\it identify} the discrete volume measure $\mu$ with the counting measure, precede the appearance of the caveat {\it ``up to Poisson-type fluctuations"} in the later literature.  Section \hyperref[subsectionaxioms]{\ref{subsectionaxioms}} contains a more thorough discussion of this point, along with appropriate quotations. 

Due to the axiom of interval finiteness (\hyperref[if]{IF}), the information encoded in a causal set can only {\it approximately} supply the causal and conformal ingredients of Malament's theorem.  It is important to emphasize that {\it causal set theory treats relativistic spacetime as an approximation of a causal set, not vice versa.}   More precisely, continuum geometry is viewed, from the causal set perspective, as a {\it large-scale smoothing} of the fundamental structure encoded by $C$, $\prec$ and $\mu$.  The extent to which a ``manifold-like" causal set {\it uniquely} determines large-scale manifold structure is, as far as I know, still an outstanding issue.\footnotemark\footnotetext{In particular, Martin and Panangaden's ``discrete metric recovery" result \cite{Martin06} involves a {\it dense} subset, and therefore has no {\it effective volume gap} for the ``size" of elements.}  The conjecture that an appropriate uniqueness result exists is called the causal set {\it hauptvermutung}, or {\it fundamental conjecture}, a historical reference to an analogous conjecture in geometric topology.\footnotemark\footnotetext{Namely, that any two triangulations of a triangulable space admit a common refinement.  John Milnor proved in 1961 that this conjecture is {\it false}.}  Sorkin discusses the causal set {\it hauptvermutung} in \cite{SorkinCausalSetsDiscreteGravity05}. 


{\bf Manifold Embedding; Sprinkling.}  Most causal sets bear little resemblance to manifolds, and are therefore irrelevant to classical spacetime structure.  Here, the quantifier ``most" is made precise by the {\it asymptotic enumeration} of Kleitman and Rothschild \cite{KleitmanRothschild75}, discussed in more detail in section \hyperref[subsectiontransitivitydeficient]{\ref{subsectiontransitivitydeficient}}.  An {\it effective realization} of the causal metric hypothesis (\hyperref[cmh]{CMH}), whereby one may {\it derive} the emergence of approximate manifold structure at appropriate scales, beginning with just the axioms of causal set theory and suitable dynamical laws, has not yet been discovered.  A useful intermediate step is to study causal sets {\it induced} by ``sprinkling" elements discretely on a pseudo-Riemannian manifold, according to a {\it Poisson process.} One may thereby obtain physically relevant causal sets for investigative purposes, without the necessity of knowing how these sets might arise dynamically.  The method of sprinkling was introduced at the dawn of causal set theory in Bombelli, Lee, Meyer, and Sorkin's inaugural paper \cite{Sorkinetal87}.  Sprinkling is illustrated schematically in figure \hyperref[sprinkling]{\ref{sprinkling}} below.  In figure \hyperref[sprinkling]{\ref{sprinkling}}a, a region of a classical spacetime manifold $X$ is shown, together with a distinguished event $x$ on $X$, and a portion of its light cone.  In figure \hyperref[sprinkling]{\ref{sprinkling}}b, a family $C$ of elements is sprinkled on $X$.  In figure \hyperref[sprinkling]{\ref{sprinkling}}c, $X$ is removed from the picture, and only the abstract causal structure of $C$, induced by $X$, remains.  Each edge in the figure represents a causal relation between its endpoints, directed upward.  This is an example of a {\it Hasse diagram}, defined more generally below. Note that the diagram omits {\it reducible} relations; i.e., relations between pairs of elements of $C$ connected by complex sequences of other relations.  This practice is permissible in causal set theory, essentially because causal sets are {\it transitive.} However, it {\it cannot} be applied to most of the alternative models of causal structure appearing in this paper, without destroying information. 

\begin{figure}[H]
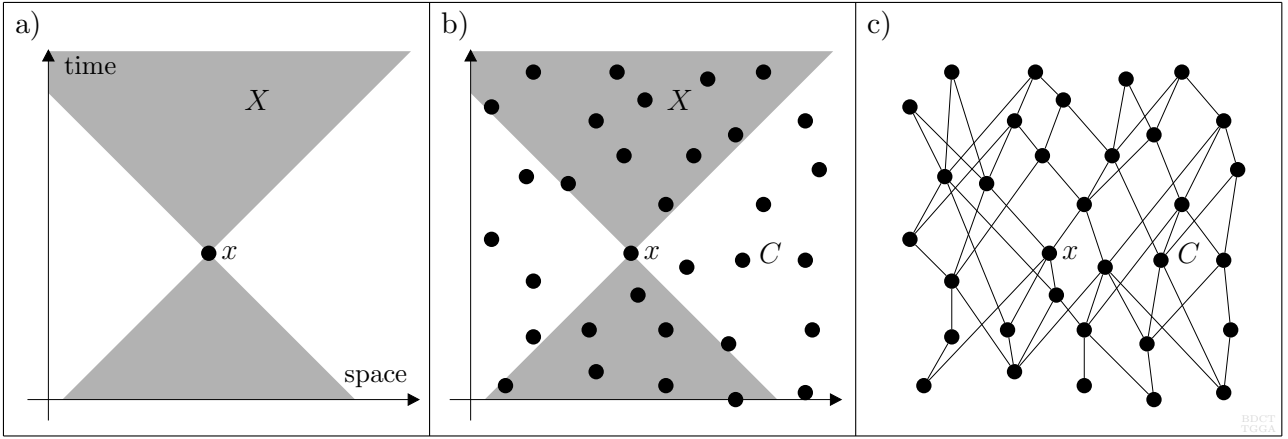


\caption{a) Classical spacetime manifold $X$; b) family $C$ of elements sprinkled on $X$; c) abstract causal structure of $C$, induced by $X$.}
\label{sprinkling}
\end{figure}
\vspace*{-.5cm}


{\bf Quantum Theory.}  The quantum theory of causal sets may be broadly divided into {\it background-dependent} and {\it background-independent} approaches.  The former involve the study of auxiliary structures, such as particles and fields, existing {\it on} a causal set, while the latter involve the quantum nature of spacetime itself.  Background dependent approaches are the easier of the two, and have been extensively developed in the existing literature.   While interesting in their own right, most of these approaches differ too much in emphasis from the approach developed in this paper to provide helpful sources.  An exception is the work of Ioannis Raptis, beginning with his 2000 paper {\it Algebraic Quantization of Causal Sets} \cite{RaptisAlgebraicQuantization00}, which provides useful context for the theory of {\it causal path algebras} appearing in section \hyperref[subsectionpathspaces]{\ref{subsectionpathspaces}}, and the {\it abstract quantum causal theory} developed in section \hyperref[subsectionquantumpathsummation]{\ref{subsectionquantumpathsummation}}, of this paper. 

Background independent quantum causal theory is much more interesting and difficult.   The familiar {\it histories approach to quantum theory} offers promising general methods in this context.  This approach may be traced back to Richard Feynman's method of {\it path integration,} first introduced in his 1948 paper {\it Space-Time Approach to Non-Relativistic Quantum Mechanics} \cite{FeynmanSOH48}, and subsequently applied with great success to a host of important problems throughout modern physics.  Chris Isham's category-theoretic treatment of histories, introduced in his 2005 paper {\it Quantising on a Category} \cite{IshamQuantisingI05}, provides part of the motivation for the theory of {\it co-relative histories} and {\it kinematic schemes} in section \hyperref[subsectionquantumcausal]{\ref{subsectionquantumcausal}} of this paper.  Also relevant is Sorkin's work on the general topic of {\it quantum measure theory,} including his 2012 paper {\it Toward a Fundamental Theorem of Quantal Measure Theory} \cite{SorkinQuantalMeasure12}, which discusses quantum-theoretic measures on configuration spaces of histories, both for causal sets and discrete lattices.  These ideas foreshadow the theory of {\it co-relative kinematics,} and {\it completions of kinematic schemes,} both discussed in section \hyperref[subsectionquantumcausal]{\ref{subsectionquantumcausal}} below. 
  

{\bf Dynamics.}  Causal set theory presently lacks dynamical laws describing the evolution of suitable manifold-like causal sets, or configuration spaces of such sets.  Nevertheless, suggestive ``toy models" have been examined.  The most prominent of these is a {\it classical stochastic model} called {\bf sequential growth dynamics}, introduced by Sorkin and his student David Rideout in their 1999 paper {\it Classical sequential growth dynamics for causal sets} \cite{SorkinSequentialGrowthDynamics99}.   In sequential growth dynamics, a causal set is ``built up one element at a time," with each ``evolutionary step" represented by a special causal set {\it morphism,} called a {\it transition}.\footnotemark\footnotetext{In fact, a transition only {\it represents} a physically more fundamental relationship called a {\it co-relative history,} as explained in section \hyperref[subsectionquantumcausal]{\ref{subsectionquantumcausal}}.}  Transitions organize the class of causal sets, considered up to isomorphism,\footnotemark\footnotetext{Throughout this paper, structured sets are usually considered only up to isomorphism, since differences between members of an isomorphism class are physically immaterial.}  into a ``higher-level" structure $\ms{K}$, which I refer to as a {\it kinematic scheme}.  The theory and terminology of kinematic schemes is discussed in more detail in section \hyperref[subsectionkinematicschemes]{\ref{subsectionkinematicschemes}} of this paper.  $\ms{K}$ has the abstract structure of a {\it multidirected set}.\footnotemark\footnotetext{In particular, it is better {\it not} to view $\ms{K}$ as a partially ordered set, since there exist ``inequivalent" transitions with the same source and target.  See section \hyperref[subsectionkinematicschemes]{\ref{subsectionkinematicschemes}} for an example of this, due to Brendan McKay.} It is analogous to a mathematical {\it category,} with its member causal sets corresponding to {\it objects,} and transitions between pairs of these causal sets corresponding to {\it morphisms.}\footnotemark\footnotetext{Of course, transitions {\it are} morphisms in the category of causal sets; the point is that they serve a similar structural role in $\ms{K}$, which is {\it not} a category.  More precisely, they {\it represent} the ``morphisms" of $\ms{K}$, which are co-relative histories, as explained in section \hyperref[subsectionquantumprelim]{\ref{subsectionquantumprelim}}.}   This analogy has been developed in a striking way by Chris Isham and his collaborators.  In section \hyperref[subsectionquantumcausal]{\ref{subsectionquantumcausal}} of this paper, I offer a somewhat different perspective regarding such constructions.  

\begin{figure}[H]
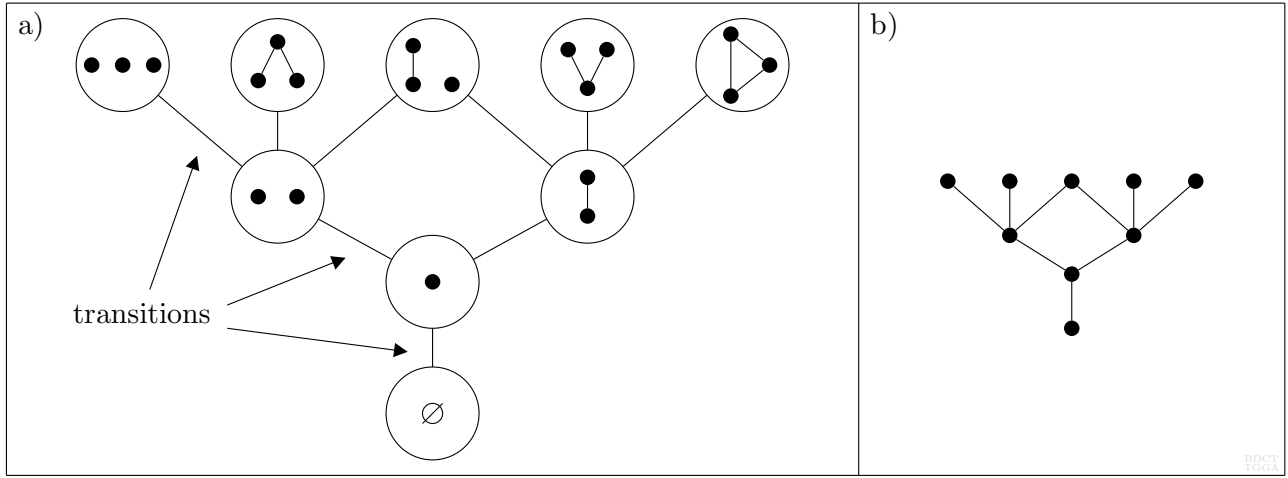


\caption{a) Portion of Sorkin and Rideout's kinematic scheme $\ms{K}$ describing sequential growth of causal sets; b) abstract structure of this portion of $\ms{K}$. }
\label{sequential}
\end{figure}

Figure \hyperref[sequential]{\ref{sequential}}a below illustrates a portion of the kinematic scheme $\ms{K}$. Individual causal sets appear in the large open nodes, and transitions are represented by the edges.  To each transition is associated an {\it initial causal set} $C_i$, and a {\it terminal causal set} $C_t$, given by adding a single element to $C_i$, along with a family of relations.  

\vspace*{-.5cm}

The kinematic scheme $\ms{K}$ only describes {\it possible} evolutionary behavior of causal sets.   The actual {\it dynamics} of sequential growth is supplied by assigning a {\it weight} or {\it phase} $\theta(\tau)$ to each transition $\tau$ in $\ms{K}$, encoding the {\it ``likelihood that $C_i$ transitions to $C_t$."}\footnotemark\footnotetext{This deliberately vague description is intended to cover a range of possible choices for the values of $\theta$, from real probabilities, to complex amplitudes, to more exotic possibilities.}   The ``likelihood" of evolution between any pair of causal sets $C$ and $C'$ in $\ms{K}$ may then be expressed in terms of the weights of all transitions along all upward-directed sequences of transitions from $C$ to $C'$ in $\ms{K}$.  In the specific context of classical sequential growth dynamics, Sorkin and Rideout take these weights to be real probabilities, but $\ms{K}$ may also be viewed in a quantum-theoretic context, as in \cite{SorkinQuantalMeasure12}.  Each upward-directed sequence of transitions from $C$ to $C'$ in $\ms{K}$ corresponds to a sequential labeling of the elements of $C'-C$.  If $C$ is empty, such a sequence defines a {\it total labeling} $\{x_0,x_1,x_2,...\}$ of $C'$, which is equivalent to a {\it linear extension} of the causal order, or alternatively, to a {\it bijective order morphism} of $C'$ into a linear suborder of the nonnegative integers.\footnotemark\footnotetext{This procedure is unambiguous if and only if $\ms{K}$ is viewed as a (higher-level) multidirected set.  In particular, a total labeling of an ``initial"  finite causal set $C_i$, together with the isomorphism class of the ``next" causal set $C_t$, generally does {\it not} determine a class of labelings of $C_t$ related by automorphisms of $C_t$.}  Such a sequence is analogous to a relativistic {\it frame of reference,} supplementing the causal structure of its terminal causal set with arbitrary nonphysical information, just as relativistic coordinates assign arbitrary time orders to pairs of spacelike-separated events.   In section \hyperref[subsectionkinematicschemes]{\ref{subsectionkinematicschemes}} of this paper, I generalize and formalize such sequences as {\it co-relative kinematics.} 


{\bf Phenomenology.} Like other approaches to fundamental spacetime structure, causal set theory cannot yet boast a robust record of phenomenological success.  However, the difficulty of wringing any definitive and experimentally accessible prediction out of a relatively young fundamental theory makes it reasonable to exercise considerable patience in evaluating the ongoing viability of the causal sets program, given its plausibility and motivation.  The most well-known phenomenological ``achievement" of causal set theory to date is an approximate prediction of the size of {\it fluctuations} of the cosmological constant, as discussed in \cite{SorkinEverPresentLambda04}.  Assuming fluctuations centered about zero, this prediction falls within the range suggested by observation.  However, this result relies on assumptions about the interpretation of volume, the emergent dimensionality of spacetime, and the large-scale structure of the universe, that one would {\it a priori} prefer to avoid.  For a fine overview of this and other topics in causal set phenomenology, I refer the reader to Sumati Surya's recent review article {\it Directions in Causal Set Quantum Gravity} \cite{Surya11}.  Surya references a wide variety of recent developments, including Dowker, Henson, and Sorkin's work on {\it diffusion in a causal set background} \cite{SorkinLorentz04}, examination of {\it causal set curvature} and analogues of the {\it Einstein-Hilbert action} by Dowker and her collaborators \cite{DowkerScalarCurvature10}, and the construction of a {\it scalar field Feynman propagator} by Johnston \cite{JohnstonFeynman09}. 

Causal set phenomenology has thus far been directed more toward working out the preliminaries of {\it how causal set theory might make contact with the experimental realm} than toward making specific, falsifiable predictions.  Most of the topics cited above involve background-dependent versions of the theory, and are therefore of only limited relevance to this paper.  An important basic phenomenological question is how nongravitational matter might arise in causal set theory.  Sorkin \cite{SorkinSequentialGrowthDynamics99} advocates {\it ``a kind of ``effective action" for a field of Ising spins living on the relations of the causal set."}   In certain cases, such a field may be {\it attributed to the underlying causal structure.}  See, for example, the analogous discussion of {\it valence fields} in section \hyperref[subsectiontopology]{\ref{subsectiontopology}} of this paper.  


\subsection{Purposes; Viewpoint and Style; Intended Audience}\label{subsectionpurpose}

Three outstanding problems of modern theoretical physics are: 1) {\it description of fundamental spacetime structure;} 2) {\it development of a suitable theory of quantum gravity;} and 3) {\it unification of physical law.}  These problems are, of course, intertwined.  Ever since the discovery of general relativity, gravity has been understood to be {\it structural} in nature, and it is difficult to imagine a successful approach to quantum gravity without a deeper structural understanding of spacetime.  Quantum theory, whatever its interpretation, is believed by most serious theorists to represent a fundamental aspect of nature, so any successful unification of physical law is expected to be quantum-theoretic.  It seems, then, that two common ingredients necessary for the solution of these three problems are {\it a suitable notion of structure} and {\it a suitable notion of quantum theory.} This paper represents an attempt to help identify such notions.  


{\bf Purposes.} Under the causal metric hypothesis (\hyperref[cmh]{CMH}), the ``suitable notion of structure" is {\it causal structure,} and the question becomes {\it how this structure should be modeled.}  The primary purpose of this paper is to address this question, beginning at the classical level.  I focus attention on {\it discrete models satisfying a local finiteness condition,} motivated by the common expectation that a finite fundamental scale exists in physics, by technical results such as Malament's theorem, and by the mathematical tractability and interest of such models.   Turning to the quantum realm, the histories approach to quantum theory, mentioned above, is a general and flexible candidate for a ``suitable notion of quantum theory," and is also particularly amenable to discrete causal theory.  An important secondary purpose of this paper, therefore, is to {\it outline how classical causal structure may be combined with the histories approach to quantum theory.}   This leads to the theory of {\it co-relative histories,} described in section \hyperref[subsectionquantumprelim]{\ref{subsectionquantumprelim}} below. The fact that the resulting version of quantum causal theory seems to have attractive properties reinforces the preceding choices. 


{\bf Viewpoint and Style.}  Causal set theory already occupies similar philosophical and theoretical ground, seeking to combine discrete causal structure with the histories approach to quantum theory.   It is therefore reasonable to take causal set theory as a starting point, and propose modifications as needed.  Other similar theories also play important roles: Finkelstein's {\it causal nets} are very similar, at the level of individual objects, to the {\it directed sets}\footnotemark\footnotetext{The term {\it directed set} has  a meaning equivalent to {\it small, simple directed graph} in this paper, which is a more general meaning than it has in the usual order-theoretic settings.  See the discussion of nonstandard terminology in section \hyperref[subsectionnotation]{\ref{subsectionnotation}}, and the definition of directed sets in section \hyperref[subsectionacyclicdirected]{\ref{subsectionacyclicdirected}} below.} used in this paper as models of classical spacetime, and Isham's category-theoretic methods supply important aspects of higher-level structure and hierarchy.  However, the overall approach developed here is closer, from a physical perspective, to Sorkin's causal set theory, than to any other existing approach.  Mathematically, this paper is significantly different from most of the causal set literature, incorporating substantial influences from modern algebra and algebraic geometry.  However, these influences are manifested less by direct use of existing algebraic notions, such as those contributing to Isham's approach, than by {\it analogy.}  For example, the multidirected structures of principal interest in this paper are not categories, but {\it resemble} categories in useful ways.  The length and style of this paper is motivated by the conviction that one cannot expect others to seriously consider an alternative approach to fundamental physics without a substantial account of its underlying motivations and potential applications.  Hence, in writing this paper, I felt compelled to venture beyond merely advocating changes to the axioms of causal set theory, at least as far as introducing the rudiments of the resulting quantum causal theory, outlined in section \hyperref[subsectionquantumcausal]{\ref{subsectionquantumcausal}} below.  Anything less would not have provided sufficient evidence that the ideas presented here actually lead to any significant improvement of existing theory.  


{\bf Intended Audience.} Every author wishes for a broad audience.   While lengthy concessions to accessibility risk exasperating expert readers, the opposite error of excessive terseness is far worse.  In writing this paper, I felt that modest expository detours, such as an extra paragraph reminding the reader what a category is, or an extension of the discussion of path summation to include comparison with the continuum analogue, were well justified.  My own background as a student of algebraic geometry naturally colors the presentation in a manner most likely to appeal to those with a taste for structure and generality, but hopefully not in an overwhelming fashion. 


\subsection{Underlying Natural Philosophy}\label{naturalphilosophy} 

A significant portion of this paper consists of analyzing the {\it physical plausibility} of various axiomatic systems for encoding  causal structure.  Einstein once remarked that the deep physical principles underlying the natural world cannot be {\it logically deduced,} but must be reached by {\it ``intuition, resting on sympathetic understanding of experience"} \cite{EinsteinEssays34}.   In recent years, Einstein has sometimes been cited as a misleading exception to the ``rule" that close grappling with experimental evidence generally yields greater success than intuition.  This rule is probably valid if theoretical success is credited solely to the {\it first} satisfactory explanation of phenomena, and if intuition is allowed to operate in a vacuum.  However, if secondary explanations, and the {\it ``sympathetic understanding of experience,"} are admitted, then intuition can lay claim to the Lagrangian and Hamiltonian formulations of classical mechanics, the histories approach to quantum theory, and many other seminal ideas besides general relativity.  Hence, in this broader context, even intuition has its merits.\footnotemark\footnotetext{The mathematical reader will recall here Gordan's reluctant endorsement of Hilbert's ``theology."}  With this in mind, I offer six philosophical principles underlying the intuitive viewpoint of this paper, together with a few contextual remarks.  I make no attempt here to be original or definitive. 

{\bf 1. Physics should seek not to prescribe what may be, but to describe what is.}\footnotemark\footnotetext{Idiotic interpretations of this statement, involving general phobia of assumptions or predictions, are easily avoided.}  This distinction is illustrated by comparing general relativity with causal theory.  The former theory views spacetime structure as {\it prescribing which pairs of events may be causally related,}\footnotemark\footnotetext{Some, such as Carlo Rovelli \cite{Rovelli04}, might argue that general relativity is merely {\it misinterpreted} along these lines.} while the latter approach views this structure as {\it describing which pairs of events are causally related.}  The descriptive philosophy avoids meaningless inconsistencies, such as ``time travel paradoxes," and unjustified assumptions, such as the presumed steady state of the universe before Hubble's observations.  The causal set axioms of transitivity (\hyperref[tr]{TR}) and interval finiteness (\hyperref[if]{IF}) are objectionably prescriptive.

{\bf 2. Mathematics and physics are distinct; each informs the other.}  Mathematical structures in physics should be chosen for their conceptual merits, rather than their familiarity or convenience.\footnotemark\footnotetext{{\it ``Concept over convenience"} recalls Gauss' {\it ``notions over notations."}} This leads not only to good physics, but also to interesting mathematics.   Mathematics returns the favor by producing concepts and methods whose physical significance is only appreciated later.  Care must be taken to distinguish between mathematical and physical properties.  An unfortunate byproduct of the long-standing success of continuum theory in physics has been the automatic and unjustified attribution of {\it mathematical} properties of the continuum, such as the {\it interpolative property,} and {\it Cauchy completeness,} to physical spacetime.  The reasoning behind the {\it independence convention} (\hyperref[ic]{IC}), introduced in section \hyperref[subsectionchains]{\ref{subsectionchains}} of this paper, provides an example of proper distinction between mathematical and physical properties; in this case, {\it irreducibility} and {\it independence} of relations between pairs of elements in a directed or multidirected set. 

{\bf 3. Basic structural concepts are crucial.}  It is instructive to consider the poverty of structural alternatives to continuum geometry available to physicists during the early 20th century, when relativity and quantum theory were being developed.  Many physically suggestive ideas from fields such as order theory, graph theory, information theory, computer science, category theory, algebra, and algebraic geometry, were not yet known.  Even group theory faced a difficult reception.\footnotemark\footnotetext{For example, Schr\"{o}dinger's {\it gruppenpest}, as related by Wigner.}  The twenty-first century scientific community is much better equipped, at least in theory, to follow up Einstein's intuition that physics is essentially structural in nature, brilliantly vindicated by general relativity, but unconsummated thereafter.   Many of the structural ideas appearing in this paper have roots in modern algebra, particularly in the work of Alexander Grothendieck.  Especially important is Grothendieck's {\it relative viewpoint} (\hyperref[rv]{RV}), discussed further in section \hyperref[settheoretic]{\ref{settheoretic}} below. 

{\bf 4. Local and global properties must be properly distinguished.}  The history of physics is littered with errors resulting from specious local-to-global reasoning and dubious extrapolation across scales.  Often such errors arise from failure to recognize the limitations of ``obvious" observations, such as the apparent motionlessness of the earth, or the apparent flatness of spacetime in the vicinity of the earth, or the apparent possibility of assigning definite values of position and momentum simultaneously to macroscopic material bodies.  Particularly troublesome are ``local" conditions adopted without recognition of their global consequences.  For example, the causal set axiom of interval finiteness (\hyperref[if]{IF}), called {\it local finiteness} in the literature, unjustifiably prescribes {\it global} structure, due to an improper distinction between local and global properties. 

{\bf 5. The nature of experimentation has a bearing on the nature of theory, not merely on the details of what theory can hope to predict.}  Besides attempting to explain specific experimental results, theorists must consider {\it general demands and prohibitions} associated with the experimental method.  For example, the {\it directed} relationship between experimental conditions and experimental results demands the application of something very similar to causal theory at our closest interface with nature.  The nature of experimentation also favors axiomatizing local rather than global properties, since the latter are often experimentally inaccessible.  For example, in the context of continuum geometry, it is reasonable to assume that classical spacetime is four-dimensional, since dimension is defined locally.  It is not reasonable, barring unlikely observational scenarios such as ``circles in the sky," to assume that classical spacetime has any particular global topology, although conclusions about global structure could conceivably be {\it derived} on dynamical grounds. 

{\bf 6. Censor the fatal, not the merely unexpected.}   A reasonable facet of theory-building is to impose conditions censoring properties so qualitatively contrary to observation that any theory exhibiting them is immediately discredited.  More succinctly, it is reasonable to ``ignore the irrelevant."  However, this must be done with great care, due to the limitations of human judgment and imagination.  For example, Planck's approach to black-body radiation, eliminating all but a discrete set of emission frequencies, to avoid the fatal ultraviolet catastrophe, is an example of justified censorship.  However, Einstein's fixing of the cosmological constant, to achieve his {\it expectation} of a steady-state universe, is not.  The causal set axioms of transitivity (\hyperref[tr]{TR}) and interval finiteness (\hyperref[if]{IF}) both censor nonfatal phenomena; the former by ignoring distinctions among certain modes of influence between pairs of events, and the latter by drastically constraining the global structure of classical spacetime. 


\subsection{Notation and Conventions; Figures}\label{subsectionnotation}

{\bf Notation and Conventions.} The mathematical objects of principal interest in this paper are three types of {\it structured sets} called {\it acyclic directed sets,} {\it directed sets} and {\it multidirected sets,} respectively.  These are defined in section \hyperref[subsectionacyclicdirected]{\ref{subsectionacyclicdirected}} below.   They may be viewed as increasingly broad generalizations of causal sets.  In almost all cases, only the {\it isomorphism class} of a structured set is significant, but most constructions are described in terms of particular representatives.  Individual structured sets are usually denoted by capital letters such as $A$, $D$ or $M$, and their individual elements by lower-case letters such as $w,x, y,$ and $z$.   Whenever a more explicit treatment is necessary, an acyclic directed set or directed set may be represented by an ordered pair $(A,\prec)$ or $(D,\prec)$, where $\prec$ is an appropriate binary relation, and a multidirected set may be represented by an ordered quadruple $(M,R,i,t)$, where $R$ is a set of relations, and where $i$ and $t$ are {\it initial} and {\it terminal element maps,} described in more detail in section \hyperref[subsectionacyclicdirected]{\ref{subsectionacyclicdirected}}. The {\it precursor symbol} $\prec$ is used for three distinct but related purposes: first, to represent a {\it binary relation} on a set, as mentioned above; second, to represent an {\it individual relation} between two elements of a set; third, to indicate that {\it at least one such relation exists.}  For an acyclic directed set or directed set, an individual relation $x\prec y$ is an {\it element} of the binary relation $\prec$.  For a multidirected set, there may exist multiple relations between a given pair of elements in either or both directions.  In this case, the expression $x\prec y$ either refers to a particular relation with initial element $x$ and terminal element $y$, or indicates that at least one such relation exists.   

Categories, and other structures filling similar hierarchical roles, such as kinematic schemes, are denoted by mathscript letters such as $\ms{C}$, for the category of causal sets, or a general category; $\ms{A}$, for the category of acyclic directed sets; $\ms{D}$, for the category of directed sets;, or $\ms{M}$, for the category of multidirected sets.  Functors, and other similar ``higher-level maps," are assigned notation {\it suggestively,} rather than in any systematic fashion; for example, the {\it transitive closure} functor on the category of directed sets is denoted by $\tn{tr}$, while the {\it relation space} functor on the category of multidirected sets is denoted by $\ms{R}$.  Familiar number systems, such as the nonnegative integers, integers, rationals, reals, and complex numbers, are denoted by the usual blackboard bold symbols $\NN, \ZZ, \QQ, \RR, \CC$.   I use ``nonnegative integers," rather than ``natural numbers," to avoid confusion about the status of zero.   The additive group of integers modulo a positive integer $n$ is denoted by $\ZZ_n$.  The term {\it space} usually means {\it structured set;} the structure involved is often {\it a priori} order-theoretic, or more generally, multidirected, rather than topological, though this structure may be used to {\it define} natural topologies.  

Appendix \hyperref[appendixindex]{A} provides a thorough index of notation.  


{\bf Nonstandard Terminology.} This paper contains notable disagreements with standard mathematical terminology.  Most of these disagreements arise because the contexts in which certain existing structures and methods appear in this paper are much different than their original contexts, thereby demanding different viewpoints.   In some cases, existing terminology is actually ``wrong," in the sense that it contradicts ``sacred notions."   For example, the condition of interval finiteness (\hyperref[if]{IF}), called {\it local finiteness} in the causal set literature, {\it is not a local condition,} and placing it in such a role leads to serious problems.   In other cases, terms whose plain sense evokes a useful general idea have acquired inconvenient proprietary meanings.  For example, the term {\it directed set} conventionally signifies a set equipped with a reflexive, transitive binary relation, satisfying a common successor property.  Since the alternative term {\it filtered set} has the same conventional meaning, but conveys it better, I reclaim the term {\it directed set} to mean simply a set equipped with a binary relation.  Similarly, I use the term (causal) {\it preorder} to mean simply a binary relation generating a causal order under the operation of transitive closure, even though the term {\it preorder} conventionally signifies reflexivity and transitivity.  The alternative term {\it quasiorder} has the same conventional meaning, and conveys it just as well.   


{\bf Highlighting and Hyperlinks.} Important technical terms, headings of definitions, axioms, lemmas, theorems, and important arguments and principles, are introduced using {\bf bold font}.  Most definitions are embedded in the text, but particularly important definitions are formally set off from the text and numbered.   {\it Italic font} is used for the definition or introduction of other technical terms, bodies of definitions, axioms, lemmas, theorems, and quotations, non-English terms and phrases, and ordinary English emphasis.  It is also used with discretion to emphasize important technical terms referenced {\it before or after} their bold-font definitions.  Hyperlinks are included throughout the text to enable easy navigation.  Axioms, conventions, and principles of sufficient importance to demand frequent reference are labeled by short acronyms.  The axiom of {\it transitivity} (\hyperref[tr]{TR}) and the {\it order extension principle} (\hyperref[oep]{OEP}) are examples.  These acronyms serve as tags for hyperlinks to the original definitions.  Their function is opposite to the common usage of acronyms as ``labor-saving devices," to the exasperation of the reader. 


{\bf Illustrative Figures.} Though many of the constructions in this paper apply to general multidirected sets, most of the accompanying figures, such as figure \hyperref[sprinkling]{\ref{sprinkling}}c in section \hyperref[subsectionapproach]{\ref{subsectionapproach}} above, depict plane diagrams of acyclic directed sets, called {\bf generalized Hasse diagrams,} analogous to Minkowski diagrams of relativistic spacetime.  In a generalized Hasse diagram, elements of an acyclic directed set are represented by nodes, and relations between pairs of elements are represented by edges between pairs of nodes.  The directions of relations are inferred via the vertical coordinates assigned to their initial and terminal nodes.  More specifically, if $x$ and $y$ are two elements of an acyclic directed set $(A,\prec)$, then by convention, $x\prec y$ in $A$ if and only if there is an edge between the corresponding nodes in a generalized Hasse diagram for $A$, such that the vertical coordinate of the node corresponding to $x$ is less than the vertical coordinate of the node corresponding to $y$.  Pairs of elements whose corresponding nodes have no edge between them are assumed to be unrelated, regardless of the coordinates of these nodes.  Heuristically, {\it causal influence flows upward along edges.}  The qualifier {\it generalized} is included because ``standard" Hasse diagrams represent only {\it irreducible} relations, in the sense of part 3 of definition \hyperref[definitionchains]{\ref{definitionchains}} below, while the diagrams appearing in this paper also admit reducible relations.  {\it Acyclic} multidirected sets may be represented by generalized Hasse diagrams in an analogous way, but directed sets and multidirected sets involving cycles cannot be so represented.  Regardless of the actual properties of the structured sets appearing in a figure, these sets are usually labeled and referenced at the same level of generality at which the corresponding constructions or examples are intended to apply.   For example, an acyclic directed set appearing in a figure may be labeled $M$, rather than $A$, if the figure is intended to apply to multidirected sets in general.  Conversely, if the purpose of the figure is to demonstrate that a certain property manifests itself even in the special case of acyclic directed sets, then the set appearing in the figure may be labeled $A$, rather than $M$.   Figures are numbered separately from definitions, lemmas and theorems, since they are auxiliary to the logical development of the paper. 


\subsection{Summary of Contents; Outline of Sections}\label{subsectionoutline}

{\bf Summary of Contents.} This paper has seven sections and one appendix.  Section \hyperref[sectionintroduction]{\ref{sectionintroduction}} is the present introduction.  Section  \hyperref[sectionaxioms]{\ref{sectionaxioms}} presents the axioms of causal set theory, and provides necessary terminology and definitions.  Sections \hyperref[sectiontransitivity]{\ref{sectiontransitivity}}, \hyperref[sectioninterval]{\ref{sectioninterval}}, and \hyperref[sectionbinary]{\ref{sectionbinary}} are devoted to careful analysis of the causal set axioms of transitivity (\hyperref[tr]{TR}), interval finiteness (\hyperref[if]{IF}), and the binary axiom (\hyperref[b]{B}).  The axiom of irreflexivity (\hyperref[ir]{IR}) is involved indirectly, since it interacts with transitivity to imply the condition of acyclicity (\hyperref[ac]{AC}). The principal conclusions of these sections are the following:
\begin{itemize}
\item The transitive causal orders of causal set theory are information-theoretically deficient as discrete models of classical spacetime. Generally nontransitive binary relations called {\it causal preorders} should be viewed as more fundamental.  In this context, causal cycles must be ruled out explicitly if they are considered undesirable.  Hence, for conservative models of classical spacetime, in which ``closed timelike curves" are forbidden, transitivity and irreflexivity should be replaced with the single axiom of acyclicity.   For more general models, allowing ``closed timelike curves," transitivity and irreflexivity should be simply discarded without replacement.   In either case, causal sets should be replaced with a more general class of directed sets. 
\item  Interval finiteness suffers from multiple pathologies.  It does not rule out locally infinite behavior, and imposes unjustified restrictions on the global structure of classical spacetime.  For these reasons, interval finiteness should be replaced with the alternative axiom of {\it local finiteness} (\hyperref[lf]{LF}).  This replacement has important consequences both for the directed sets used to model classical spacetime, and for the multidirected sets arising in quantum causal theory. 
\item The useful practice of encoding spacetime events as elements of a directed set, formalized by the binary axiom (\hyperref[b]{B}), may be generalized to take advantage of natural directed and multidirected structures whose elements correspond to more complex physical entities than individual spacetime events, such as causal relations between pairs of events, or sequences of such relations.  For example, viewing relations between elements of a multidirected set $M$ as elements of a directed set $\ms{R}(M)$, called its {\it relation space,} enables a path-summation approach to quantum causal theory with no direct analogue in ordinary element space.   

\end{itemize}
Section 6 demonstrates how the developments of sections \hyperref[sectiontransitivity]{\ref{sectiontransitivity}}, \hyperref[sectioninterval]{\ref{sectioninterval}}, and \hyperref[sectionbinary]{\ref{sectionbinary}} may be applied to the quantum theory of spacetime.  Combining Grothendieck's relative viewpoint (\hyperref[rv]{RV}) with the histories approach to quantum theory leads to the theory of {\it co-relative histories} and {\it kinematic schemes.}  Applying the theory of relation space, developed in section \hyperref[sectionbinary]{\ref{sectionbinary}}, leads to the derivation of {\it causal Schr\"{o}dinger-type equations,} which describe the dynamics of quantum spacetime in the discrete causal setting.  An example of such an equation is the following:
\[\psi_{R;\theta}^-(r)=\theta(r)\sum_{r^-\prec r}\psi_{R;\theta}^-(r^-).\]
The notation appearing in this equation is explained in sections \hyperref[sectionbinary]{\ref{sectionbinary}} and \hyperref[subsectionquantumcausal]{\ref{subsectionquantumcausal}}.  Briefly, $\psi_{R;\theta}^-$ is a causal analogue of Schr\"{o}dinger's wave function, $R$ is a subset of relation space analogous to a region of spacetime, $\theta$ is a {\it phase map} analogous to a Lagrangian, and $r$ and $r^-$ are relations.    

Section \hyperref[sectionconclusions]{\ref{sectionconclusions}} gathers together the conclusions of preceding sections, and proposes specific changes to the axioms of causal set theory.  This section also includes brief discussions of omitted material, outlines directions for future research, and describes potential connections with other areas of physics and mathematics.  Appendix \hyperref[appendixindex]{A} provides a detailed index of notation.  


{\bf Topical Outline.} Following is a hyperlinked topical outline of sections 2-7:


{\bf Section \hyperref[sectionaxioms]{\ref{sectionaxioms}}: Axioms and Definitions.} 

{\bf \hyperref[subsectioncausalmetric]{\ref{subsectioncausalmetric}}: }{\bf The Causal Metric Hypothesis.}  
\begin{itemize}
\item \hyperref[cmh]{Causal metric hypothesis (CMH): {\it ``The properties of the physical universe are manifestations of}} \hyperref[cmh]{{\it causal structure."}  Combines ``physics is structural" with familiar building block for structure.} 
\item \hyperref[ccmh]{Classical causal metric hypothesis (CCMH): {\it ``Classical spacetime arises from the structure of a directed set."}}  
\item \hyperref[strongform]{Scope of the CMH; e.g., strong form: {\it ``All physical phenomena may be explained in terms of}} \hyperref[strongform]{{\it causal structure."}} \hyperref[strongform]{Too ambitious?}   
\item \hyperref[technicalimplementation]{Technical implementations of the CMH.}  
\item \hyperref[quantumcmh]{Quantum causal metric hypothesis (QCMH). Revisited in section \hyperref[subsectionkinematicschemes]{\ref{subsectionkinematicschemes}}.}
\end{itemize}


{\bf \hyperref[subsectionaxioms]{\ref{subsectionaxioms}}: }{\bf The Axioms of Causal Set Theory.} 
\begin{itemize}
\item \hyperref[subsectionaxioms]{Two formulations: irreflexive formulation and partial order formulation.}  
\item \hyperref[irreflexive]{Irreflexive formulation: axioms.}  
\item \hyperref[defcausalset]{Formal definition of a causal set.}  
\item \hyperref[partialorder]{Partial order formulation: disadvantages.}  
\item \hyperref[axiomsinliterature]{Treatment in the literature.}
\end{itemize}


{\bf \hyperref[subsectionacyclicdirected]{\ref{subsectionacyclicdirected}}: }{\bf Acyclic Directed Sets; Directed Sets; Multidirected Sets.} 
\begin{itemize}
\item \hyperref[subsectionacyclicdirected]{Generalizations of causal sets.}  
\item \hyperref[acyclicdirected]{Acyclic directed sets: conservative models of classical spacetime.} 
\item \hyperref[graph]{Graph-theoretic viewpoint.}   
\item \hyperref[directedsets]{Directed sets: can model causal analogues of closed timelike curves.}
\item \hyperref[multidirectedsets]{Multidirected sets: necessary for quantum theory.} 
\end{itemize}


{\bf \hyperref[subsectionchains]{\ref{subsectionchains}}: }{\bf Chains; Antichains; Irreducibility; Independence.} 
\begin{itemize}
\item \hyperref[chains]{Chains; antichains; mathematical properties such as reducibility and irreducibility.}
\item \hyperref[conditionsmulti]{Conditions on multidirected sets: generalizing transitivity, interval finiteness, etc.} 
\item \hyperref[independence]{Independence: a physical property.} 
\item \hyperref[ic]{Independence convention (IC): {\it ``Relations are independent unless stated otherwise."}}
\item  \hyperref[interpolativecomparison]{Comparison to interpolative case; e.g., domain theory.}
\end{itemize}


{\bf \hyperref[subsecpredecessors]{\ref{subsecpredecessors}}: }{\bf Domains of Influence; Predecessors and Successors; Boundary and Interior.} 
\begin{itemize}
\item \hyperref[relpastfuture]{Relativistic notions; e.g., light cone; chronological and causal pasts and futures.}  
\item \hyperref[newbehavior]{New behavior in causal theory due to irreducibility and independence.} 
\item \hyperref[domains]{Predecessors and successors; domains of influence in multidirected sets.} 
\item \hyperref[predecessors]{Direct pasts and futures; maximal past; minimal future.} 
\item \hyperref[boundaryint]{Boundary and interior of a multidirected set.}
\end{itemize}


{\bf \hyperref[settheoretic]{\ref{settheoretic}}: }{\bf Order Theory; Category Theory; Influence of Grothendieck.}  
\begin{itemize}
\item  \hyperref[isomclassesfinite]{Counting (isomorphism classes of) finite acyclic directed sets.}  
\item \hyperref[isomclassescountable]{Counting (isomorphism classes of) countable acyclic directed sets.}  
\item \hyperref[orderext]{Linear orders; order extensions.}  
\item \hyperref[oep]{Order extension principle (OEP): {\it ``Every acyclic binary relation extends to a linear order."}}
\item \hyperref[cat]{Category theory.  Two roles: framework and analogy.}
\item \hyperref[ishamtopos]{Isham's topos theory;} \hyperref[differentnotions]{different notions suggested by causal theory.}  
\item \hyperref[groth]{Influence of Grothendieck: relative viewpoint (RV); hidden structure (HS).} 
\end{itemize}


\vspace*{.5cm}

{\bf Section \hyperref[sectiontransitivity]{\ref{sectiontransitivity}}: Transitivity, Independence, and the Causal Preorder.}

{\bf \hyperref[subsectionmodes]{\ref{subsectionmodes}}: }{\bf Independent Modes of Influence.}  
\begin{itemize}
\item \hyperref[transproblem]{Basic problem: transitive relations fail to distinguish among independent modes of influence.}  
\item \hyperref[infoanalogies]{Na\"{i}ve analogies from information theory.}  
\item \hyperref[finkelstein]{Nontransitive relations in the literature: Finkelstein's causal nets.}
\item \hyperref[raptisquant]{Raptis' algebraic quantization of causal sets. } 
\item \hyperref[sorkincurrent]{Sorkin's current view.}  
\item \hyperref[independencefundamental]{Independence at the fundamental scale.}
\end{itemize}


{\bf \hyperref[subsectiontransitivitydeficient]{\ref{subsectiontransitivitydeficient}}: }{\bf Six Arguments that Transitive Binary Relations are Deficient.}  
\begin{itemize}
\item 1) \hyperref[cauchyarg]{Cauchy surface analogy.}  
\item 2) \hyperref[futurearg]{Elements need not ``know their futures."}  
\item 3) \hyperref[irredindarg]{Irreducibility and independence are distinct.}  
\item 4) \hyperref[configpath]{Configuration space pathologies (Kleitman-Rothschild).}
\item 5) \hyperref[commcat]{Analogy with commutativity in a category.}  
\item 6) \hyperref[otherimprovements]{Nontransitivity facilitates other improvements; e.g. local finiteness.} 
\end{itemize}


{\bf \hyperref[subsectionpreorder]{\ref{subsectionpreorder}}: }{\bf The Causal Preorder.}  
\begin{itemize}
\item \hyperref[nontransgen]{Nontransitive binary relation generating the causal order (both may have cycles).} 
\item \hyperref[finkelsteinconnect]{Similar to Finkelstein's ``causal connection" relation.} 
\item  \hyperref[notusualpreorder]{Not a preorder in usual sense.} 
\end{itemize}


{\bf \hyperref[subsectiontransitiveclosure]{\ref{subsectiontransitiveclosure}}: }{\bf Transitive Closure; Skeleton; Degeneracy; Functoriality.}  
\begin{itemize}
\item \hyperref[transclosure]{Two operations: transitive closure adds reducible relations;} \hyperref[skeleton]{skeleton deletes reducible relations.}  
\item \hyperref[skelnottransred]{Skeleton distinct from transitive reduction.}  
\item \hyperref[causordtranspreorder]{Causal order is transitive closure of causal preorder.} 
\item \hyperref[degenconfig]{Degeneracy; configuration space implications.} 
\item \hyperref[transfunctorial]{Functorial properties: transitive closure a functor, adjoint to inclusion; skeleton not a functor.}  
\end{itemize}


\vspace*{.5cm}

{\bf Section \hyperref[sectioninterval]{\ref{sectioninterval}}: Interval Finiteness versus Local Finiteness.}

{\bf \hyperref[subsectiontopology]{\ref{subsectiontopology}}: }{\bf Local Conditions; Topology.} 
\begin{itemize}
\item \hyperref[localcond]{Local conditions: detectable in arbitrarily small neighborhoods.}
\item  \hyperref[relnotionstop]{Relativistic notions: topology and the metric.} 
\item  \hyperref[causallocality]{Causal locality.}  
\item  \hyperref[gentopology]{General topology;} \hyperref[hiddentop]{topological hidden structure: ``extra elements;"} \hyperref[toplocfin]{topological local finiteness.} 
\item  \hyperref[fourtopspaces]{Four topologies:} \hyperref[discretetop]{discrete topology;} \hyperref[inttop]{interval topology;} \hyperref[conttop]{continuum topology;} \hyperref[startop]{star topology.}  
\item \hyperref[cardvalscalar]{Cardinal-valued scalar fields.}  
\end{itemize}


{\bf \hyperref[subsectionintervalfiniteness]{\ref{subsectionintervalfiniteness}}: }{\bf Interval Finiteness versus Local Finiteness.} 
 \begin{itemize}
\item \hyperref[intcriticisms]{Initial criticisms of interval finiteness.} 
\item \hyperref[lf]{Local finiteness (LF).} 
\item  \hyperref[locfintoplocfinstar]{Lemma 4.2.1: {\it ``Local finiteness coincides with topological local finiteness in the star topology."}} 
\item  \hyperref[varlocfin]{Variations on local finiteness for directed sets:} \hyperref[lft]{Local finiteness in the transitive closure (LFT).} \hyperref[lfs]{Local finiteness in the skeleton (LFS).}   
\item \hyperref[intfinlocfinincomp]{Interval finiteness and local finiteness are incomparable.} 
\item \hyperref[interactiontrans]{Interaction with transitivity;} \hyperref[interactionmeasure]{interaction with the measure axiom.}  
\end{itemize}


{\bf \hyperref[relativeacyclicdirected]{\ref{relativeacyclicdirected}}: }{\bf Relative Multidirected Sets over a Fixed Base.}
\begin{itemize}
\item \hyperref[infintmeaning]{Infinite intervals: physically meaningful?}
\item \hyperref[causetrelz]{Causal sets as relative directed sets over $\ZZ$.} 
\item \hyperref[theoremcausalsetrelint]{Theorem 4.3.1: {\it ``Any countable, interval finite acyclic directed set is relative over $\ZZ$."}} 
\item \hyperref[relconseq]{Connection to sequential growth dynamics.}  
\item \hyperref[relacdirarb]{Relative multidirected sets over an arbitrary base.} 
\item \hyperref[democracy]{Democracy of bases: why restrict to relative directed sets over $\ZZ$?} 
\item \hyperref[infkleitroth]{Temporal versus spatial size: ``countably infinite Kleitman-Rothschild-type pathology."} 
\end{itemize}


{\bf \hyperref[subsectionintervalfinitenessdeficient]{\ref{subsectionintervalfinitenessdeficient}}: }{\bf Eight Arguments against Interval Finiteness and Similar Conditions.} 
\begin{itemize}
\item 1) \hyperref[intfinnonloc]{Interval finiteness is causally nonlocal.}  
\item 2) \hyperref[nonlocininttop]{Interval finiteness is nonlocal in the interval topology.}  
\item 3) \hyperref[toplocfinintnofinnbhd]{Topological local finiteness in the interval topology does not imply the existence of} \\\hspace*{.45cm}\hyperref[toplocfinintnofinnbhd]{finite interval neighborhoods.}  
\item 4) \hyperref[intnotcapture]{Open intervals fail to capture local multidirected structure.}  
\item 5) \hyperref[intnotimplyloc]{Interval finiteness does not imply local finiteness.}  
\item 6) \hyperref[locnotimplyint]{It does not follow from local finiteness.}  
\item 7) \hyperref[fatalsorkin]{It permits fatal local behavior under Sorkin's version of the CMH.}  
\item 8) \hyperref[unjustglobal]{It imposes unjustified global restrictions on classical spacetime structure.} 
\end{itemize}


{\bf \hyperref[subsectionarglocfin]{\ref{subsectionarglocfin}}: }{\bf Six Arguments for Local Finiteness.} 
\begin{itemize}
\item 1) \hyperref[locfincausloc]{It is causally local.}  
\item 2) \hyperref[locfintoploc]{It is topologically local.}  
\item 3) \hyperref[locfincaptures]{It captures and isolates local multidirected structure.}  
\item 4) \hyperref[compatsorkin]{It is compatible with Sorkin's version of the CMH.} 
\item 5) \hyperref[notunjustglobal]{It does not impose unjustified global restrictions.}  
\item 6) \hyperref[naturalnontrans]{It is natural in the nontransitive setting.}  
\end{itemize}


{\bf \hyperref[subsectionhierarchyfiniteness]{\ref{subsectionhierarchyfiniteness}}: }{\bf Hierarchy of Finiteness Conditions.} 
\begin{itemize}
\item  \hyperref[eltfin]{Other finiteness conditions:} \hyperref[eltfin]{element finiteness (EF);} \hyperref[loceltfin]{local element finiteness (LEF);} \\ \hyperref[relfin]{relation finiteness (RF);} \hyperref[parrelfin]{pairwise relation finiteness (PRF);} \hyperref[chainfin]{chain finiteness (CF);} \\\hyperref[antichainfin]{antichain finiteness (AF).}  
\item \hyperref[discussfincond]{Discussion regarding finiteness conditions for classical spacetime models.}  
\item \hyperref[theoremhierarchyfiniteness]{Theorem 4.6.1: logical implications among finiteness conditions for acyclic directed sets.}  
\item \hyperref[theoremhierarchyfinitenessmulti]{Theorem 4.6.2: logical implications among finiteness conditions for multidirected sets.}  
\end{itemize}


\vspace*{.5cm}

{\bf Section \hyperref[sectionbinary]{\ref{sectionbinary}}: The Binary Axiom: Events versus Elements.}

{\bf \hyperref[subsectionrelation]{\ref{subsectionrelation}}: }{\bf Relation Space over a Multidirected Set.}
\begin{itemize}
\item \hyperref[spacesinphys]{Importance of ``spaces" other than ordinary spacetime in physics.}  
\item \hyperref[defirelationspace]{Definition of relation space.} 
\item \hyperref[thmrelspacefunctor]{Theorem 5.1.2: {\it ``Passage to relation space is a functor $\ms{R}$ with special properties.}}  \item \hyperref[acyclicinterpolativediscussion]{Discussion of acyclic directed and interpolative cases.} 
\item \hyperref[abeltspace]{Inverse problem: abstract element space.} 
\item \hyperref[theoremelementfunctor]{Theorem 5.1.4: {\it ``Passage to abstract element space is a functor $\ms{E}$."}}
\item \hyperref[theoreminterior]{Theorem 5.1.5: {\it ``$\ms{R}$ preserves information except at the boundary."}}
\item \hyperref[cauchyrel]{Cauchy surfaces: impermeability.} \hyperref[eltspaceperm]{Maximal antichains of elements are generally permeable.}  
\item \hyperref[rideoutperm]{The literature on permeability: Major, Rideout, Surya.}  
\item \hyperref[theoremrelimpermeable]{Theorem 5.1.7: {\it ``Maximal antichains in relation space are impermeable."}} 
\item\hyperref[preferrelspace]{Preferability of relation space for spatial notions.}  
\item \hyperref[analogymorphism]{Analogy between $\ms{R}(M)$ and morphism categories.}  
\item \hyperref[starrevisited]{The star model revisited.}
\item \hyperref[inducedtwoelt]{Induced binary relation on two-element subsets.}  
 \end{itemize}


{\bf \hyperref[subsectionpowerset]{\ref{subsectionpowerset}}: }{\bf Power Spaces.} 
\begin{itemize}
\item \hyperref[powerspaces]{Built from ``relations between subsets of arbitrary size."} 
\item  \hyperref[inducedpowerspaces]{Induced power spaces;} \hyperref[holisticpowerspaces]{holistic power spaces.} 
\item \hyperref[higherinduced]{Higher induced relations;} \hyperref[splicerel]{splice relations.} 
\item  \hyperref[causalatoms]{Causal atoms;} \hyperref[atomictop]{atomic topologies;} \hyperref[causalatomicdec]{causal atomic decomposition;} \hyperref[causalatomicres]{causal atomic resolution.}    
\item \hyperref[classicalhol]{Top-down causation, classical holism;} \hyperref[degreeshol]{Degrees of holism.}  
\item \hyperref[twistor]{Analogies with twistor theory;} \hyperref[shapedynamics]{shape dynamics.} 
\end{itemize}

\newpage
{\bf \hyperref[subsectionpathspaces]{\ref{subsectionpathspaces}}: }{\bf Causal Path Spaces.} 
\begin{itemize}
\item \hyperref[causalpaths]{Causal paths.} 
\item \hyperref[pathspacephys]{Path spaces in mathematics and physics.} 
\item \hyperref[pathmorph]{Paths as morphisms.} \hyperref[linsource]{Linear directed sets: sources of paths.}  
\item \hyperref[pathsets]{Causal path sets;} \hyperref[pathprod]{products of paths; causal path spaces.} 
\item  \hyperref[concatprod]{Concatenation product for directed sets;} \hyperref[dirprod]{directed product for multidirected sets;} 
\\\hyperref[spliceprod]{splice products: more general.} 
\item \hyperref[catsemicat]{Categories and semicategories;} \hyperref[causalpathsemicat]{causal path semicategories.}  
\item \hyperref[causalpathalg]{Causal path algebras;} \hyperref[conventionalpath]{conventional applications of path algebras;}  \hyperref[raptisincidence]{Raptis' incidence algebra.}
\end{itemize}


{\bf \hyperref[subsectionpathsummation]{\ref{subsectionpathsummation}}: }{\bf Path Summation over a Multidirected Set.}
\begin{itemize}
\item \hyperref[pathfunctcont]{Path functionals; motivation from continuum theory;} \hyperref[lagrangehamilton]{Lagrangian and Hamiltonian;} \\\hyperref[lagrangehamilton]{Hamilton's principle.} 
\item  \hyperref[pathfunctac]{Path functionals for multidirected sets;} \hyperref[pathsumac]{path summation over a multidirected set.}  
\end{itemize}


\vspace*{.5cm}

{\bf Section \hyperref[subsectionquantumcausal]{\ref{subsectionquantumcausal}}: Quantum Causal Theory.}

{\bf \hyperref[subsectionquantumprelim]{\ref{subsectionquantumprelim}}: }{\bf Quantum Preliminaries; Iteration of Structure; Co-Relative Histories.} 
\begin{itemize}
\item \hyperref[superposition]{Superposition; path integration; histories approach to quantum theory in general.}  
\item  \hyperref[historiesquantumcausal]{Histories approach to quantum causal theory;} \hyperref[backgrounddependentqct]{background dependent quantum causal theory;}  
\\\hyperref[backgroundindependentqct]{background independent quantum causal theory.}  
\item \hyperref[ishamquantcat]{The literature:} \hyperref[ishamquantcat]{Isham's quantization on a category;} \hyperref[sorkinquantal]{Sorkin's quantum measure theory.}  
\item \hyperref[iteration]{Iteration of structure (IS): {\it ``Multidirected sets whose elements are directed sets."}} 
\item \hyperref[transitions]{Transitions.} \hyperref[benderrobinson]{Generally too specific to be fundamental:} \hyperref[benderrobinson]{Bender and Robinson's rigidity result.} 
\item \hyperref[symmetry]{Symmetry considerations;} \hyperref[galois]{causal Galois groups.} 
\item \hyperref[labelcorelative]{Co-relative histories: induce higher level multidirected structure.} \hyperref[corelmulti]{McKay's example.} 
\item \hyperref[deficorelative]{Formal definition of co-relative histories;} \hyperref[corelativesubtle]{category-theoretic subtleties.}
 \end{itemize}


{\bf \hyperref[subsectionquantumpathsummation]{\ref{subsectionquantumpathsummation}}: }{\bf Abstract Quantum Causal Theory via Path Summation.}  
\begin{itemize}
\item \hyperref[adaptinghistories]{Adapting the histories approach;} \hyperref[backgrounddepadapt]{background-dependent adaptations;} \\ \hyperref[backgroundindepadapt]{ingredients for background independent approach.}  
\item \hyperref[appliestoboth]{Path summation principle (PS): {\it ``The same abstract theory applies to both background-dependent}}  \hyperref[appliestoboth]{{\it and background-independent cases."}} 
\item \hyperref[contpathintegral]{Feynman's continuum path integral;} \hyperref[feynpost]{Feynman's postulates;} \hyperref[feynamp]{Feynman's quantum amplitude.}  
\item \hyperref[causalpathint]{Causal analogues of Feynman's path integral;} \hyperref[dependsimperm]{importance of impermeability;} \\\hyperref[abstractamp]{abstract quantum amplitudes.}  
\end{itemize}


{\bf \hyperref[subsectionschrodinger]{\ref{subsectionschrodinger}}: }{\bf Schr\"{o}dinger-Type Equations.}  
\begin{itemize}
\item \hyperref[schrodhistory]{Schr\"{o}dinger-type equations via the histories approach in general.} 
\item \hyperref[contwavefunc]{Continuum wave functions;} \hyperref[conschrodequ]{continuum Schr\"{o}dinger equation via path integrals.}  
\item  \hyperref[abstractpathfunct]{Abstract chain functionals;} \hyperref[abstractwavefunct]{abstract wave functions.}  
\item \hyperref[causalschrodequ]{Causal Schr\"{o}dinger-type equations;} \hyperref[causalfeynamp]{causal analogues of Feynman's inner product formulas.}
\end{itemize}  


{\bf \hyperref[subsectionkinematicschemes]{\ref{subsectionkinematicschemes}}: }{\bf Kinematic Schemes.} 
\begin{itemize}
\item \hyperref[pathsumbackground]{Path summation in the background independent context.}
\item  \hyperref[kinversusdyn]{Kinematics versus dynamics;} \hyperref[kinpreschemes]{kinematic preschemes;} 
\\\hyperref[underlyingdirclass]{Underlying directed sets and multidirected sets.} 
\item \hyperref[kinschemes]{Kinematic schemes;} \hyperref[hereditary]{hereditary property (H);} \hyperref[weakaccessibility]{weak accessibility (WA).}
\item \hyperref[quantumcausalmetric]{Quantum causal metric hypothesis (QCMH): {\it ``The properties of quantum spacetime arise from a kinematic scheme of directed sets."}} 
\item\hyperref[pathsumkinscheme]{Path summation over a kinematic scheme;} \hyperref[corelkin]{co-relative kinematics.}  
\item \hyperref[possequkinscheme]{Positive sequential kinematic scheme;} \hyperref[relsorkinrideout]{comparison to Sorkin and Rideout's kinematic scheme} \hyperref[relsorkinrideout]{describing sequential growth of causal sets.}  
\item \hyperref[genkinematics]{Generational kinematics: analogous to relativistic frames of reference.} 
\item \hyperref[completions]{Completions of kinematic schemes;} \hyperref[kinschemecountablyinf]{kinematic schemes of countable acyclic directed sets.}  
\item \hyperref[morphkinscheme]{``Functors" of kinematic schemes;} \hyperref[numberanalogies]{analogies with familiar number systems.} 
\item  \hyperref[universalkinschemes]{Universal kinematic schemes; kinematic spaces.}  
\end{itemize}


\vspace*{.5cm}

{\bf Section \hyperref[sectionconclusions]{\ref{sectionconclusions}}: Conclusions.}

{\bf \hyperref[subsectionalternative]{\ref{subsectionalternative}}: }{\bf New Axioms, Perspectives, and Technical Methods.} 
\begin{itemize}
\item \hyperref[axiomaticanalysis]{Summary of axiomatic analysis.} 
\item \hyperref[suggestedalt]{Suggested alternative axioms;} \hyperref[conservativealt]{conservative alternative;} \hyperref[radicalalt]{radical alternatives.} 
\item \hyperref[summaryperspectives]{Summary of new perspectives and technical methods.}
 \end{itemize}


{\bf \hyperref[subsectionomitted]{\ref{subsectionomitted}}: }{\bf Omitted Topics and Future Research Directions.}
\begin{itemize}
\item \hyperref[remainingax]{Further remarks on countability, irreflexivity, and acyclicity.} 
\item \hyperref[covariance]{Covariance.} 
\item \hyperref[algstructure]{Algebraic structure and hierarchy.} 
\item \hyperref[phasetheory]{Phase theory: what is the target object of the phase map?}  
\item \hyperref[randomgraph]{Random graph dynamics; phase transitions.} 
\item \hyperref[holalt]{Alternatives to power spaces.} 
\item \hyperref[othertheories]{Connections with other physical theories;} \hyperref[othermath]{connections with other mathematical topics.}   
\end{itemize}


{\bf \hyperref[subsectionacknowledgements]{\ref{subsectionacknowledgements}}: }{\bf Acknowledgements; Personal Notes.}
\begin{itemize}
\item \hyperref[ack]{Acknowledgements.} 
\item \hyperref[personal]{Personal notes.} 
\end{itemize}
  
\newpage

\section{Axioms and Definitions}\label{sectionaxioms}

Causal set theory is based on Rafael Sorkin's version of the {\it causal metric hypothesis} (\hyperref[cmh]{CMH}), summarized by the phrase, {\it ``order plus number equals geometry."}  The general philosophy of the causal metric hypothesis is that the observed properties of the physical universe, including the structure of spacetime, and the dynamical behavior of matter and energy, are ultimately just manifestations of cause and effect.  The axioms of causal set theory represent one of many possible ways to distill from this broad idea a specific quantitative approach to fundamental physics.   In this section, I introduce these axioms, in a sufficiently general context to support their analysis in succeeding sections.  I also define other important structures and methods used throughout the remainder of this paper. 

In section \hyperref[subsectioncausalmetric]{\ref{subsectioncausalmetric}} below, I explain the general reasoning behind the causal metric hypothesis, and place Sorkin's specific version in a broader context.  In section \hyperref[subsectionaxioms]{\ref{subsectionaxioms}}, I present the axioms of causal set theory, using the {\it irreflexive formulation,} and explain why this formulation is structurally preferable to the alternative {\it partial order formulation.}  In section \hyperref[subsectionacyclicdirected]{\ref{subsectionacyclicdirected}}, I introduce three increasingly general analogues of causal sets, which I call {\it acyclic directed sets,} {\it directed sets,} and {\it multidirected sets.}  Acyclic directed sets are equivalent to {\it small, simple, acyclic directed graphs,} but are viewed in order-theoretic, rather than graph-theoretic, terms.  They serve as ``conservative" models of classical spacetime structure.   Directed sets, not necessarily acyclic, appear in the study of more general models of classical spacetime, analogous to solutions of general relativity admitting closed timelike curves.  They are equivalent to {\it small, simple directed graphs.}  The reader should be aware that this is a more general meaning than the standard one.  Multidirected sets arise naturally in the theory of {\it configuration spaces of directed sets,} and play an important role in the version of quantum causal theory developed in section \hyperref[subsectionquantumcausal]{\ref{subsectionquantumcausal}} of this paper.   They are equivalent to {\it small directed multigraphs,} or equivalently, {\it small quivers.} 

In section \hyperref[subsectionchains]{\ref{subsectionchains}}, I introduce {\it chains} and {\it antichains} in directed sets and multidirected sets.  I also discuss distinctions among mathematical and physical properties associated with chains, and more specifically, with individual relations between pairs of elements.  Particularly important is the distinction between {\it irreducibility,} a mathematical property, and {\it independence,} a physical property.  I state the {\it independence convention} (\hyperref[ic]{IC}), which is crucial to the interpretation of directed sets and multidirected sets throughout this paper.  In section \hyperref[subsecpredecessors]{\ref{subsecpredecessors}}, I discuss {\it domains of influence}, which are generalizations of chronological and causal pasts and futures in general relativity.   They are distinct from the {\it domains} appearing in the related field of domain theory.  In particular, I discuss {\it direct pasts and futures,} {\it maximal pasts,} and {\it minimal futures,} which are vital for analyzing the axioms of transitivity (\hyperref[tr]{TR}) and interval finiteness (\hyperref[if]{IF}) in sections \hyperref[sectiontransitivity]{\ref{sectiontransitivity}} and \hyperref[sectioninterval]{\ref{sectioninterval}} below.  In section \hyperref[settheoretic]{\ref{settheoretic}}, I introduce general structural ideas from order theory and category theory, such as the {\it order extension principle} (\hyperref[oep]{OEP}), and Grothendieck's {\it relative viewpoint} (\hyperref[rv]{RV}). 


\subsection{The Causal Metric Hypothesis}\label{subsectioncausalmetric}

{\bf Background and Definition.} Causal theory is founded on a single fundamental idea called the {\it causal metric hypothesis}.  The special case of causal set theory represents one possible way of formalizing Sorkin's specific version of this hypothesis. The philosophical content of the causal metric hypothesis is that the observed properties of the physical universe arise from causal relationships between pairs of events, or more generally, between pairs of families of events.\footnotemark\footnotetext{This caveat is included to allow for the possibility of {\it classical holism,} though I focus on classically reductionist models in this paper.  See section \hyperref[subsectionpowerset]{\ref{subsectionpowerset}} for further discussion.}  

\hspace*{.3cm} CMH.\refstepcounter{textlabels}\label{cmh} {\bf Causal Metric Hypothesis.} {\it The properties of the physical universe are manifestations \\ \hspace*{1.45cm} of causal structure.}  

The causal metric hypothesis fleshes out the longstanding idea, going back to Gauss, Riemann, Einstein, Kaluza and Klein, Weyl, and many others, that physics is {\it essentially structural in nature,} by proposing the familiar relationship between cause and effect as the fundamental building block of this structure.  Focusing on the classical case, the principal philosophical difference between the causal metric hypothesis and the viewpoint of general relativity is one of {\it description versus prescription,} as described in section \hyperref[naturalphilosophy]{\ref{naturalphilosophy}}.  Whereas general relativity treats spacetime geometry as a {\it constraint} on causal structure, the causal metric hypothesis treats it as an emergent manifestation thereof.  The causal-theoretic rejoinder to John Wheeler's statement that {\it ``spacetime tells matter how to move; matter tells spacetime how to curve"} \cite{WheelerQuantum98}, is {\it ``things happen; `spacetime' and `matter' are ways of describing them."}   

The viewpoint afforded by the causal metric hypothesis must be expressed mathematically before its physical consequences may be explored in a precise manner.  Still working in the classical context, the basic mathematical structure associated with a {\it particular instance} of cause and effect is an {\it ordered pair of abstract elements,} the first representing the cause, and the second representing the effect.  In causal theory, the {\it arrow of time} arises, ultimately, from many instances of this primitive order.  For an entire collection $D$ of events, or families of events, one must consider a corresponding collection $\prec$ of ordered pairs of elements of $D$.  Viewing $D$ as a set, the collection $\prec$ is, by definition, a subset of the Cartesian product $D\times D$; i.e., a {\it binary relation} on $D$.  In this paper, the pair $(D,\prec)$ is called a {\it directed set.}  A more precise {\it classical version} of the causal metric hypothesis may then be stated as follows:

\hspace*{.3cm} CCMH.\refstepcounter{textlabels}\label{ccmh} {\bf Classical Causal Metric Hypothesis.} {\it The properties of classical spacetime arise \\ \hspace*{1.7cm} from the structure of a directed set.}  

This is still a very general idea, whose practical application requires narrowing the focus to ``physically relevant directed sets." This involves nontrivial choices.  First, the apparently {\it unidirectional} nature of time suggests, at least at a na\"{i}ve level, that the binary relation $\prec$ should be {\it acyclic;} i.e., that events do not contribute to their own causes, either directly or indirectly.  An acyclic binary relation {\it generates a partial order}, in a sense described in sections \hyperref[subsectionaxioms]{\ref{subsectionaxioms}}, \hyperref[subsectionpreorder]{\ref{subsectionpreorder}}, and \hyperref[subsectiontransitiveclosure]{\ref{subsectiontransitiveclosure}}.  This lends plausibility to the appearance of {\it order} in Sorkin's version of the causal metric hypothesis, without reference to geometric notions.   General relativity permits {\it closed timelike curves,} so acyclicity is a significant assumption.  Second, experimental evidence over the last century reveals the fundamental {\it discreteness} of numerous physical quantities at ``small" scales.  Causal set theory takes the bold step of incorporating this discreteness at the {\it classical} level, via a {\it discrete measure} $\mu$ that, roughly speaking, ``counts fundamental volume units."  This lends plausibility to the appearance of {\it number} in Sorkin's version of the causal metric hypothesis, without reference to ``quantizing spacetime."   Since familiar geometric notions are expected to emerge only at relatively large scales in causal set theory, the persistence of volume as a meaningful concept down to the fundamental scale is again a significant assumption.    


\refstepcounter{textlabels}\label{strongform}

{\bf Scope of the Causal Metric Hypothesis.} The proper scope of the causal metric hypothesis is debatable.   A conservative approach is to view causal structure as merely a {\it ``replacement for relativistic spacetime,"} without attempting to explain ``particles" and ``fields" by means of the same structure.  Versions of causal theory following this approach are {\it theories of gravity,} rather than {\it unified theories.}  At the opposite extreme is the {\bf strong form of the causal metric hypothesis,} which seeks to explain {\it all physical phenomena} in terms of causal structure.  This is the most ambitious and optimistic form of the causal metric hypothesis, but also the most pleasing at a philosophical level, since it removes any possibility of tension between ``material bodies" and ``background structure," whether static or dynamical.  In this sense, the strong form of the causal metric hypothesis achieves perfect {\it background independence.}  


\refstepcounter{textlabels}\label{technicalimplementation}

{\bf Technical Implementations.} Many different technical implementations of the causal metric hypothesis are possible.  For example, rather than assigning the {\it same} volume to each element of a directed set, except possibly for small statistical fluctuations, as in Sorkin's approach, one might choose to take account of {\it local causal structure,} as described in section \hyperref[sectioninterval]{\ref{sectioninterval}}, in the assignment of volume.  Such an approach accords with the general philosophy that familiar properties of spacetime emerge from causal structure at appropriate scales, without necessarily admitting meaningful extension down to the fundamental scale.  For example, it would be absurd, for models as general as causal sets, to expect properties such as {\it dimension} and {\it curvature} to retain their usual geometric meanings in this regime.\footnotemark\footnotetext{Causal dynamical triangulations {\it does} assume a fundamental dimension, but this is a much less general approach than causal set theory.}  


\refstepcounter{textlabels}\label{quantumcmh}

{\bf Quantum Version.}  Finally, there is a {\it quantum theoretic version} of the causal metric hypothesis (\hyperref[qcmh]{QCMH}), discussed further in section \hyperref[subsectionkinematicschemes]{\ref{subsectionkinematicschemes}} below, in which the role of classical causal structure is superseded by {\it higher-level multidirected structures} on configuration spaces of directed sets, called {\it kinematic schemes.}   In this context, the quantum causal metric hypothesis states that {\it the properties of quantum spacetime arise from the structure of a kinematic scheme of directed sets.}   As mentioned in section \hyperref[subsectionapproach]{\ref{subsectionapproach}}, this approach is partly motivated by Isham's {\it quantization on a category} \cite{IshamQuantisingI05}, and Sorkin's {\it quantum measure theory} \cite{SorkinQuantalMeasure12}.


\subsection{The Axioms of Causal Set Theory}\label{subsectionaxioms}

The causal set literature contains two {\it almost equivalent} formulations of causal set theory, which I refer to as the {\it irreflexive formulation,} and the {\it partial order formulation.}  In this section, I introduce and discuss both.  Subsequently, the irreflexive formulation is assumed throughout the paper, unless stated otherwise.  I occasionally borrow convenient terminology from the partial order formulation, however. 

\refstepcounter{textlabels}\label{irreflexive} 

{\bf Irreflexive Formulation.} The irreflexive formulation of causal set theory is defined in terms of irreflexive binary relations, analogous to the familiar {\it less than} relation on the integers.  This formulation may be expressed by the following six axioms:

\refstepcounter{textlabels}\label{b}

\hspace*{.4cm}B. \hspace*{.25cm}{\bf Binary Axiom}: {\it Classical spacetime may be modeled as a set $C$, whose elements
 \\ \hspace*{1.1cm} represent spacetime events, together with a binary relation $\prec $ on $C$, whose elements
 \\ \hspace*{1.1cm} represent causal relations between pairs of spacetime events.}
 
 \refstepcounter{textlabels}\label{m}
 
 \hspace*{.4cm}M. \hspace*{.2cm}{\bf Measure Axiom}: {\it The volume of a spacetime region corresponding to a subset  
\\ \hspace*{1.1cm} $S$ of $C$ is equal to the cardinality of $S$ in fundamental units, up to Poisson-type 
\\ \hspace*{1.1cm} fluctuations.}

\refstepcounter{textlabels}\label{c}
 
 \hspace*{.4cm}C. \hspace*{.25cm}{\bf Countability}: {\it $C$ is countable.}
 
 \refstepcounter{textlabels}\label{tr}

\hspace*{.25cm}TR. \hspace*{.1cm}{\bf Transitivity}: {\it Given three elements $x,y,$ and $z$ in $C$, if $x\prec y\prec z$, then $x\prec z$.}

\refstepcounter{textlabels}\label{if}

\hspace*{.3cm}IF. \hspace*{.18cm}{\bf Interval Finiteness}: {\it For every pair of elements $x$ and $z$ in $C$, the open interval \\\hspace*{1.1cm}$\llangle x,z\rrangle:=\{y\in C\hspace*{.1cm}|\hspace*{.1cm} x\prec y\prec z\}$ has finite cardinality.}  

 \refstepcounter{textlabels}\label{ir}
 
\hspace*{.3cm}IR. \hspace*{.2cm}{{\bf Irreflexivity}: {\it Elements of $C$ are not self-related with respect to $\prec $; i.e., $x\nprec x$.} 

The measure axiom may be expressed more precisely in terms of a {\it discrete measure} $\mu:\ms{P}(C)\rightarrow\RR^+$, where $\ms{P}(C)$ is the {\it power set} of $C$; i.e., the set of all subsets of $C$, and where $\RR^+$ is the set of positive real numbers.\footnotemark\footnotetext{Use of the real numbers here is merely for convenience; any ``sufficiently large" extension of the positive integers suffices.  No essential properties of the continuum are necessary.} Transitivity and irreflexivity together imply the condition of {\it acyclicity}, discussed further below, which rules out causal analogues of closed timelike curves.  Interval finiteness is called {\it local finiteness} in the literature, but it is not a local condition in any suitable sense.  I reserve the term {\it local finiteness} for a different, genuinely local condition (\hyperref[lf]{LF}), introduced in section \hyperref[subsectionintervalfiniteness]{\ref{subsectionintervalfiniteness}} below.  Cardinality conditions rarely appear explicitly in the causal set literature, but countability is usually implicit.\footnotemark\footnotetext{For example, Sorkin and Rideout's theory of sequential growth dynamics involves specific enumerations of causal sets.  More generally, interval finite but uncountable ``causal sets" are ``physically ridiculous." } 

For ease of reference, I include a formal definition of causal sets in terms of the above axioms:\\

\begin{defi}\label{defcausalset} A {\bf causal set} is a countable set $C$, equipped with a transitive, interval finite, irreflexive binary relation $\prec$, physically interpreted according to the binary axiom and the measure axiom.  
\end{defi} 

It is sometimes necessary to make the binary relation $\prec$ on a causal set $C$ explicit.  In such cases, a causal set may be introduced as a pair $(C,\prec)$.  Alternatively, it may be denoted by the single symbol $C$, with the understanding that this is an abbreviation.  A subset $S$ of a causal set $C=(C,\prec)$ inherits from $C$ a binary relation, called the {\bf subset relation}, also denoted by $\prec$, where $x\prec y$ in $S$ if and only if $x\prec y$ in $C$.   It is often useful to view the class of causal sets, together with the class of structure-preserving maps between pairs of causal sets, as a {\it category} $\ms{C}$, called the  {\bf category of causal sets}.  Basic category-theoretic notions are outlined in section \hyperref[settheoretic]{\ref{settheoretic}}.  For the present, it is sufficient to think of a category as a collection of {\it objects}, in this case causal sets, together with a collection of {\it morphisms} between pairs of objects, in this case {\it structure-preserving maps} between pairs of causal sets.   To be explicit, a morphism between causal sets $C=(C,\prec)$ and $C'=(C',\prec')$ is a set map $\phi:C\rightarrow C'$ that {\it preserves structure,} in the sense that $\phi(x)\prec' \phi(y)$ in $C'$ whenever $x\prec y$ in $C$.  Here, $C$ is called the {\bf source} of $\phi$, and $C'$ is called the {\bf target} of $\phi$.  The {\bf index} of $\phi$ is the supremum of the cardinalities of its fibers, where the {\bf fiber} $\phi^{-1}(x')$ of $\phi$ over $x'$ in $C'$ is the set of elements of $C$ that map to $x'$ under $\phi$.   An {\bf isomorphism} is an invertible morphism.  In almost all cases, only the isomorphism class of a causal set is significant.   A bijective morphism is generally {\it not} an isomorphism, since the target may have ``extra relations," extending the binary relation of the source.  Such bijective morphisms play an important role in sections \hyperref[settheoretic]{\ref{settheoretic}} and \hyperref[relativeacyclicdirected]{\ref{relativeacyclicdirected}}.   A {\bf monomorphism} is an injective morphism.   I sometimes refer to an injective morphism $\phi:C\rightarrow C'$ as {\it embedding its source as a subobject of its target.}


{\bf Partial Order Formulation.}\refstepcounter{textlabels}\label{partialorder}  A common alternative formulation of causal set theory uses {\it interval finite partial orders} in the place of irreflexive binary relations.  A {\bf partial order} $\preceq$ on a set $P$ is a {\it reflexive, antisymmetric, transitive, binary relation} on $P$, where {\bf reflexivity} means that $x\preceq x$ for every $x$ in $P$, and {\bf antisymmetry} means that if $x\preceq y$ and $y\preceq x$ for two elements $x$ and $y$ in $P$, then $x=y$.  Transitivity and antisymmetry together imply that $P$ has no cycles {\it except for} the reflexive cycles $x\preceq x$.  As the notation suggests, a partial order $\preceq$ is analogous to the familiar {\it less than or equal to} relation on the integers.  Interval finiteness is an extra condition, not part of the definition of a partial order.   Interval finite partial orders are special cases of {\bf discrete orders,} which are partial orders in which every nonextremal element has a {\it maximal predecessor} and {\it minimal successor.}\footnotemark\footnotetext{See the quote of Sorkin in section \hyperref[subsectionmodes]{\ref{subsectionmodes}} for an example of this terminology.  The reader should also be aware, however, that the term {\it discrete order} is sometimes assigned other, mutually contradictory meanings, such as a discrete {\it linear} order, or a ``trivial" order involving only reflexive relations.} Predecessors and successors are discussed in more detail in section \hyperref[subsecpredecessors]{\ref{subsecpredecessors}} below. 

The irreflexive and partial order formulations of causal set theory are {\it equivalent at the level of objects} in the following sense: a causal set $(C,\prec)$ may be viewed as a countable interval finite partially ordered set, by taking $x\preceq y$ if $x\prec y$ or $x=y$; conversely, a countable interval finite partially ordered set $(P,\preceq)$ may be viewed as a causal set, by taking $x\prec y$ if $x\preceq y$ and $x\neq y$.  

The partial order formulation of causal set theory has the superficial advantage of familiarity.  For example, in general relativity, the {\it causal relation} is usually taken to be reflexive,\footnotemark\footnotetext{In my view, this is simply an unfortunate choice of definition.} while the {\it chronological relation} is taken to be irreflexive.\footnotemark\footnotetext{{\it Neither} the irreflexive binary relation $\prec$ nor the partial order $\preceq$ corresponds to the chronological relation, which ``excludes the entire boundaries of light cones."}  Also, partial orders are more popular  than irreflexive relations in many mathematical contexts.  On structural grounds, however, the partial order formulation is inconvenient.  Partial orders are technically not {\it acyclic,} due to the existence of reflexive relations $x\le x$.  Hence, the {\it categories} of causal sets and interval finite partially ordered sets are not equivalent, since the latter category admits ``structure-destroying morphisms" that wrap nontrivial relations around reflexive cycles.  These inconveniences, of course, do not reflect any essential difference in physically relevant {\it information content} between the two formulations.  Though I use the irreflexive formulation throughout this paper, I sometimes abuse terminology and {\it refer} to the binary relation on a causal set as a partial order, via the object-level equivalence mentioned above.  I also apply familiar techniques from order theory to causal sets, such as the {\it order extension principle} (\hyperref[oep]{OEP}), introduced in section \hyperref[settheoretic]{\ref{settheoretic}} below.  


{\bf Treatment in the Literature.}\refstepcounter{textlabels}\label{axiomsinliterature} The binary axiom (\hyperref[b]{B}), the measure axiom (\hyperref[m]{M}), and countability (\hyperref[c]{C}), do not appear as {\it axioms} in the causal set literature.  However, the {\it content} of the binary axiom and measure axiom appear explicitly, and countability is implicit in the focus and methods.  Treatment of the remaining axioms depends on whether the irreflexive formulation or the partial order formulation is being used.  Here, I reproduce a few representative quotations to illustrate this treatment. 

In their inaugural paper {\it Space-Time as a Causal Set} \cite{Sorkinetal87}, Bombelli, Lee, Meyer, and Sorkin use the partial order formulation of causal set theory, with the exceptions that an explicit statement of countability, and the nuance of  {\it Poisson-type fluctuations,} do not appear:
\begin{quotation}\noindent{\it ``...when we measure the volume of a region of space-time, we are merely indirectly counting the number of ``point events" it contains... ...volume is number, and macroscopic causality reflects a deeper notion of order in terms of which all the ``geometrical" structures of spacetime must find their ultimate expression... ...Before proceeding any further, let us put the notion of a causal set into mathematically precise language.   A {\it partially ordered set}... ...is a set... ...provided with an order relation, which is transitive... ...noncircular... ...\tn{[and]} reflexive.  A partial ordering is {\it locally finite} if every ``Alexandroff set"... ...contains a finite number of elements... ...a {\it causal set} is then by definition a locally finite, partially ordered set."} (page 522)
\end{quotation}
Here, {\it ``noncircular"} means {\it acyclic} (\hyperref[ac]{AC}) {\it except for reflexive cycles.}  An {\it ``Alexandroff set"} is an {\it open interval} $\llangle x,z\rrangle:=\{y\in C\hspace*{.1cm}|\hspace*{.1cm} x\prec y\prec z\}$; hence, {\it ``local finiteness"} means {\it interval finiteness} (\hyperref[if]{IF}) in this context.  Later papers are split between the irreflexive formulation and the partial order formulation.  For example, in {\it Classical sequential growth dynamics for causal sets} \cite{SorkinSequentialGrowthDynamics99}, Sorkin and Rideout use the irreflexive formulation (see page 2).   Later papers also amend the interpretation of volume to allow for Poisson-type fluctuations.  For example, in {\it Everpresent $\Lambda$} \cite{SorkinEverPresentLambda04}, Ahmed, Dodelson, Greene, and Sorkin write, 
\begin{quotation}\noindent{\it ``In order to do justice to local Lorentz invariance, the correspondence between number and volume... ...must be subject to Poisson-type fluctuations..."} (page 2)
\end{quotation}
Absence of explicit cardinality conditions in the causal set literature may be attributed to the fact that countability is {\it automatic} from the viewpoint of {\it recovering a spacetime manifold from a discrete subset.}  More fundamentally, allowing uncountable ``causal sets" greatly  exacerbates the existing ``imbalance" of the category of causal sets toward ``large spatial and small temporal size," as indicated by the {\it Kleitman-Rothschild pathology,} discussed in section \hyperref[subsectiontransitivitydeficient]{\ref{subsectiontransitivitydeficient}}, and its ``countable analogue," discussed in section \hyperref[relativeacyclicdirected]{\ref{relativeacyclicdirected}} below.  These pathologies arise, ultimately, because of the axioms of transitivity (\hyperref[tr]{TR}) and interval finiteness (\hyperref[if]{IF}).


\subsection{Acyclic Directed Sets; Directed Sets; Multidirected Sets}\label{subsectionacyclicdirected}

It is useful to study causal sets in the context of more general structured sets, which I refer to as {\it acyclic directed sets,} {\it directed sets,} and {\it multidirected sets}, respectively.  Acyclic directed sets represent a ``blank canvas for conservative models of classical spacetime;" i.e., models without causal analogues of closed timelike curves.  Directed sets and multidirected sets are even more general.   While {\it possibly} unnecessary for modeling classical spacetime, directed sets and multidirected sets arise unavoidably in the theory of configuration spaces of classical spacetime models in causal theory.  As elaborated in section \hyperref[subsectionquantumcausal]{\ref{subsectionquantumcausal}} of this paper, such configuration spaces are of central importance in the histories approach to quantum causal theory. 


\refstepcounter{textlabels}\label{acyclicdirected}

{\bf Acyclic Directed Sets.} In this paper, the term {\it acyclic directed set} means merely a set $A$ equipped with an acyclic binary relation.  The meaning of acyclicity is already evident from the foregoing discussion of the causal metric hypothesis and the axioms of causal set theory, but it is convenient, for future reference, to spell out the notion here.  A {\bf cycle} in a set equipped with a binary relation is a sequence of relations with the same initial and terminal element: $x=x_0\prec x_1\prec...\prec x_{n-1}\prec x_n=x$.  A ``trivial example" of a cycle is a {\it reflexive relation} $x\prec x$.   The set $A$ is called acyclic if its binary relation is acyclic; i.e., has no cycles.  It is useful to state this condition as an axiom: 

\refstepcounter{textlabels}\label{ac}

\hspace*{.3cm} AC. {\bf Acyclicity}: {\it The binary relation on $A$ is acyclic.}

It is also useful, though redundant, to include a formal definition of acyclic directed sets:\\

\begin{defi} An {\bf acyclic directed set} is a set $A$ equipped with an acyclic binary relation $\prec$.  
\end{defi} 

Irreflexivity (\hyperref[ir]{IR}) alone does not imply acyclicity; for example, cycles of the form $x\prec y\prec x$ are allowed under irreflexivity.  However, irreflexivity and transitivity (\hyperref[tr]{TR}) together {\it do} imply acyclicity.  This distinction is important in this paper, since most of the binary relations considered here are not assumed to be transitive.   

An acyclic directed set may be introduced explicitly as a pair $(A,\prec)$, or it may be denoted in abbreviated fashion by a single symbol $A$.  {\it Transitive} acyclic directed sets are equivalent {\it as objects} to partially ordered sets, extending the correspondence introduced in section \hyperref[subsectionaxioms]{\ref{subsectionaxioms}} in the countable case.  A subset $S$ of an acyclic directed set $(A,\prec)$ inherits a natural {\bf subset relation}, given by restricting $\prec$ to $S$.  The {\bf category of acyclic directed sets} is the category $\ms{A}$ whose objects are acyclic directed sets, and whose morphisms $(A,\prec)\rightarrow (A',\prec')$ are set maps $\phi:A\rightarrow A'$, such that $\phi(x)\prec' \phi(y)$ in $A'$ whenever $x\prec y$ in $A$.  Related notions, such as sources, targets, fibers, indices, isomorphisms, and monomorphisms, generalize from the case of causal sets in an obvious way.   Because of the choice to define the category $\ms{C}$ of causal sets using the irreflexive formulation of causal set theory, causal sets may be viewed as objects of either $\ms{C}$ or $\ms{A}$ without ambiguity concerning their morphism classes.  Technically, this means that $\ms{C}$ {\it embeds into $\ms{A}$ as a full subcategory.}  Heuristically, it means that pairs of causal sets share the same relationships whether viewed as objects of $\ms{C}$ or $\ms{A}$.\refstepcounter{textlabels}\label{conventionads}

\refstepcounter{textlabels}\label{graph}

Acyclic directed sets are equivalent to {\it small, simple, acyclic directed graphs}, under the correspondence sending elements to vertices and relations to directed edges.  Here {\it small} means that the vertex class of the graph under consideration is a set, rather than a {\it proper class,} and {\it simple} means that for any pair of vertices $x$ and $y$, there is at most one edge from $x$ to $y$.  In this case, the equivalence extends to the level of categories.  Both sides of this equivalence are useful: the set-theoretic side for its convenient terminology, and the graph-theoretic side for the visual perspective afforded by generalized Hasse diagrams.   


\refstepcounter{textlabels}\label{directedsets}

{\bf Directed Sets.}  As mentioned above, general relativity admits solutions involving {\it closed timelike curves,} whose causal analogues are the {\it cycles} discussed above.  In fact, such curves appear in rather generic situations, such as the Kerr black hole.  Acyclic directed sets are inadequate to model causal analogues of such spacetimes; more general objects, which I refer to as {\it directed sets,} are needed.  In this paper, the term {\it directed set} means merely a set $D$ equipped with a binary relation.   However, the reader should keep in mind that the term {\it directed set} has a different conventional meaning in the context of order theory, as explained in the discussion of nonstandard terminology in section \hyperref[subsectionnotation]{\ref{subsectionnotation}} above.  

It is useful to include a formal definition of this term as it is used in this paper:

\refstepcounter{textlabels}\label{defdirected}
\vspace*{.2cm}
\begin{defi} A {\bf directed set} is a set $D$ equipped with a binary relation $\prec$.
\end{defi} 

A directed set may be introduced explicitly as a pair $(D,\prec)$, or it may be denoted in abbreviated fashion by a single symbol $D$.   A subset $S$ of a directed set $(D,\prec)$ inherits a natural {\bf subset relation}, given by restricting $\prec$ to $S$.  The {\bf category of directed sets} is the category $\ms{D}$ whose elements are directed sets, and whose morphisms $(D,\prec)\rightarrow (D',\prec')$ are set maps $\phi:D\rightarrow D'$, such that $\phi(x)\prec' \phi(y)$ in $D'$ whenever $x\prec y$ in $D$.  Related notions, such as sources, targets, fibers, indices, isomorphisms, and monomorphisms, generalize from the case acyclic directed sets in an obvious way.  The category $\ms{A}$ of acyclic directed sets embeds into $\ms{D}$ as a full subcategory.\footnotemark\footnotetext{It is tempting to invent suggestive terminology for special classes of directed sets in the context of causal theory.  The best term, of course, is ``causal sets," which is already appropriated.   Finkelstein \cite{Finkelstein88} uses the term {\it causal nets.}  Benincasa and Dowker \cite{DowkerScalarCurvature10}, following Riemann, use the term {\it discrete manifolds} in a general context, including discrete causal models as a special case.  Other possible choices are {\it causal graphs} and {\it discrete chronofolds,} though the latter is the name of an Apple smartphone application.}

Like acyclic directed sets, directed sets may also be understood in graph-theoretic terms, as {\it small, simple, directed graphs.}  Note that the definition of a {\it simple} directed graph allows for the coexistence of ``reciprocal edges" $x\prec y$ and $y\prec x$, but prohibits multiple edges between $x$ and $y$ in the same direction.  Besides their use as models of classical spacetime, directed sets arise naturally in more abstract contexts, in which their elements represent {\it more general entities} than structureless spacetime events.  Examples include the {\it relation spaces} and {\it power spaces} introduced in section \hyperref[sectionbinary]{\ref{sectionbinary}}. 


\refstepcounter{textlabels}\label{multidirectedsets}

{\bf Multidirected Sets.} The most general structured sets enjoying broad use in this paper are {\it multidirected sets,} which arise in the study of configuration spaces of acyclic directed sets and directed sets.  Special configuration spaces, called kinematic schemes, arise naturally in the histories approach to quantum causal theory, developed in section \hyperref[subsectionquantumcausal]{\ref{subsectionquantumcausal}} below.  The multidirected sets of principal interest in this paper arise by {\it``decategorifying"} kinematic schemes; i.e., demoting them to a lower level of algebraic hierarchy by forgetting the internal structure of their member sets.  Possible {\it classical} applications of multidirected sets are mentioned very briefly in section \hyperref[subsectionomitted]{\ref{subsectionomitted}} below.  

\refstepcounter{textlabels}\label{defmultidirected}
\vspace*{.2cm}
\begin{defi} A {\bf multidirected set} consists of a set of elements $M$, a set of relations $R$, and {\bf initial} and {\bf terminal element maps} $i:R\rightarrow M$ and  $t:R\rightarrow M$, assigning to each relation initial and terminal elements.  
\end{defi} 

A multidirected set for which each ordered pair $(x,y)$ of elements has at most one relation $r$ satisfying the conditions that $i(r)=x$ and $t(r)=y$, may be viewed as a directed set, whose binary relation $\prec$ is defined by setting $x\prec y$ whenever there exists a relation $r$ such that $i(r)=x$ and $t(r)=y$.  However, the structure of a general multidirected set {\it cannot} be expressed by a single binary relation, since an ordered pair $(x,y)$ of elements of $M$ generally does not {\it uniquely identify} a relation from $x$ to $y$.  In this context, the expression $x\prec y$ either refers to a {\it particular} relation with initial element $x$ and terminal element $y$, or indicates that {\it at least one such relation exists.}  A multidirected set may be introduced as a quadruple $(M,R,i,t)$, or it may be denoted in abbreviated fashion by a single symbol $M$.  A {\bf subobject} of a multidirected set $M=(M,R,i,t)$ consists of a subset of $M$, together with subset of $R$, chosen in such a way that the initial and terminal element maps $i$ and $t$ map the latter subset into the former.  Subsets of causal sets, acyclic directed sets, and directed sets, together with their subset relations, are all special cases of subobjects of multidirected sets.  

\newpage

Abstractly, multidirected sets are structurally analogous to {\it small categories;} i.e., categories whose object classes are sets.  As in the case of directed sets, the elements of a multidirected set are often taken to represent entities more complex than events.  From a graph-theoretic perspective, multidirected sets are equivalent to {\it small directed multigraphs,} or equivalently, {\it small quivers}.\footnotemark\footnotetext{This term deliberately evokes the ``quiver of arrows" carried by an archer.}  Use of the term {\it quiver} often has {\it algebraic connotations;} for example, the {\it concatenation algebra} over a multidirected set, introduced in section \hyperref[subsectionpathspaces]{\ref{subsectionpathspaces}} below, is a special type of {\it quiver algebra}.\footnotemark\footnotetext{The use of multidirected sets as {\it algebraic substrata} goes back at least to {\it Gabriel's theorem} in 1972, which classifies {\it connected quivers of finite type} and their representations in terms of {\it Dynkin diagrams} and {\it root systems.}}  As I have defined it, a multidirected set is precisely what Abrams and Pino call a ``directed graph" in \cite{AbramsPathAlgebra05}. The terminology of {\it multidirected sets} is the most convenient choice for this paper, for at least five reasons: 1) it is descriptive; 2) it is congenial to the causal set viewpoint; 3) it is relatively neutral in its connotations; 4) it is amenable to generalized order-theoretic notions and terminology; and 5) it eliminates potential confusion about proper classes.  The {\bf category of multidirected sets} is the category $\ms{M}$ whose objects are multidirected sets, and whose morphisms are pairs of maps of elements and relations that respect initial and terminal element maps.  More precisely, a morphism $\phi$ between two multidirected sets $(M,R,i,t)$ and $(M',R',i',t')$ consists of {\it two} maps: a {\bf map of elements} $\phi_{\tn{\fsz{elt}}}:M\rightarrow M'$, and a {\bf map of relations} $\phi_{\tn{\fsz{rel}}}:R\rightarrow R'$, satisfying the conditions that 
$\phi_{\tn{\fsz{elt}}}(i(r))=i'(\phi_{\tn{\fsz{rel}}}(r))$ and $\phi_{\tn{\fsz{elt}}}(t(r))=t'(\phi_{\tn{\fsz{rel}}}(r))$ for each relation $r$ in $R$.\footnotemark\footnotetext{The student of algebraic geometry will recall that a morphism of algebraic schemes also consists of two maps: a map of topological spaces, and a map of structure sheaves.  Of course, the latter map is ``in the opposite direction."  Nevertheless, there is a loose analogy between $R$ and the {\it tangent sheaf} in geometry.}   Morphisms of multidirected sets play a prominent role in section \hyperref[subsectionrelation]{\ref{subsectionrelation}}, where I use them to analyze the properties of the {\it relation space functor} and the {\it abstract element space functor.}   Related notions, such as sources, targets, fibers, indices, isomorphisms, and monomorphisms, may be defined by applying familiar notions to {\it both} maps $\phi_{\tn{\fsz{elt}}}$ and $\phi_{\tn{\fsz{rel}}}$. 

Figure \hyperref[figmultidirected]{\ref{figmultidirected}} below illustrates the differences among {\it causal sets,} {\it acyclic directed sets,} {\it directed sets,} {\it acyclic multidirected sets,} and {\it multidirected sets.}  Figure \hyperref[figmultidirected]{\ref{figmultidirected}}a shows a causal set.  Note the transitivity of the binary relation.  Figure \hyperref[figmultidirected]{\ref{figmultidirected}}b shows an acyclic directed set.  Here, only {\it some} of the relations ``implied by transitivity" are present; for example, there is a relation $w\prec z$ corresponding to the pair of relations $w\prec x\prec z$, but no relation $w\prec y$ corresponding to the pair of relations $w\prec x\prec y$.  Figure \hyperref[figmultidirected]{\ref{figmultidirected}}c shows a directed set.  Note that reflexive relations, such as $t\prec t$, and {\it reciprocal relations,} such as $u\prec v$ and $v\prec u$, are allowed.  Figure \hyperref[figmultidirected]{\ref{figmultidirected}}d shows an acyclic multidirected set.   While acyclicity is defined above only for binary relations, a straightforward generalization to multidirected sets is given in section \hyperref[subsectionchains]{\ref{subsectionchains}} below.  Informally, multiple independent relations between a given pair of elements are allowed in this case, but only in one direction.  For example, there are two independent relations from $w$ to $x$ and from $x$ to $z$.  Figure \hyperref[figmultidirected]{\ref{figmultidirected}}e shows a general multidirected set.  Figures \hyperref[figmultidirected]{\ref{figmultidirected}}c and \hyperref[figmultidirected]{\ref{figmultidirected}}e illustrate the fact that cycles cannot be represented by generalized Hasse diagrams.  Whenever cycles are present, the directions of relations are indicated explicitly by arrows.  

\begin{figure}[H]
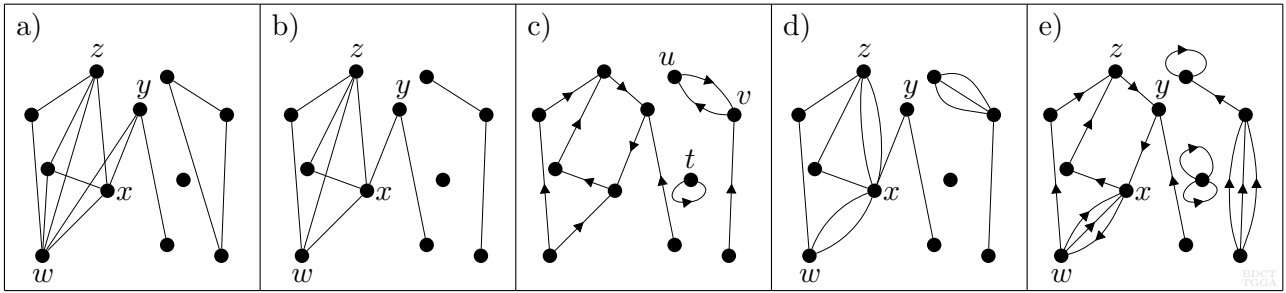


\caption{a) Causal set;  b) acyclic directed set; c) directed set; d) acyclic multidirected set; e) multidirected set with cycles.}
\label{figmultidirected}
\end{figure}
\vspace*{-.5cm}


\subsection{Chains; Antichains; Irreducibility; Independence}\label{subsectionchains}

Many important properties of multidirected sets may be described in terms of sequences of consecutive elements and relations, called {\it chains.}  For example, a multidirected set $M$ is acyclic if and only if no chain in $M$ has the same initial and terminal element.  Subsets of multidirected sets admitting no chains between pairs of elements are also important; these are called {\it antichains}.  Both chains and antichains are formally introduced in definition \hyperref[definitionchains]{\ref{definitionchains}} below.  Chains are used in definition \hyperref[deficonditionsmulti]{\ref{deficonditionsmulti}} to generalize familiar conditions on directed sets, originally defined in terms of binary relations, to the case of multidirected sets.  

When a multidirected set is assigned a physical interpretation, for example, as a model of information flow or causal structure, it is vital to distinguish between its {\it absolute mathematical properties,} and the {\it physical characteristics subjectively ascribed to it.} An important example is the distinction between mathematical {\it reducibility} or {\it irreducibility} of a relation between a pair of elements $x$ and $y$, and physical {\it dependence} or {\it independence} of the information encoded by this relation, with respect to the information encoded by other chains between $x$ and $y$.  Definition \hyperref[definitionindependence]{\ref{definitionindependence}} introduces the property of independence in the more general context of families of chains between pairs of elements.   This leads to the {\it independence convention} (\hyperref[ic]{IC}), which is crucial to the physical interpretation of directed sets and multidirected sets throughout the remainder of the paper. 


{\bf Chains; Antichains; Mathematical Properties.}\refstepcounter{textlabels}\label{chains}  Definition \hyperref[definitionchains]{\ref{definitionchains}} introduces chains and antichains.  The reader should note that chains are special cases of {\it paths,} discussed in more detail in section \hyperref[subsectionpathspaces]{\ref{subsectionpathspaces}} below. \\

\begin{defi}\label{definitionchains} Let $M=(M,R,i,t)$ be a multidirected set, and let $x$ and $y$ be elements of $M$.  
\begin{enumerate}
\item A {\bf chain}  $\gamma$ in $M$ is a sequence of elements and relations of the form $...\prec x_0\prec x_1\prec...$ in $M$, where the notation $x_n\prec x_{n+1}$ refers to a particular relation $r$ in $R$ such that $x_n=i(r)$ and $x_{n+1}=t(r)$.  The {\bf chain set} $\tn{Ch}(M)$ of $M$ is the set of all chains in $M$.  
\item A {\bf chain of length} $n$, or {\bf $n$-chain}, from  $x$ to $y$ in $M$, is a chain of the form $x=x_0\prec x_1\prec...\prec x_{n-1}\prec x_n=y$. The {\bf set of $n$-chains} $\tn{Ch}_n(M)$ in $M$ is the subset of $\tn{Ch}(M)$ consisting of chains of length $n$.   A {\bf complex chain} is a chain of length at least two. 
\item\refstepcounter{textlabels}\label{irred} A relation $r$ in $R$ is called {\bf reducible} if there exists a complex chain from its initial element to its terminal element.  Such a chain is called a {\bf reducing chain} for $r$.   If $r$ is not reducible, it is called {\bf irreducible}.   $M$ itself is called {\bf irreducible} if all its relations are irreducible. 
\item An {\bf antichain} $\sigma$ in $M$ is a subset of $M$ admitting no chain in $M$ between any pair of its elements. 
\end{enumerate}
\end{defi}

In the context of classical causal structure, a chain is analogous to a relativistic {\it world line,} while an antichain corresponds to a {\it spacelike section} of spacetime; i.e., a {\it Cauchy surface}.  A $0$-chain in $M$ is just an element of $M$, and a $1$-chain in $M$ is just an element of the relation set $R$ of $M$.   $R$ is the underlying set of the {\it relation space} $\ms{R}(M)$ over $M$, studied in detail in section \hyperref[subsectionrelation]{\ref{subsectionrelation}} below.  Similarly, {\it chain spaces} over $M$, whose underlying sets are the chain sets $\tn{Ch}_n(M)$ and $\tn{Ch}(M)$, are introduced in section \hyperref[subsectionpathspaces]{\ref{subsectionpathspaces}}.  A chain in $M$ may include a given element or relation more than once; when this occurs, it indicates the existence of a cycle in $M$, as spelled out in definition \hyperref[deficonditionsmulti]{\ref{deficonditionsmulti}} below.   Any relation $r$ in $M$ belonging to a cycle is reducible, via any chain beginning and ending with $r$.  The reason for defining chains in terms of {\it elements and relations,} with the awkward use of the expression $x_n\prec x_{n+1}$ to denote a {\it particular} relation, is to include individual elements in the definition, as $0$-chains.   For most other purposes, it is more convenient to define chains purely in terms of relations. 


\refstepcounter{textlabels}\label{conditionsmulti}

{\bf Conditions on Multidirected Sets.}  Conditions on directed sets, such as transitivity (\hyperref[tr]{TR}), interval finiteness (\hyperref[if]{IF}), irreflexivity (\hyperref[ir]{IR}), and acyclicity (\hyperref[ac]{AC}), defined in sections \hyperref[subsectionaxioms]{\ref{subsectionaxioms}} and \hyperref[subsectionacyclicdirected]{\ref{subsectionacyclicdirected}} above in terms of binary relations, are also useful in the more general context of multidirected sets.   It is therefore necessary to re-express these conditions in terms of relation sets and initial and terminal element maps. Chains are particularly convenient for this purpose.\\

\begin{defi}\label{deficonditionsmulti} Let $M=(M,R,i,t)$ be a multidirected set.  
\begin{enumerate}
\item $M$ is {\bf transitive} if every complex chain in $M$ is a reducing chain for a relation $r$ in $R$. 
\item Let $x$ and $z$ be two elements of $M$.  The {\bf open interval} $\llangle x,z\rrangle$ in $M$ is the subset of $M$ consisting of all elements $y$ admitting a chain from $x$ to $y$ and a chain from $y$ to $z$.  $M$ is {\bf interval finite} if every open interval in $M$ is finite. 
\item A relation $r$ in $R$ is called a {\bf reflexive relation} at $x$ in $M$ if $i(r)=t(r)=x$.  $M$ is {\bf irreflexive} if $R$ includes no reflexive relations.  
\item A {\bf cycle} in $M$ is a chain $x_0\prec x_1\prec...\prec x_{n-1}\prec x_n$ such that $x_0=x_n$.  $M$ is {\bf acyclic} if it contains no cycles.  
\end{enumerate}
\end{defi}

Other important conditions on multidirected sets, such as local finiteness (\hyperref[lf]{LF}), are examined and compared in section \hyperref[sectioninterval]{\ref{sectioninterval}} below. 


\refstepcounter{textlabels}\label{independence}

{\bf Independence: A Physical Property.} Definitions \hyperref[definitionchains]{\ref{definitionchains}} and \hyperref[deficonditionsmulti]{\ref{deficonditionsmulti}} above are {\it purely mathematical,} making no reference to the physical interpretation of the multidirected set $M$.  Definition \hyperref[definitionindependence]{\ref{definitionindependence}} below introduces the notion of {\it independence} of a family of chains in a multidirected set, which is a physical property subjectively associated with such a family.  The mathematical reader should be careful to observe that this notion of independence is different than the notions of independence appearing in mathematical fields such as linear algebra and matroid theory, and that this difference is not merely one of distinction between physical and mathematical properties.   In such mathematical contexts, independence is, broadly speaking, an {\it absolute, ``internal" property, signifying absence of redundancy,} while in the present physical context, it is a {\it relative property, signifying uniqueness of information content.}  For example, the linear independence of a set $S$ of vectors in a vector space $V$ does not depend on the complement $V-S$; in particular, one cannot ``spoil" the linear independence of $S$ by adding new dimensions to $V$.  Further, $S$ has no ``monopoly" on the information about $V$ it contains; any basis for the span of $S$ in $V$ contains the same information about $V$.  By contrast, independence of a family $\Gamma$ of chains in a multidirected set $M$ means that $\Gamma$ encodes, possibly in a redundant fashion, {\it at least some information} that cannot be recovered from any subobject of $M$ not containing $\Gamma$.   Definition \hyperref[definitionindependence]{\ref{definitionindependence}} makes this idea precise for the special case of a family of chains between a particular pair of elements of $M$. \\

\begin{defi}\label{definitionindependence}  Let $M=(M,R,i,t)$ be a multidirected set, interpreted as a model of information flow or causal structure, and let $x$ and $y$ be elements of $M$.  
\begin{enumerate}
\item A family $\Gamma$ of chains between $x$ and $y$ in $M$ is called {\bf dependent} if there exists another such family $\Gamma'$, not containing $\Gamma$, encoding all information or causal influence encoded by $\Gamma$.  
\item In particular, a chain $\gamma$ from $x$ to $y$ in $M$ is called {\bf dependent} if there exists a family $\Gamma'$  of chains from $x$ to $y$, not including $\gamma$, encoding all information or causal influence encoded by $\gamma$. 
\item If a chain or family of chains is not dependent, it is called {\bf independent}. 
\end{enumerate}
\end{defi}

Dependence and independence may be easily generalized to the case of families of chains between pairs of {\it subsets} of $M$, but definition \hyperref[definitionindependence]{\ref{definitionindependence}} is sufficient for the purposes of this paper. Independence is constrained, but not determined, by the mathematical structure of $M$.  Of principal interest is the {\it dependence or independence of individual relations;} i.e., $1$-chains.    For example, {\it irreducible relations in directed sets are necessarily independent,} since no other chains exist between their initial and terminal elements.  Reducible relations, on the other hand, may be interpreted as either dependent or independent. For multidirected sets, multiple relations may share the same initial and terminal elements, so even irreducible relations may be interpreted as dependent in this context.  These ambiguities illustrate the need to fix {\it conventions} regarding dependence and independence.  Causal set theory implicitly treats reducible relations as {\it dependent}, as reflected in the practice of representing causal sets by standard Hasse diagrams.  However, {\it this convention leads to information-theoretic shortcomings in causal set theory,} as explained in section \hyperref[sectiontransitivity]{\ref{sectiontransitivity}}.  Since the multidirected sets of principal interest in this paper satisfy a local finiteness condition (\hyperref[lf]{LF}) rendering dependent relations information-theoretically superfluous, I adopt the following convention:

 \refstepcounter{textlabels}\label{ic}

\hspace*{.3cm} IC. {\bf Independence Convention}: {\it Every relation in a multidirected set is interpreted as \\ 
\hspace*{1.1cm}independent unless stated otherwise.}

\refstepcounter{textlabels}\label{interpolativecomparison}

It is reasonable to ask why one would even {\it consider} including dependent relations in a physical model.  In the locally finite case, dependent relations only add redundancy, but nontrivial {\it locally infinite} multidirected sets exist in which {\it every relation is both reducible and dependent.}  The principal example is, of course, relativistic spacetime, which exhibits this behavior due to the {\bf interpolative property} of the continuum.  Under the interpolative property, every relation $x\prec z$ admits an {\bf interpolating element} $y$ such that $x\prec y\prec z$.\footnotemark\footnotetext{For the continuum itself, the interpolative property may be realized by {\it averaging}.}  The interpolative property has been studied since antiquity; for example, it forms the crux of Zeno's {\it dichotomy paradox,} mentioned in section \hyperref[subsectionapproach]{\ref{subsectionapproach}} above.  In a more modern context, the interpolative property is closely related to the use of {\it Cauchy surfaces} to separate ``past" and ``future" regions of spacetime in relativistic cosmology.  Suitable analogues of Cauchy surfaces for multidirected sets are introduced in section \hyperref[subsectionrelation]{\ref{subsectionrelation}} below.   {\it Domain theory} is a branch of causal theory involving interpolative causal structures.\footnotemark\footnotetext{As noted in the introduction to section \hyperref[sectionaxioms]{\ref{sectionaxioms}}, the domains studied in domain theory are distinct from the {\it domains of influence} introduced in section \hyperref[subsecpredecessors]{\ref{subsecpredecessors}}.}  A useful reference for domain theory is the paper of Martin and Panangaden \cite{Martin06}, cited in section \hyperref[subsectionapproach]{\ref{subsectionapproach}} in the discussion of Malament's metric recovery theorem.  

The independence convention makes physical sense only if one asserts the freedom to use relations between pairs of elements to encode {\it actual influence or information flow}.  The causal set axiom of transitivity (\hyperref[tr]{TR}) removes this freedom by {\it prescribing} the existence of a relation between two elements in a causal set whenever there is a chain between them.  This eliminates the option of representing {\it dependent} influence solely by means of complex chains, thereby limiting the variety of causal structures that can be modeled.  This is the root of the {\it Kleitman-Rothschild configuration space pathology,} revisited in greater depth in section \hyperref[sectiontransitivity]{\ref{sectiontransitivity}} below.   However, a greater problem is the way in which the presence of dependent relations clouds the structural picture in causal set theory, hampering important perspectives and technical methods. 


\subsection{Domains of Influence; Predecessors and Successors; Boundary and Interior}\label{subsecpredecessors}

\refstepcounter{textlabels}\label{relpastfuture}

{\bf Domains of Influence; Predecessors and Successors.} An iconic feature of Minkowski spacetime diagrams in special relativity is the {\it light cone} of an event $x$, which is the union of its past and future.  In terms of causal structure, the light cone of $x$ is the set of all events in Minkowski spacetime that can either influence $x$ or be influenced by $x$.  In general relativity, the light cone is superseded in this role by the {\it total domain of influence} $J(x)$ of $x$.  The uneasy distinction between which events {\it can} influence $x$ or be influenced by $x$, and which events actually {\it do} exert or experience such influence, arises from the {\it prescriptive} nature of spacetime geometry in general relativity, as mentioned in section  \hyperref[naturalphilosophy]{\ref{naturalphilosophy}} above.  This distinction disappears under the causal metric hypothesis (\hyperref[cmh]{CMH}).   More specific domains of influence are defined by taking subsets of $J(x)$.  For example, the {\it chronological future} $I^+(x)$ of $x$ is the set of events that may be reached from $x$ by a {\it future-directed timelike curve,} while the {\it causal future} $J^+(x)$ of $x$ is the set of events that may be reached from $x$ by a {\it future-directed causal curve;} i.e., a future-directed curve that is either timelike or null.  Chronological pasts $I^-(x)$ and causal pasts $J^-(x)$ are defined in a similar way.\footnotemark\footnotetext{This is standard relativistic notation.  In the causal context, I replace the letter $J$ by $D$, for ``domain."   Distinct analogues of chronological pasts and futures are not needed in this paper; see, however, the discussion of the {\it interval topology} in section \hyperref[subsectiontopology]{\ref{subsectiontopology}} below.}   Relativistic domains of influence play a central role in Malament's metric recovery theorem, discussed in section \hyperref[subsectionapproach]{\ref{subsectionapproach}} above.   

\refstepcounter{textlabels}\label{newbehavior}

Despite the conceptual tension between potential and actual influence in general relativity, the mathematical classification of relativistic domains of influence is comparatively simple, due to the reducibility and dependence of relations between events in relativistic spacetime, discussed in section \hyperref[subsectionchains]{\ref{subsectionchains}} above.  For more general directed sets and multidirected sets, including those of principal interest in discrete causal theory, the possibilities of irreducibility and independence introduce qualitatively new features, giving rise to more nuanced domains of influence.  These include {\it direct pasts and futures,}  {\it maximal pasts,} and {\it minimal futures,}  introduced in definition \hyperref[predecessors]{\ref{predecessors}} below.  More complicated domains of influence may be defined in terms of {\it filtrations} of the past and future of a general subset of a multidirected set, but the simple pointwise definitions given here suffice for the purposes of this paper.  The physically suggestive notions of ``past" and ``future" are useful for general multidirected sets, even though the classical spacetime models of principal interest in this paper are merely directed sets. \\

\begin{defi}\label{domains} Let $M=(M,R,i,t)$ be a multidirected set, and let $w$, $x$, and $y$ be elements of $M$.
\begin{enumerate}
\item If there exists a chain from $w$ to $x$ in $M$, then $w$ is called a {\bf predecessor} of $x$.   The set $D^-(x)$ of all predecessors of $x$ is called the {\bf past} of $x$.  
\item If there exists a chain from $x$ to $y$ in $M$, then $y$ is called a {\bf successor} of $x$.  The set $D^+(x)$ of all successors of $x$ is called the {\bf future} of $x$.  
\item The {\bf total domain of influence} $D(x)$ of $x$ in $M$ is the union $D^-(x)\cup D^+(x)$ of the past and future of $x$. 
\item  A {\bf domain of influence} of $x$ is a subset of $D(x)$.  
\end{enumerate}
\end{defi}

Particularly important domains of influence are defined in terms of {\it direct predecessors} and {\it successors, maximal predecessors,} and {\it minimal successors.}\\

\begin{defi}\label{predecessors} Let $M=(M,R,i,t)$ be a multidirected set, and let $x$ be an element of $M$.  Let $w$ be a predecessor of $x$, and let $y$ be a successor of $x$.  
\begin{enumerate}
\item If there exists a relation $r$ in $R$ such that $i(r)=w$ and $t(r)=x$, then $w$ is called a {\bf direct predecessor} of $x$.  If $r$ is irreducible, then $w$ is called a {\bf maximal predecessor} of $x$. 
\item The sets of direct predecessors and maximal predecessors of $x$ are denoted by $D_0^-(x)$ and $D_{\text{\tn{\fsz{max}}}}^-(x)$, and are called the {\bf direct past} and the {\bf maximal past} of $x$, respectively. 
\item If there exists a relation $r$ in $R$ such that $i(r)=x$ and $t(r)=y$, then $y$ is called a {\bf direct successor} of $x$.   If $r$ is irreducible, then $y$ is called a {\bf minimal successor} of $x$. 
\item The sets of direct successors and minimal successors of $x$ are denoted by $D_0^+(x)$ and $D_{\text{\tn{\fsz{min}}}}^+(x)$, and are called the {\bf direct future} and the {\bf minimal future} of $x$, respectively. 
\end{enumerate}
\end{defi}

It follows immediately from definition \hyperref[predecessors]{\ref{predecessors}} that $D_{\text{\tn{\fsz{max}}}}^-(x)\subset D_0^-(x)\subset D^-(x)$, and that $D_{\text{\tn{\fsz{min}}}}^+(x)\subset D_0^+(x)\subset D^+(x)$.  The direct past and the past of $x$ coincide in the transitive case, by the definition of transitivity, and similarly for the direct future and the future of $x$.   In the interpolative case, maximal pasts and minimal futures are always empty.   For multidirected sets including cycles, pasts and futures are not necessarily disjoint.  For example, if $M$ includes a reflexive relation $x\prec x$, then the element $x$ belongs to its own past and future.  The physical interpretation of direct pasts and futures requires the independence convention (\hyperref[ic]{IC}), or a suitable alternative, since without such a convention the physical significance of reducible relations is ambiguous.  Figure \hyperref[successors]{\ref{successors}} below illustrates the future, direct future, and minimal future of an element $x$ in a multidirected set, which in this case is a nontransitive acyclic directed set with $35$ elements.   

\begin{figure}[H]
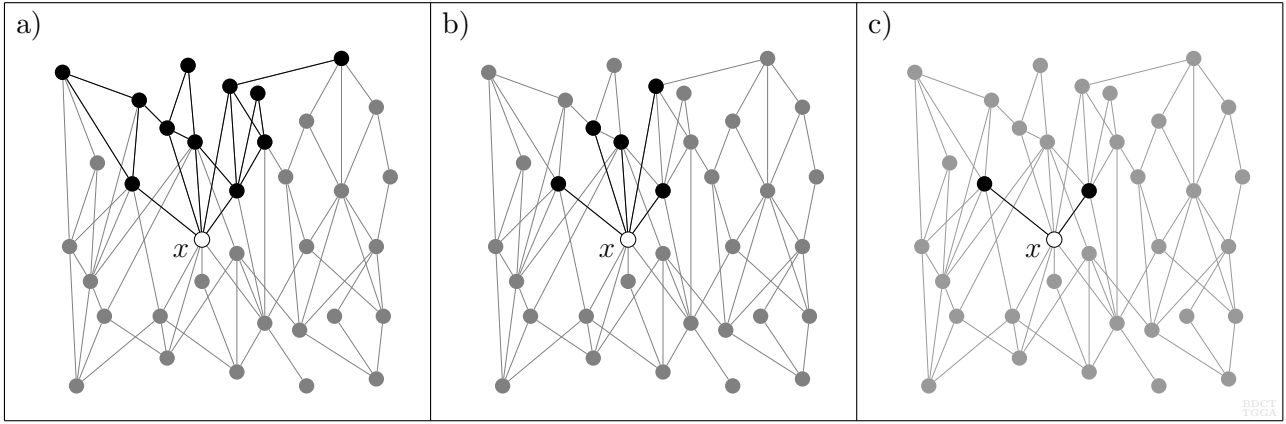


\caption{a) Future of an element $x$; b) direct future of $x$; c) minimal future of $x$.}
\label{successors}
\end{figure}
\vspace*{-.5cm}


\refstepcounter{textlabels}\label{boundaryint}

{\bf Boundary and Interior.} An element $x$ in a multidirected set $M$ is called an {\bf extremal element} if either its past or its future is empty.   A ``Big Bang-like" or ``Big Crunch-like" event might be modeled as an extremal element. ``Free agency" might also be represented in this way.  In the context of classical spacetime structure, the set of all extremal elements serves as a {\it temporal boundary.}   Spatial boundaries, by contrast, are not fundamental in causal theory.  The {\it interior} of a multidirected set is the complement of its boundary.  In the context of classical spacetime structure, the interior is where ``ordinary physics" occurs: every event in the interior has causes, and produces effects.  These concepts are made precise in definition \hyperref[defboundary]{\ref{defboundary}}.  \\ 

\begin{defi}\label{defboundary} Let $M=(M,R,i,t)$ be a multidirected set.
\begin{enumerate}
\item The {\bf boundary} $\partial M$ of $M$ is the subset of all extremal elements of $M$; i.e., elements with either empty past or empty future.  
\item The {\bf interior} $\tn{Int}(M)$ of $M$ is the subset of all nonextremal elements of $M$; i.e., the complement of the boundary of $M$. 
\end{enumerate}
\end{defi}

Figure \hyperref[boundary]{\ref{boundary}}a below illustrates the boundary of a multidirected set $M$.  Figure \hyperref[boundary]{\ref{boundary}}b illustrates the interior of $M$.

It is sometimes useful to regard domains of influence of elements in a multidirected $M$, as well as its boundary and interior, as {\it subobjects} of $M$, rather than merely structureless subsets of $M$.  This requires specifying sets of relations to complement the sets of elements specified in definitions \hyperref[predecessors]{\ref{predecessors}} and \hyperref[defboundary]{\ref{defboundary}}. Unless stated otherwise, {\it all} relations between pairs of elements in the specified subset are included;\footnotemark\footnotetext{Of course, the only possible relations in the boundary $\partial M$ are relations from minimal to maximal elements.}  the resulting subobjects are called {\bf full subobjects.}  For example, the interior $\tn{Int}(M)$ of $M=(M,R,i,t)$, viewed as a subobject of $M$, consists of all nonextremal elements of $M$, {\it together with} all relations in $M$ between pairs of nonextremal elements.  This convention is important for interpreting some of the results of later sections, such as theorem \hyperref[theoreminterior]{\ref{theoreminterior}}.   Figure \hyperref[boundary]{\ref{boundary}}c below illustrates the interior of a multidirected set $M$, viewed as a full subobject.   The domains of influence illustrated in figure \hyperref[successors]{\ref{successors}} above are also shown as full subobjects.

\begin{figure}[H]
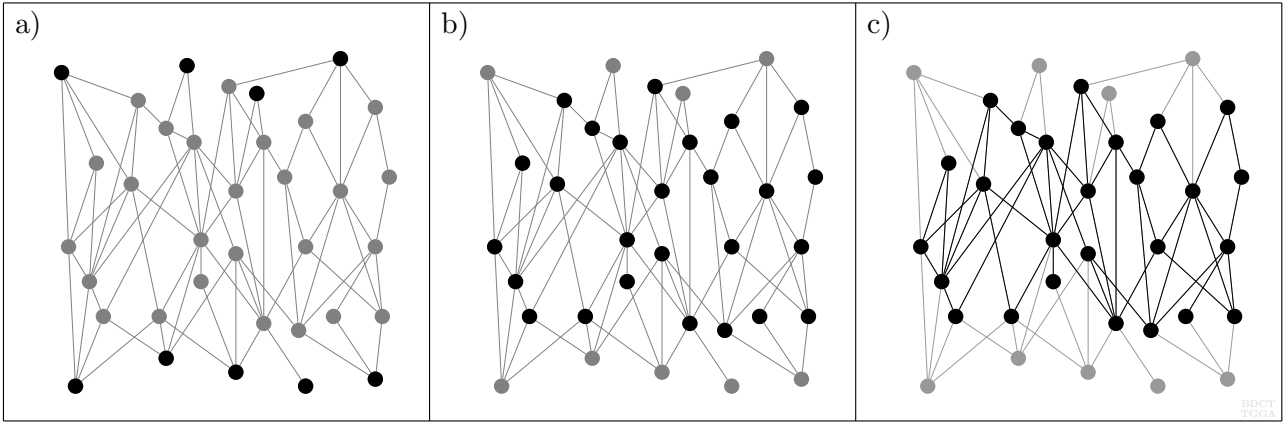


\caption{a) Boundary $\partial M$ of a multidirected set $M$; b) interior $\tn{Int}(M)$ of $M$; c) $\tn{Int}(M)$ as a full subobject of $M$.}
\label{boundary}
\end{figure}
\vspace*{-.5cm} 
  

\subsection{Order Theory; Category Theory; Influence of Grothendieck}\label{settheoretic}

Two broad structural paradigms used in this paper are {\it generalized order theory} and {\it category theory.}  Generalized order theory includes all the structured sets considered in this paper, including ordered sets, partially ordered sets, acyclic directed sets, directed sets, and multidirected sets.  In this section, I discuss a few order-theoretic subtleties involving the cardinalities of certain isomorphism classes of acyclic directed sets, and the properties of order extensions and linear orders.  These subtleties are relevant to the theory of {\it relative multidirected sets over a fixed base,} introduced in section \hyperref[relativeacyclicdirected]{\ref{relativeacyclicdirected}}, and the theory of {\it kinematic schemes,} introduced in section \hyperref[subsectionkinematicschemes]{\ref{subsectionkinematicschemes}} below.  Category theory, conceived in a pure mathematical context, plays a dual role in this paper, both as a general mathematical framework, and as a novel structural analogy to causal theory.   Important aspects of this analogy already appear in Chris Isham's {\it topos-theoretic approach} to quantum gravity.  Causal theory returns the favor to pure mathematics by suggesting structural paradigms similar to, but distinct from, category theory.  This section concludes with a discussion of how the mathematical work of Alexander Grothendieck influences the conceptual framework of this paper, focusing on the {\it relative viewpoint} (\hyperref[rv]{RV}), and the notion of {\it hidden structure} (\hyperref[hs]{HS}). 


{\bf Counting Finite Acyclic Directed Sets; Countable Acyclic Directed Sets.}\refstepcounter{textlabels}\label{isomclassesfinite}   I begin this section by establishing the cardinalities of a few distinguished classes of acyclic directed sets, considered up to isomorphism.  These classes serve as the ``object" classes of the ``smallest physically relevant kinematic schemes," considered in section \hyperref[subsectionquantumcausal]{\ref{subsectionquantumcausal}} of this paper.   This type of computation may be easily generalized to determine the cardinalities of isomorphism classes of multidirected sets with cardinality bounds on their sets of elements and relations, as well as the cardinalities of isomorphism classes of {\it power spaces,} discussed in section \hyperref[subsectionpowerset]{\ref{subsectionpowerset}} below. 

\refstepcounter{textlabels}\label{isomclassescountable}

Since a finite set admits only a finite number of binary relations, the class $\ms{A}_{\tn{\fsz{fin}}}$ of finite acyclic directed sets, considered up to isomorphism, is a set of cardinality $\aleph_0$, where $\aleph_0$ is the unique countably infinite cardinal.  The class $\ms{A}_{\aleph_0}$ of countable acyclic directed sets, considered up to isomorphism, is a set of continuum cardinality.  To see this, first note that the unit continuum interval $[0,1]$, and hence the continuum itself, may be placed in bijective correspondence with a subclass of $\ms{A}_{\aleph_0}$.  One way of doing this is illustrated in figure \hyperref[encodingpi]{\ref{encodingpi}} below. 

\begin{figure}[H]
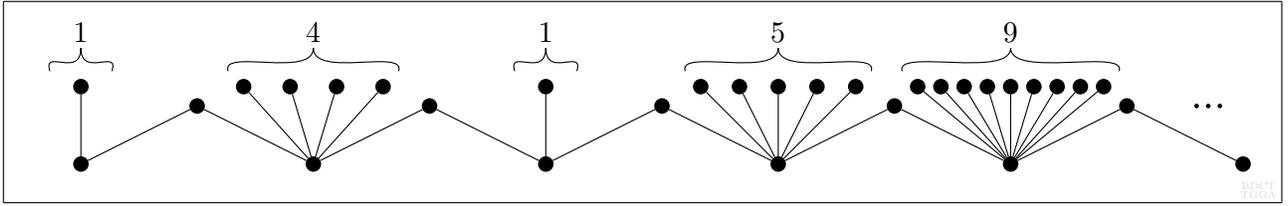


\caption{Portion of a locally finite causal set encoding the decimal $\pi-3=.14159...$}
\label{encodingpi}
\end{figure}
\vspace*{-.5cm}

Anticipating the introduction of {\it local finiteness} (\hyperref[lf]{LF}) in section \hyperref[subsectionintervalfiniteness]{\ref{subsectionintervalfiniteness}}, the construction illustrated in figure \hyperref[encodingpi]{\ref{encodingpi}} involves only the class $\ms{C}_{\tn{\fsz LF}}$ of isomorphism classes of {\it locally finite causal sets}.  The larger classes $\ms{A}_{\aleph_0}$, $\ms{C}$, and $\ms{A}_{\aleph_0,\tn{\fsz LF}}$ of isomorphism classes of countable acyclic directed sets, causal sets, and countable locally finite acyclic directed sets, respectively, are therefore {\it at least} as large as the continuum.   Now let $S_{\aleph_0}$ be a {\it fixed countably infinite set}.  The class of binary relations on $S_{\aleph_0}$ includes a representative of every infinite member of $\ms{A}_{\aleph_0}$; the finite members may be ignored since they are countable in number.  Hence, $\ms{A}_{\aleph_0}$ corresponds bijectively with a subset of the power set of $S_{\aleph_0}\times S_{\aleph_0}$, which has cardinality $2^{\aleph_0}$, the cardinality of the continuum.\footnotemark\footnotetext{Under the {\it continuum hypothesis,} $2^{\aleph_0}$ is equal to the smallest uncountable cardinal $\aleph_1$.}   Therefore, the four classes $\ms{C}_{\tn{\fsz LF}}$, $\ms{A}_{\aleph_0}$, $\ms{C}$, and $\ms{A}_{\aleph_0,\tn{\fsz LF}}$ also have cardinality {\it at most} as large as the continuum.  


\refstepcounter{textlabels}\label{orderext}

{\bf Linear Orders; Order Extensions.} {\it Linear orders} are, in a sense, the simplest nontrivial orders.  For this reason, it is often useful to {\it compare} multidirected sets, particularly acyclic directed sets, to linearly ordered sets.   The theory of {\it order extensions} provides a means of formalizing such comparisons, by extending a partial order to a linear order.   Sorkin and Rideout's theory of sequential growth dynamics of causal sets \cite{SorkinSequentialGrowthDynamics99}, already mentioned in section \hyperref[subsectionapproach]{\ref{subsectionapproach}} above, illustrates the importance of linear extensions, and their corresponding morphisms, in causal set theory.  In particular, a {\it total labeling} $\{x_0,x_1,x_2,...\}$ of a causal set $(C,\prec)$ is equivalent to a bijective morphism from $(C,\prec)$ into a linearly ordered subset of the nonnegative integers $\NN$.   In this paper, it is necessary to compare acyclic directed sets to more general linearly ordered sets, particularly those with {\it discrete linear orders.}  Such sets are {\it locally isomorphic} to the integers, except possibly at extremal elements.  This topic is discussed in more detail in section \hyperref[relativeacyclicdirected]{\ref{relativeacyclicdirected}} below. 

Let $(A,\prec)$ be an acyclic directed set.  The transitive closure of $A$ is equivalent to a partially ordered set, via the equivalence introduced in sections \hyperref[subsectionaxioms]{\ref{subsectionaxioms}} and \hyperref[subsectionacyclicdirected]{\ref{subsectionacyclicdirected}} above.   Regarding the binary relation $\prec$ on $A$ as a subset of the Cartesian product $A\times A$, an {\bf extension} of $\prec$ is a subset $\prec'$ of $A\times A$ containing $\prec$.   In terms of individual relations between pairs of elements of $A$, $\prec'$ is an extension of $\prec$ if and only if each relation $x\prec y$ implies the relation $x\prec'y$.   Abusing terminology in order to remain in the acyclic context, a {\bf linear order} on $A$ is an transitive acyclic binary relation $\prec''$ on $A$, such that either $x\prec''y$ or $y\prec''x$ for every pair of {\it distinct} elements $x$ and $y$ in $A$. Equivalently, a linear order is a {\it total} order, in the sense that it admits no extension. If $(A,\prec)$ is any acyclic directed set, an extension of $\prec$ to a linear order $\prec''$ on $A$ is equivalent to a bijective morphism $(A,\prec)\rightarrow (A,\prec'')$.  

The existence of linear extensions of acyclic binary relations is surprisingly subtle.  The transitive analogue of the following {\it order-extension principle} is sometimes used as an {\it axiom} of modern set theory:

\refstepcounter{textlabels}\label{oep}

\hspace*{.3cm} OEP. {\bf Order Extension Principle:} {\it Every acyclic binary relation $\prec$ on a set $A$ \\ 
\hspace*{1.4cm} may be extended to a linear order on $A$.}

In Zermelo-Fraenkel set theory, the order extension principle may be {\it proven} using the power set axiom and Zorn's lemma.  The {\it power set axiom} states that the power set $\ms{P}(A)$ of a set $A$ is itself a set, rather than a {\it proper class}; i.e., a class ``too large" to be a set.\footnotemark\footnotetext{As von Neumann and others have pointed out, ``size" is a poor way to understand the distinction between sets and proper classes, but no great sophistication is needed in the present context.}  {\it Zorn's lemma} is an {\it axiom} closely related to the {\it axiom of choice;} it states that any partially ordered set satisfying the property that every chain is bounded above has at least one maximal element.  Zorn's lemma may be applied to the class of extensions of a transitive acyclic binary relation $\prec$ on $A$, ordered by inclusion, which is {\it itself} a set, by an elementary application of the power set axiom.  The union of the elements of any chain of extensions of $\prec$ serves as an upper bound for the chain, and the maximal element guaranteed by Zorn's lemma is a linear order on $A$, generally {\it not} unique.   


{\bf Category Theory.}\refstepcounter{textlabels}\label{cat} In sections \hyperref[subsectionaxioms]{\ref{subsectionaxioms}} and \hyperref[subsectionacyclicdirected]{\ref{subsectionacyclicdirected}} of this paper, I introduced the specific categories of causal sets, acyclic directed sets, directed sets, and multidirected sets.  Here, I briefly discuss categories in general, and outline the use of category theory in this paper.  A {\bf category} $\ms{C}$ consists of a class of objects $\tn{Obj}(\ms{C})$, and a class of morphisms $\tn{Mor}(\ms{C})$, such that every morphism $\gamma$ has an initial object $\mbf{i}(\gamma)$ and a terminal object $\mbf{t}(\gamma)$, every object $C$ has an identity morphism $\gamma_C$,\footnotemark\footnotetext{The property of $\gamma_C$ distinguishing it as an identity morphism is that it is {\it absorbed under composition;} i.e., for any morphisms $\gamma:B\rightarrow C$ and $\gamma':C\rightarrow D$, $\gamma_C\circ\gamma=\gamma$ and $\gamma'\circ\gamma_C=\gamma'$.} and morphisms compose associatively.    For many categories, objects possess {\it internal structure,} and morphisms are chosen to {\it preserve structure} in some way.  For example, morphisms of multidirected sets  preserve initial and terminal elements. However, there are also important cases in which the objects of a category do {\it not} possess internal structure, and all the structure of the category resides in its morphisms.   An example is the category-theoretic approach to {\it groups}, in which group elements are represented by endomorphisms of a single structureless object.

\refstepcounter{textlabels}\label{ishamtopos}Category theory plays two distinct roles in this paper, the first conventional, and the second relatively novel.  First, category theory serves as a {\it general structural paradigm;} i.e., a convenient method of organizing information about mathematical objects and relationships between pairs of objects. Category theory is used in this way in many, if not most, mathematical papers involving modern algebra.  Second, structures of particular interest in causal theory, including multidirected sets and kinematic schemes, {\it closely resemble categories.}  This leads to interesting interplay between causal theory and category theory.  Chris Isham and his collaborators have carried this analogy very far in a particular direction, organizing families of causal sets, as well as many other types of physical models, into special categories called {\it topoi,} in an effort to build up a general {\it topos-theoretic} framework for physics.  Though Isham's work is one of the major motivations for this paper, my perspective on the role of category theory in physics is somewhat different.  Consider the following two questions:

\hspace*{.3cm}{\bf Question 1:} {\it How can existing structural notions in mathematics, such as category theory, 
\\\hspace*{.3cm}be applied to physics?}

\hspace*{.3cm}{\bf Question 2:} {\it Does existing mathematical theory supply adequate structural notions for
 \\\hspace*{.3cm} theoretical physics, and if not, what more can physics teach us about structural notions?}

 \refstepcounter{textlabels}\label{differentnotions}Isham's program seems to focus mainly on the first question, with the principal ``existing structural notion" being topos theory.  To be sure, Isham considers the second question as well; for example, in his recent article {\it Topos Methods in the Foundations of Physics} \cite{Isham11}, he points out the dangers of taking for granted} the real and complex numbers in the context of quantum gravity, proposes replacing the notion of probabilities with {\it ``generalized truth values,"} and suggests that {\it ``conventional quantum formalism is inadequate to the task of quantum gravity."}  All these points are well taken.  However, Isham advocates topos-theoretic solutions to these issues, rather than new, physically motivated, structural paradigms.

My own view is that category theory and topos theory may not be ideally suited to fundamental physics, and that causal theory suggests {\it essentially different} structural notions that are are worth studying in their own right.  Category theory and topos theory arose from the fields of homological algebra and algebraic geometry between 1940 and 1960, in response to specific mathematical issues only distantly related to fundamental spacetime structure.  These issues include the essential uniqueness of certain {\it cohomology theories,} and the assignment of algebraic structure to generalized topological spaces.  While sufficiently rich and general to {\it accommodate} any conceivable physical theory, categories and topoi may not always be {\it optimal} or {\it natural} for this purpose.  Indeed, the clear physical relevance of essentially different structures playing similar roles, such as kinematic schemes in causal theory, suggests that alternative structural notions may be more appropriate in certain physical contexts.  Such alternatives already appear, in a variety of mathematical settings, as ``enriched" or ``categorified"  graphs or quivers, but are seldom accorded broad structural significance in their own right.   However, the legacy of interaction between mathematics and physics suggests than any concept of fundamental physical significance should be taken very seriously, even from a purely mathematical perspective.\footnotemark\footnotetext{One of many instances of the continued mathematical potency of physically motivated methods, setting aside the question of their actual physical success, is the interplay between string theory and algebraic geometry/topology, for example, in {\it Gromov-Witten theory,} {\it ``Monstrous Moonshine,"} {\it mirror symmetry,} and so on.}

This paper contains several examples of ``category-like" structures, arising naturally in causal theory, for which na\"{i}ve category-theoretic intuition {\it leads in the wrong direction.}  Examples include {\it``objects without identity morphisms,"} and {\it ``morphisms which do not compose,"} appearing repeatedly in the theory of multidirected sets, relation spaces, and power spaces; {\it ``morphisms without initial and terminal objects,"} appearing in the theory of {\it causal path semicategories,} discussed in section \hyperref[subsectionpathspaces]{\ref{subsectionpathspaces}}; and {\it ``morphisms whose compositions are families of morphisms rather than single morphisms,"} appearing in the theory of {\it co-relative histories,} discussed in section \hyperref[subsectionquantumprelim]{\ref{subsectionquantumprelim}}.   These examples provide evidence that the analysis of basic structural questions is still very interesting and open-ended in theoretical physics, not merely an exercise in choosing and applying familiar mathematical tools.  


{\bf Influence of Grothendieck.}\refstepcounter{textlabels}\label{groth} Many of the structural ideas appearing in this paper, both conventional and novel, have {\it algebraic} roots in the work of  Alexander Grothendieck, the father of modern algebraic geometry.  It was Grothendieck, along with his student Verdier, who introduced topos theory around 1960. Most of the topics to which Grothendieck applied his methods play little direct role in discrete causal theory,\footnotemark\footnotetext{{\it Noncommutative} geometry provides methods for studying the {\it causal path algebras} introduced in section \hyperref[subsectionpathspaces]{\ref{subsectionpathspaces}}. Ioannis Raptis discusses the application of noncommutative geometry to {\it incidence algebras} of causal sets.  However, noncommutative geometry took its modern shape beginning around 1980, after Grothendieck's most productive period.} but his {\it broad structural concepts} may be applied in a variety of useful ways in this context. Two such concepts used repeatedly in this paper are the {\it relative viewpoint,} and the principle of {\it hidden structure.}\footnotemark\footnotetext{The relative viewpoint is well-known as such.  {\it Hidden structure} is my own terminology.}

\refstepcounter{textlabels}\label{rv} 

\hspace*{.3cm}RV. \hspace*{.1cm}{\bf Relative Viewpoint.} {\it It is often more natural, and more informative, to study relationships 
\\\hspace*{1.1cm}between objects of a certain type, than to study such objects individually.} 

Grothendieck applied the relative viewpoint to his work on {\it commutative algebraic varieties and schemes,} which are geometric objects defined locally by the vanishing of polynomials.  In this context, one fixes a {\it base scheme} $Z$, and considers the {\it category of schemes over} $Z$; i.e., the category whose objects are morphisms of schemes $X\rightarrow Z$.  The signature achievement of this method was Grothendieck's proof of the {\it Grothendieck-Riemann-Roch theorem.}\footnotemark\footnotetext{This theorem may be viewed as a special case of the {\it Atiyah-Singer index theorem.}  It has become so prominent in string theory that at least one string theorist has proposed renaming it the ``Atiyah-Singer string index theorem."}  It has since become ubiquitous in modern algebra.  The first significant application of the relative viewpoint in this paper comes in section \hyperref[relativeacyclicdirected]{\ref{relativeacyclicdirected}}, where I use it to study {\it relative multidirected sets over a fixed base.}  For example, causal sets may be viewed as relative directed sets over the integers. In section \hyperref[subsectionrelation]{\ref{subsectionrelation}}, the relative viewpoint underlies the theory of {\it relation space,} which leads to the technical methods of abstract quantum causal theory introduced in section \hyperref[subsectionquantumpathsummation]{\ref{subsectionquantumpathsummation}}.  In section \hyperref[subsectionquantumprelim]{\ref{subsectionquantumprelim}}, the relative viewpoint serves as the conceptual foundation of the theory of {\it co-relative histories,} which formalize the idea of ``spacetime evolution."  Although the relative viewpoint often involves the study of morphisms in some category, more nuanced versions are sometimes useful.  For example, co-relative histories are {\it not} morphisms in the category of directed sets. The relative viewpoint is generally distinct from the long-standing physical and metaphysical philosophy of {\it relationism,} but nontrivial connections between the two ideas arise in the special context of causal theory, particularly in the theory of relation space. 

\refstepcounter{textlabels}\label{hs}

\hspace*{.3cm}HS. \hspace*{.1cm}{\bf Hidden Structure.} {\it Many objects presented as structured sets may be more profitably viewed 
\\\hspace*{1.1cm}as enlarged sets with ``extra elements," naturally induced by their native structures.}

Grothendieck used the idea of hidden structure to construct algebraic schemes, by adding new elements of positive dimension to the sets of zero-dimensional points of algebraic varieties.  Since schemes are more natural and better behaved than varieties in many respects, the ``promotion of varieties to schemes" may be viewed as a belated recognition of hidden structure latent in varieties, unnoticed for more than two millennia.  In section \hyperref[subsectiontopology]{\ref{subsectiontopology}} of this paper, I treat the {\it relations} in a multidirected set $M=(M,R,i,t)$ as elements of a topological space called the {\it star model} of $M$, which plays an important role in the study of {\it causal locality.}   As a set, the star model is the {\it union} of $M$ and its relation set $R$.  In section \hyperref[subsectionrelation]{\ref{subsectionrelation}}, I elevate the relation set $R$ to the {\it relation space} $\ms{R}(M)$ over $M$, already mentioned above in the context of the relative viewpoint.  Hidden structure is further exploited in the context of power spaces, causal path spaces, and kinematic schemes, discussed in sections \hyperref[sectionbinary]{\ref{sectionbinary}} and \hyperref[subsectionquantumcausal]{\ref{subsectionquantumcausal}} below. 

Much of Grothendieck's mathematical output came during a period of relative estrangement between fundamental physics and abstract mathematics, which continued until the string theory revolutions, Alain Connes' work on noncommutative geometry, and other developments during the 1980's and 1990's.   With a few exceptions,\footnotemark\footnotetext{For example, Roger Penrose introduced twistor theory in the middle 1960's.} physicists during this interregnum relied on mathematical notions conceived in the century from Riemann to Weyl.   Even today, the structural ideas of Grothendieck and his contemporaries are not as broadly appreciated in the physics community as one might wish.\footnotemark\footnotetext{As late as 2005, Chris Isham felt compelled to essentially {\it apologize} for using methods as abstract as category theory.  See \cite{IshamQuantisingI05}, pages 3, 10.}  This unfortunate state of affairs may be attributed partly to a legacy of physically unilluminating exposition emanating from certain mathematical sources,\footnotemark\footnotetext{For example, the {\it Bourbaki} group, to which Grothendieck briefly contributed.  One of the principal scribes for {\it Boubaki} was Jean Dieudonn\'{e}, who also penned Grothendieck's monumental {\it \'{E}l\'{e}ments de G\'{e}om\'{e}trie Alg\'{e}brique} (EGA).} and partly to ``pragmatism" in the physics community during the heyday of experimental particle physics.

\newpage

\section{Transitivity, Independence, and the Causal Preorder}\label{sectiontransitivity}

Transitivity (\hyperref[tr]{TR}) is deeply ingrained in ordinary thought as part of the {\it meaning} of classical causal structure.  For example, the relation between grandparent and grandchild, though usually indirect, is recognized as causal, due to transitivity, as illustrated by its role in the so-called {\it grandfather paradox.} Challenging this convention would only incite terminological controversy, to no good purpose.  Hence, rather than disputing the notion of a transitive causal order, I introduce a more fundamental {\it nontransitive} binary relation, which I refer to as the {\it causal preorder}, from which the transitive causal order may be derived.  Here, it is expedient to use the term {\it order} in a generalized sense, as explained at the beginning of section \hyperref[settheoretic]{\ref{settheoretic}} above.  In particular, ``causal orders" may contain cycles, analogous to closed timelike curves in general relativity.  The causal preorder encodes only {\it independent} influences between pairs of events, and {\it generates} the causal order under the operation of {\it transitive closure.}  Using the independence convention (\hyperref[ic]{IC}), a causal preorder may be unambiguously represented by a directed set in the context of discrete causal theory. 

In section \hyperref[subsectionmodes]{\ref{subsectionmodes}} below, I present the basic problem of transitive binary relations: they fail to distinguish among classical causal structures involving {\it multiple independent modes of influence} between pairs of events.  I discuss two of the few treatments of this issue appearing in the literature, due to David Finkelstein and Ioannis Raptis, and relate Rafael Sorkin's current view of the subject, via private communication.  In section \hyperref[subsectiontransitivitydeficient]{\ref{subsectiontransitivitydeficient}}, I give six arguments that transitivity represents a major information-theoretic deficiency in causal set theory.  In section \hyperref[subsectionpreorder]{\ref{subsectionpreorder}}, I introduce the causal preorder.  Finally, in section \hyperref[subsectiontransitiveclosure]{\ref{subsectiontransitiveclosure}}, I introduce the {\it transitive closure functor,} which converts a causal preorder into a transitive causal order, and the {\it skeleton operation,} a ``partial inverse" of the transitive closure, which annihilates reducible relations.  

Throughout section \hyperref[sectiontransitivity]{\ref{sectiontransitivity}}, I focus on the special case of directed sets, since the intended context is the structure of classical spacetime.   In particular, every consideration of significant importance in this section manifests itself already in the acyclic directed case.  The principal conclusion is that {\it causal sets are information-theoretically inadequate to model classical spacetime structure in discrete causal theory, even under the assumption of acyclicity.}  It is safe, in any case, to omit multidirected sets from the present discussion, since the existence of multiple independent ``modes of influence" is obvious from the very beginning in the settings in which such sets arise in discrete causal theory, such as in the {\it ``decategorification"} of a kinematic scheme.  These concepts are clarified and expanded in section \hyperref[subsectionquantumcausal]{\ref{subsectionquantumcausal}} of this paper. 


\subsection{Independent Modes of Influence}\label{subsectionmodes}

\refstepcounter{textlabels}\label{transproblem}

{\bf The Basic Problem.}  Recognition of a nontransitive binary relation underlying the transitive causal order is motivated by the possibility of multiple independent ways in which one event may influence another, called {\bf modes of influence}.   A simple example of this is illustrated in figure  \ref{transitivity} below.  Consider three events, labeled $x$, $y$, and $z$.  If $x$ influences $z$, this influence may be either exclusively independent of $y$, as illustrated in \ref{transitivity}a, exclusively mediated by $y$, as illustrated in \ref{transitivity}b, or it may involve {\it both independent influence and influence mediated by} $y$, as illustrated in \ref{transitivity}c.  In accordance with the independence convention (\hyperref[ic]{IC}), the relations indicated by solid lines in the figure all represent independent influences; the dashed line in figure \ref{transitivity}b indicates the {\it hypothetical} relation $x\prec z$ prescribed under the axiom of transitivity (\hyperref[tr]{TR}).  Causal set theory, which does {\it not} employ the independence convention, and which {\it does} impose transitivity, requires the presence of this relation, and therefore cannot distinguish between the scenarios illustrated in figures \ref{transitivity}b and \ref{transitivity}c. {\it A priori}, this is a serious information-theoretic deficiency.  This situation extends in an obvious way to any relation in a multidirected set admitting a reducing chain.  

\begin{figure}[H]
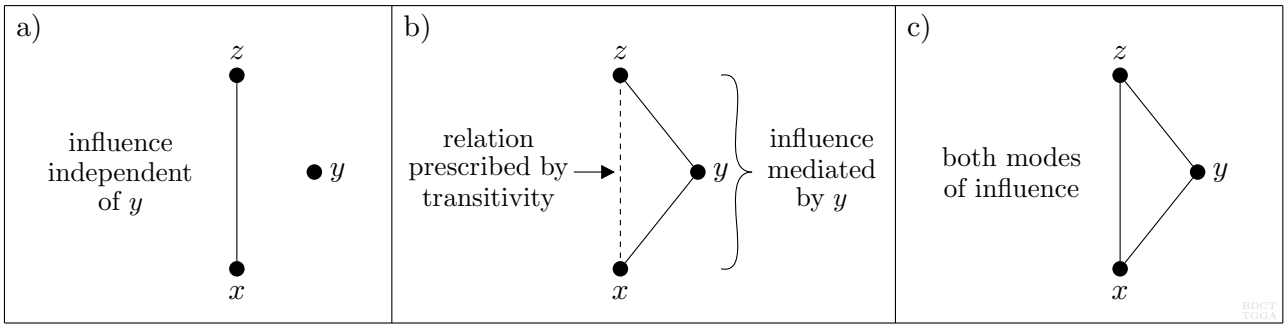


\caption{Modes of influence between related elements in a three-element directed set.}
\label{transitivity}
\end{figure}
\vspace*{-.5cm}

\refstepcounter{textlabels}\label{infoanalogies}

{\bf Na\"{i}ve Analogies from Information Theory.} A cryptographic analogy is illustrative of this deficiency.  Consider the problem of sending a secret message from $x$ to a remote receiver at $z$, either independently, as in figure \hyperref[transitivity]{\ref{transitivity}}a, or via an intermediary at $y$, as in figure \hyperref[transitivity]{\ref{transitivity}}b.  The intermediary can prevent third-party tampering, but may not be trustworthy.  One possible strategy is to send part of the message independently, and the other part via the intermediary, as in figure \hyperref[transitivity]{\ref{transitivity}}c, in such a way that its meaning can only be deciphered by combining both parts.  Transitive relations are clearly deficient for describing these three scenarios, since they fail to distinguish between the scenarios illustrated in figures \hyperref[transitivity]{\ref{transitivity}}b and \hyperref[transitivity]{\ref{transitivity}}c.  Transitive relations can encode {\it which} pairs of events communicate, but cannot encode {\it how} they communicate.  This problem only becomes worse for larger families of events, in which there may be many independent modes of influence between a given pair of events.   The consequences of ignoring such independence may be illustrated by imagining an academic conference, at which the chair announces that only {\it one} member need return the following year, since the information brought to the second meeting by all the other members is sure to be redundant!


\refstepcounter{textlabels}\label{finkelstein}

{\bf Finkelstein's Causal Nets.} The foregoing ideas are so elementary, and the reasonable alternatives so few, that the same basic steps have surely been followed many times previously.  However, it is surprisingly difficult to find explicit treatments of this issue in the literature, at least in the context of classical spacetime structure.   Here I briefly detail a few instances that come as close as any I could find to addressing this problem.  David Finkelstein, in his 1988 paper {\it ``Superconducting" Causal Nets} \cite{Finkelstein88}, writes,  
\begin{quotation}
{\noindent\it``...the \tn{[transitive]} causal relation $x\mbf{C}y$ is not local, but may hold for events as far apart as the birth and death of the universe.  Since we have committed ourselves to local variables, we abandon $\mbf{C}$ for a local causal relation $\mbf{c}$..." }(page 476)
\end{quotation}
and later,
\begin{quotation}
{\noindent\it``We call $\mbf{c}$ the (causal) connection relation, understanding that this connection is... ...immediate ($x\mbf{c}y$ and $y\mbf{c}z$ do not necessarily imply $x\mbf{c} z$).  Events in a continuum theory have no immediate causal relations..."}(page 477)
\end{quotation}
and again,
\begin{quotation}
{\noindent\it``Every causal set $\mbf{C}$ may also be regarded as a causal net $\mbf{c}$, with the connection $x\mbf{c}y$ defined to hold if and only if \tn{[the relation $x\mbf{C}y$ holds]} and no event $z$ exists with $x\mbf{C}z\mbf{C}y$... ... but \tn{[causal nets]} are more general than causal sets."}(page 477)
\end{quotation}

These remarks suggest that Finkelstein's principal objections to transitive binary relations were\footnotemark\footnotetext{I say ``were," rather than ``are," because Finkelstein suggested to me, via private communication \cite{FinkelsteinPrivate13}, that his views on the topic have changed considerably since 1988.} based on issues of {\it locality} rather than independence, although the greater generality afforded by dropping the axiom of transitivity (\hyperref[tr]{TR}), noted in the final quote above, is information-theoretically meaningful only if reducible relations can encode independent influence.  Finkelstein's locality-based critique of transitivity is valid, and in my opinion, unanswerable, though the structural problems in causal set theory involving locality do not arise from transitivity alone.  This topic is elaborated in section \hyperref[sectioninterval]{\ref{sectioninterval}} below.   Due to Finkelstein's use of a local causal relation, his causal nets share a greater abstract similarity with the directed sets used in this paper to model classical spacetime than Sorkin's causal sets do, at least at the level of individual objects.  However, the overall approach developed here is much closer to causal set theory, particularly in the quantum-theoretic context. 


\refstepcounter{textlabels}\label{raptisquant}

{\bf Raptis' Algebraic Quantization of Causal Sets.}  A few relatively recent papers examining the use of nontransitive causal relations in the study of fundamental spacetime structure may also be found in the literature.  Ioannis Raptis, in his 2000 paper {\it Algebraic Quantization of Causal Sets} \cite{RaptisAlgebraicQuantization00}, deprecates transitive binary relations as unsuitable for modeling {\it quantum} spacetime, writing that {\it ``the physical connection between events in the quantum deep should be one connecting nearest-neighboring events."}  He cites Finkelstein's paper \cite{Finkelstein88} as the source of this idea.   However, Raptis uses {\it irreducible} relations in his ``quantum causal sets,"  thereby sidestepping the issue of independence.  Also, the objections to transitivity outlined above apply not only in the {\it ``quantum deep,"} but also at the classical level. 


\refstepcounter{textlabels}\label{sorkincurrent}

{\bf Sorkin's Current View.} An interesting {\it historical} point, regarding the alternative between transitive and nontransitive causal relations in the study of fundamental spacetime structure, is that Finkelstein \cite{Finkelstein88} {\it credits Sorkin} with suggesting the idea of the nontransitive local ``causal connection relation" $\mbf{c}$ appearing in Finkelstein's paper, via private communication in 1987!  This indicates that Sorkin has considered, and to some extent encouraged, the development of {\it both} approaches.  Sorkin recently expressed to me his current view of the matter in the following remarks, reproduced here with his permission:
\begin{quotation}
{\noindent\it``I suppose the main reason \tn{[not to express causal structure in terms of nontransitive binary relations]} is that \tn{[doing so appears]} to add nothing to the theory, insofar as it tries to explain how a Lorentzian spacetime can emerge from an underlying discrete order.  In the continuum one has well defined causal relations like \tn{[the relativistic ``causal" relation]} $J^+$, and they are automatically transitive, whence suited to correspond to a discrete transitive relation of precedence. But what in the continuum would correspond to a more general... ...not-necessarily transitive relation...?"} \cite{SorkinPrivate13}
\end{quotation}
These remarks suggest, at least to me, that the choice involved is largely one of focus and ambition.  If one seeks merely to {\it recover} relativistic spacetime, at a suitable level of approximation, from a binary relation on a ``discrete set," then it may indeed {\it ``add nothing to the theory"} to consider the independence of reducible relations.\footnotemark\footnotetext{However, if one relaxes the measure axiom to admit volume dependence on {\it local causal structure,} as discussed at the end of section \hyperref[subsectionintervalfiniteness]{\ref{subsectionintervalfiniteness}}, then independence becomes metrically relevant.}  There is something humorous about attaching the adjective ``merely" to a research objective which would represent a monumental advance in theoretical physics, should it succeed.  However, if one does aspire to go further, and to recover ``matter and energy" along with spacetime, the possible role of independence becomes clearer.  Sorkin goes on to suggest this very idea, while expressing open-minded reservations about its viability:
\begin{quotation}
{\noindent ``[the question is] \it whether such additional structure could play a useful kinematical or dynamical role in the theory, perhaps as a kind of ``matter" living on the \tn{[causal set]} (a rudimentary scalar field).  At present that looks unlikely to me, but of course I'd never say it was impossible."} \cite{SorkinPrivate13}
\end{quotation}
My own view is that, given the extreme simplicity of binary relations, compared, for example, to continuum manifolds, any approach to fundamental physics based on binary relations should {\it a priori} attempt to leverage every bit of natural structure available.  This is particularly true if one can ``obtain more structure by assuming less;" in this case, obtaining a ``scalar field" by replacing the axioms of transitivity and irreflexivity with the weaker axiom of acyclicity, or omitting these axioms altogether.   I do not know the details of Sorkin's reasons for doubting that such a generalization can play a {\it ``useful kinematical or dynamical role"} in discrete causal theory, but sections \hyperref[sectioninterval]{\ref{sectioninterval}}, \hyperref[sectionbinary]{\ref{sectionbinary}}, and \hyperref[subsectionquantumcausal]{\ref{subsectionquantumcausal}} of this paper present, in my belief, strong reasons to believe that it can.   Hence, I view the omission of nontransitive binary relations in causal set theory as an unfortunate example of {\it censoring the merely unexpected,} in the sense of section \hyperref[naturalphilosophy]{\ref{naturalphilosophy}} above. 
 

\refstepcounter{textlabels}\label{independencefundamental}

{\bf Independence at the Fundamental Scale.} Regardless of the opinions of the experts, the na\"{i}ve information-theoretic analogies given above make the deficiency of transitive binary relations in discrete causal theory seem more obvious than it actually is.  For example, the intuition of ``breaking a message into multiple parts" is dubious in the context of fundamental spacetime structure, though the notion of independent modes of influence does not depend on this intuition.  Arguments for transitivity may also be made by drawing analogies with relativistic spacetime geometry, but adopting any such argument endangers both the correctness and the explanatory power of the resulting theory.  The safe position is to abstain from transitivity, since this leads to a more general theory.  For this reason, I do not discuss {\it specific} defenses of this axiom.  However, I  emphasize that any such defense must establish that {\it only one of the two scenarios} illustrated in figures \hyperref[transitivity]{\ref{transitivity}}b and \hyperref[transitivity]{\ref{transitivity}}c above is physically relevant at the fundamental scale, since otherwise the need to distinguish between them is unavoidable.   

Hence, to rescue transitivity, one must argue either that independent modes of influence along a relation and a reducing chain {\it never exist} at the fundamental scale, or, alternatively, that they {\it always exist.}  The latter argument is absurd, since it leads to the conclusion that, whenever one event influences another, however long the chain of intervening influences, {\it independent direct influence} must also propagate between the two events; e.g., between the {\it ``birth and death of the universe,"} as Finkelstein puts it.   In particular, this means that the independence convention (\hyperref[ic]{IC}) makes no sense in the transitive paradigm, as already noted at the end of section \hyperref[subsectionchains]{\ref{subsectionchains}} above.  Eliminating this choice leaves two possibilities: either independent modes of influence along a relation and a reducing chain never exist at the fundamental scale, or transitive binary relations are deficient for modeling the fundamental structure of classical spacetime in discrete causal theory.


\subsection{Six Arguments that Transitive Binary Relations are Deficient}\label{subsectiontransitivitydeficient}

I now present six arguments that the scenario illustrated in figure \hyperref[transitivity]{\ref{transitivity}}c above, in which independent modes of influence exist along a relation and a reducing chain between two events, {\it should not be ruled out at any scale.}  As discussed at the end of section \hyperref[subsectionmodes]{\ref{subsectionmodes}}, this leads immediately to the conclusion that transitive binary relations are deficient for modeling the fundamental structure of classical spacetime in discrete causal theory.  Some of the arguments presented below are direct, while others merely make note of suggestive structural analogies.   These arguments are supplemented by several of the general principles discussed in section \hyperref[naturalphilosophy]{\ref{naturalphilosophy}} above.  

\refstepcounter{textlabels}\label{cauchyarg}

\begin{figure}[H]
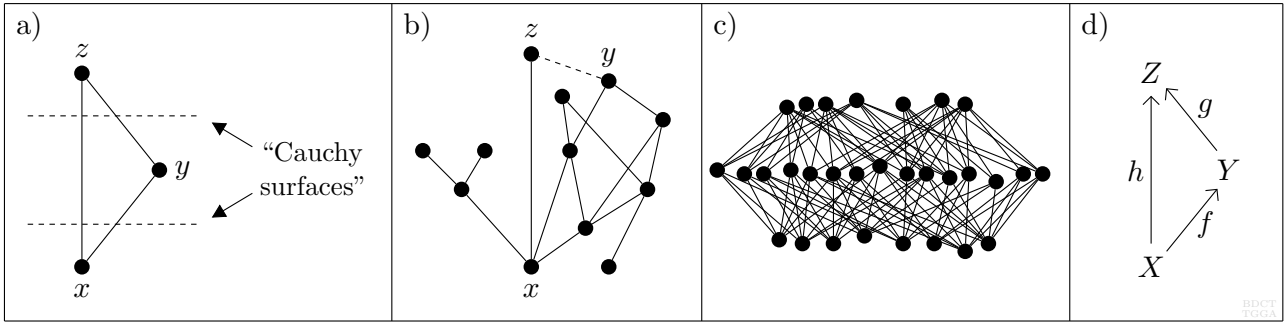


\caption{a) ``Cauchy surfaces" in the directed set first appearing in figure 3.1.1c; b) hypothetical relation in the ``distant future" of an element $x$; c) Kleitman-Rothschild-type partial order (irreducible relations shown); d) category-theoretic analogue of figure 3.1.1b or 3.1.1c.}
\label{transitivityarguments}
\end{figure}

{\bf 1.} {\bf Multiple independent modes of influence between pairs of events are ubiquitous in conventional physics.}  An important example of this is illustrated by the widespread use of {\it Cauchy surfaces} in continuum theory, both in the classical and quantum settings.  In the context of general relativity, a Cauchy surface is a spacelike hypersurface uniquely intersecting {\it every causal path} between its past and future, in a given region of spacetime.  The prominence of Cauchy surfaces is based on the assumption that different causal paths from past to future can carry independent information.  Extension of this assumption to the fundamental scale allows the scenario illustrated in figure \hyperref[transitivity]{\ref{transitivity}}c above as a special case.  Figure \hyperref[transitivityarguments]{\ref{transitivityarguments}}a below illustrates ``Cauchy surfaces" intersecting the two independent paths from $x$ to $z$ appearing in figure \hyperref[transitivity]{\ref{transitivity}}c.  These ``Cauchy surfaces" may be viewed as primitive examples of {\it maximal antichains of relations.}   Such maximal antichains play a major role in the theory of {\it relation space,} introduced in section \hyperref[sectionbinary]{\ref{sectionbinary}} of this paper, and in the quantum causal theory developed in section \hyperref[subsectionquantumcausal]{\ref{subsectionquantumcausal}}.

\vspace*{-.5cm}

\refstepcounter{textlabels}\label{futurearg}

{\bf 2.} {\bf Influences exerted by an event should not be constrained by details of its future.}  This is particularly true in the acyclic case, in which such constraints may legitimately be viewed as pathological, rather than merely dubious.  In particular, given events $x$, $y$, and $z$, together with the knowledge that $x$ influences both $y$ and $z$, the question of whether or not $x$ {\it directly} influences $z$, independently of $y$, should not depend on the existence of a relation $y\prec z$ lying in the {\it future} of $x$.  For example, figure   \hyperref[transitivityarguments]{\ref{transitivityarguments}}b above illustrates an acyclic directed set in which the existence of a reducing chain for the relation $x\prec z$ depends on the existence of a relation $y\prec z$ in the ``distant future" of $x$.  It is intuitively obvious that the presence or absence of this relation should impose no ``retroactive" constraint on influences emanating from $x$.  

\refstepcounter{textlabels}\label{irredindarg}

{\bf 3.} {\bf Irreducibility and independence of relations between pairs of elements are {\it a priori} distinct conditions.}  This logical distinction was pointed out in section \hyperref[subsectionchains]{\ref{subsectionchains}} above.  Irreducibility is an absolute mathematical condition, while independence is a subjective physical condition.  In a directed set, irreducible relations between pairs of elements are necessarily independent, but the converse is false.  The question then becomes whether or not there are compelling physical reasons to equate the two conditions in the context of discrete causal theory.  As far as I know, the answer to this question is negative.  It is therefore natural to encode discrete causal structure by means of independent relations between pairs of elements, following the independence convention (\hyperref[ic]{IC}).  

\refstepcounter{textlabels}\label{configpath}

{\bf 4.} {\bf Configuration spaces of transitive binary relations are physically pathological.}  A classic example of such a pathology occurs in the acyclic directed case, in which the distribution of finite partial orders is asymptotically dominated by objects manifestly unsuitable as models of classical spacetime.  This has been known since before the advent of causal set theory, due to Kleitman and Rothschild's 1975 paper {\it Asymptotic Enumeration of Partial Orders on a Finite Set} \cite{KleitmanRothschild75}.  This paper shows that a ``generic large" partially ordered set of cardinality $n$ has just three generations, of cardinalities roughly $n/4$, $n/2$, and $n/4$, respectively, with the number of irreducible relations beginning at a typical nonmaximal element depending linearly on $n$.  A Kleitman-Rothschild-type partial order is illustrated in figure \hyperref[transitivityarguments]{\ref{transitivityarguments}}c above. In the context of classical spacetime structure, such a partial order represents a universe of {\it large spatial and negligible temporal size,} with an unreasonably large luminal velocity.  {\it A priori,} this raises the concern that the histories approach to quantum causal set theory might be dominated by such objects.  This is sometimes referred to as an {\it entropy problem.}\footnotemark\footnotetext{This terminology comes from statistical thermodynamics, in which the large-scale state of a system is ``determined" by the multiplicities of microscopic states sharing the same large-scale properties.}  Avoiding problems of this nature was one of the principal motivations for Sorkin and Rideout's theory of sequential growth dynamics, as explained on page 4 of \cite{SorkinSequentialGrowthDynamics99}.\footnotemark\footnotetext{Sorkin and Rideout write, {\it ``Maybe this is not so different from \tn{[Feynman's path integral],} where the smooth paths, which form a set of measure zero... ...dominate the sum over histories..."}  If this is so, the most reasonable explanation might be that {\it both situations are cluttered by physically irrelevant structure;} in the first case, the continuum; in the second case, dependent relations due to transitivity.}  It is certainly possible, and in a sense, easy, to avoid such problems dynamically, but it is more promising to have a class of models that does not present such troubling features in the first place.  Often, at least in mathematics, generic behavior of this type indicates some form of {\it degeneracy,} or {\it lack of naturality} in the class of structures under consideration.  In contrast, the asymptotic behavior of (generally nontransitive) acyclic directed sets seems to be much more reasonable, as indicated by Brendan McKay's 1989 paper {\it On the shape of a random acyclic digraph} \cite{McKayDigraphs89}, and Stephan Wagner's 2013 paper {\it Asymptotic Enumeration of Extensional Acyclic Digraphs} \cite{WagnerDigraphs13}.  In particular, the configuration space of finite acyclic directed sets is not asymptotically dominated by {\it either} ``near antichains," such as the Kleitman-Rothschild orders, or by ``near-chains."  

\refstepcounter{textlabels}\label{commcat}

{\bf 5.} {\bf Structural notions from mathematics motivate independent modes of influence}.  For example, there is a useful and far-reaching analogy between plane diagrams of multidirected sets, including the generalized Hasse diagrams appearing in the acyclic directed case, and category-theoretic diagrams of objects and morphisms. This analogy is illustrated by the triangular diagram in figure \hyperref[transitivityarguments]{\ref{transitivityarguments}}d above, where $X$, $Y$, and $Z$ are objects of a category, and where $f$, $g$, and $h$ are morphisms.   This diagram may be viewed as analogous to {\it either} of the two physical scenarios illustrated in figures \ref{transitivity}b and \ref{transitivity}c above, depending on whether or not it {\it commutes}; i.e., whether or not the morphism $h$ is equal to the composition $g\circ f$.  If the diagram commutes, then $h$ is ``merely a consequence of $f$ and $g$," encoding no independent information.  This case is analogous to the scenario illustrated in figure \ref{transitivity}b.  If the diagram does not commute, then $h$ involves information independent of $f$ and $g$.  This case is analogous to the scenario illustrated in figure \ref{transitivity}c.  

In a sense, then, considering only transitive binary relations in discrete causal theory is analogous to considering only categories in which {\it every diagram commutes}, which is a very restrictive condition.  The reader familiar with ``concrete" categories, in which objects possess underlying sets, may argue in favor of such ``commutativity" in the case of directed sets, on the grounds that the unique map of singleton sets from $\{x\}$ to $\{z\}$ is equal to the composition of unique maps from $\{x\}$ to $\{y\}$ and from $\{y\}$ to $\{z\}$.  This observation is true but irrelevant; the appropriate analog of a ``morphism" in this case is not a set map.  Indeed, there is a unique set map between {\it any} pair of singletons in a directed or multidirected set.  However, as noted at the end of section \hyperref[settheoretic]{\ref{settheoretic}} above, there are many important examples of categories whose objects have no internal structure, but which nonetheless admit multiple independent morphisms between pairs of objects.  I have already cited the example of groups, whose elements may be represented by endomorphisms of a single structureless object.  Analogous but more general constructions, with important applications in this paper, arise in the theory of {\it semicategories,} discussed in section \hyperref[subsectionpathspaces]{\ref{subsectionpathspaces}} below. 

\refstepcounter{textlabels}\label{otherimprovements}

{\bf 6.} {\bf Nontransitive relations lead naturally to other improvements in causal theory.}  As pointed out by Finkelstein,  nontransitive relations enable an improved treatment of {\it causal locality.}  This notion is discussed in detail in section \hyperref[subsectiontopology]{\ref{subsectiontopology}} below.  The same basic reasoning leads to superior notions of local conditions for multidirected sets in general.  Of particular importance is {\it local finiteness} (\hyperref[lf]{LF}), introduced in section \hyperref[subsectionintervalfiniteness]{\ref{subsectionintervalfiniteness}}, which replaces the pathological causal set axiom of interval finiteness (\hyperref[if]{IF}).  Building on these improvements is the theory of {\it relation space,} introduced in section \hyperref[subsectionrelation]{\ref{subsectionrelation}}, which is almost unrecognizable in the transitive paradigm.  In particular, the use of maximal antichains of relations as analogues of Cauchy surfaces, which avoids the generic {\it permeability problem} presented by maximal antichains of elements, depends crucially on the independence convention (\hyperref[ic]{IC}), and therefore makes sense only in the nontransitive context.  Similarly, much of the algebraic machinery associated with the theory of path summation over a multidirected set, introduced in section \hyperref[subsectionpathsummation]{\ref{subsectionpathsummation}}, is seriously hampered if dependent relations are forced into the picture.  Ultimately, the causal analogues of Feynman's path integral and Schr\"{o}dinger's equation, appearing in sections \hyperref[subsectionquantumpathsummation]{\ref{subsectionquantumpathsummation}} and \hyperref[subsectionschrodinger]{\ref{subsectionschrodinger}}, find full expression only in the nontransitive setting. 


\subsection{The Causal Preorder}\label{subsectionpreorder}

\refstepcounter{textlabels}\label{nontransgen}
\refstepcounter{textlabels}\label{finkelsteinconnect}
\refstepcounter{textlabels}\label{adscausalpreorder}
\refstepcounter{textlabels}\label{notusualpreorder}

To remedy the shortcomings of transitive binary relations in modeling classical spacetime structure in discrete causal theory, I propose that the transitive causal order should be viewed as a {\it derivative construct,} generated by a more fundamental, generally nontransitive, binary relation called the {\it causal preorder.}  Referring again to Finkelstein \cite{Finkelstein88}, the causal preorder corresponds to the {\it ``causal connection relation"} $\mbf{c}$, while the causal order corresponds to the {\it ``transitive causal relation"} $\mbf{C}$.  The reader should be aware that the term {\it preorder} already has an inconvenient proprietary definition in the context of order theory, as explained in the discussion of nonstandard terminology in section \hyperref[subsectionnotation]{\ref{subsectionnotation}}.  For the duration of this paper, I reclaim this term as follows: \\

\begin{defi}\label{defpreorder} Let $D=(D,\prec)$ be a directed set, viewed as a model of causal structure under the independence convention \tn{(\hyperref[ic]{IC})}.  In this context, the binary relation $\prec$ on $D$ is called the {\bf causal preorder} on $D$.  
\end{defi}

The independence convention gives an unambiguous physical meaning to every relation appearing in a causal preorder: each relation encodes independent influence.  Dependent influence, meanwhile, is encoded by complex chains. For example, the causal preorder illustrated in figure \hyperref[transitivity]{\ref{transitivity}}b above does {\it not} include the relation $x\prec z$, since $x$ does not influence $z$ independently of $y$, while the causal preorder illustrated in figure \hyperref[transitivity]{\ref{transitivity}}c {\it does} include the relation $x\prec z$.  The causal {\it orders} are the same in both cases.


\subsection{Transitive Closure; Skeleton; Degeneracy; Functoriality}\label{subsectiontransitiveclosure}

{\bf Transitive Closure; Skeleton.} Two useful operations on directed sets are the {\it transitive closure} functor and the {\it skeleton} operation.  The transitive closure functor is particularly important because it {\it generates causal orders from causal preorders.}  These two operations realize the opposite extremes of transitivity and irreducibility for binary relations.  The transitive closure is defined by adding reducible relations between pairs of elements, while the skeleton is defined by deleting such relations.  A variety of generalizations to the case of multidirected sets are possible, but are not needed in this paper.\\

\refstepcounter{textlabels}\label{transclosure}
\refstepcounter{textlabels}\label{skeleton}

\begin{defi}\label{deftransclosure}Let $D=(D,\prec)$ be a directed set.  
\begin{enumerate}
\item The {\bf transitive closure} of $D$ is the directed set $\text{\tn{tr}}(D):=(D,\prec_{\text{\tn{\fsz{tr}}}})$ whose binary relation $\prec_{\text{\tn{\fsz{tr}}}}$ is defined by setting $x\prec_{\text{\tn{\fsz{tr}}}}y$ if and only if there is a chain from $x$ to $y$ in $D$.  
\item The {\bf skeleton} of $D$ is the acyclic directed set $\text{\tn{sk}}(D):=(D,\prec_{\text{\tn{\fsz{sk}}}})$ whose binary relation $\prec_{\text{\tn{\fsz{sk}}}}$ is defined by setting $x\prec_{\text{\tn{\fsz{sk}}}}y$ if and only if $x\prec y$ is an irreducible relation in $D$.  
\end{enumerate}
\end{defi}

\refstepcounter{textlabels}\label{skelnottransred}

The skeleton $\tn{sk}(D)$ of any directed set $D$ is automatically acyclic, since any relation in a cycle admits a reducing chain given by going around the cycle. The skeleton is generally not the same as the {\it transitive reduction,} familiar from graph theory, which {\it preserves accessibility.}  However, the two coincide for finite acyclic directed sets.  Since every relation is itself a chain, the transitive closure of a directed set $D$ has at least as many relations as $D$, with equality occurring if and only if $D$ is transitive.  Similarly, since every irreducible relation is a relation, the skeleton of $D$ has at most as many relations as $D$, with equality occurring if and only if $D$ is irreducible.    The transitive closure of a directed set is closely related to its {\it chain space,} discussed in a more general context in section \hyperref[subsectionpathspaces]{\ref{subsectionpathspaces}} below.  This relationship arises from the fact that every relation in $\tn{tr}(D)$ corresponds to a chain in $D$.  This suggests one way to generalize the transitive closure to multidirected sets: by {\it adding a relation corresponding to each complex chain.}  Figure \ref{transitiveskel} below illustrates the transitive closure and the skeleton of a directed set. 

\begin{figure}[H]
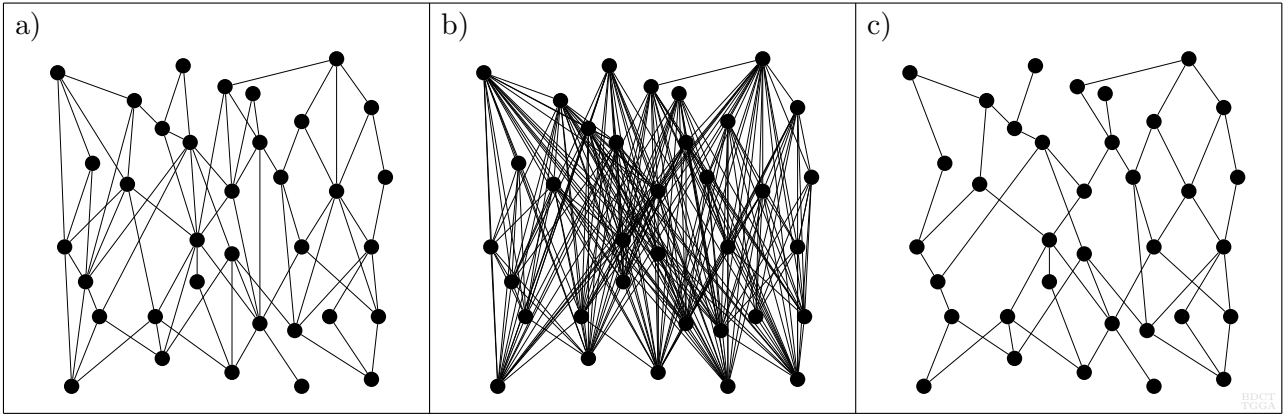


\caption{a) Directed set $D$; b) transitive closure $\tn{tr}(D)$ of $D$; c) skeleton $\tn{sk}(D)$ of $D$.}
\label{transitiveskel}
\end{figure}
\vspace*{-.5cm}


\refstepcounter{textlabels}\label{causordtranspreorder}

Definition \hyperref[defipreorderorder]{\ref{defipreorderorder}} below gives the precise relationship between the causal preorder on a directed set and the corresponding causal order.  I emphasize again that the term {\it order} is used here in a generalized sense.  In particular, the transitive closure of a causal preorder including cycles also includes cycles, and is therefore not an order in the usual sense. \\

\begin{defi}\label{defipreorderorder} Let $D$ be a directed set, viewed as a model of causal structure, with causal preorder $\prec$. Then the binary relation $\prec_{\text{\tn{\fsz{tr}}}}$ on the transitive closure $(D,\prec_{\text{\tn{\fsz{tr}}}})$ is called the {\bf causal order} on $D$.  
\end{defi}


\refstepcounter{textlabels}\label{degenconfig}

{\bf Degeneracy.}  In general, many different directed sets share the same transitive closure and skeleton, up to isomorphism.  For example, the transitive closure and skeleton of the directed set $D$ illustrated in figure \hyperref[transitiveskel]{\ref{transitiveskel}}a above differ by about $200$ reducible relations.  Adding distinct families of these relations to $\tn{sk}(D)$ produces distinct directed sets; hence, there are roughly $2^{200}$ {\it information-theoretically distinct} directed sets sharing $\tn{tr}(D)$ and $\tn{sk}(D)$ as their transitive closures and skeletons, respectively.  The {\bf transitive degeneracy class} of a directed set $D$ is the set of isomorphism classes of directed sets with transitive closure isomorphic to $\tn{tr}(D)$, and the {\bf skeletal degeneracy class} of $D$ is the set of isomorphism classes of directed sets with skeleton isomorphic to $\tn{sk}(D)$.   The two classes coincide for {\it finite acyclic directed sets,} since the transitive closure of such a set may be recovered from its skeleton, and vice versa.  The {\bf transitive} and {\bf skeletal degeneracies} of $D$ are the cardinalities of the corresponding degeneracy classes.   

Degeneracy plays an important role in the structure of configuration spaces of directed sets.  In particular, degeneracy distinguishes configuration spaces of acyclic directed sets from the corresponding configuration spaces of partial orders, whose elements correspond to transitive degeneracy classes of acyclic directed sets.   This clarifies the source of the {\it Kleitman-Rothschild pathology} for finite partial orders, discussed in section \hyperref[subsectiontransitivitydeficient]{\ref{subsectiontransitivitydeficient}} above, in which ``near-antichains" dominate asymptotically: enumeration of partial orders counts only {\it one} member of each degeneracy class.  Referring again to the directed set illustrated in figure \hyperref[transitiveskel]{\ref{transitiveskel}}a, a Kleitman-Rothschild-type order of the same cardinality has roughly $80$ reducible relations, and hence a transitive degeneracy of roughly $2^{80}$, about $120$ powers of $2$ less than the ``more physically realistic" directed set in the figure.  This illustrates how enumeration of partial orders {\it relatively overcounts} ``near-antichains" by many of orders of magnitude, even for small cardinalities.  


\refstepcounter{textlabels}\label{transfunctorial}

{\bf Functorial Properties.}  The transitive closure and the skeleton are {\it idempotent} operations on the category $\ms{D}$ of directed sets.  This means that repeated application of these operations produces nothing new: $\tn{tr}\big(\tn{tr}(D)\big)=\tn{tr}(D)$ and $\tn{sk}\big(\tn{sk}(D)\big)=\tn{sk}(D)$.  In the finite acyclic directed case, the two operations are ``roughly inverse" to each other, but this is not true in general; for example, the skeleton of the rational numbers $\QQ$, with its usual order, has {\it no relations}, since every relation in $\QQ$ is reducible.  Similarly, the skeleton of a reflexive relation is a singleton with no relations, since a reflexive relation is reduced by any chain ``going around it multiple times."  Taking morphisms into account, the transitive closure extends to a {\it functor} from the category $\ms{D}$ of directed sets to its subcategory $\ms{D}_{\tn{\fsz TR}}$ of transitive directed sets, essentially because chains map to chains under morphisms.  There is an inclusion functor $\ms{D}_{\tn{\fsz TR}}\rightarrow \ms{D}$ in the opposite direction, and the two functors are {\it adjoint.} This means that for any directed set $(D,\prec)$, and any transitive directed set $(D',\prec')$, there is a natural identification of morphism sets
\refstepcounter{textlabels}\label{transadj}
\refstepcounter{textlabels}\label{skelnotfunctor}
\[\tn{Mor}_{\ms{D}}\big(D,D'\big)\cong \tn{Mor}_{\ms{D}_{\tn{\fsz TR}}}\big(\tn{tr}(D),D'\big),\]
given by identifying morphisms with the same underlying set maps.  The skeleton operation does {\it not} define a functor, since irreducible relations do not always map to irreducible relations under morphisms.   For example, the inclusion $\ZZ\rightarrow\QQ$, with the usual orders, is a morphism in $\ms{D}$, but does not induce a morphism $\tn{sk}(\ZZ)\rightarrow\tn{sk}(\QQ)$.

\newpage

\section{Interval Finiteness versus Local Finiteness}\label{sectioninterval}

Interval finiteness (\hyperref[if]{IF}), called {\it local finiteness} in the causal set literature, is described by Rafael Sorkin  as {\it``a formal way of saying that a} [causal set] {\it is discrete"} (\cite{SorkinCausalSetsDiscreteGravity05}, page 309). The discrete {\it topology} is useless in causal theory, since it ignores causal structure, but order-theoretic discreteness and measure-theoretic discreteness are central to the emergence of geometry under Sorkin's version of the causal metric hypothesis (\hyperref[cmh]{CMH}).  Local finiteness of some form is an obvious requirement in this setting, at least if one wishes to avoid manifestly nonphysical scenarios, such as the instantaneous emergence of an infinite volume of spacetime from a single event.  Unfortunately, interval finiteness permits this very scenario.  The need for a better local finiteness condition, and for better treatment of local conditions in general, brings topology back into the picture.  Many different topological spaces may be associated with a directed or multidirected set, with each topology inducing its own local conditions.  The suitability of these conditions, and of the topologies that induce them, must be judged on {\it physical} grounds.  In particular, for directed sets, viewed as models of classical spacetime under the classical causal metric hypothesis (\hyperref[ccmh]{CCMH}), {\it local} should mean {\it causally local} in a suitable sense.   Interval finiteness is not even close to being a causally local condition, and related topological conditions, such as {\it topological local finiteness in the interval topology,} are little better.   As an alternative, I propose the axiom of {\it local finiteness} (\hyperref[lf]{LF}).  Besides being causally local, this axiom admits a topological description that facilitates a better and more comprehensive understanding of discrete causal structure.  

In section \hyperref[subsectiontopology]{\ref{subsectiontopology}} below, I discuss local conditions and topology in general terms.   I define {\it causal locality} for physical systems, in essential agreement with Finkelstein's \cite{Finkelstein88} and Raptis' \cite{RaptisAlgebraicQuantization00} notions of locality.  I briefly review standard topological material.  After this, I discuss an important application of the principle of {\it hidden structure} (\hyperref[hs]{HS}), introduced in section \hyperref[settheoretic]{\ref{settheoretic}} above, in which multidirected sets induce {\it extra elements} in associated topological spaces.  I then discuss four different topological spaces associated with a multidirected set $M=(M,R,i,t)$.  The most useful of these I call the {\it star model} of $M$; its underlying set is the union of $M$ and its relation set $R$, and its topology is called the {\it star topology.}  In section \hyperref[subsectionintervalfiniteness]{\ref{subsectionintervalfiniteness}}, I discuss the shortcomings of interval finiteness, and propose local finiteness as an alternative.  I prove that local finiteness coincides with {\it topological local finiteness for the star topology.}  I show that interval finiteness and local finiteness are {\it incomparable conditions;} i.e., neither implies the other.   In section \hyperref[relativeacyclicdirected]{\ref{relativeacyclicdirected}}, I present the theory of {\it relative multidirected sets over a fixed base,} a generalized order-theoretic application of Grothendieck's {\it relative viewpoint} (\hyperref[rv]{RV}). This leads to reinterpretation of interval finite acyclic directed sets, including causal sets, as relative directed sets over the integers. The alternative axiom of local finiteness identifies a much richer class of relative directed sets.   In section \hyperref[subsectionintervalfinitenessdeficient]{\ref{subsectionintervalfinitenessdeficient}}, I present eight arguments against interval finiteness as a local finiteness condition for multidirected sets, focusing on the case of directed sets in the context of classical spacetime structure.   In section \hyperref[subsectionarglocfin]{\ref{subsectionarglocfin}}, I present six arguments in favor of local finiteness as a suitable alternative.  Finally, in section \hyperref[subsectionhierarchyfiniteness]{\ref{subsectionhierarchyfiniteness}}, I introduce a few other finiteness conditions, and prove some results on the hierarchy of finiteness conditions in causal theory. 


\subsection{Local Conditions; Topology}\label{subsectiontopology}

\refstepcounter{textlabels}\label{localcond}

{\bf Local Conditions.} A condition on a physical system or mathematical structure is called {\bf local} if it may be checked by examining {\it arbitrarily small neighborhoods of events or elements.}  In general mathematical settings, the notion of a small neighborhood is usually made precise in terms of a {\it topology,} as described below.  In classical continuum physics, including general relativity, neighborhoods may be defined in {\it metric} terms; i.e., in terms of ``distances" in space or spacetime.  Under the causal metric hypothesis (\hyperref[cmh]{CMH}), the roles are reversed: local conditions are more fundamental than metric notions in this context.  This requires a suitable definition of locality for multidirected sets.  


\newpage
\refstepcounter{textlabels}\label{relnotionstop}

{\bf Relativistic Notions.} Before proceeding, I briefly review a few relativistic notions for contextual purposes.  The word {\it metric} has a different meaning in differential geometry, and hence in general relativity, than in the abstract mathematical setting of a {\it metric space.}  In the latter setting, a metric is just a {\it global distance function} on a set $S$.  Such a metric induces a {\it metric topology} on $S$.  For example, the {\it Euclidean metric}, which generalizes the Pythagorean theorem, induces the {\it Euclidean metric topology} on  $\RR^n$, which in turn induces a natural topology on any $n$-dimensional continuum manifold $X$, via its coordinate charts.  However, $X$ generally does not inherit any preferred notion of conformal structure or scale from $\RR^n$. A {\it pseudo-Riemannian metric} $g$ on $X$, meanwhile, is a special additional structure on a {\it smooth} continuum manifold $X$, assigning to the tangent spaces of $X$ smoothly varying inner products.  The metric $g$ endows $X$ with specific conformal and scale data.  One {\it may} express topological properties of $X$ in terms of $g$, but it is generally {\it not} possible to express metric or conformal structure in topological terms.  Results such as the theorems of Hawking \cite{Hawking76} and Malament \cite{Malament77}, cited in section \hyperref[subsectionapproach]{\ref{subsectionapproach}} above, give special conditions under which this may be done; modern results such as those of Martin and Panangaden \cite{Martin06}, dispense with the smoothness assumption.  Other useful topologies may also be defined on pseudo-Riemannian manifolds.  One example is the {\it Alexandrov topology,} mentioned below in the context of the {\it interval topology} on a multidirected set. 


\refstepcounter{textlabels}\label{causallocality}

{\bf Causal Locality.} For directed sets, viewed as models of classical spacetime under the classical causal metric hypothesis (\hyperref[ccmh]{CCMH}), no pseudo-Riemannian metric exists at the fundamental level, and topological notions must be treated more abstractly.  Since the appropriate choice of topology depends on physical considerations in this context, I begin by specifying a {\it physical} notion of what it means for a condition to be local in a causal sense.\\

\begin{defi}\label{definitioncausallocality} A condition on a physical system is {\bf causally local} if it may be checked by examining all independent causes and effects associated with each event. 
\end{defi}

This definition accords with Finkelstein's \cite{Finkelstein88} and Raptis' \cite{RaptisAlgebraicQuantization00} notions of locality.  The same intuition is useful in the more general case of multidirected sets, although the physical interpretation is generally different in this context.  Use of the terms {\it physical system} and {\it event} in definition \hyperref[definitioncausallocality]{\ref{definitioncausallocality}} is intended to convey the idea that the physical notion of causally locality {\it precedes any specific choice of mathematical model.}  However, this definition is only applicable to models for which the independence convention (\hyperref[ic]{IC}) makes sense.\footnotemark\footnotetext{In the context of classical spacetime, this includes the general class of {\it discrete manifolds,} as described by Benincasa and Dowker \cite{DowkerScalarCurvature10}, following Riemann.}  For example, interpolative models such as domains require modified definitions. 


\refstepcounter{textlabels}\label{gentopology}

{\bf General Topology.} In mathematics, local conditions are formalized by the theory of {\it topology.}  A {\bf topology} $\ms{T}$ on a set $S$, with or without any additional structure, is a special family of subsets of $S$, which may be viewed as generalizations of open intervals on the real line.  Elements of $\ms{T}$ are called {\it open sets} of $S$.  An open set containing a given element $x$ of $S$ is called a {\it neighborhood} of $x$.  To qualify as a topology, $\ms{T}$ must satisfy the axioms that the ``total set" $S$ and the empty set $\oslash$ are open, that finite intersections of open sets are open, and that arbitrary unions of open sets are open.  A {\it basis} for $\ms{T}$ is a family of open sets such that every open set of $S$ is a union of sets in the basis.  A {\it subbasis} for $\ms{T}$ is a {\it generating family} of open sets for $\ms{T}$, meaning that any topology on $S$ containing the subbasis contains all of $\ms{T}$.  More concretely, this means that every open set of $S$, with the possible exceptions of $S$ and $\oslash$, is a union of finite intersections of sets in the subbasis.  A {\it topological space} is a set $S$ together with a topology $\ms{T}$ on $S$.  A morphism in the category of topological spaces is called a {\it continuous map}; an isomorphism is called a {\it homeomorphism}.  This paper does not make explicit use of topological morphisms, however.\footnotemark\footnotetext{The alternative notion of {\it Scott continuity} is often used in order theory, and particularly in domain theory.}

A set $S$ may be endowed with multiple different topologies, which may be compared in the following way: a topology $\ms{T}$ on $S$ is called {\it finer} than an alternative topology $\ms{T}'$ on $S$ if every element of $\ms{T}'$ contains an element of $\ms{T}$.  If the converse is true, then $\ms{T}$ is called {\it coarser} than $\ms{T}'$.  If neither condition is true, then $\ms{T}$ and $\ms{T}'$ are called {\it incomparable}.  Topologies most suitable for encoding useful local conditions are those that incorporate a careful balance between fineness and coarseness, with special respect for any auxiliary structure on $S$, such as a binary relation.  If a topology is too fine, then arbitrarily small neighborhoods of an element $x$ are generally too small to capture all the behavior one would like to classify as occurring {\it near} $x$.  If a topology is too coarse, it generally does not adequately {\it resolve} structure near $x$; i.e., every neighborhood of $x$ is so large that it contains behavior ``irrelevant to $x$" in some sense.  Among the topologies associated with multidirected sets discussed later in this section, the {\it discrete topology} is too fine, while the {\it interval topology} is too coarse. 


\refstepcounter{textlabels}\label{hiddentop}

{\bf Topological Hidden Structure.}  Not every interesting topological space that may be associated with a multidirected set $M=(M,R,i,t)$ involves a topology on its underlying set.  Also of interest are topologies on {\it enlarged sets} containing $M$, whose {\it extra elements} facilitate the analysis of $M$ itself.   The {\it continuum model} $M_\RR$ of $M$, and the {\it star model} $M_\star$ of $M$, both defined below, are examples of such enlarged sets.   The continuum model of $M$ contains an uncountable number of extra elements, drawn from an {\it extrinsic} source; namely, the unit continuum interval $[0,1]$.   This limits its relevance, since the resulting topological space is not actually {\it induced} by $M$.  By contrast, the extra elements appearing in the star model of $M$  correspond to the elements of the relation set $R$ of $M$, and are therefore an {\it intrinsic} part of the multidirected structure. These elements represent {\it topological hidden structure} associated with $M$.  

As mentioned in section \hyperref[settheoretic]{\ref{settheoretic}} above, the principle of hidden structure (\hyperref[hs]{HS}) is inspired by Alexander Grothendieck's approach to commutative algebraic geometry, in which an algebraic variety is augmented by the addition of extra elements corresponding to its closed subsets, thereby producing an algebraic scheme.    In the context of discrete causal theory, the star model $M_\star$ of a multidirected set $M$ represents a primitive example of an analogous, and equally useful, type of construction, in which $M$ is augmented by the addition of extra elements corresponding to its relations, chains etc.  Going in the opposite direction, it is sometimes useful to adjoin to $M$ extra elements ``smaller than ordinary elements."   This situation arises, for example, in the theory of {\it abstract element space,} introduced in section \hyperref[subsectionrelation]{\ref{subsectionrelation}} below.  Similar ideas appear in the theory of {\it causal atoms,} introduced in section \hyperref[subsectionpowerset]{\ref{subsectionpowerset}}.  Causal atoms may be viewed as ``generalized elements" of a multidirected set.  Ordinary elements are sometimes  ``smaller than any causal atom" in the cyclic case.


\refstepcounter{textlabels}\label{toplocfin}

{\bf Topological Local Finiteness.} I now turn to the specific topic of local finiteness conditions.  Each topology on a set $S$ is equipped with its own notion of local finiteness.  Since it is necessary, in the context of discrete causal theory, to consider the case of topologies on enlarged sets containing a multidirected set, I formalize topological local finiteness in the following generalized way:\\

\begin{defi}\label{topologicallocalfiniteness} Let $S$ and $\overline{S}$ be sets, with $S$ a subset of $\overline{S}$, and let $\overline{\ms{T}}$ be a topology on $\overline{S}$.  Then $S$ is called {\bf topologically locally finite} with respect to the pair $(\overline{S},\overline{\ms{T}})$ if every element of $S$ is contained in an element of $\overline{\ms{T}}$ of finite cardinality.
\end{defi}

Topological local finiteness may be informally reduced to the statement that {\it every element has a finite neighborhood.} However, the definition of ``neighborhood" depends on the choice of topology, and neighborhoods are generally subsets of the enlarged set $\overline{S}$, rather than subsets of $S$ itself.\footnotemark\footnotetext{More generally, one may define a local condition {\it at a particular element} $x$ in $S$, then say that $S$ satisfies the condition (globally) if it satisfies the condition at every element. For example, a generic spacetime in general relativity is smooth almost everywhere, but is {\it not} globally smooth because of the singularity theorems.}


\refstepcounter{textlabels}\label{fourtopspaces}

{\bf Four Illustrative Topologies.}  {\it Many} different topological spaces may be associated with a multidirected set $M=(M,R,i,t)$. Some involve topologies on the underlying set of $M$, while others involve topologies on an enlarged set containing the underlying set of $M$.   Here, I discuss just four illustrative examples of such spaces.  Their topologies are called the {\it discrete}, {\it interval}, {\it continuum}, and {\it star topologies}.\footnotemark\footnotetext{The interval and continuum topologies are each analogous to the Euclidean metric topology on the real line, but in different ways. The star topology is distinct from the identically named notion appearing in the study of computer networks, although the underlying idea is similar.}  The discrete topology and the interval topology are topologies on the underlying set of $M$, while the continuum topology and the star topology are topologies on enlarged sets $M_\RR$ and $M_\star$ containing the underlying set of $M$, called the {\it continuum model} and the {\it star model} of $M$, respectively.  

In topological terms, the goals of this section are modest, and indeed primitive, compared to what appears in the graph-theoretic and order-theoretic literature.  The four topological spaces discussed here are chosen specifically to facilitate and clarify the study of local finiteness in causal theory.  The star topology on the star model of a multidirected set $M$ is the most useful of these four spaces for defining causally local properties in the context of classical spacetime structure.  In this setting, {\it topological local finiteness} for the star topology coincides with both {\it causal local finiteness} in the sense of definition \hyperref[definitioncausallocality]{\ref{definitioncausallocality}} above, as shown in lemma \hyperref[localfinitenesslemma]{\ref{localfinitenesslemma}} below, and the axiom of {\it local finiteness} (\hyperref[lf]{LF}), introduced in section \hyperref[subsectionintervalfiniteness]{\ref{subsectionintervalfiniteness}} below, which is {\it chosen} to generalize classical causal local finiteness.   For a more detailed topological perspective in the special case of causal sets, the sources cited in Sumati Surya's review article \cite{Surya11} provide an excellent starting point.   From a broader perspective, it is instructive to study the various topologies arising in domain theory, such as those discussed by Martin and Panangaden in \cite{Martin06}.    

Figure \hyperref[topologies]{\ref{topologies}} below illustrates small neighborhoods of an element $x$ in a multidirected set $M$, for each of the four topologies listed above.  The unusual appearance of the diagrams in figures \hyperref[topologies]{\ref{topologies}}c and \hyperref[topologies]{\ref{topologies}}d is due to the fact that these diagrams represent the enlarged sets $M_\RR$ and $M_\star$; the precise meaning of these diagrams is explained below.   As usual, these figures only involve acyclic directed sets, but the associated definitions apply to multidirected sets in general.  The reader should keep in mind that ``strange things can happen" when cycles are admitted; for example, an open interval of the form $\llangle x,x \rrangle$ may be nonempty in this context. 

\begin{figure}[H]
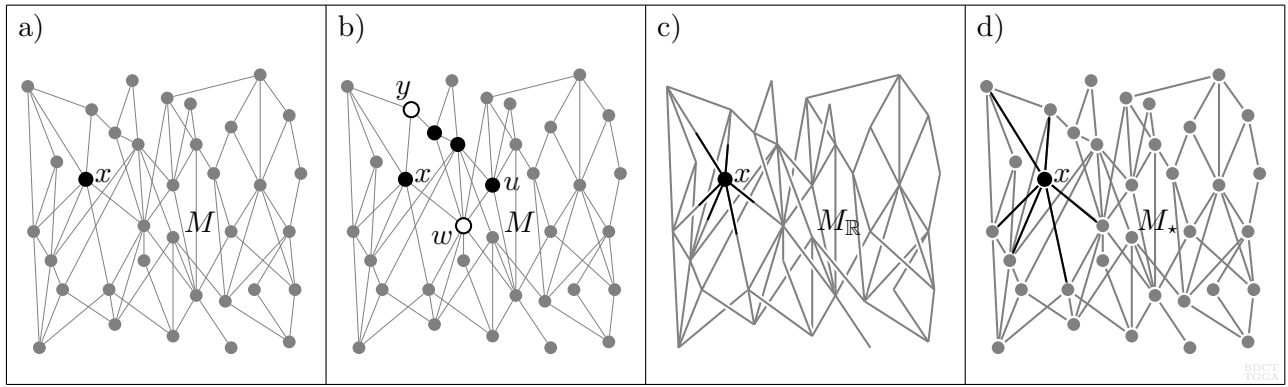


\caption{Small neighborhood of an element $x$ in various topologies: a) discrete topology on $M$; b) interval topology on $M$; c) continuum topology on $M_\RR$; d) star topology on $M_\star$.}
\label{topologies}
\end{figure}
\vspace*{-.5cm}

\refstepcounter{textlabels}\label{discretetop}

{\bf Discrete Topology.} The discrete topology $\ms{T}_{\tn{\fsz{dis}}}$ on a multidirected set $M$ takes the individual elements of $M$ to be open sets. Since arbitrary unions of open sets are open, {\it every} subset of $M$ is open in the discrete topology.  The discrete topology is too fine to be useful in causal theory.  Indeed, as illustrated in figure \hyperref[topologies]{\ref{topologies}}a, every element $x$ of $M$ defines a singleton neighborhood of itself.  Hence, one can decipher {\it nothing} about local multidirected structure near $x$ by examining arbitrarily small neighborhoods of $x$ in the discrete topology on $M$. 


\refstepcounter{textlabels}\label{inttop}

{\bf Interval Topology.} The interval topology $\ms{T}_{\tn{\fsz{int}}}$ on a multidirected set $M$ is perhaps the most natural topology to study in the special case when $M$ is a causal set.  In particular, it is the obvious topology to use for examining the local attributes of the axiom of interval finiteness (\hyperref[if]{IF}).  A convenient subbasis for the interval topology on $M$ is the family of pasts and futures $D^\pm(x)$ of elements $x$ of $M$.  An open interval $\llangle w,y\rrangle$, for two elements $w$ and $y$ in $M$, may be expressed as the intersection $D^+(w)\cap D^-(y)$ of subbasis elements.\footnotemark\footnotetext{One reason for defining the interval topology in this way, instead of using open intervals directly, is to avoid the nuisance of explicitly identifying ``half-open intervals," which include extremal elements, as members of the subbasis.}  As noted above, ``strange properties" arise in the cyclic case; for example, the open interval $\llangle x,x \rrangle$ is nonempty whenever $x$ lies on a cycle.  The reader should note that the term {\it interval topology} has many alternative meanings besides the one used here, even in the specific context of partially ordered sets.\footnotemark\footnotetext{Consider, for example, the ``interval topologies" discussed by Marcel Ern\'{e} in \cite{ErneTopologies80}.}  The interval topology is one possible causal analogue of the {\it Alexandrov topology} on a pseudo-Riemannian manifold $X$ in the context of general relativity,\footnotemark\footnotetext{This is the source of the term {\it ``Alexandroff set,"} appearing in the quote of Bombelli, Lee, Meyer and Sorkin in section \hyperref[subsectionaxioms]{\ref{subsectionaxioms}} above. The term {\it Alexandrov topology} has a slightly different meaning in modern mathematics, but the above usage is too deeply rooted to amend, appearing in classic relativity texts such as Hawking and Ellis \cite{HawkingEllis73}.}  which has subbasis consisting of the {\it chronological} pasts and futures $I^\pm(x)$ of spacetime events $x$ in $X$.  The Alexandrov topology is generally coarser than the manifold topology on $X$ induced by the Euclidean metric topology, but the two coincide for {\it strongly causal spacetimes,} which is one reason why the interval topology is {\it a priori} interesting in causal set theory.   Nontrivial choices are involved in abstracting the Alexandrov topology to the discrete causal context,\footnotemark\footnotetext{For example, the chronological pasts and futures $I^\pm(x)$ exclude the boundaries of the corresponding causal pasts and futures $J^\pm(x)$ in the relativistic context.  One way to abstract this condition to the discrete causal context would be to exclude $y$ from $D^{+}(x)$ whenever $x\prec y$ is an {\it irreducible} relation, and similarly for the past.} but these choices do not significantly affect the conclusions necessary for this paper.     

Although the interval topology respects multidirected structure in an obvious sense, it is nonetheless a poor choice for defining local properties of multidirected sets.  For example, in the context of classical spacetime, the interval topology on a directed set generally fails to capture and isolate local causal structure near a given event.  This is partly because the interval topology is too coarse, but it also has other deficiencies.  These are elaborated in section \hyperref[subsectionintervalfiniteness]{\ref{subsectionintervalfiniteness}}, and illustrated in the examples appearing in section \hyperref[subsectionintervalfinitenessdeficient]{\ref{subsectionintervalfinitenessdeficient}} below.  Figure \hyperref[topologies]{\ref{topologies}}b above illustrates a small neighborhood of an element $x$ in the interval topology on $M$, consisting of the four elements indicated by the dark nodes.   In this case, the chosen neighborhood is itself an open interval.  Observe that this neighborhood contains elements, such as $u$, that are {\it causally unrelated} to $x$.  This is a generic problem for the interval topology.  For future reference, note that {\it interval finiteness is not the same as topological local finiteness for the interval topology,} since interval finiteness requires that {\it every} open interval must be finite, not merely that arbitrarily small intersections of open intervals must be finite.  Also note that an open set in the interval topology does not always contain an open interval; i.e., open intervals do not form a basis for the interval topology.   Note, by contrast, that the Alexandrov sets $I^+(w)\cap I^-(y)$ {\it do} form a basis for the Alexandrov topology in the relativistic context. 


\refstepcounter{textlabels}\label{conttop}

{\bf Continuum Topology.} A ``concrete," though cumbersome, enlarged set containing a multidirected set $M=(M,R,i,t)$, may be constructed by identifying each element of the relation set $R$ of $M$ with a {\it continuum line segment,} then gluing these segments together at their endpoints in an appropriate way.  This approach leads to a topology, called the {\it continuum topology,} on the resulting quotient space, called the {\it continuum model} of $M$.  More precisely, define the {\bf relation set} $R(x)$ at an element $x$ in $M$ to be the subset of $R$ consisting of all relations $r$ such that either $i(r)=x$, or $t(r)=x$, or both.  Identify every element of $R$ with a copy of  the closed unit interval $[0,1]$ in $\RR$, equipped with its Euclidean metric topology.  Label the initial and terminal points of each interval with the initial and terminal elements of the corresponding relation.  Define the {\bf continuum model} $M_\RR$ of $M$ to be the quotient space given by identifying all endpoints of intervals with matching labels.  In particular, the set of intervals with at least one endpoint labeled by $x$ corresponds to $R(x)$.  The quotient topology $\ms{T}_\RR$ on $M_\RR$ is called the {\bf continuum topology}. 

Although the continuum topology involves a great deal of {\it extrinsic structure,} it has the advantage of providing natural families of neighborhoods that capture and isolate local multidirected structure near each element $x$ of $M$.  In particular, for every $0< \rho<1$, define the {\bf (open) star at $x$ of radius $\rho$} to be the {\it star-shaped subset} $\tn{St}_\rho(x)$ of $M_\RR$ with {\it limbs of length $\rho$} glued at $x$; i.e., the subset 
 \[\tn{St}_\rho(x):=\{t\in M_\RR\hspace*{.1cm}|\hspace*{.1cm} d(t,x)<\rho\},\]
 where $d$ is the Euclidean metric on each unit interval.  Any neighborhood $U$ of $x$ in the continuum topology on $M_\RR$ contains $\tn{St}_\rho(x)$ for sufficiently small values of $\rho$.   Hence, given such a neighborhood $U$, one may ascertain the cardinality of the relation set $R(x)$ at $x$, and hence the local multidirected structure near $x$, by examining an appropriate star $\tn{St}_\rho(x)$ contained in $U$.  Note that if $r$ is a reflexive relation $x\prec x$, and if $\rho\ge 1/2$, then $\tn{St}_\rho(x)$ contains the {\it entire interval corresponding to} $r$, so {\it``star-shaped"} is merely a suggestive term, carried over from the acyclic case.   Figure \hyperref[topologies]{\ref{topologies}}c above illustrates the star $\tn{St}_{1/2}(x)$ at an element $x$ in the continuum topology on $M_\RR$, with limbs represented by the dark segments.  The figure is deliberately drawn to suggest a {\it one-dimensional subspace of three-dimensional Euclidean space.} This  is intended to emphasize the difference between $M$ and $M_\RR$, particularly the large number of extra elements of $M_\RR$.  
 

\refstepcounter{textlabels}\label{startop}

{\bf Star Topology.} The best features of the continuum topology may be retained without the burden of extrinsic structure, as I now explain.  Define the {\bf star model} $M_\star$ of $M=(M,R,i,t)$ to be the union $M_\star:=M\cup R$ of (the underlying set of) $M$ with its relation set $R$. For each element $x$ in $M$, define the {\bf star at $x$ in $M_\star$} to be the subset
\[\tn{St}(x):=\{x\}\cup R(x),\]
of $M_\star$.  Define the {\bf star topology} $\ms{T}_\star$ on $M_\star$ to be the topology with subbasis $\{\tn{St}(x)\hspace*{.1cm}|\hspace*{.1cm} x\in M\}$.  It is useful to compare $\tn{St}(x)$ with the corresponding open stars $\tn{St}_\rho(x)$, defined above in the context of the continuum topology.  In particular, note that while there exist uncountably many open stars $\tn{St}_\rho(x)$ of different radii,  $\tn{St}(x)$ is unique.  Unlike the extra elements of $M_\RR$, which originate ultimately from the continuum interval $[0,1]$, an external source {\it a priori} unrelated to $M$, the extra elements $R$ of $M_\star$ are an intrinsic, uniquely-defined part of the structure of $M=(M,R,i,t)$.  Variations of the star topology may be defined in terms of {\it irreducible stars,} {\it past stars,} {\it future stars,} and so on.  

Figure \hyperref[topologies]{\ref{topologies}}d above illustrates the star $\tn{St}(x)$ at an element $x$ in the star topology on $M_\star$.  As in the case of the continuum topology, the appearance of this figure warrants some comment.  The elements of the relation set $R$ of $M=(M,R,i,t)$ now have a {\it dual role:} they serve both as {\it relations} between pairs of elements of $M$ in the usual sense, and as {\it elements} of $M_\star$ in their own right.  I have schematically highlighted this difference by drawing the edges in the ``Hasse diagram" for $M_\star$ slightly darker than the corresponding edges in figures  \hyperref[topologies]{\ref{topologies}}a and \hyperref[topologies]{\ref{topologies}}b, and by slightly separating the edges from the nodes representing the elements of $M$.  I represent $R$ in a different diagrammatic fashion in section \hyperref[subsectionrelation]{\ref{subsectionrelation}} below, where it serves as the underlying set of the {\it relation space} $\ms{R}(M)$ over $M$.  


\refstepcounter{textlabels}\label{cardvalscalar}

{\bf Cardinal-Valued Scalar Fields in the Star Topology.}  It is interesting to consider the question of how much information about a multidirected set $M=(M,R,i,t)$ may be recovered from the star topology on $M_\star$.   The star topology detects the {\it valences} $v(x)$ at each element $x$ of $M$; i.e., the number of relations incident at $x$.\footnotemark\footnotetext{Reflexive relations $x\prec x$ are counted twice.}  These valences form a {\bf cardinal-valued scalar field} $v_M$ on $M$, called the {\bf valence field} of $M$.  An arbitrary cardinal-valued scalar field on an arbitrary set $S$ is generally {\it not} a valence field for any multidirected structure on $S$.  For example, there is no multidirected structure on a singleton $S=\{x\}$ such that $v_S(x)=3$.\footnotemark\footnotetext{The inverse problem of recovering a ``multidirected set" from an arbitrary ``valence field" suggests consideration of  generalized multidirected sets in which the initial and terminal element maps are only partially defined.  This idea is actually somewhat interesting, due to an analogy involving the {\it causal path semicategories} studied in section \hyperref[subsectionpathspaces]{\ref{subsectionpathspaces}}.  However, I do not explore it in this paper.}  The valence field $v_M$ of a multidirected set $M$ may be decomposed into past and future components $v_{M}^\pm$, where $v_M^-(x)$ is the cardinality of the set of relations terminating at $x$, and $v_M^+(x)$ is the cardinality of the set of relations beginning at $x$.  Unlike the valence field itself, the past and future valence fields generally {\it cannot} be detected by the star topology.  For each relation $r$ in the relation set $R$ of $M$, one may consider linear combinations of $v_M(i(r))$ and $v_M(t(r))$ or of $v_M^\pm(i(r))$ and $v_M^\pm(t(r))$; these are analogues of {\it timelike vector fields} in an obvious sense.  Under the axiom of locally finiteness, introduced in section \hyperref[subsectionintervalfiniteness]{\ref{subsectionintervalfiniteness}} below, all such fields take nonnegative integer values, and their arithmetic properties are of significant interest. 


\subsection{Interval Finiteness versus Local Finiteness}\label{subsectionintervalfiniteness}

\refstepcounter{textlabels}\label{intcriticisms}

{\bf Initial Criticisms of Interval Finiteness.} The causal set axiom of interval finiteness (\hyperref[if]{IF}) is utterly inadequate as a local finiteness condition for classical spacetime structure under Sorkin's version of the causal metric hypothesis (\hyperref[cmh]{CMH}).  Physically, it permits the very type of pathological behavior an appropriate local finiteness condition should rule out.   It is not {\it causally local} in the sense of definition \hyperref[definitioncausallocality]{\ref{definitioncausallocality}} above, since the direct causes and effects associated with a given element generally cannot be isolated in an open interval.  Mathematically, it is not a local condition for the interval topology, since it involves arbitrarily large neighborhoods, nor is it a consequence of the topological local finiteness condition induced by the interval topology.  Worse, open intervals do not even {\it capture} local multidirected structure in any suitable sense, let alone isolate such structure. These shortcomings, and others, are discussed in more detail below. 


\refstepcounter{textlabels}\label{locfinintro}

{\bf Local Finiteness.}  Sorkin's version of the causal metric hypothesis (\hyperref[cmh]{CMH}) virtually demands {\it some} type of local finiteness condition, since the associated measure axiom (\hyperref[m]{M}) assigns an infinite spacetime volume to an infinite subset of a directed set.  In this context, local infinities produce absurd behavior of the worst sort from a physical perspective.  An example of such behavior occurs when an element $x$ in a directed set has an {\it infinite number of maximal predecessors or minimal successors,} corresponding to the instantaneous collapse or expansion of an infinite volume of spacetime.  Certain other types of ``locally infinite" behavior may be interpreted as artifacts of transitivity, but this particular type is a fatal threat to any theory permitting it.   One could try to formulate a different version of the causal metric hypothesis, in an effort to ameliorate this problem without imposing a local finiteness condition.  For example, one might consider altering the measure axiom to admit discrete measures assigning {\it arbitrarily small volumes} to subsets of a directed set; i.e., measures without an {\it effective volume gap.}  However, this strategy would lead far away from the basic insight that ``number" can serve as a proxy for volume in the context of Malament's metric recovery theorem.   Axiomatic changes of such significance should be made for physical reasons, not for the purpose of rescuing other axioms.\footnotemark\footnotetext{As discussed at the end of this section, there are legitimate reasons to consider {\it relaxing} the measure axiom, to allow volume to depend on local causal structure.  However, this requires a satisfactory treatment of local conditions, which open intervals cannot supply.} 

How, then, {\it should} local finiteness be axiomatized?  The crucial {\it physical} requirement is that local finiteness should coincide with {\it causal local finiteness} in the context of classical spacetime structure.   Modeling classical spacetime as a directed set $D$, and applying the independence convention (\hyperref[ic]{IC}), the definition of causal local finiteness translates to the condition that {\it the relation set $R(x)$ at each element $x$ of $D$ is finite.}  Since the same condition is useful for general multidirected sets, I state the axiom in this context.

\refstepcounter{textlabels}\label{lf}

\hspace*{.3cm} LF. {\bf Local Finiteness}: {\it For every element $x$ of $M$, the relation set $R(x)$ at $x$ is finite.}

In the special case of acyclic directed sets, the cardinality of $R(x)$ is the same as the number of {\it direct predecessors and successors} of $x$.  Hence, local finiteness may be stated in terms of {\it elements} in this case.  For directed sets including cycles, this correspondence is altered slightly, by the possibility of reflexive relations, and relations in both directions between a given pair of elements.  Finally, for multidirected sets, one may imagine situations, such as the existence of an {\it infinite number of reflexive relations at a singleton,}\footnotemark\footnotetext{This is actually not so strange; it is a causal analogue of the category-theoretic representation of an infinite group.} in which local finiteness cannot be determined by counting predecessors and successors.  For these reasons, it is best in the general case to express local finiteness in terms of relation sets.  Local finiteness may be detected by either the continuum topology or the star topology.  In fact, as mentioned above, local finiteness is the same condition as topological local finiteness for a multidirected set $M$ with respect to the star topology on the star model $M_\star$ of $M$.  This is the content of lemma \hyperref[localfinitenesslemma]{\ref{localfinitenesslemma}}. 
\vspace*{.2cm}

\refstepcounter{textlabels}\label{locfintoplocfinstar}

\begin{lem}\label{localfinitenesslemma} Local finiteness \tn{(\hyperref[lf]{\tn{LF}})} for a multidirected set $M=(M,R,i,t)$ coincides with topological local finiteness of $M$ with respect to the pair $(M_\star,\ms{T}_\star)$.  
\end{lem}
\begin{proof} Local finiteness means that the relation set $R(x)$ is finite for each element $x$ of $M$.  This is true if and only if the star $\tn{St}(x)$ at $x$ is finite for each $x$ in $M$.  Since $\tn{St}(x)$ is a neighborhood of $x$ in $M_\star$ with respect to $\ms{T}_\star$, this immediately implies topological local finiteness of $M$ with respect to $(M_\star,\ms{T}_\star)$.  Conversely, the intersection of two stars $\tn{St}(x)\cap\tn{St}(y)$ is either a family of relations between $x$ and $y$, or empty.  Hence, the smallest neighborhood of $x$ in the star topology on $M_\star$ is $\tn{St}(x)$.   Therefore, topological local finiteness of $M$ with respect to $(M_\star,\ms{T}_\star)$, implies that $\tn{St}(x)$, and hence $R(x)$, is finite for every $x$ in $M$.   
\end{proof}

Another equivalent condition is that the valence field $v_M$ of $M$ takes values in $\NN$.   


\refstepcounter{textlabels}\label{varlocfin}

{\bf Variations on Local Finiteness for Directed Sets.}  The axiom of local finiteness (\hyperref[lf]{LF}) may be modified in a number of different ways.  For example, in the general context of multidirected sets, one may first collapse all relations of the form $x\prec y$, for each pair of elements $x$ and $y$ in $M$, into a single relation, before testing for local finiteness.   This leads to the condition of {\it local element finiteness} (\hyperref[lef]{LEF}), introduced in section \hyperref[subsectionhierarchyfiniteness]{\ref{subsectionhierarchyfiniteness}} below.  In the special case of directed sets, it is sometimes useful to consider alternative finiteness conditions involving the addition or removal of {\it reducible relations.}  Though inferior in some ways to local finiteness for describing classical spacetime structure, such conditions have a sufficiently important auxiliary role to warrant mention.  The two extreme cases of such conditions admit convenient descriptions in terms of the transitive closure and the skeleton.  

\refstepcounter{textlabels}\label{lft}

\hspace*{.3cm} LFT. {\bf Local Finiteness in the Transitive Closure}: $\tn{tr}(D)$ is locally finite.

\refstepcounter{textlabels}\label{lfs}

\hspace*{.3cm} LFS.\hspace*{.05cm} {\bf Local Finiteness in the Skeleton}: $\tn{sk}(D)$ is locally finite.

Local finiteness in the transitive closure is not a causally local condition, since it requires the entire past and future of each element of $D$ to be finite.  This does {\it not} imply that $D$ itself has finite ``temporal size;" for example, a disjoint union of chains of each nonnegative integer length is locally finite in the transitive closure.  However, this condition is still too restrictive of global structure to serve as a suitable replacement for interval finiteness in the context of classical spacetime.  Local finiteness in the skeleton, meanwhile, is far too weak a condition to restrict attention to physically plausible directed sets under Sorkin's version of the causal metric hypothesis (\hyperref[cmh]{CMH}).   In particular, this condition permits any interpolative directed set, including the continuum, Minkowski spacetime, domains, etc., since the skeleta of such sets have no relations.  It also permits ``pathologically complex cycles" in ``otherwise simple" directed sets, since the skeleton operation destroys cycles.  Local finiteness in the skeleton is interesting primarily in concert with other axioms.  For example, if one insists on retaining interval finiteness, local finiteness in the skeleton cures some of its deficiencies. 


\refstepcounter{textlabels}\label{intfinlocfinincomp}

{\bf Incomparability of Interval Finiteness and Local Finiteness.} Interval finiteness (\hyperref[if]{IF}) and local finiteness (\hyperref[lf]{LF}) are incomparable conditions; i.e., neither condition implies the other.  This is true even in the special case of acyclic directed sets.  More comprehensive {\it pairwise} comparisons of various finiteness conditions, for both acyclic directed sets and multidirected sets, are given in theorems \hyperref[theoremhierarchyfiniteness]{\ref{theoremhierarchyfiniteness}} and \hyperref[theoremhierarchyfinitenessmulti]{\ref{theoremhierarchyfinitenessmulti}} of section \hyperref[subsectionhierarchyfiniteness]{\ref{subsectionhierarchyfiniteness}} below. In this section, I merely present a few illustrative counterexamples. 

\begin{figure}[H]
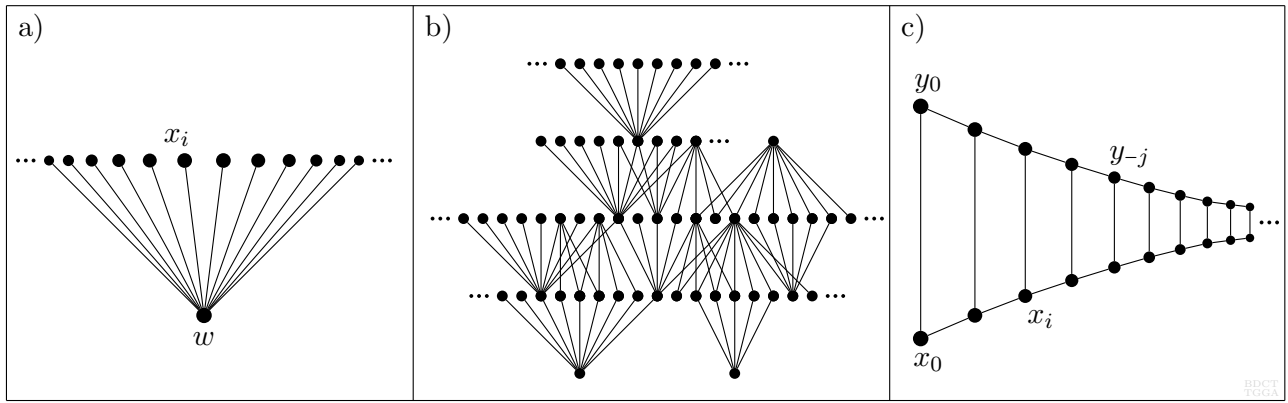


\caption{a) the infinite bouquet is interval finite but not locally finite, even in the skeleton; b) absurd interval-finite behavior; c) Jacob's ladder is locally finite but not interval finite.}
\label{jacobs}
\end{figure}

To see that interval finiteness does not imply local finiteness, consider the causal set with elements $\{w\}\cup\{x_i\hspace*{.1cm}|\hspace*{.1cm}i\in\ZZ\}$, and relations $w\prec x_i$ for all $i$, illustrated in figure \hyperref[jacobs]{\ref{jacobs}}a below.  For future reference, I call this causal set the {\it infinite bouquet}.  The infinite bouquet possesses no nonempty open intervals, since its longest chain has length one, and is therefore interval finite.  However, it is not locally finite, even in the skeleton, because the element $w$ has an infinite number of minimal successors.   The example of the infinite bouquet illustrates how interval finiteness permits the worst possible type of locally infinite behavior under Sorkin's version of the causal metric hypothesis (\hyperref[cmh]{CMH}).  This type of pathology is not limited to causal sets of small ``temporal size."   For example, figure \hyperref[jacobs]{\ref{jacobs}}b illustrates part of the skeleton of a physically absurd causal set with more than two generations. This example demonstrates how interval finiteness permits all manner of locally infinite behavior, restricted only by the condition that such behavior is not {\it complemented nonlocally} in such a way as to create infinite intervals.  As mentioned above, such behavior includes both instantaneous expansion, and instantaneous collapse, of an infinite volume of spacetime, under the measure axiom (\hyperref[m]{M}).  Dismissing implausible speculation about the Big Bang, the Big Crunch, eternal inflation, vacuum collapse, and other universal cataclysms, such scenarios have no reasonable place in discrete causal theory. 

To see that local finiteness does not imply interval finiteness, consider the acyclic directed set with elements $\{x_0, x_1,...\}\cup\{...,y_{-1},y_0\}$, and relations 
\[x_i\prec x_{i+1}, \hspace*{.5cm} x_i\prec y_{-i}, \hspace*{.5cm} y_{-j}\prec y_{-j+1},\]
illustrated in figure \hyperref[jacobs]{\ref{jacobs}}c.  I refer to this set as {\it Jacob's ladder}.\footnotemark\footnotetext{This is a reference to Genesis 28:12, in which the patriarch Jacob dreams of a ladder reaching from earth to heaven.}  Each element $x_i$, with the exception of $x_0$, has exactly one direct predecessor and two direct successors, and each element $y_{-j}$, with the exception of $y_0$, has exactly two direct predecessors and one direct successor.  The elements $x_0$ and $y_0$ are unique global minimal and maximal elements of Jacob's ladder, with two direct successors, and two direct predecessors, respectively.  Hence, Jacob's ladder is locally finite.  However, for every choice of nonnegative integers $i$ and $j$, there exist arbitrarily long chains of the form
\[x_i\prec x_{i+1}\prec...\prec x_{i+n}\prec y_{-i-n}\prec y_{-i-n+1}\prec ...\prec y_{-j}.\]
Thus, every open interval of the form $\llangle x_i,y_{-j}\rrangle$ in Jacob's ladder is infinite.  

\vspace*{-.5cm}


\refstepcounter{textlabels}\label{interactiontrans}

\begin{figure}[H]
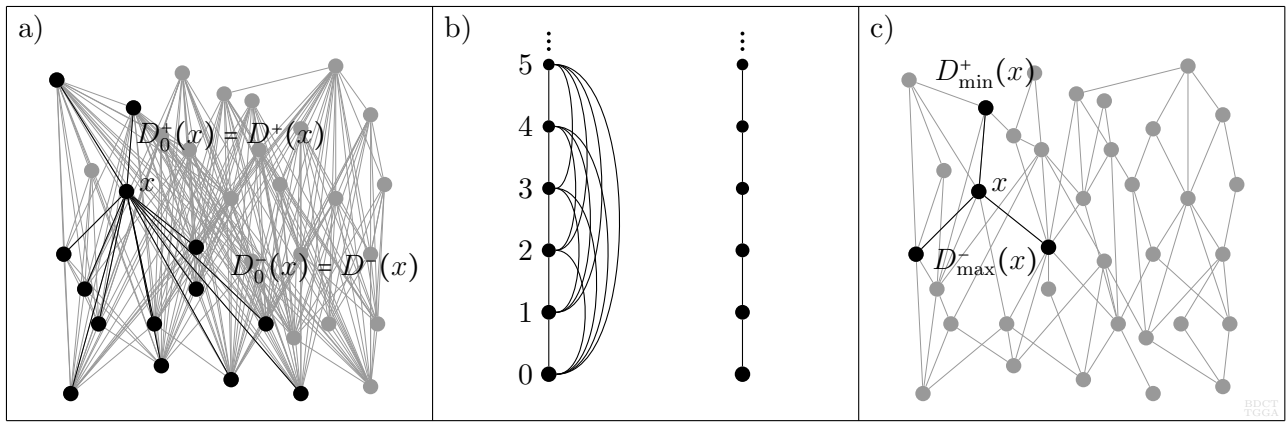


\caption{a) Transitivity obscures the advantages of local finiteness; b) $\NN$ is locally infinite, but locally finite in the skeleton; c) local finiteness in the skeleton is a better condition than interval finiteness in many cases.}
\label{interactiontransitivity}
\end{figure}

{\bf Interaction with Transitivity.} The foregoing criticisms of interval finiteness (\hyperref[if]{IF}) might lead the reader to question why this axiom was ever proposed in the context of discrete causal theory.  A {\it plausible} explanation is that it is an artifact of continuum theory, and the associated transitive paradigm, bolstered by the prominence of Alexandrov subsets and the Alexandrov topology in general relativity.   For a transitive directed set $D$, viewed as a model of classical spacetime structure, local finiteness (\hyperref[lf]{LF}) is not a reasonable alternative to interval finiteness, since the direct past $D_0^-(x)$ and direct future $D_0^+(x)$ of an element $x$ in $D$ coincide with its total past $D^-(x)$ and total future $D^+(x)$, respectively.   This situation is illustrated in figure \hyperref[interactiontransitivity]{\ref{interactiontransitivity}}a below.   In this context, the independence convention (\hyperref[ic]{IC}) does not make sense physically, since ``most" relations terminating at $x$ come from its ``distant past," and ``most" relations beginning at $x$ reach to its ``distant future."   Hence, the relation set $R(x)$ at $x$ is dominated by redundant information, and fails to isolate local causal structure in any suitable sense.  From a different point of view, the condition of local finiteness is generally {\it too restrictive} under the transitive paradigm, since transitivity creates ``innocuous" locally infinite behavior.  For example, the set $\NN$ of nonnegative integers is locally infinite at each element, but is locally finite in the skeleton, as illustrated in figure \hyperref[interactiontransitivity]{\ref{interactiontransitivity}}b.  Since $\NN$ may be recovered from its skeleton by taking the transitive closure, this locally infinite behavior is ``no worse," in an information-theoretic sense, than the simple structure of $\tn{sk}(\NN)$.  Any transitive cosmological model in discrete causal theory admitting unbounded chains is locally infinite for essentially the same reason, regardless of its actual information-theoretic complexity. 

\vspace*{-.5cm}

Even in the transitive case, however, better local finiteness conditions than interval finiteness are available.   For example, restricting attention to transitive directed sets recoverable from their skeleta,\footnotemark\footnotetext{This condition rules out objects like $\QQ$, which is locally finite in the skeleton because of the interpolative property.  It also rules out cycles, but may be easily amended to permit them.} local finiteness in the skeleton (\hyperref[lfs]{LFS}) is a better condition than interval finiteness, since it is causally local, and since it rules out pathologies such as the infinite bouquet.  Figure \hyperref[interactiontransitivity]{\ref{interactiontransitivity}}c above illustrates the set $D_{\tn{max}}^-(x)$ of maximal predecessors and the set $D_{\tn{min}}^+(x)$ of minimal successors of an element $x$ in a directed set $D$.  The cardinalities of such sets, as $x$ ranges over $D$, determine whether or not $D$ is locally finite in the skeleton.  The most obvious way to ensure that a directed set is recoverable from its skeleton is simply to impose the condition of {\it irreducibility;} i.e., to {\it begin} with ``skeletal objects."   In particular, this approach makes information-theoretic sense for causal sets, since reducible relations are already regarded as redundant in this context.  However, imposing irreducibility reduces the condition of local finiteness in the skeleton to local finiteness. Further, the independence convention (\hyperref[ic]{IC}) is automatic in the irreducible directed case.  Hence, one recovers the same basic conditions favored in the nontransitive case, only applied to a smaller class of objects; namely, the class of irreducible locally finite directed sets.  Despite these restrictions, this class is {\it still} much different, and vastly richer in general, than the class of causal sets.  


\newpage

\refstepcounter{textlabels}\label{interactionmeasure}

{\bf Interaction with the Measure Axiom.}  The clear picture of local structure provided by local finiteness (\hyperref[lf]{LF}) and the star topology suggests reconsideration of the measure axiom (\hyperref[m]{M}) in the context of classical spacetime structure.  Following this axiom, the volume measure $\mu$ on a causal set $C$ assigns ``approximately" one fundamental unit of spacetime volume to each element $x$ of $C$, {\it regardless of the details of local causal structure near $x$.}  While it would certainly be simplest, and most attractive, if ``order plus number" really {\it did} ``equal geometry," an {\it exact relationship} of this nature already seems to be in tension with known bounds on Lorentz invariance violation.  As discussed in section \hyperref[subsectionaxioms]{\ref{subsectionaxioms}} above, this is the reason for the caveat {\it ``up to Poisson-type fluctuations,"} appearing in the later causal set literature.  Sharpening the picture by dropping transitivity and adopting local finiteness raises the possibility that {\it relations,} as well as elements, might play some role in the emergent notion of volume.  Any particular proposal along these lines would require solid motivation, to avoid the conceptual trap of postulating increasing arbitrary and complex structures to ``rescue" a preconceived result.   However, it does seem reasonable, and even natural, to consider at least {\it local} causal structure in the interpretation of volume. 


\subsection{Relative Multdirected Sets over a Fixed Base}\label{relativeacyclicdirected}

\refstepcounter{textlabels}\label{infintmeaning}

{\bf Infinite Intervals.} What, if any, significance should be attached to the infinite intervals appearing in locally finite directed sets such as Jacob's ladder, and more generally, in multidirected sets?  Should such intervals be viewed as large-scale pathologies that one would prefer to avoid in a physical theory, or should they be taken seriously as structures of possible physical import?  In the context of classical spacetime structure, I know of no {\it positive physical evidence} for the latter view, but there seem to be good structural grounds for considering it.  The theory of {\it relative multidirected sets over a fixed base,} introduced in this section, provides an excellent perspective on why censoring infinite intervals seems arbitrary and unjustified.  The terminology comes from Grothendieck's {\it relative viewpoint} (\hyperref[rv]{RV}), introduced in section \hyperref[settheoretic]{\ref{settheoretic}} above.  


\refstepcounter{textlabels}\label{grothrelview}
\refstepcounter{textlabels}\label{causetrelz}

{\bf Causal Sets as Relative Directed Sets over the Integers.} Every nonempty causal set $C=(C,\prec)$ admits a bijective morphism into a linear suborder of the integers, either $\ZZ$ itself, the nonnegative integers $\NN$, the nonpositive integers $-\NN$, or a {\it finite simplex} $[n]:=\{0,1,...,n\}$ for some nonnegative integer $n$.   Since $\NN$, $-\NN$, and $[n]$ all embed into $\ZZ$, every causal set may be presented as a {\it relative directed set over} $\ZZ$; i.e., a directed set together with a distinguished morphism into $\ZZ$.  Generally, there are many such morphisms for a given causal set.  Such a morphism is equivalent to an {\it extension of the partial order $\prec$ on} $C$ to an order isomorphic to that of its linear image.  The existence of such an extension does {\it not} follow immediately from the order extension principle (\hyperref[oep]{OEP}), since a linear extension of an interval finite partial order is generally not interval finite.   However, such an extension may be explicitly constructed.   The proof I give here is essentially due to Joel David Hamkins \cite{Hamkins13}, though expressed in different language.  It uses a special case of a technique I refer to as {\it atomic accretion,} in which an order-theoretic analogue of a convex set, called a {\it causal atom} in the context of classical spacetime structure, is built up by a sequential process.   I revisit the theory of causal atoms in section \hyperref[subsectionpowerset]{\ref{subsectionpowerset}} below.  It seems that this simple extension result was first recognized in the context of pure order theory, in the work of Marcel Ern\'{e} circa 1979.\footnotemark\footnotetext{The earliest published references I can find are secondary remarks citing the unpublished lecture notes \cite{Erne79}, which Ern\'{e} was kind enough to send me copies of.}\\

\refstepcounter{textlabels}

\begin{theorem}\label{theoremcausalsetrelint} Any nonempty, countable, interval finite, acyclic directed set admits a morphism into the integers.  
\end{theorem}
\begin{proof} The proof has two steps.  The first step is an induction argument establishing that any nonempty, countable, interval finite, acyclic directed set $A=(A,\prec)$ admits a {\it sequential atomic accretion} $\{x_i\}_{i\in I}$, where $I$ is an interval in $\ZZ$ containing $0$.  This means that every finite, two-sided truncation $\{x_m,...,x_n\}$ of $\{x_i\}_{i\in I}$ defines a {\it causal atom}  in $A$; i.e., a subset of $A$ containing the intersection of its past and future.   The second step is merely the observation that such an atomic accretion defines a morphism into $\ZZ$.  

\begin{enumerate}

\item {\it $A$ admits a sequential atomic accretion.}  First note that $A$ admits an {\it enumeration} $A=\{y_0,y_1,...\}$ by hypothesis, since $A$ is countable.  A sequential atomic accretion $\{x_i\}_{i\in I}$ of $A$ may be constructed from this enumeration by induction.  By acyclicity, any singleton in $A$ is a causal atom, so set $x_0=y_0$.   If $A$ has only one element, the accretion process terminates at this stage, and the result is obvious.  Otherwise, one must proceed to consider the succeeding elements $y_1,y_2,...$ in the enumeration.

I explicitly describe the step involving $y_1$ for illustrative purposes.  If $x_0$ precedes $y_1$ in $(A,\prec)$, then $\llangle x_0,y_1\rrangle$ is a finite interval in $(A,\prec)$ by interval finiteness.   If this interval is empty, set $x_1=y_1$.  If not, choose $x_1$ to be any minimal element in $\llangle x_0,y_1\rrangle$.   In either case, the set $\{x_0,x_1\}$ is a causal atom in $A$.   If the interval $\llangle x_0,y_1\rrangle$ contains no other elements, set $x_2=y_1$.   Otherwise, choose $x_2$ to be a minimal element in $\llangle x_0,y_1\rrangle-\{x_1\}$.  In either case, the set $\{x_0,x_1,x_2\}$ is a causal atom in $A$.   Continuing in this fashion, the interval $\llangle x_0,y_1\rrangle$ is exhausted after a finite number of steps.   A symmetric argument applies if $x_0$ succeeds $y_1$; in this case, the elements in the interval $\llangle y_1,x_0\rrangle$ are chosen to be $x_{-1}$, $x_{-2}$, and so on.  If $y_1$ is unrelated to $x_0$, then $y_1$ may be chosen to be either $x_1$ or $x_{-1}$, since in this case the intersection of the past and future of $\{x_0,y_1\}$ is empty.   All these cases lead to an accretion process that terminates after a finite number of steps, yielding a causal atom containing $y_0$ and $y_1$.  

Now assume, by induction, that $\alpha:=\{x_m,...,x_n\}$ is a causal atom containing $y_0,...,y_k$, and consider $y_{k+1}$.  If $\alpha=A$, or if $y_{k+1}\in\alpha$, there is nothing to show.  Suppose that $y_{k+1}$ belongs to the future of $\alpha$, but not to $\alpha$ itself.  Since $\alpha$ is a causal atom, $y_{k+1}$ does not belong to the past of $\alpha$.  Consider the generalized open interval $\llangle \alpha, y_{k+1}\rrangle$ consisting of elements in the future of $\alpha$ and the past of $y_{k+1}$, but not belonging to $\alpha$ or equal to $y_{k+1}$.  If $\llangle \alpha, y_{k+1}\rrangle$ is empty, set $x_{n+1}=y_{k+1}$.  Otherwise, choose $x_{n+1}$ to be a minimal element in $\llangle \alpha,y_1\rrangle$.  In either case, $\alpha\cup\{x_{n+1}\}$ is a causal atom.   Continuing in this fashion, the interval $\llangle \alpha,y_1\rrangle$ is exhausted after a finite number of steps.   A symmetric argument applies if $y_{k+1}$ precedes $\alpha$. If $y_1$ is unrelated to any element of $\alpha$, then $y_1$ may be chosen to be either $x_{n+1}$ or $x_{m-1}$.   All these cases lead to an accretion process that terminates after a finite number of steps, yielding a causal atom containing $y_0,...,y_{k+1}$.   By induction, the accretion extends to all of $A$.  

\item {\it The set map $A\rightarrow\ZZ$ sending $x_i$ to $i$ is a morphism.}  At each step in the accretion process, the image $i\in\ZZ$ of the ``next element" $x_i$ either precedes or succeeds all previously-defined images.  The corresponding elements form a causal atom, so $x_i$ cannot precede some of these elements while succeeding others.  Hence, the accretion process automatically respects the binary relation $\prec$ on $A$.  

\end{enumerate}

\end{proof}

Every causal set is a nonempty, countable, interval finite, acyclic directed set, and therefore admits a morphism into $\ZZ$ by the theorem.  In fact, the proof of the theorem yields more: if $A$ is finite, with cardinality $n+1$, the construction described in the proof yields a family of bijective morphisms into $[n]$, determined by the choice of enumeration and the choices of maximal or minimal elements in generalized open intervals.  If $A$ is infinite, bounded below,\footnotemark\footnotetext{There are many different notions of boundedness for acyclic directed sets, both local and global.  In this case, {\it bounded below} means that every chain in $A$ has a minimal element.  This is a relatively ``weak" boundedness condition.} and has a finite number of minimal elements, then the construction yields, after appropriate ``shifts," a family of bijective morphisms into $\NN$.   If $A$ is infinite, bounded above, and has a finite number of maximal elements, then the construction yields a family of bijective morphisms into $-\NN$.  If $A$ is infinite and unbounded above and below, the construction yields a family of bijective morphisms into $\ZZ$.  The remaining cases may involve bijective morphisms into either $\NN,-\NN,$ or $\ZZ$, depending on the details.  For example, a countably infinite set with no relations admits bijective morphisms into all three of these targets. 

\refstepcounter{textlabels}\label{relconseq}

A directed set $(D,\prec)$, together with a {\it particular} morphism into $\ZZ$, bijective or otherwise, is called a {\bf relative directed set over} $\ZZ$.  Hence, every causal set may be viewed as a relative directed set over $\ZZ$, generally in many different ways.  Conversely, every transitive relative directed set $D$ of finite index\footnotemark\footnotetext{Here, {\it index} refers to the index of the morphism $D\rightarrow\ZZ$; i.e., the supremum of the cardinalities of its fibers.} over $\ZZ$ is a causal set, since every interval in $D$ is contained in the preimage of an interval in $\ZZ$, and since $(\ZZ,<)$ itself is irreflexive and interval finite.  The axioms of causal set theory may therefore be reduced from six to four, with countability (\hyperref[c]{C}), interval finiteness (\hyperref[if]{IF}), and irreflexivity (\hyperref[ir]{IR}) replaced by the existence of a morphism of finite index into $\ZZ$.  This axiomatic streamlining is one advantage of the theory of relative multidirected sets over a fixed base.  A more concrete application appears in Sorkin and Rideout's theory of sequential growth dynamics \cite{SorkinSequentialGrowthDynamics99}, in which a total labeling $\{x_0,x_1,x_2,...\}$ of a causal set $C$ exhibits $C$ as a relative directed set over $\ZZ$, via the inclusion $\NN\rightarrow\ZZ$. 

 

\refstepcounter{textlabels}\label{relacdirarb}

{\bf Relative Multidirected Sets over an Arbitrary Base.}   More generally, let $M''=(M'',R'',i'',t'')$ be a fixed multidirected set.  A {\bf relative multidirected set over} $M''$ is a multidirected set $M=(M,R,i,t)$, together with a morphism $M\rightarrow M''$.  The {\bf category of relative multidirected sets over} $M''$ is the category whose objects are multidirected sets over $M''$, and whose morphisms are commutative triangles of the form

\begin{pgfpicture}{0cm}{0cm}{15cm}{2cm}
\begin{pgfmagnify}{1.03}{1.03}
\pgfputat{\pgfxy(8.5, .25)}{\pgfbox[center,center]{$M''$}}
\pgfputat{\pgfxy(7.5, 1.75)}{\pgfbox[center,center]{$M$}}
\pgfputat{\pgfxy(9.5, 1.75)}{\pgfbox[center,center]{$M'$}}
\pgfsetendarrow{\pgfarrowlargepointed{3pt}}
\pgfxyline(7.9,1.75)(9.1,1.75)
\pgfxyline(7.6,1.4)(8.2,.5)
\pgfxyline(9.4,1.4)(8.8,.5)
\end{pgfmagnify}\end{pgfpicture}

where the vertical arrows are the distinguished morphisms defining $M$ and $M'$ as multidirected sets over $M''$.   In this context, $M''$ is called the {\bf base}, while $M$ and $M'$ are called {\bf sources}.  Of particular interest in the study of classical spacetime structure is the case in which $M$, $M'$ and $M''$ are three directed sets $(D,\prec)$, $(D',\prec')$, and $(D'',\prec'')$, respectively, where the underlying sets coincide, where the morphisms are the identity set maps, and where the binary relations $\prec'$ and $\prec''$ are successive extensions of $\prec$.  In this case, the morphisms represent {\it generalized frames of reference,} as mentioned in section \hyperref[subsectionapproach]{\ref{subsectionapproach}} above. 


\refstepcounter{textlabels}\label{democracy}

{\bf Democracy of Bases.}  Consider now the following question, which bears directly on the plausibility of the axiom of interval finiteness (\hyperref[if]{IF}) in discrete causal theory:

\hspace*{.3cm}{\bf Question: }{\it Why should the causal structure of classical spacetime be modeled exclusively in terms   
\\\hspace*{2.25cm} of relative directed sets over $\ZZ$?}

\begin{figure}[H]
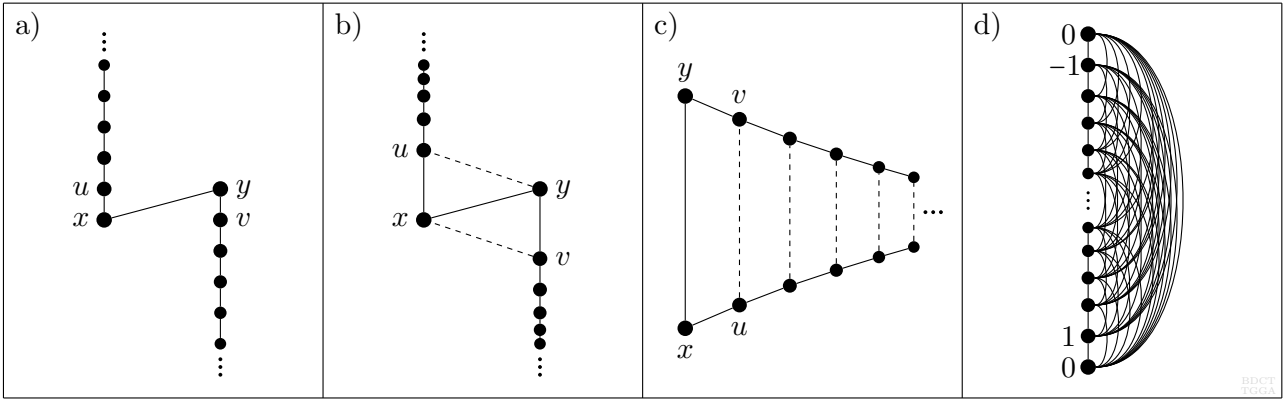


\caption{a) Skeleton of a causal set $(C,\prec)$; b) adding relations $v\prec x$ and $y\prec u$, and taking the transitive closure, defines a bijective morphism $C\rightarrow\ZZ$; c) transitive closure of Jacob's ladder gives an alternative extension, which is not interval finite; d) this extension defines a bijective morphism $C\rightarrow\NN\coprod-\NN$.}
\label{alternativeextensions}
\end{figure}

Even causal sets, and in particular, connected\footnotemark\footnotetext{Connected, that is, in a graph-theoretic sense.  The point of highlighting this property is that the existence of bijective morphisms into ``strange" bases for sufficiently ``sparse" causal sets, such as a countable set with no relations, is pedestrian.} causal sets, generally admit bijective morphisms into linearly ordered bases {\it not contained in the integers,} and it is difficult to argue that such bases should be ignored.  For example, consider the causal set $C=(C,\prec)$ whose skeleton is depicted in figure \hyperref[alternativeextensions]{\ref{alternativeextensions}}a below.  $C$ consists of two infinite chains, one bounded below, with minimal element $x$; the other bounded above, with maximal element $y$; these chains are connected  by a relation $x\prec y$.  Approaching $C$ from the viewpoint of sequential growth, one wishes to provide a ``complete kinematic account of the evolution of $C$;" i.e., to exhibit $C$ bijectively as a relative directed set over a linearly ordered base.  The addition of relations $v\prec x$ and $y\prec u$, as shown in figure \hyperref[alternativeextensions]{\ref{alternativeextensions}}b, uniquely determines a linear order on $C$, since every pair of elements is now connected by a chain.  This linear order defines a bijective morphism into the integers.  However, an alternative linear order on $C$, extending $\prec$, may be defined by taking the transitive closure of the acyclic directed set illustrated in figure  \hyperref[alternativeextensions]{\ref{alternativeextensions}}c, which I introduced as {\it Jacob's ladder} in section \hyperref[subsectionintervalfiniteness]{\ref{subsectionintervalfiniteness}} above.  This linear order defines a bijective morphism from $C$ into a linearly ordered acyclic directed set, depicted in figure \hyperref[alternativeextensions]{\ref{alternativeextensions}}d, which is {\it not interval finite.} Rather, it is isomorphic to $\NN\coprod-\NN$, where every element of $\NN$ is taken to precede every element of $-\NN$.  

\vspace*{-.5cm}

Both of the linearly ordered sets $\ZZ$ and  $\NN\coprod-\NN$ are examples of {\it discrete linearly ordered sets,} already mentioned in section \hyperref[subsectionaxioms]{\ref{subsectionaxioms}} above, in the context of the partial order formulation of causal set theory.  To review, a discrete linearly ordered set is a linearly ordered set in which every nonextremal element has a maximal predecessor and minimal successor.  Hence, except at extremal elements, every discrete linearly ordered set is {\it locally isomorphic} to $\ZZ$, in either the interval topology, or the irreducible star topology.   Discrete linearly ordered sets may be viewed as ``one-dimensional manifolds with boundary over the integers."  The fact that discrete linearly ordered sets are defined in terms of {\it local properties} makes them particularly attractive as bases for relative directed sets in the context of classical spacetime structure, since it is better not to {\it prescribe global structure.}  The relationship between $\ZZ$ and an arbitrary discrete linearly ordered set may be viewed as analogous to the relationship between $\RR^4$, in Newtonian mechanics, and pseudo-Riemannian manifolds, in general relativity, which are only {\it locally} isomorphic to $\RR^4$.  


\refstepcounter{textlabels}\label{infkleitroth}

{\bf Countable Analogue of the Kleitman-Rothschild Pathology.} Order theory provides notions of ``size" more nuanced than cardinality; namely, the isomorphism classes of linearly ordered sets.   In this sense, Cantor's first transfinite ordinal $\omega$ is ``larger" than $\NN$, and the discrete linearly ordered set $\NN\coprod-\NN$ is ``larger" still.   Restrictions on ``minimal linearly ordered  bases" of relative acyclic directed sets, imposed by finiteness conditions such as interval finiteness (\hyperref[if]{IF}) or local finiteness (\hyperref[lf]{LF}), may be viewed as restrictions on {\it temporal size} in this refined sense.   For example, theorem \hyperref[theoremcausalsetrelint]{\ref{theoremcausalsetrelint}} above demonstrates that interval finiteness limits acyclic directed sets to ``temporal size at most $\ZZ$," while local finiteness permits ``much larger temporal sizes."  This is another way of expressing how severely interval finiteness restricts the global structure of causal sets.  As mentioned in section \hyperref[subsectiontransitivitydeficient]{\ref{subsectiontransitivitydeficient}} above, large finite causal sets are known to be biased toward large spatial and small temporal size, with the three-generation Kleitman-Rothschild orders dominating asymptotically.  Restriction to temporal size at most $\ZZ$ creates an analogous bias in the case of countably infinite interval finite acyclic directed sets, since ``spatial size is unrestricted" in this case, except by countability.  Figure \hyperref[semiordinalspatial]{\ref{semiordinalspatial}} below illustrates the skeleton of a causal set of ``spatial size greater than $\ZZ$."  A better justification of this vague statement may be given in terms of the theory of {\it relation space,} introduced in section \hyperref[subsectionrelation]{\ref{subsectionrelation}} below.  In particular, the three connected subgraphs indicated by the braces in the figure may be viewed as  ``antichains of size $\ZZ$ in relation space."  

\begin{figure}[H]
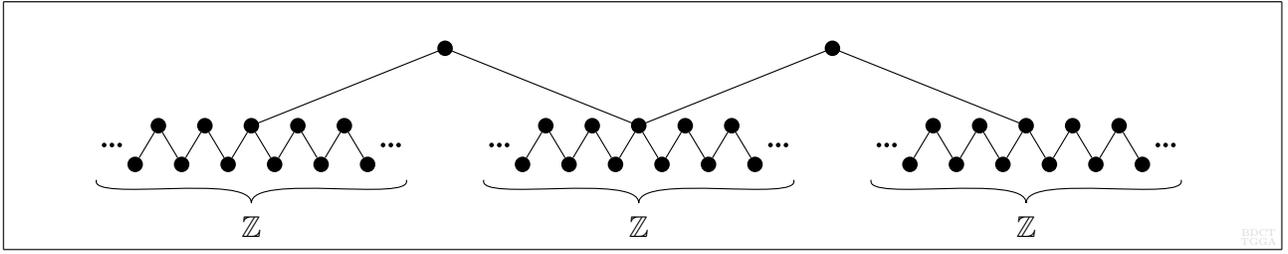


\caption{Skeleton of a causal set of ``spatial size greater than $\ZZ$."}
\label{semiordinalspatial}
\end{figure}
\vspace*{-.5cm}


\subsection{Eight Arguments against Interval Finiteness and Similar Conditions}\label{subsectionintervalfinitenessdeficient}

I have now assembled the necessary background to fully expose the shortcomings of interval finiteness (\hyperref[if]{IF}) as a local finiteness axiom for directed sets, viewed as models of classical spacetime structure under Sorkin's version of the causal metric hypothesis (\hyperref[cmh]{CMH}).  Some of these shortcomings have already been mentioned in previous sections.  The following arguments also show that interval finiteness cannot be repaired by working only with arbitrarily small open intervals, or intersections of open intervals.   In particular, topological local finiteness for the interval topology is also an unsuitable axiom.   The examples appearing below involve only acyclic directed sets.  This makes it clear that the arguments apply even under conservative assumptions. 

\refstepcounter{textlabels}\label{intfinnonloc}

\begin{figure}[H]
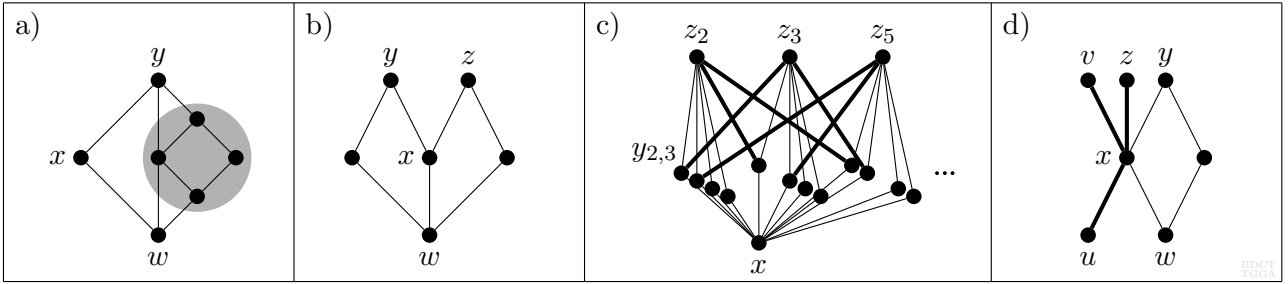


\caption{a) The only open interval containing $x$ also contains irrelevant directed structure; b) the singleton $\{x\}$ is open in the interval topology, but does not contain an open interval; c) every element lies in a finite open set in the interval topology, but not in a finite open interval; d) the open interval $\llangle w,y\rrangle$ contains $x$, but does not capture local directed structure near $x$.}
\label{intervaldeficiency}
\end{figure}

{\bf 1.} {\bf Interval finiteness is not a causally local condition in the context of classical spacetime structure.} Referring to definition \hyperref[definitioncausallocality]{\ref{definitioncausallocality}} above, this means that interval finiteness generally cannot be checked by examining the direct causes and effects associated with each element in a directed set.  For example, consider the element $x$ in the acyclic directed set illustrated in figure \hyperref[intervaldeficiency]{\ref{intervaldeficiency}}a below.  The only open interval containing $x$ is the interval $\llangle w,y\rrangle$, which also contains all of the directed structure highlighted by the grey circle, which is causally unrelated to $x$.   This is also the smallest open set containing $x$ in the interval topology.  

\refstepcounter{textlabels}\label{nonlocininttop}

{\bf 2.} {\bf Interval finiteness is not a local condition for the interval topology.}   In particular, interval finiteness is not the same as topological local finiteness for the interval topology, as already mentioned in section \hyperref[subsectiontopology]{\ref{subsectiontopology}} above.  This is intuitively obvious, since interval finiteness involves open intervals of arbitrary size.  More generally, {\it no} condition stated solely in terms of open intervals, even ``small" open intervals, is local in the interval topology, because open intervals do not form a basis for the topology: a neighborhood containing an element need not contain an {\it open interval} containing the element.  For example, consider the element $x$ in the acyclic directed set illustrated in figure \hyperref[intervaldeficiency]{\ref{intervaldeficiency}}b. The only nonempty open intervals in this set are $\llangle w,y\rrangle$ and $\llangle w,z\rrangle$, both of which contain $x$.   Their intersection is the singleton $\{x\}$, which is therefore an open set in the interval topology, but {\it not} an open interval.  Of course, interval finiteness is even worse in this regard, since it is defined in terms of {\it all} open intervals. 

\refstepcounter{textlabels}\label{toplocfinintnofinnbhd}

{\bf 3.} {\bf Topological local finiteness for the interval topology does not imply that each element is contained in a finite open interval.}  This illustrates the fact that interval finiteness is {\it not even close} to having a reasonable topological interpretation as a local finiteness condition.  It is already clear from figure \hyperref[intervaldeficiency]{\ref{intervaldeficiency}}b that an element in a multidirected set may have neighborhoods in the interval topology which are smaller than any nonempty open interval.  In the worst case scenario, a minimal nonempty open interval containing a given element may be {\it infinite} even though the element has a finite neighborhood in the interval topology!  For example, consider the acyclic directed set $A=(A,\prec)$ with elements
\[\{x\}\cup\{y_{p,q}\hspace*{.1cm}|\hspace*{.1cm}p,q \tn{ prime, } q\neq p\}\cup\{z_l\hspace*{.1cm}|\hspace*{.1cm}l \tn{ prime}\},\]
and relations
\[x\prec y_{p,q},\hspace*{.3cm} y_{p,q}\prec z_p,\hspace*{.3cm}\tn{and}\hspace*{.3cm}y_{p,q}\prec z_q.\]
A portion of $A$ is illustrated in figure \hyperref[intervaldeficiency]{\ref{intervaldeficiency}}c, with the elements $y_{p,q}$ grouped together for each $p$, and with the relations $y_{p,q}\prec z_q$ indicated by dark segments.  The only nonempty open intervals in $A$ are the infinite intervals $\llangle x,z_{l}\rrangle=\{y_{p,q}\hspace*{.1cm}|\hspace*{.1cm} p=l\tn{ or }q=l \}$ for each $l$.  The intersection of two such intervals $\llangle x,z_{l}\rrangle$ and $\llangle x,z_{l'}\rrangle$ is the two-point set $\{y_{l,l'},y_{l',l}\}$.
Meanwhile, the open set $D^+(y_{p,q})$ in $A$ is the two-point set $\{z_p\}\cup\{z_q\}$, and the open set $D^-(y_{p,q})$ is the singleton $\{x\}$ for all $p,q$.  Hence, every element of $A$ is contained in a finite open subset of $A$ in the interval topology, but {\it no element of $A$ lies in a finite open interval.}  

\vspace*{-.5cm}

\refstepcounter{textlabels}\label{intnotcapture}

{\bf 4.} {\bf Open intervals fail to even capture local multidirected structure, let alone isolate it.}  Heuristically, this means that open intervals are not only {\it imprecise,} but also {\it inaccurate.}  Despite being ``too large" to {\it resolve} local multidirected structure, open intervals generally do not even contain adequate information about local multidirected structure near each of  their elements.   For example, consider the element $x$ lying in the open interval $\llangle w,y \rrangle$ in the acyclic directed set illustrated in figure \hyperref[intervaldeficiency]{\ref{intervaldeficiency}}d above.  Three relations involving $x$; namely, $u\prec x$, $x\prec v$, and $x\prec z$, represented in the figure by dark segments, also involve elements outside of $\llangle w,y\rrangle$.  Hence, $\llangle w,y \rrangle$ contains relatively little local information relevant to $x$.   Open sets in the interval topology are even worse in general, since they are generally smaller than any open interval. 

\refstepcounter{textlabels}\label{intnotimplyloc}

{\bf 5.} {\bf Interval finiteness does not imply local finiteness.}  This was already demonstrated in section \hyperref[subsectionintervalfiniteness]{\ref{subsectionintervalfiniteness}} above, with the infinite bouquet (figure \hyperref[jacobs]{\ref{jacobs}}a) as a counterexample.  Since open intervals are generally ``too large" to resolve local multidirected structure, one might hope that requiring even such ``large" structures to be finite would at least constrain local causal structure, viewed as ``small," to be finite as well.   The fact that this is not true illustrates how badly open intervals fail to capture local causal structure.   Topological local finiteness in the interval topology is even worse in this regard, since it is a weaker condition.  

\refstepcounter{textlabels}\label{locnotimplyint}

{\bf 6.} {\bf Interval finiteness does not follow from local finiteness.} Again, this was demonstrated in section \hyperref[subsectionintervalfiniteness]{\ref{subsectionintervalfiniteness}} above, with Jacob's ladder (figure \hyperref[jacobs]{\ref{jacobs}}c) as a counterexample.  This fact is not surprising, since conditions stated in terms of arbitrary open intervals are not local even in the interval topology.  However, topological local finiteness in the interval topology does not follow from local finiteness either.  To see this, ``augment" Jacob's ladder by replacing the relation $x_0\prec y_0$ with a $2$-chain $x_0\prec z_{0}\prec y_{0}$.   Then the only open interval containing $z_0$ is the infinite interval $\llangle x_0,y_0\rrangle$.  

\refstepcounter{textlabels}\label{fatalsorkin}

{\bf 7.} {\bf Interval finiteness permits physically fatal local behavior under Sorkin's version of the causal metric hypothesis.}  In particular, it permits instantaneous collapse, or instantaneous expansion, of an infinite volume of spacetime, as explained in section \hyperref[subsectionintervalfiniteness]{\ref{subsectionintervalfiniteness}} above.  Mixing finite and infinite local behavior is dubious even without taking the measure axiom (\hyperref[m]{M}) into account, but is particularly problematic when the measure axiom is assumed to hold.   Topological local finiteness in the interval topology is equally problematic in this regard.  

\refstepcounter{textlabels}\label{unjustglobal}

{\bf 8.} {\bf Interval finiteness imposes unjustified restrictions on the global structure of classical spacetime.}  This was made precise for causal sets by theorem \hyperref[theoremcausalsetrelint]{\ref{theoremcausalsetrelint}} above, which demonstrates that every causal set may be exhibited as a relative directed set over the integers.  Nothing we know about cosmology justifies this restriction.  Topological local finiteness in the interval topology does not impose such a restriction, since it is at least a local condition.  However, the absence of this particular pathology is not enough to rescue this condition.


\subsection{Six Arguments for Local Finiteness}\label{subsectionarglocfin}

I now briefly review some of the principal arguments in favor of local finiteness as a replacement for interval finiteness (\hyperref[lf]{LF}).   These arguments have all appeared, explicitly or implicitly, in the foregoing analysis.

\refstepcounter{textlabels}\label{locfincausloc}
  
{\bf 1.} {\bf Local finiteness is a causally local condition in the context of classical spacetime structure.}  Referring to definition \hyperref[definitioncausallocality]{\ref{definitioncausallocality}} above, this means that local finiteness may be checked by examining the independent causes and effects associated with each element in a directed set $D$.   These causes and effects are represented by the relation sets $R(x)$ at each element $x$ in $D$, whose cardinality determines local finiteness. 

\refstepcounter{textlabels}\label{locfintoploc}

{\bf 2.} {\bf Local finiteness is a local condition in the topological sense.} More precisely, local finiteness coincides with the condition of topological local finiteness for the star topology on the star model $M_\star$ of a multidirected set $M$, as proven in lemma \hyperref[localfinitenesslemma]{\ref{localfinitenesslemma}} above. 

\refstepcounter{textlabels}\label{locfincaptures}

{\bf 3.} {\bf Local finiteness both captures and isolates local multidirected structure.}  This means that consideration of the entire relation set $R(x)$ at each element $x$ in a multidirected set $M$ is both necessary and sufficient to determine local finiteness.  This follows immediately from the definition of local finiteness. 

\refstepcounter{textlabels}\label{compatsorkin}

{\bf 4.} {\bf Local finiteness is compatible with Sorkin's version of the causal metric hypothesis.}  In particular, local finiteness allows a minimum finite spacetime volume to be assigned to each element in a directed set, without the danger of fatal local behavior.  Indeed, since the set $D_0^-(x)\cup D_0^+(x)$ of direct predecessors and successors of any element $x$ in {\it any} locally finite multidirected set $(M,\prec)$ is finite, this set has a finite volume for {\it any} measure $\mu:\ms{P}(M)\rightarrow\RR^+$. 

\refstepcounter{textlabels}\label{notunjustglobal}

{\bf 5.} {\bf Local finiteness does not impose unjustified restrictions on the global structure of classical spacetime.} For example, in the acyclic case, local finiteness admits relative directed sets over a wide variety of discrete linearly ordered sets, not only suborders of $\ZZ$.   

\refstepcounter{textlabels}\label{naturalnontrans}

{\bf 6.} {\bf Local finiteness is a natural condition for nontransitive relations under the independence convention.}  Under the independence convention (\hyperref[ic]{IC}), each element $r$ of the relation set $R(x)$ at an element $x$ of a multidirected set $M$ possesses independent information-theoretic significance.  In this context, local finiteness is determined precisely by the information ``directly relevant to $x$," as $x$ ranges over $M$.  


\subsection{Hierarchy of Finiteness Conditions}\label{subsectionhierarchyfiniteness}

To conclude this section, I briefly introduce a few more finiteness conditions for multidirected sets, and prove two elementary theorems relating these conditions for acyclic directed sets and multidirected sets, respectively.  The corresponding conditions for directed sets may be easily inferred from the second of these two theorems; namely, theorem \hyperref[theoremhierarchyfinitenessmulti]{\ref{theoremhierarchyfinitenessmulti}}. 

\newpage

Let $M=(M,R,i,t)$ be a multidirected set.  Consider the following finiteness conditions on $M$: 

\refstepcounter{textlabels}\label{eltfin}

\hspace*{.3cm} EF.\hspace*{.2cm} {\bf Element Finiteness:} {\it $M$ has finite cardinality.}\label{ef}

\refstepcounter{textlabels}\label{loceltfin}

\hspace*{.3cm} LEF. {\bf Local Element Finiteness:} {\it Every element of $M$ has a finite number of direct predecessors 
\\\hspace*{1.3cm} and successors.}\label{lef}

\refstepcounter{textlabels}\label{relfin}

\hspace*{.3cm} RF.\hspace*{.2cm} {\bf Relation Finiteness:} {\it $R$ has finite cardinality; i.e., $M$ has a finite number of \\
\hspace*{1.3cm} relations.}\label{rf}

\refstepcounter{textlabels}\label{parrelfin}

\hspace*{.3cm} PRF. {\bf Pairwise Relation Finiteness:} {\it Every pair of elements in $M$ has a finite number of 
\\\hspace*{1.3cm} relations between them.}\label{prf}

\refstepcounter{textlabels}\label{chainfin}

\hspace*{.3cm} CF.\hspace*{.2cm} {\bf Chain Finiteness:} {\it Every chain in $M$ has finite length.}\label{cf}

\refstepcounter{textlabels}\label{antichainfin}

\hspace*{.3cm} AF.\hspace*{.2cm} {\bf Antichain Finiteness:} {\it Every antichain in $M$ has finite cardinality.}\label{af}

\refstepcounter{textlabels}\label{discussfincond}

Local element finiteness coincides with local finiteness for directed sets, and pairwise relation finiteness follows directly from the definition in this case.  The remaining four conditions are all too restrictive to serve as axioms for directed sets as models of classical spacetime structure under the causal metric hypothesis (\hyperref[cmh]{CMH}).  Element finiteness is too restrictive because it prescribes a finite classical spacetime, and relation finiteness is too restrictive because it prescribes a classical spacetime with finite connected components.  Chain finiteness is too restrictive because it prescribes a classical spacetime that is {\it locally temporally finite,} in the sense that every sequence of events has a first cause and final effect.  Finally, antichain finiteness is too restrictive because it prescribes a classical spacetime that is spatially finite ``in any given frame of reference."   Each of these prescriptions rules out interesting cosmological models. \\

\begin{theorem}\label{theoremhierarchyfiniteness} For acyclic directed sets, the conditions of element finiteness \tn{(\hyperref[ef]{\tn{EF}})}, relation finiteness \tn{(\hyperref[rf]{\tn{RF}})}, local finiteness in the transitive closure \tn{(\hyperref[lft]{\tn{LFT}})}, local finiteness \tn{(\hyperref[lf]{\tn{LF}})}, local finiteness in the skeleton \tn{(\hyperref[lfs]{\tn{LFS}})}, chain finiteness \tn{(\hyperref[cf]{\tn{CF}})}, interval finiteness \tn{(\hyperref[if]{\tn{IF}})}, and antichain finiteness \tn{(\hyperref[af]{\tn{AF}})}, are related as follows, where arrows indicate logical implications:

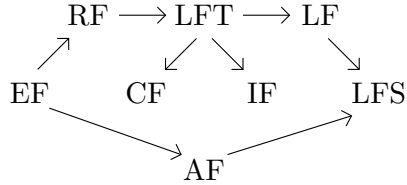
\begin{pgfpicture}{0cm}{0cm}{15cm}{2.3cm}
\begin{pgfmagnify}{1.03}{1.03}
\begin{pgftranslate}{\pgfpoint{1cm}{-.5cm}}
\pgfputat{\pgfxy(4.75,1.5)}{\pgfbox[center,center]{\tn{EF}}}
\pgfputat{\pgfxy(5.5,2.5)}{\pgfbox[center,center]{\tn{RF}}}
\pgfputat{\pgfxy(6.25,1.5)}{\pgfbox[center,center]{\tn{CF}}}
\pgfputat{\pgfxy(7,2.5)}{\pgfbox[center,center]{\tn{LFT}}}
\pgfputat{\pgfxy(8.5,2.5)}{\pgfbox[center,center]{\tn{LF}}}
\pgfputat{\pgfxy(9.25,1.5)}{\pgfbox[center,center]{\tn{LFS}}}
\pgfputat{\pgfxy(7,.5)}{\pgfbox[center,center]{\tn{AF}}}
\pgfputat{\pgfxy(7.75,1.5)}{\pgfbox[center,center]{\tn{IF}}}
\pgfsetendarrow{\pgfarrowlargepointed{3pt}}
\pgfxyline(4.85,1.8)(5.25,2.2)
\pgfxyline(6.9,2.2)(6.5,1.8)
\pgfxyline(5.9,2.5)(6.5,2.5)
\pgfxyline(7.5,2.5)(8.1,2.5)
\pgfxyline(8.6,2.2)(9,1.8)
\pgfxyline(7.1,2.2)(7.5,1.8)
\pgfxyline(5,1.3)(6.7,.7)
\pgfxyline(7.3,.7)(8.9,1.2)
\end{pgftranslate}
\begin{colormixin}{15!white}
\begin{pgfmagnify}{.6}{.6}
\pgfputat{\pgfxy(27,.2)}{\pgfbox[center,center]{\tiny{BDCT}}}
\pgfputat{\pgfxy(27,0)}{\pgfbox[center,center]{\tiny{TGGA}}}
\end{pgfmagnify}
\end{colormixin}
\end{pgfmagnify}\end{pgfpicture}

Moreover, these are the only pairwise logical implications among these conditions. 

\end{theorem}
\begin{proof} Let $A=(A,\prec)$ be an acyclic directed set.  I break the proof into two parts, first showing that the claimed logical  implications are valid, then showing that no other pairwise logical implications exist among these conditions.  Some of the statements proven here are more general than actually necessary to establish the theorem.   

\begin{enumerate}

\item {\it The claimed logical implications are valid.} 

EF $\Rightarrow$ RF: For {\it any} directed set, relations may be identified with pairs of elements; hence, the number of relations is bounded by the square of the number of elements.  

RF $\Rightarrow$ LFT:  Suppose that $A$ has $n$ relations.  The cardinality of the relation set $R(x)$ at any element $x$ in $\tn{tr}(A)$ is bounded by the total number of relations in $\tn{tr}(A)$, which is bounded by the total number of chains in $A$.  Since chains may be identified with families of relations in the acyclic case, this number is bounded by the cardinality of the power set of the set of relations of $A$, which is $2^{n}$. 

LFT $\Rightarrow$ CF: I prove the contrapositive.  Suppose that $A$ has an infinite chain $\gamma$, and let $x$ be an element of $A$ lying on $\gamma$.  Then $\tn{tr}(A)$ includes a relation between $x$ and every other element of $\gamma$, so the relation set at $x$ in $\tn{tr}(A)$ is infinite.

LFT $\Rightarrow$ IF: I prove the contrapositive.  Suppose that $A$ is not interval finite, and let $\llangle w,y \rrangle $ be an infinite open interval in $A$.  Every element in this interval is a direct successor of $w$ in $\tn{tr}(A)$, so $A$ is not locally finite in the transitive closure.  

LFT $\Rightarrow$ LF:  Every relation in $A$ corresponds to a relation in $\tn{tr}(A)$, so any element with a finite relation set in $\tn{tr}(A)$ also has a finite relation set in $A$.  

LF $\Rightarrow$ LFS:  Every relation in $\tn{sk}(A)$ corresponds to a relation in $A$, so any element with a finite relation set in $A$ also has a finite relation set in $\tn{sk}(A)$. 

EF $\Rightarrow$ AF: The number of elements in any antichain in {\it any} multidirected set is bounded by the total number of elements.  If the latter is finite, then so is the former. 

AF $\Rightarrow$ LFS: I prove the contrapositive.  Suppose that $A$ is not locally finite in the skeleton.  Then there exists an element in $A$ with either an infinite number of maximal predecessors or an infinite number of minimal successors, which comprise an infinite antichain in $A$ by the definition of irreducibility. 

\item {\it No other pairwise logical implications exist among these conditions.}  Note that disproving a particular implication also disproves every ``upstream implication."  For example, since element finiteness (EF) implies relation finiteness (RF), disproving the implication LFT$\Rightarrow$ RF automatically disproves the implication LFT$\Rightarrow$ EF. 

RF $\nRightarrow$ AF:  A countably infinite set with no relations is a counterexample.

LFT $\nRightarrow$ RF, AF:  A countably infinite number of disjoint $1$-chains is a counterexample.

LF $\nRightarrow$ CF, IF, AF:  Augment Jacob's ladder by replacing each relation $x_i\prec y_{-i}$ with a $2$-chain $x_i\prec z_{i}\prec y_{-i}$.  The resulting acyclic directed set is a counterexample for all three conditions. 

LFS $\nRightarrow$ CF, IF, AF, LF:  The transitive closure of the acyclic directed set in the previous step is a counterexample for all four conditions.   

AF $\nRightarrow$ CF, IF, LF:  Cantor's first transfinite ordinal is a counterexample for all three conditions.  

IF $\nRightarrow$ CF, AF, LFS:  The disjoint union of the infinite bouquet and $\ZZ$ is a counterexample for all three conditions. 

CF $\nRightarrow$ AF, IF, LFS:  The {\it mirror bouquet,} with elements $x$, $\{y_i\}_{i\in\ZZ}$, and $z$, and relations $x\prec y_i$ and $y_i\prec z$ for all $i$, is a counterexample for all three conditions. 

\end{enumerate}

\end{proof}

The main purpose of theorem \hyperref[theoremhierarchyfiniteness]{\ref{theoremhierarchyfiniteness}} is to compare finiteness conditions for conservative models of classical spacetime structure in the discrete causal context.  Theorem \hyperref[theoremhierarchyfinitenessmulti]{\ref{theoremhierarchyfinitenessmulti}}, meanwhile, is intended to apply primarily to the abstract structure of configuration spaces of directed sets, particularly kinematic schemes, arising in the histories approach to quantum causal theory developed in section \hyperref[subsectionquantumcausal]{\ref{subsectionquantumcausal}} below.\\

\newpage

\begin{theorem}\label{theoremhierarchyfinitenessmulti} For multidirected sets, the conditions of element finiteness \tn{(\hyperref[ef]{\tn{EF}})}, local element finiteness \tn{(\hyperref[lef]{\tn{LEF}})}, relation finiteness \tn{(\hyperref[rf]{\tn{RF}})}, pairwise relation finiteness \tn{(\hyperref[prf]{\tn{PRF}})}, local finiteness \tn{(\hyperref[lf]{\tn{LF}})}, chain finiteness \tn{(\hyperref[cf]{\tn{CF}})}, interval finiteness \tn{(\hyperref[if]{\tn{IF}})}, and antichain finiteness \tn{(\hyperref[af]{\tn{AF}})}, are related as follows, where arrows indicate logical implications:

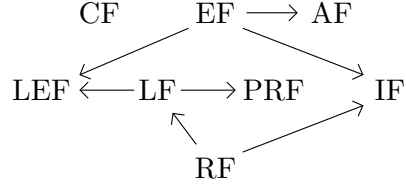
\begin{pgfpicture}{0cm}{0cm}{15cm}{2.3cm}
\begin{pgfmagnify}{1.03}{1.03}
\begin{pgftranslate}{\pgfpoint{2cm}{0cm}}
\pgfputat{\pgfxy(4.75,2)}{\pgfbox[center,center]{\tn{CF}}}
\pgfputat{\pgfxy(6.25,2)}{\pgfbox[center,center]{\tn{EF}}}
\pgfputat{\pgfxy(7.75,2)}{\pgfbox[center,center]{\tn{AF}}}
\pgfputat{\pgfxy(4,1)}{\pgfbox[center,center]{\tn{LEF}}}
\pgfputat{\pgfxy(5.5,1)}{\pgfbox[center,center]{\tn{LF}}}
\pgfputat{\pgfxy(7,1)}{\pgfbox[center,center]{\tn{PRF}}}
\pgfputat{\pgfxy(8.5,1)}{\pgfbox[center,center]{\tn{IF}}}
\pgfputat{\pgfxy(6.25,0)}{\pgfbox[center,center]{\tn{RF}}}
\pgfsetendarrow{\pgfarrowlargepointed{3pt}}
\pgfxyline(6.65,2)(7.3,2)
\pgfxyline(5.9,1.8)(4.5,1.2)
\pgfxyline(6.6,1.8)(8.15,1.2)
\pgfxyline(6.6,.2)(8.15,.8)
\pgfxyline(6,.3)(5.7,.7)
\pgfxyline(5.2,1)(4.5,1)
\pgfxyline(5.8,1)(6.5,1)
\end{pgftranslate}
\begin{colormixin}{15!white}
\begin{pgfmagnify}{.6}{.6}
\pgfputat{\pgfxy(27,.2)}{\pgfbox[center,center]{\tiny{BDCT}}}
\pgfputat{\pgfxy(27,0)}{\pgfbox[center,center]{\tiny{TGGA}}}
\end{pgfmagnify}
\end{colormixin}
\end{pgfmagnify}\end{pgfpicture}

Moreover, these are the only pairwise logical implications among these conditions. 

\end{theorem}
\begin{proof} Let $M=(M,R,i,t)$ be a multidirected set.  As in theorem \hyperref[theoremhierarchyfiniteness]{\ref{theoremhierarchyfiniteness}}, I first prove the claimed logical implications, then prove that no other pairwise logical implications exist among these conditions. 

\begin{enumerate}

\item {\it The claimed logical implications are valid.}

EF $\Rightarrow$ LEF: The number of direct predecessors and successors of each element in $M$ is bounded by the total number of elements of $M$.  If the latter number is finite, then so is the former.  

EF $\Rightarrow$ AF: The proof of this implication in theorem \hyperref[theoremhierarchyfiniteness]{\ref{theoremhierarchyfiniteness}} is for general multidirected sets.  

EF $\Rightarrow$ IF: The number of elements in any open interval in $M$ is bounded by the total number of elements of $M$.  If the latter number is finite, then so is the former. 

RF $\Rightarrow$ LF: The number of relations in the relation set $R(x)$ at any element $x$ in $M$ bounded by the cardinality of the relation set $R$ of $M$.  If the latter number is finite, then so is the former. 

RF $\Rightarrow$ IF: I prove the contrapositive.  Suppose that $M$ possesses an infinite interval $\llangle w,y\rrangle$, and let $\{x_i\}_{i\in\ZZ}$ be an infinite subfamily of $\llangle w,y\rrangle$.  By the definition of an open interval, there exist chains $\gamma_i$ in $M$ beginning at $w$ and terminating at $x_i$ for each $i$.  The terminal relations in these chains are all distinct, since they have distinct terminal elements.  These relations form an infinite subset of the relation set $R$ of $M$. 

LF $\Rightarrow$ LEF: The number of elements directly related to an element $x$ of $M$  is bounded by the cardinality of the relation set $R(x)$ at $x$.  If $M$ is locally finite, then the latter number is finite by definition, so the former number is also finite. 

LF $\Rightarrow$ PRF: The number of relations between two elements $x$ and $y$ in $M$ is bounded by the cardinalities of both the relation sets $R(x)$ and $R(y)$.  If either of the latter numbers is finite, then so is the former.   

\item {\it No other pairwise logical implications exist among these conditions.}  All the counterexamples from theorem \hyperref[theoremhierarchyfiniteness]{\ref{theoremhierarchyfiniteness}} remain valid, since multdirected sets are more general than acyclic directed sets.  The remaining necessary counterexamples are as follows:
 
EF $\nRightarrow$ CF, PRF: A multidirected set with two elements and a countably infinite number of relations between them in both directions is a counterexample for both conditions. 

LEF $\nRightarrow$ CF, AF, IF, PRF:  The disjoint union of the previous counterexample with the augmented version of Jacob's ladder, in which each relation $x_i\prec y_{-i}$ is replaced with a chain $x_i\prec z_{i}\prec y_{-i}$, is a counterexample for all four conditions. 

PRF $\nRightarrow$ CF, LEF, AF, IF:  The disjoint union of the infinite bouquet and the augmented version of Jacob's ladder from the previous counterexample, is a counterexample for all four conditions.  

CF $\nRightarrow$ LEF, PRF:  Replace each relation in the infinite bouquet with a countably infinite number of relations.  The resulting multidirected set is a counterexample for both conditions. 

AF $\nRightarrow$ LEF, PRF:  The disjoint union of $\ZZ$ with a multidirected set consisting of a pair of elements with a countably infinite number of relations between them is a counterexample for both conditions.  

IF $\nRightarrow$ LEF, PRF:  The previous counterexample suffices. 

RF $\nRightarrow$ CF: A reflexive relation $x\prec x$ is a counterexample. 

\end{enumerate}

\end{proof}


\newpage

\section{The Binary Axiom: Events versus Elements}\label{sectionbinary}

The motivation for isolating the physical interpretation of causal set elements as {\it spacetime events} in the binary axiom (\hyperref[b]{B}), is to distinguish the causal set viewpoint from interesting alternative viewpoints regarding similar generalized order-theoretic structures, in which elements may be assigned {\it different physical interpretations}.   In this section, I describe a variety of such interpretations, and explore some of their applications.  This section is not intended to advocate any actual change in the axioms of causal set theory, beyond those changes already suggested in sections \hyperref[sectiontransitivity]{\ref{sectiontransitivity}} and \hyperref[sectioninterval]{\ref{sectioninterval}} above, although possible alterations of the binary axiom are briefly outlined in section \hyperref[subsectionpowerset]{\ref{subsectionpowerset}}, in the context of {\it classical holism.}  Instead, this section provides an expanded structural perspective, introduces new technical methods, and begins the process of exploiting the conclusions of sections \hyperref[sectiontransitivity]{\ref{sectiontransitivity}} and \hyperref[sectioninterval]{\ref{sectioninterval}}.  Mathematically, it has an {\it algebraic} flavor, with multiple appearances of Grothendieck's relative viewpoint (\hyperref[rv]{RV}), and the principle of hidden structure (\hyperref[hs]{HS}).

In section \hyperref[subsectionrelation]{\ref{subsectionrelation}} below, I show how {\it relations} in a multidirected set $M=(M,R,i,t)$ may be interpreted as {\it elements} of an induced directed set $\ms{R}(M)=(R,\prec)$, called the {\it relation space} over $M$.  The {\it induced (binary) relation} $\prec$ on $\ms{R}(M)$ is defined by setting $r\prec s$ if and only if the terminal element of $r$ coincides with the initial element of $s$ in $M$.  Relation spaces are called {\it line digraphs} in the context of graph theory.  They possess special information-theoretic properties, such as {\it impermeability of maximal antichains}, proven in theorem \hyperref[theoremrelimpermeable]{\ref{theoremrelimpermeable}}.  This property is crucial to the theory of {\it path summation} over a multidirected set; in particular, it implies the information-theoretic adequacy of the {\it causal Schr\"{o}dinger-type equations} derived via path summation in section \hyperref[subsectionschrodinger]{\ref{subsectionschrodinger}}.  The theoretical importance of relation space motivates the consideration of analogous spaces whose elements have ``more complicated internal structures" than relations. In section \hyperref[subsectionpowerset]{\ref{subsectionpowerset}}, I discuss a special class of such spaces, called {\it power spaces}, which are power sets endowed with multidirected structures.  I focus on the case of power spaces {\it induced} by the structure of a multidirected set $M=(M,R,i,t)$.   From a conservative perspective, power spaces  provide a convenient tool for organizing the information contained in a multidirected set.  An example is the theory of {\it causal atoms,} already used in the proof of theorem \hyperref[theoremcausalsetrelint]{\ref{theoremcausalsetrelint}} above. More ambitiously, power spaces may be used to encode causal structure that {\it cannot be described} by relations between pairs of elements representing individual events.  This prompts the consideration of {\it top-down causation} and {\it classical holism} in discrete causal theory. 

In section \hyperref[subsectionpathspaces]{\ref{subsectionpathspaces}}, I discuss the theory of {\it causal path spaces} over multidirected sets, which abstract and generalize different aspects of relation spaces.  Paths have special significance in theoretical physics, representing the flow of information or causal influence.  In the context of discrete causal theory, a {\it causal path} is simply a {\it directed path} in a multidirected set $M$; i.e., a morphism from a {\it linear directed set} into $M$.   I usually drop the qualifier ``causal" when referring to individual paths.     Causal path spaces are amenable to a wide array of modern algebraic  tools, providing powerful technical methods for approaching basic physical problems.  In particular, the theory of {\it causal path semicategories} and {\it causal path algebras} leads to a precise algebraic implementation of the histories approach to quantum causal theory, developed in section \hyperref[subsectionquantumcausal]{\ref{subsectionquantumcausal}}.  Relations between pairs of elements in a multidirected set $M$ generalize to yield a family of multidirected structures called {\it splice relations} on causal path spaces over $M$.  Two of the simplest and most important splice relations are the {\it concatenation relation} and the {\it directed product relation,}  which underlie product operations on special causal path semicategories and causal path algebras of particular interest in the quantum setting.  In section \hyperref[subsectionpathsummation]{\ref{subsectionpathsummation}}, I outline the theory of path summation over a multidirected set.  Path sums combine the values of special maps called {\it path functionals,} leading ultimately to the generalized quantum amplitudes, wave functions, and Schr\"{o}dinger-type equations studied in section \hyperref[subsectionquantumcausal]{\ref{subsectionquantumcausal}}.   Path functionals over multidirected sets play a role similar to the role of the {\it classical action} and the {\it Lagrangian} in continuum theory. 


\subsection{Relation Space over a Multidirected Set}\label{subsectionrelation}

\refstepcounter{textlabels}\label{spacesinphys}
\refstepcounter{textlabels}\label{relspace}

{\bf Relation Space Preliminaries.} Physicists are well-accustomed to working in terms of ``spaces" other than ordinary spacetime.   Some of these spaces, such as  {\it Hilbert spaces} in quantum theory, {\it configuration spaces} and {\it phase spaces} in mechanics and thermodynamics, {\it measure spaces} in physical applications of probability, and the various {\it Lie groups,} {\it bundles,} and {\it sheaves} arising in field theories, correspond to spacetime only in the sense that the same algebraic, topological, and geometric intuitions and techniques are useful for studying both.  Other spaces, such as {\it frequency spaces} in electronics, {\it momentum spaces} in classical and quantum mechanics, {\it twistor spaces} in Roger Penrose's twistor theory, and the {\it conformal boundary} in the AdS/CFT correspondence, represent {\it the same or very similar information} as the ordinary space or spacetime to which they correspond.\footnotemark\footnotetext{Such spaces may often be studied in the context of an appropriate {\it duality} theory, such as {\it Pontryagin duality,} {\it Tannaka-Klein duality,} or {\it Maldacena duality.}  The rudiments of such a duality theory for multidirected sets and their relation spaces appear in definition \hyperref[defabstractelementset]{\ref{defabstractelementset}} and theorem \hyperref[theoreminterior]{\ref{theoreminterior}} below.} The {\it relation space} $\ms{R}(M)$ over a multidirected set $M=(M,R,i,t)$, introduced in definition \hyperref[defirelationspace]{\ref{defirelationspace}} below, falls approximately into this latter class.  

The relation space $\ms{R}(M)$ over $M$ is a directed set whose underlying set is the relation set $R$ of $M$, identified in section \hyperref[subsectionchains]{\ref{subsectionchains}} with the set $\tn{Ch}_1(M)$ of $1$-chains in $M$, and used in section \hyperref[subsectiontopology]{\ref{subsectiontopology}} in the definition of the continuum and star topologies.   The binary relation $\prec$ on $\ms{R}(M)$, called the {\it induced relation,} arises naturally from the multidirected structure of $M$, as described in definition \hyperref[defirelationspace]{\ref{defirelationspace}} below.   $\ms{R}(M)$  represents the {\it same information} as $M$ itself, {\it except on the boundary} $\partial M$ of $M$; i.e., its subset of extremal elements.  This statement is made precise in theorem \hyperref[theoreminterior]{\ref{theoreminterior}}.  $\ms{R}(M)$  is therefore ``approximately equivalent" to $M$ in an information-theoretic sense, but provides a superior viewpoint in several important ways.  For example, {\it passage to relation space reduces multidirected structure to directed structure,} since $\ms{R}(M)$ is merely a directed set.  Further, $\ms{R}(M)$ does not suffer from the generic problem of {\it permeability of maximal antichains in multidirected sets,} discussed later in this section.  Relation space methods seem to be best suited to the discrete, nontransitive context, which makes them particularly useful for studying the spacetime models of principal interest in this paper. 

Relation space represents another application of Grothendieck's {\it relative viewpoint} (\hyperref[rv]{RV}), with the {\it ``objects"} involved being the elements of a multidirected set $M=(M,R,i,t)$, and the {\it ``relationships between them"} being the elements of its relation set $R$.  In the special case of directed sets, viewed as models of classical spacetime, these ``objects" and ``relationships" represent spacetime events and independent causal relations between pairs of events, respectively.  For a multidirected set arising from the abstract structure of a kinematic scheme, {\it ``objects"} are directed sets, viewed as {\it classical universes,} and {\it ``relationships"} are {\it co-relative histories,} discussed in detail in section \hyperref[subsectionquantumprelim]{\ref{subsectionquantumprelim}} below. 


{\bf Definition and First Properties of Relation Space.}  The induced binary relation $\prec$ on the relation space $\ms{R}(M)$ over a multidirected set $M=(M,R,i,t)$ is defined by taking one relation (i.e., element of $R$) to precede another if and only if the terminal element (in $M$) of the first relation coincides with the initial element of the second.  Since $\prec$ is defined in terms of {\it ordered pairs} of elements of $R$, the relation space $\ms{R}(M)$ is automatically a {\it directed set;} i.e., no pair of (induced) relations in $\ms{R}(M)$ share the same initial and terminal elements (in $R$).  This is the precise meaning of the above statement that {\it passage to relation space reduces multidirected structure to directed structure.}\\  

\begin{defi}\label{defirelationspace} Let $M=(M,R,i,t)$ be a multidirected set, and let $r$ and $s$ be elements of its relation set $R$.
\begin{enumerate}
\item The {\bf induced relation} $\prec$ on $R$ is the binary relation defined by setting $r\prec s$ if and only if $t(r)=i(s)$. 
\item The directed set $\ms{R}(M)=(R, \prec)$ is called the {\bf relation space over} $M$.   
\end{enumerate}
\end{defi}

In the parlance of graph theory, $\ms{R}(M)$ is called the {\it line digraph} of $M$, where $M$ is viewed as a small directed multigraph.  Here, {\it line} means {\it relation,} and {\it digraph} means {\it directed graph.}  The first appearance of line digraphs in the literature seems to be Frank Harary and Robert Norman's 1960 paper {\it Some Properties of Line Digraphs} \cite{HararyLineDigraphs60}. As far as I know, these objects have received little or no attention in the context of fundamental spacetime structure, although they do appear in recent articles on physically relevant topics, such as the {\it control theory of network dynamics} \cite{NepuszEdgeDynamics12}. 

Figure \hyperref[relationspace]{\ref{relationspace}} below illustrates the construction of the relation space $\ms{R}(M)$ over a multidirected set $M=(M,R,i,t)$. Figure \hyperref[relationspace]{\ref{relationspace}}a shows the induced relation between two elements $r$ and $s$ of $R$, with initial and terminal elements $x$ and $y$, and $y$ and $z$, in $M$, respectively. Figure \hyperref[relationspace]{\ref{relationspace}}b shows the global structure of $\ms{R}(M)$ as a directed set.  In this particular example, $M$ is an acyclic directed set, so $\ms{R}(M)$ is an {\it irreducible} acyclic directed set, as proven in theorem \hyperref[theoremrelationfunctor]{\ref{theoremrelationfunctor}} below.   In this figure, I introduce the convention of representing elements of relation space by {\it square nodes,} while elements of ordinary element space are represented by circular nodes in the usual way.  

\begin{figure}[H]
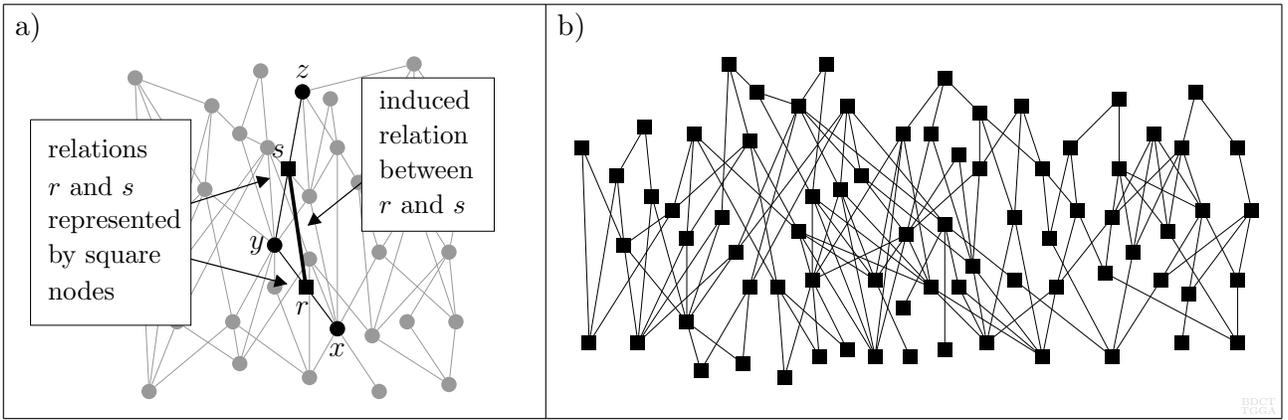


\caption{a) Induced relation between two relations $r$ and $s$ in a multidirected set $M=(M,R,i,t)$; b) relation space $\ms{R}(M)$ over $M$, with induced relation $\prec$.}
\label{relationspace}
\end{figure}
\vspace*{-.5cm}

Theorem \hyperref[theoremrelationfunctor]{\ref{theoremrelationfunctor}} establishes the basic properties of relation space. 

\refstepcounter{textlabels}\label{thmrelspacefunctor}
\vspace*{.2cm}
\begin{theorem}\label{theoremrelationfunctor} Passage to relation space defines a functor $\ms{R}$ from the category $\ms{M}$ of multidirected sets to the category $\ms{D}$ of directed sets.  This functor sends acyclic multidirected sets to irreducible acyclic directed sets, and preserves local finiteness.   Moreover, a cycle of length $n$ in a multidirected set induces a cycle of length $n$ in its relation space.
\end{theorem}
\begin{proof} Let $M=(M,R,i,t)$ be a multidirected set, and let $\ms{R}(M)=(R,\prec)$ be its relation space.  The proof proceeds in five steps:

\begin{enumerate}

\item {\it A morphism $\phi:M\rightarrow M'$ in $\ms{M}$ induces a morphism $\ms{R}(\phi):\ms{R}(M)\rightarrow \ms{R}(M')$ in $\ms{D}$.}  Let $M'=(M',R',i',t')$ be a second multidirected set, and let $\ms{R}(M')=(R',\prec')$ be its relation space.  Recall that a morphism $\phi:M\rightarrow M'$ consists of a {\it pair of maps} $\phi_{\tn{\fsz elt}}:M\rightarrow M'$ and $\phi_{\tn{\fsz rel}}:R\rightarrow R'$, respecting $i,t,i'$ and $t'$.   Define $\ms{R}(\phi)$ to be the map $\phi_{\tn{\fsz rel}}$, as a map of sets, and let $r\prec s$ be a relation in $\ms{R}(M)$.  Then by definition, $t(r)=i(s)$.  Since $\phi$ is a morphism, 
\[t'(\phi_{\tn{\fsz rel}}(r))=\phi_{\tn {\fsz elt}}(t(r))=\phi_{\tn {\fsz elt}}(i(s))=i'(\phi_{\tn{\fsz rel}}(s)).\]

Hence, $\phi_{\tn{\fsz rel}}(r)\prec'\phi_{\tn{\fsz rel}}(s)$ in $\ms{R}(M')$, so $\ms{R}(\phi)$ is a morphism in $\ms{D}$.

\item {\it $\ms{R}$ is a functor.}  It remains to show that $\ms{R}$ preserves identity morphisms and compositions.  The identity morphism $\tn{Id}_{(M,R,i,t)}$ consists of the identity maps $\tn{Id}_M$ and $\tn{Id}_R$ on the {\it sets} $M$ and $R$, respectively, the latter of which coincides with the induced morphism $\ms{R}(\tn{Id}_{(M,R,i,t)})$ on the relation space $\ms{R}(M)$, by definition.  Hence, $\ms{R}$ preserves identities.  Similarly, given two morphisms $\phi:(M,R,i,t)\rightarrow (M',R',i',t')$ and $\psi:(M',R',i',t')\rightarrow (M'',R'',i'',t'')$, the map $(\psi\circ\phi)_{\tn{\fsz rel}}$ is by definition equal to the composition $\psi_{\tn{\fsz rel}}\circ\phi_{\tn{\fsz rel}}$, so $\ms{R}$ preserves compositions. 

\item {\it If $M$ is acyclic, then so is $\ms{R}(M)$.}  I prove the contrapositive.  Suppose that $r_0\prec r_1\prec...\prec r_n\prec r_0$ is a cycle in $\ms{R}(M)$.   Then $t(r_j)=i(r_{j+1})$ for $0\le j<n$, and $t(r_n)=i(r_0)$, by the definition of $\ms{R}(M)$.  Therefore  $i(r_0)\prec i(r_1)\prec...\prec i(r_n)\prec i(r_0)$ is a cycle in $M$.  

\item {\it If $M$ is acyclic, then $\ms{R}(M)$ is irreducible.} Again, I prove the contrapositive.  Suppose that $r\prec s$ is a relation in $\ms{R}(M)$ and $r=r_0\prec r_1\prec...\prec r_n=s$ is a reducing chain.   Then $t(r)$ coincides with the both $i(s)$ and $i(r_1)$.  If $n=2$, then $t(r_1)$ coincides with $i(s)$, so $r_1$ is a reflexive relation.  If $n>2$, then $t(r_1)$ precedes $i(s)$, so $i(r_1)\prec...\prec i(s)=i(r_1)$ is a cycle in $M$.  

\item {\it If $M$ is locally finite, then so is $\ms{R}(M)$.} I prove the contrapositive.  Suppose that $\ms{R}(M)$ is not locally finite, and let $r$ be an element of $\ms{R}(M)$ with infinite relation set.  Since $\ms{R}(M)$ is a directed set, either the set of direct predecessors of $r$, or the set of direct successors of $r$, is infinite.  Assume the former condition, and let $\{r_j\}_{j\in\ZZ}$ be an infinite subfamily of direct predecessors of $r$.  By the definition of $\ms{R}(M)$, the initial element $i(r)$ of $r$ coincides with each terminal element $t(r_j)$ for all $j$.  Therefore, the relation set $R(i(r))$ in $M$ is infinite, so $M$ is not locally finite.  A symmetric argument applies if the set of direct successors of $r$ is infinite. 

\item {\it A cycle of length $n$ in $M$ induces a cycle of length $n$ in $\ms{R}(M)$.}  Let $x_0\prec x_1\prec...\prec x_n= x_0$ be a cycle of length $n$ in $M$, where the distinguished relation $x_j\prec x_{j+1}$ is labeled as $r_j$.\footnotemark\footnotetext{Of course, there are generally other relations between these two elements, since $M$ is multidirected, but $r_j$ is the relation belonging to the cycle under consideration.}  Then $r_0\prec r_1\prec...\prec r_{n-1}\prec r_0$ is a cycle of length $n$ in $\ms{R}(M)$ by the definition of relation space. 
\end{enumerate} 
\end{proof}

\refstepcounter{textlabels}
\label{acyclicinterpolativediscussion}

The relation space $\ms{R}(A)$ over an acyclic directed set $A=(A,\prec)$ satisfies a stronger condition than irreducibility.  For example, no acyclic directed set with multiple $2$-chains between a given pair of elements is the relation space of an acyclic directed set $A$, since the existence of such chains implies multidirected structure on $A$.   Harary and Norman \cite{HararyLineDigraphs60} give a precise criterion identifying when a digraph is the line digraph of a multidigraph, but this result is not needed for the purposes of this paper.  

As mentioned above, the relation space functor $\ms{R}$ is well-suited to the discrete context.   Preservation of local finiteness in theorem \hyperref[theoremrelationfunctor]{\ref{theoremrelationfunctor}} is one indication of this.  It is interesting to observe how strongly this contrasts with the behavior of $\ms{R}$ in the interpolative context.   For example, the relation space over the real line is not only noninterpolative, but irreducible.  To see this, observe that the relation space $\ms{R}(\RR)$ may be identified, as a set, with the half plane above the diagonal $\{(x,x)\hspace*{.1cm}\big|\hspace*{.1cm} x\in\RR\}$ in $\RR^2$.  In the induced relation on $\ms{R}(\RR)$, each element $(x,y)$ directly precedes all elements of the form $(y,z)$, and these relations are irreducible. Similar considerations apply to Minkowski spacetime $X$; every relation directly precedes a ``future light cone" of relations in a section of $X\times X$, and these relations are irreducible.


\refstepcounter{textlabels}\label{abeltspace}

{\bf Abstract Element Space.} I turn now to explaining the sense in which the relation space $\ms{R}(M)$ over a multidirected set $M=(M,R,i,t)$ represents the same information as $M$, except on the boundary $\partial M$ of $M$.   A conceptually useful way to approach this topic is by constructing an {\it approximate inverse} to the relation space functor $\ms{R}$, called the {\it abstract element space functor,} denoted by $\ms{E}$.   Heuristically, one imagines that the elements of some directed set $R=(R,\prec)$ are {\it actually relations,} then constructs in a canonical way a multidirected set $\ms{E}(R)=(M,R,i,t)$, called the {\it abstract element space over} $R$, whose relation space is isomorphic to $(R,\prec)$.\footnotemark\footnotetext{The functor $\ms{E}$ may be applied to  {\it any} directed set; the letter ``$R$" is used for suggestive purposes.  However, not every directed set is the relation space of a multidirected set, as proven by Harary and Norman \cite{HararyLineDigraphs60}.} As demonstrated in theorem \hyperref[theoreminterior]{\ref{theoreminterior}} below, $\ms{E}(R)$ has the property that {\it any multidirected set whose relation space is isomorphic to $R$ has interior isomorphic to} $\tn{Int}(\ms{E}(R))$. Abstract element space is constructed via the following procedure:\\

\begin{defi}\label{defabstractelementset} Let $R=(R,\prec)$ be a directed set.  Construct a multidirected set $\ms{E}(R)=(M,R,i,t)$ by means of the following procedure:
\begin{enumerate}
\item Define abstract sets $M^-:=\{x_r\hspace*{.1cm}|\hspace*{.1cm}r\in R\}$ and $M^+:=\{y_r\hspace*{.1cm}|\hspace*{.1cm}r\in R\}$. 
\item Form the quotient set $M:=M^-\coprod M^+/(y_r\sim x_s \tn{ iff } r\prec s \tn{ in } R)$.  Denote the equivalence classes of $x_r$ and $y_r$ by $\widetilde{x_r}$ and $\widetilde{y_r}$, respectively. 
\item Define maps $i$ and $t$ from $M$ to $R$ by setting $i(r)=\widetilde{x_r}$ and $t(r)=\widetilde{y_r}$ for each $r\in R$. 
\end{enumerate}
The multidirected set $\ms{E}(R)=(M,R,i,t)$ is called the {\bf abstract element space} over $R$. 
\end{defi}

Figure \hyperref[abstractelt]{\ref{abstractelt}} below illustrates the construction of the abstract element space $\ms{E}(R)$ over a directed set $R$.   The elements of $R$ are represented by circular nodes in figure \hyperref[abstractelt]{\ref{abstractelt}}a, to emphasize that $R$ need not be a relation space, although applying the relation space functor $\ms{R}$ to $\ms{E}(R)$ recovers $R$ in this particular example.  The same elements are represented by directed edges in figure \hyperref[abstractelt]{\ref{abstractelt}}c, where they are ``realized" as relations.   The elements of $M^\pm$ and $\ms{E}(R)$ are represented by open circular nodes. 

\begin{figure}[H]
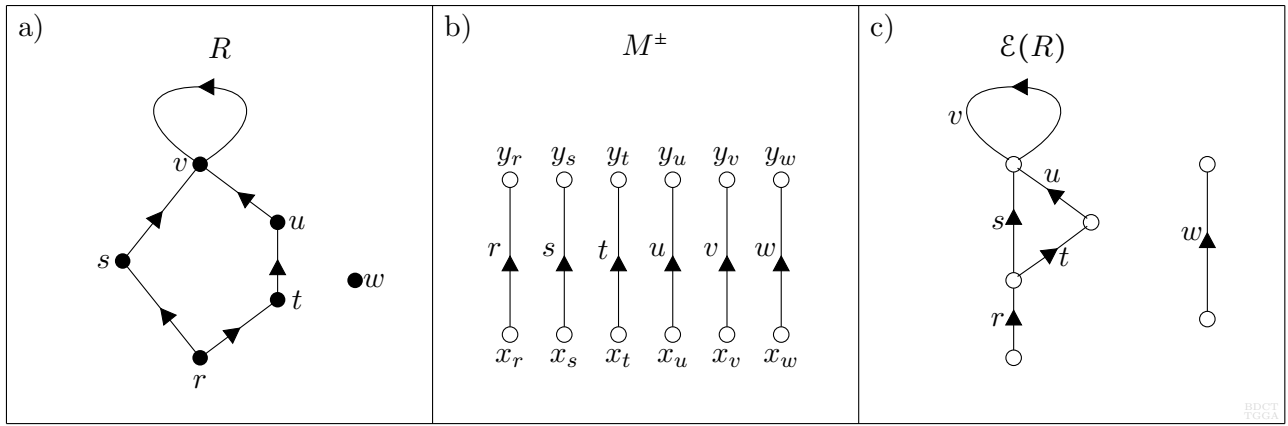


\caption{a) A directed set $R$; b) abstract sets $M^\pm$ (open nodes); arrows anticipate relations in $\ms{E}(R)$; c) abstract element space $\ms{E}(R)$; note that appropriate pairs of elements of $M^\pm$ have been identified according to definition \hyperref[defabstractelementset]{\ref{defabstractelementset}}.2.}
\label{abstractelt}
\end{figure}
\vspace*{-.2cm}

In general, $R$ cannot be recovered from $\ms{E}(R)$ by application of $\ms{R}$.  This is true even for small finite irreducible acyclic directed sets, as illustrated in figure \hyperref[einducescycles]{\ref{einducescycles}} below.  

\begin{figure}[H]
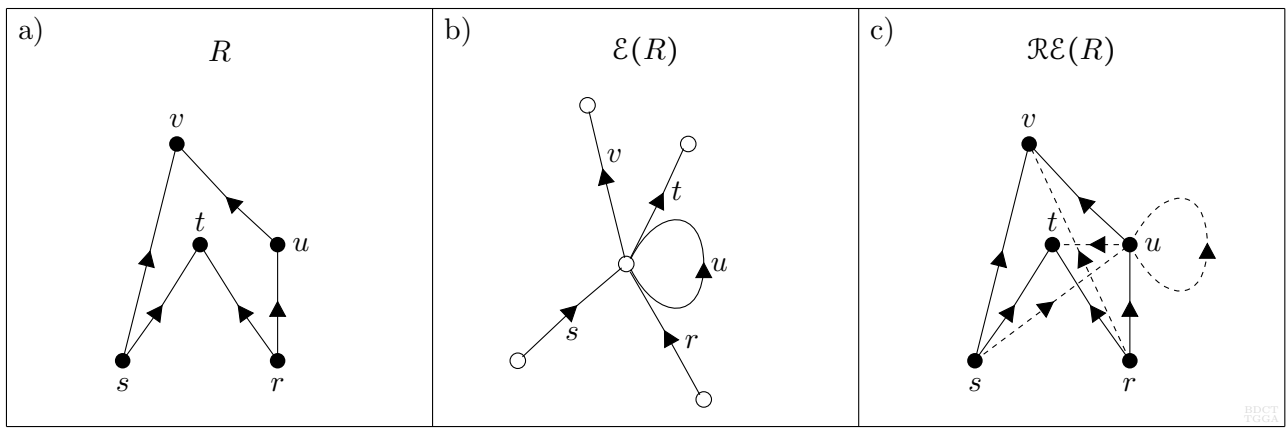


\caption{a) Irreducible acyclic directed set $R$; b) abstract element space $\ms{E}(R)$; c) relation space of $\ms{E}(R)$; note ``extra relations."}
\label{einducescycles}
\end{figure}
\vspace*{-.2cm}

In certain important cases, the directed set $R=(R,\prec)$ in definition \hyperref[defabstractelementset]{\ref{defabstractelementset}} arises by application of the relation space functor $\ms{R}$ to some ``preexisting" multidirected set.  In such cases, there are {\it two different notions} of initial and terminal elements for an element $r$ of $R$.   The images of $r$ under the initial and terminal element maps of the preexisting multidirected set are called {\bf concrete initial and terminal elements} of $r$, while the equivalence classes $\widetilde{x_r}$ and $\widetilde{y_r}$, given by the construction in definition \hyperref[defabstractelementset]{\ref{defabstractelementset}}, are called the {\bf abstract initial and terminal elements} of $r$.   Theorem \hyperref[theoreminterior]{\ref{theoreminterior}} below shows that {\it concrete interior elements correspond bijectively to abstract interior elements,} while {\it each concrete boundary element corresponds to a family of abstract boundary elements.}  Hence, any multidirected set with relation space isomorphic to $R$ may be obtained from $\ms{E}(R)$, up to isomorphism, by identifying appropriate families of abstract boundary elements.   

It is interesting to note that the abstract element space over the continuum is a multidirected set consisting of $|\RR|$ reflexive relations on a singleton.  In particular, all notion of linear order is lost after applying $\ms{E}$.  To see this, note that any pair $r$ and $s$ of real numbers have a common successor, whose abstract initial element coincides with the abstract terminal elements of $r$ and $s$, by the definition of $\ms{E}(\RR)$.  Thus, all terminal elements in $\ms{E}(\RR)$ coincide.  Similarly, all initial elements in $\ms{E}(\RR)$ coincide.  An analogous argument applies to Minkowski spacetime.   More generally, the same argument applies to any directed set satisfying the property that each pair of elements has a common predecessor and a common successor; the abstract element spaces over such sets depend only on their cardinality.\footnotemark\footnotetext{I thank Johnny Feng \cite{Feng13} for pointing out this obvious fact to me.   Using conventional terminology, such sets are often called ``upward directed" and ``downward directed," respectively.}   It is also useful to examine the interaction of $\ms{E}$ with the transitive closure functor $\tn{tr}$ and the skeleton operation $\tn{sk}$.  The integers $\ZZ$ and rational numbers $\QQ$ provide instructive examples. Recall that $\tn{tr}(\ZZ)=\ZZ$ and $\tn{tr}(\QQ)=\QQ$, since the binary relations on $\ZZ$ and $\QQ$ are transitive, but $\tn{sk}(\ZZ)$ is an irreducible unbounded chain, while $\tn{sk}(\QQ)$ is a set of cardinality $\aleph_0$ with no relations.  The abstract element spaces $\ms{E}(\ZZ)$ and $\ms{E}(\QQ)$ both consist of $\aleph_0$ reflexive relations on a singleton, but $\ms{E}(\tn{sk}(\ZZ))\cong\tn{sk}(\ZZ)$, while $\ms{E}(\tn{sk}(\QQ))$ consists of $\aleph_0$ disjoint $1$-chains.   These examples illustrate another reason for abstaining from transitivity in modeling discrete spacetime structure, since redundant reducible relations introduce useless clutter under application of $\ms{E}$.  

Theorem \hyperref[theoremelementfunctor]{\ref{theoremelementfunctor}} establishes the basic properties of abstract element space. 

\refstepcounter{textlabels}\label{thmelspacefunctor}
\vspace*{.2cm}
\begin{theorem}\label{theoremelementfunctor} Passage to abstract element space defines a functor $\ms{E}$ from the category $\ms{D}$ of directed sets to the category $\ms{M}$ of multidirected sets.   This functor preserves local finiteness, but fails to preserve acyclicity, even for irreducible directed sets.  
\end{theorem}
\begin{proof} Let $R=(R,\prec)$ and $R'=(R',\prec')$ be directed sets, with abstract element spaces $\ms{E}(R)=(M,R,i,t)$ and $\ms{E}(R')=(M',R',i',t')$, respectively.  Let $\phi:R\rightarrow R'$ be a morphism.  Recall that for elements $r$ and $r'$ in $R$ and $R'$, respectively,
\[i(r)=\widetilde{x_r}, \hspace*{.5cm} t(r)=\widetilde{y_r}, \hspace*{.5cm} i'(r')=\widetilde{x_{r'}}, \hspace*{.5cm}\tn{and}\hspace*{.5cm} t'(r')=\widetilde{y_{r'}},\]
by part 3 of definition \hyperref[defabstractelementset]{\ref{defabstractelementset}}.  The induced morphism $\ms{E}(\phi)$ consists of a map of elements $\phi_{\tn{\fsz elt}}:M\rightarrow M'$, and a map of relations $\phi_{\tn{\fsz rel}}:R\rightarrow R'$.  Define $\phi_{\tn{\fsz elt}}$ by setting $\phi_{\tn{\fsz elt}}(\widetilde{x_r})=\widetilde{x_{\phi(r)}}$ and $\phi_{\tn{\fsz elt}}(\widetilde{y_r})=\widetilde{y_{\phi(r)}}$.  Define $\phi_{\tn{\fsz rel}}$ to coincide with the given morphism $\phi:R\rightarrow R'$.   The proof proceeds in four steps:

\begin{enumerate}
\item {\it The map of elements $\phi_{\tn{\fsz elt}}:M\rightarrow M'$ is well-defined.} A given element $z\in M$ may be the abstract initial element of many different relations in $R$, and the abstract terminal element of many others.  It is necessary to show that the images of these elements under $\phi_{\tn{\fsz elt}}$ all coincide in $M'$. For example, suppose that $z=\widetilde{x_r}=\widetilde{x_{s}}$ for two different relations $r$ and $s$ in $R$, and consider the corresponding images $\widetilde{x_{\phi(r)}}$ and $\widetilde{x_{\phi(s)}}$ in $M'$ under $\phi_{\tn{\fsz elt}}$.  In this case, $z$ is nonminimal in $M$, since extremal elements in abstract element space, viewed as equivalence classes, are singletons.  Therefore, there exists an element $q$ in $R$ whose terminal element $t(q)$ in $M$ is $z$; i.e., $q$ directly precedes both $r$ and $s$ in $R$.   Since $\phi$ is a morphism, $\phi_{\tn{\fsz rel}}(q)$ directly precedes both $\phi(r)$ and $\phi(s)$ in $R'$.   Hence, the abstract terminal element $\widetilde{y_{\phi(q)}}$ of $q$ coincides with both the abstract initial elements $\widetilde{x_{\phi(r)}}$ and $\widetilde{x_{\phi(s)}}$, which therefore coincide with each other.  A symmetric argument applies if $z=t(r)=t(s)$.  Finally, if $z=t(r)=i(s)$, then $r$ directly precedes $s$ in $R$, so $\phi(r)$ directly precedes $\phi(s)$ in $R'$, and therefore $\widetilde{y_{\phi(r)}}=\widetilde{x_{\phi(s)}}$. 

\item {\it The pair $\ms{E}(\phi):=(\phi_{\tn{\fsz elt}},\phi_{\tn{\fsz rel}})$ is a morphism of multidirected sets.}  It is necessary to show that $\phi_{\tn{\fsz elt}}$ and $\phi_{\tn{\fsz rel}}$ respect initial and terminal element maps in the sense that
\[\phi_{\tn{\fsz elt}}\big(i(r)\big)=i'\big(\phi_{\tn{\fsz rel}}(r)\big)\hspace*{.5cm}\tn{and}\hspace*{.5cm} \phi_{\tn{\fsz elt}}\big(t(r)\big)=t'\big(\phi_{\tn{\fsz rel}}(r)\big),\]
for every $r$ in $R$.   Writing this in terms of equivalence classes, and recalling that $\phi_{\tn{\fsz rel}}$ is just the given morphism $\phi$, this becomes 
\[\phi_{\tn{\fsz elt}}(\widetilde{x_r})=\widetilde{x_{\phi(r)}}\hspace*{.5cm}\tn{and}\hspace*{.5cm} \phi_{\tn{\fsz elt}}(\widetilde{y_r})=\widetilde{y_{\phi(r)}},\]
which is the {\it definition} of $\phi_{\tn{\fsz elt}}$.

\item {\it $\ms{E}$ is a functor.}  If $\phi:R\rightarrow R$ is the identity morphism on $R$, then the map of relations $\phi_{\tn{\fsz rel}}$ in $\ms{E}(\phi)=(\phi_{\tn{\fsz elt}},\phi_{\tn{\fsz rel}})$ is the identity morphism on $R$ by definition, and the map of elements $\phi_{\tn{\fsz elt}}$ sends $\widetilde{x_r}$ to $\widetilde{x_{\phi(r)}}=\widetilde{x_r}$.  A symmetric statement applies to abstract terminal elements.  Hence, $\ms{E}$ preserves identities.  Similarly, given two morphisms $\phi:R\rightarrow R'$ and $\psi:R'\rightarrow R''$, where $R''=(R'',\prec'')$ is a third directed set, the map of relations $(\psi\circ\phi)_{\tn{\fsz rel}}$ in the induced morphism $\ms{E}(\psi\circ\phi)$ is the composition $\psi\circ\phi=\psi_{\tn{\fsz rel}}\circ\phi_{\tn{\fsz rel}}$ by definition, while the map of elements $(\psi\circ\phi)_{\tn{\fsz elt}}$ sends $\widetilde{x_r}$ to $\widetilde{x_{\psi\circ\phi(r)}}=\psi_{\tn{\fsz elt}}(\widetilde{x_{\phi(r)}})=\psi_{\tn{\fsz elt}}\circ\phi_{\tn{\fsz elt}}(\widetilde{x_r})$, and similarly for terminal elements.  Hence, $\ms{E}$ preserves compositions.  

\item {\it If $R$ is locally finite, then so is $\ms{E}(R)$.}  I prove the contrapositive.  Suppose that $\ms{E}(R)$ is not locally finite, and let $x$ be an element of $\ms{E}(R)$ with infinite relation set.  Then either $x$ is related to an infinite number of other elements of $\ms{E}(R)$, or there exist an infinite number of relations between $x$ and some other individual element of $\ms{E}(R)$.  First assume that $x$ has an infinite number of direct successors $\{y_n\}_{n\in\ZZ}$ in $\ms{E}(R)$.   Since $R$ is the relation set of $\ms{E}(R)$, there exist distinct $r_n$ in $R$ for each $n$, such that $x=\widetilde{x_{r_n}}$ and $y=\widetilde{y_{r_n}}$ for each $n$.   Since the abstract initial elements $\widetilde{x_{r_n}}$ of the relations $r_n$ coincide in $\ms{E}(R)$, there exists a relation $r^-$ in $R$ such that $\widetilde{y_{r^-}}=\widetilde{x_{r_n}}$ in $\ms{E}(R)$ for each $n$.  But this implies that $\{r_n\}_{n\in\ZZ}$ constitutes an infinite family of direct successors of $r^-$ in $R$, so $R$ is not locally finite.  A symmetric argument applies if $x$ has an infinite number of directed predecessors.   Next, assume that there exist an infinite number of relations of the form $x\prec y$ for some other element $y$ in $\ms{E}(R)$.  Then there exists an infinite family $\{r_n\}_{n\in\ZZ}$ of elements of $R$ such that $x=\widetilde{x_{r_n}}$ and $y=\widetilde{y_{r_n}}$ for all $n$.   Since the abstract initial elements of each $r_n$ coincide in $\ms{E}(R)$, and similarly for the abstract terminal elements, there exist elements $r^\pm$ of $R$ such that $\widetilde{y_{r^-}}=\widetilde{x_{r_n}}$ and $\widetilde{x_{r^+}}=\widetilde{y_{r_n}}$ in $\ms{E}(R)$ for each $n$.   Hence, $\{r_n\}_{n\in\ZZ}$ simultaneously constitutes an infinite family of direct successors of $r^-$ and an infinite family of direct predecessors of $r^+$ in $R$, so again $R$ is not locally finite.  A symmetric argument applies if there exist an infinite number of relations of the form $w\prec x$ for some other element $w$ in $\ms{E}(R)$.  

\item {\it $\ms{E}$ fails to preserve acyclicity, even in the irreducible case.}  The irreducible acyclic directed set illustrated in figure \hyperref[einducescycles]{\ref{einducescycles}} above is a counterexample. 

\end{enumerate}
\end{proof}


{\bf Approximate Information Preservation by $\ms{R}$.}  The relation space functor $\ms{R}$, and the abstract element space functor $\ms{E}$, interact in interesting and useful ways.  Theorem  \hyperref[theoreminterior]{\ref{theoreminterior}} below renders precise the assertion that the relation space $\ms{R}(M)$ of a multidirected set $M=(M,R,i,t)$ represents the same information as $M$, except on the boundary $\partial M$ of $M$.   This theorem has the same substance as theorem 3 of Harary and Norman \cite{HararyLineDigraphs60}.    This result has important physical applications, particularly in the quantum theoretic context.  For example, the replacement of certain ``badly-behaved" multidirected sets with their ``well-behaved" relation spaces facilitates the development of a suitable version of the histories approach to quantum causal theory, expressed via path summation, as described in section \hyperref[subsectionquantumcausal]{\ref{subsectionquantumcausal}} of this paper.\\

\refstepcounter{textlabels}\label{thmrelpreserves}

\begin{theorem}\label{theoreminterior} Any pair of multidirected sets with isomorphic relation spaces have isomorphic interiors.  
\end{theorem}
\begin{proof} The proof actually establishes the more specific statement that if $R=(R,\prec)$ is a directed set, and if $M$ is any multidirected set whose relation space $\ms{R}(M)$ is isomorphic to $R$, then $\tn{Int}(M)\cong\tn{Int}(\ms{E}(R))$.   Since the relation space $\ms{R}(M)$ of $M$ is isomorphic to $(R,\prec)$, $M$ may be expressed, up to isomorphism, as a quadruple $(M,R,i,t)$, where the ``$R$" appearing in the quadruple coincides, as a set, with the given directed set $(R,\prec)$, and where the  maps $i$ and $t$ identify initial and terminal elements in $M$; i.e., {\it concrete} initial and terminal elements.   Throughout the proof, I use the ``tilde notation," appearing in part 2 of definition \hyperref[defabstractelementset]{\ref{defabstractelementset}} above, to denote elements of $\tn{Int}(\ms{E}(R))$; i.e., {\it abstract} initial and terminal elements.   

The interior $\tn{Int}(M)$ of $M$, viewed as a subobject of $M$, is the multidirected set whose elements are nonextremal elements of $M$, whose relations are elements of $R$ having nonextremal initial and terminal elements, and whose initial and terminal element maps are the restrictions of $i$ and $t$.   In particular, any relation between two elements in $\tn{Int}(M)$ belongs to $\tn{Int}(\ms{R}(M))$, by the definition of the induced relation on $\tn{Int}(\ms{R}(M))$.  Analogous statements apply to $\tn{Int}(\ms{E}(R))$.  Therefore, an isomorphism $\phi$ between $\tn{Int}(M)$ and $\tn{Int}(\ms{E}(R))$, if it exists, consists of a map of elements $\phi_{\tn{\fsz elt}}:\tn{Int}(M)\rightarrow \tn{Int}(\ms{E}(R))$, and a map of relations $\phi_{\tn{\fsz rel}}:\tn{Int}(\ms{R}(M))\rightarrow \tn{Int}(R)$.  I define the latter map $\phi_{\tn{\fsz rel}}$ to be the restriction of the given isomorphism $\phi:\ms{R}(M)\rightarrow R$ to the interior of $\ms{R}(M)$.  Using the fact that any element of $\tn{Int}(M)$ may be written as either an initial element or a terminal element, I define the former map $\phi_{\tn{\fsz elt}}$ by setting $\phi_{\tn{\fsz elt}}(i(r))=\widetilde{x_{\phi_{\tn{\fsz rel}}(r)}}$ and $\phi_{\tn{\fsz elt}}(t(r))=\widetilde{y_{\phi_{\tn{\fsz rel}}(r)}}$.  The proof then proceeds in four steps.
\begin{enumerate}
\item {\it The map $\phi_{\tn{\fsz elt}}$ is well-defined.}  This is similar to part 1 of the proof of theorem \hyperref[theoremelementfunctor]{\ref{theoremelementfunctor}} above.  Suppose that $z=i(r)=i(s)$ for two elements $r$ and $s$ of $R$.   Then it is necessary to show that the images $\widetilde{x_{\phi_{\tn{\fsz rel}}(r)}}$ and $\widetilde{x_{\phi_{\tn{\fsz rel}}(s)}}$ belong to $\tn{Int}(\ms{E}(R))$ and are equal.  Since $z$ belongs to $\tn{Int}(M)$, there exists an element $q$ of $R$ whose such that $t(q)=z$ in $M$.  Therefore, $q$ directly precedes both $r$ and $s$ in $\ms{R}(M)$, so $\phi_{\tn{\fsz rel}}(q)$ directly precedes both $\phi_{\tn{\fsz rel}}(r)$ and $\phi_{\tn{\fsz rel}}(s)$ in $R$.  Hence, the elements $\widetilde{x_{\phi_{\tn{\fsz rel}}(r)}}$ and $\widetilde{x_{\phi_{\tn{\fsz rel}}(s)}}$ coincide in $\ms{E}(R)$, by the definition of abstract element space.  A symmetric argument applies if $z=t(r)=t(s)$.  Finally, if $z=t(r)=i(s)$, then $\phi_{\tn{\fsz rel}}(r)$ directly precedes $\phi_{\tn{\fsz rel}}(s)$ in $R$, so $\widetilde{y_{\phi_{\tn{\fsz rel}}(r)}}=\widetilde{x_{\phi_{\tn{\fsz rel}}(s)}}$ in $\ms{E}(R)$, again by the definition of abstract element space. 

\item {\it The pair of maps $(\phi_{\tn{\fsz elt}},\phi_{\tn{\fsz rel}})$ define a morphism of multidirected sets.}  The map $\phi_{\tn{\fsz rel}}$ simply identifies each relation in $\tn{Int}(\ms{R}(M))$ with the corresponding relation in $\tn{Int}(R)$, via the hypothesized isomorphism $\phi:\ms{R}(M)\cong R$ of directed sets.  If $r$ belongs to $\tn{Int}(\ms{R}(M))$, then its concrete initial and terminal elements $i(r)$ and $t(r)$ in $M$ map to the abstract initial and terminal elements $\widetilde{x_{\phi_{\tn{\fsz rel}}(r)}}$ and $\widetilde{y_{\phi_{\tn{\fsz rel}}(r)}}$ of its image $\phi_{\tn{\fsz rel}}(r)$ in $\tn{Int}(\ms{E}(R))$, by the definition of $\phi_{\tn{\fsz elt}}$.  Hence, $(\phi_{\tn{\fsz elt}},\phi_{\tn{\fsz rel}})$ respects initial and terminal element maps.  

\item {\it The map $\phi_{\tn{\fsz elt}}$ is bijective.}  Suppose that $z=t(r)=i(s)$ and $w=t(\overline{r})=i(\overline{s})$ are two elements of $\tn{Int}(\ms{R}(M))$ whose images under $\phi_{\tn{\fsz elt}}$ coincide, where $r,s,\overline{r}$ and $\overline{s}$ are elements of $R$.   Then by definition, $\widetilde{x_{\phi_{\tn{\fsz rel}}(s)}}=\widetilde{x_{\phi_{\tn{\fsz rel}}(\overline{s})}}=\widetilde{y_{\phi_{\tn{\fsz rel}}(r)}}=\widetilde{y_{\phi_{\tn{\fsz rel}}(\overline{r})}}$ in $\tn{Int}(M)$.  Therefore, $\phi_{\tn{\fsz rel}}(r)$ and $\phi_{\tn{\fsz rel}}(\overline{r})$ both directly precede $\phi_{\tn{\fsz rel}}(s)$ and $\phi_{\tn{\fsz rel}}(\overline{s})$ in $\tn{Int}(R)$.   Since $\phi_{\tn{\fsz rel}}$ is an isomorphism of directed sets, $r$ and $\overline{r}$ both directly precede $s$ and $\overline{s}$ in $\tn{Int}(\ms{R}(M))$.  Hence, $t(r)=t(\overline{r})=i(s)=i(\overline{s})$, so $z=w$, and $\phi_{\tn{\fsz elt}}$ is injective.  Next, observe that any element in $\tn{Int}(\ms{E}(R))$ may be written as $\widetilde{x_{\phi_{\tn{\fsz rel}}(s)}}$ for some relation $s$ in $\tn{Int}(\ms{R}(M))$.  By definition of $\phi_{\tn{\fsz elt}}$, this element lifts to the element $i(s)$ in $\tn{Int}(M)$.  Thus, $\phi_{\tn{\fsz elt}}$ is surjective.  

\item {\it The pair of inverse maps $(\phi_{\tn{\fsz elt}}^{-1},\phi_{\tn{\fsz rel}}^{-1})$ define a morphism of multidirected sets.} Since $\phi_{\tn{\fsz rel}}=\phi$ is an isomorphism, any element of $\tn{Int}(R)$ may be expressed as $\phi_{\tn{\fsz rel}}(r)$ for a unique element $r$ of $\ms{R}(M)$.  The image of this element under the inverse relation map $\phi_{\tn{\fsz rel}}^{-1}$ is just $r$.  By definition, the inverse element map $\phi_{\tn{\fsz elt}}^{-1}$ carries the abstract initial and terminal elements of $\phi_{\tn{\fsz rel}}(r)$ in $\tn{Int}(\ms{E}(R))$ to the concrete initial and terminal elements $i(r)$ and $t(r)$ in $\tn{Int}(M)$.  Hence, the pair $(\phi_{\tn{\fsz elt}}^{-1},\phi_{\tn{\fsz rel}}^{-1})$ respects initial and terminal element maps.  
\end{enumerate}
\end{proof}


\refstepcounter{textlabels}\label{cauchyrel}

\begin{figure}[H]
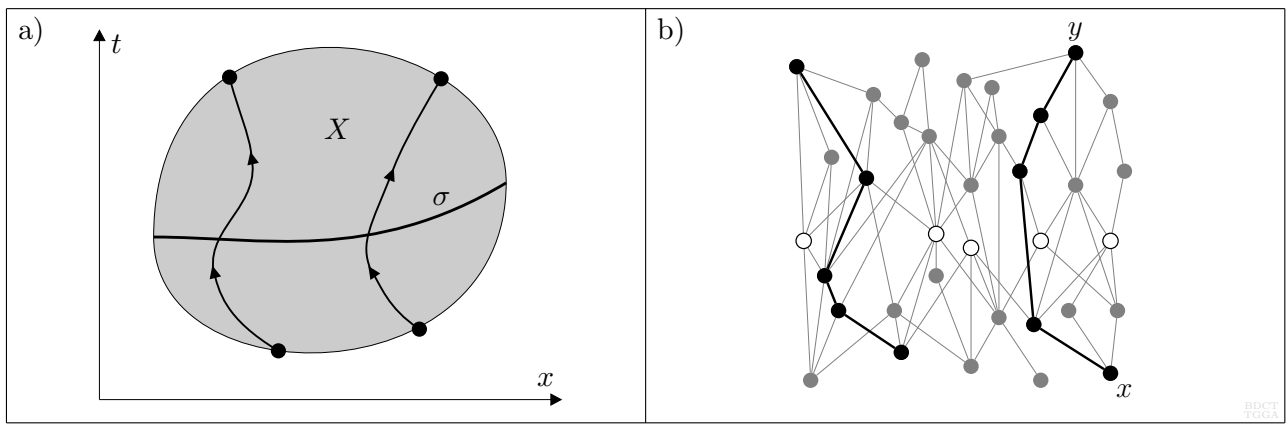


\caption{a) Cauchy surface in relativistic spacetime; b) maximal antichain in a multidirected set $M$ (open nodes); dark paths are permeating chains.}
\label{permeability}
\end{figure}

{\bf Cauchy Surfaces; Impermeability.} I now discuss an information-theoretic device of great importance in continuum theory, both in the classical and quantum settings, whose adaptation to causal set theory has hitherto been hampered by a particular technical problem.   A {\bf Cauchy surface} is a subset $\sigma$ of a region $X$ of continuum spacetime, admitting a {\it unique point of intersection with every causal path in $X$ between its past and future.}  A Cauchy surface may be viewed as a {\it generalized spacelike section} of spacetime, representing a single ``instant in time."  From an information-theoretic viewpoint, a Cauchy surface filters the flow of information from past to future through $X$. Foliation of spacetime via Cauchy surfaces is an important technique in relativistic cosmology, and also plays a central role in various approaches to quantum gravity, including loop quantum gravity, noncommutative geometry, and shape dynamics.  Figure \hyperref[permeability]{\ref{permeability}}a below illustrates a Cauchy surface $\sigma$ in relativistic spacetime, with two representative causal paths passing through $\sigma$ from past to future.

\vspace*{-.5cm}

\refstepcounter{textlabels}\label{eltspaceperm}

The na\"{\i}ve analogue of a Cauchy surface in a multidirected set $M=(M,R,i,t)$  is a {\bf maximal antichain} in $M$; i.e., a subset $\sigma$ of $M$ such that no chain in $M$ connects any pair of elements of $\sigma$, but every element in the complement of $\sigma$ is connected to an element of $\sigma$ by a chain.  This is illustrated in figure \hyperref[permeability]{\ref{permeability}}b above, which shows the acyclic directed case.   The maximal antichain is represented by the five open nodes in the figure.  The dark-colored chains, and the labels $x$ and $y$, are for future reference. 

A technical problem associated with maximal antichains in multidirected sets is that they are generally {\bf permeable}, meaning that there generally exist chains extending from the past of a maximal antichain $\sigma$ to its future {\it without intersecting} $\sigma$.  Two such paths are illustrated in figure \hyperref[permeability]{\ref{permeability}}b.  This problem is particularly severe in the transitive case, with numerous relations, mostly reducible, passing directly from the ``distant past" to the ``distant future" of $\sigma$.   In causal set theory, reducible relations are taken to be dependent, so one might na\"{i}vely hope that the problem is only a formal one in this context, disappearing upon passage to the skeleton.  This hope fails, as illustrated by the permeating chain from $x$ to $y$ in figure \hyperref[permeability]{\ref{permeability}}b, which involves only irreducible relations.  The problem of permeability therefore {\it does not depend on the physical interpretation of reducible relations;} i.e., on whether or not one accepts the independence convention (\hyperref[ic]{IC}).

\refstepcounter{textlabels}\label{rideoutperm}

Seth Major, David Rideout, and Sumati Surya examine maximal antichains as analogues of Cauchy surfaces in causal set theory in their 2006 paper {\it Spatial Hypersurfaces in Causal Set Cosmology} \cite{RideoutSpatialHypersurface06}.   Recognizing the permeability problem, they attempt to address it by {\it thickening} maximal antichains to ``capture" permeating chains.\footnotemark\footnotetext{Surya mentions the same problem in her 2011 review article \cite{Surya11}.}   While such constructions carry some intrinsic interest, they are not particularly natural, and do not really solve the problem.  A better approach is to pass to relation space, and work with {\it maximal antichains of relations.}  Remarkably, no additional device is needed: the permeability problem disappears in relation space.  This is made precise by theorem \hyperref[theoremrelimpermeable]{\ref{theoremrelimpermeable}} below, which states that {\it maximal antichains in relation space are impermeable.} 

\begin{figure}[H]
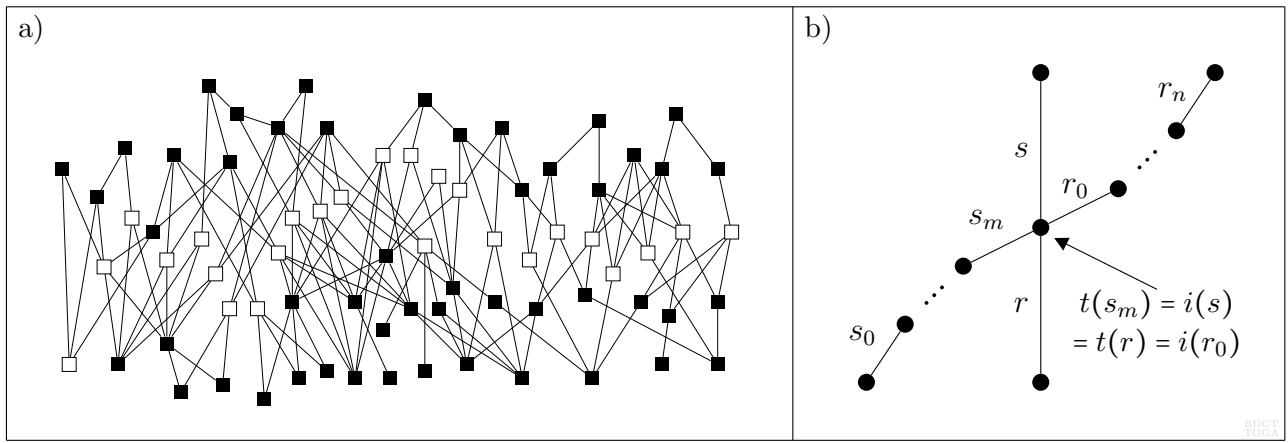


\caption{a) Maximal antichain in relation space; b) maximal antichains in relation space are impermeable.}
\label{figimpermeability}
\end{figure}

Figure \hyperref[figimpermeability]{\ref{figimpermeability}}a below illustrates a typical maximal antichain in relation space.  The underlying multidirected set from which this relation space is derived is the familiar example appearing in figure \hyperref[permeability]{\ref{permeability}}b above.  Figure  \hyperref[figimpermeability]{\ref{figimpermeability}}b is used in the proof of theorem \hyperref[theoremrelimpermeable]{\ref{theoremrelimpermeable}} below. 

\vspace*{-.2cm}

\refstepcounter{textlabels}\label{thmrelspaceimperm}
\addtocounter{theorem}{1}
\begin{theorem}\label{theoremrelimpermeable} Maximal antichains in relation space are impermeable.   
\end{theorem}
\begin{proof}  Let $M=(M,R,i,t)$ be a multidirected set with relation space $\ms{R}(M)=(R,\prec)$, and let $\sigma$ be a maximal antichain in $\ms{R}(M)$.  Note that by the definition of an antichain, the past and future of $\sigma$ are disjoint.  Indeed, if $r\in \ms{R}(M)$ belongs to both the past and future of $\sigma$, then by definition, there exist elements $r^-$ and $r^+$ in $\sigma$, together with chains from $r^-$ to $r$ and from $r$ to $r^+$. These chains join together to form a chain from $r^-$ to $r^+$, contradicting the hypothesis that $\sigma$ is an antichain.  

The proof is by contradiction.  Suppose that $\sigma$ is permeable, and let $\gamma$ be a chain in $\ms{R}(M)$ permeating $\sigma$.  By definition, $\gamma$ may be expressed as a sequence of elements and relations in $\ms{R}(M)$ with initial element in the past of $\sigma$ and terminal element  in the future of $\sigma$.   Hence, there is at least one relation $r\prec s$ in $\gamma$ with initial element $r$ in the past of $\sigma$, and terminal element $s$ in the future of $\sigma$.\footnotemark\footnotetext{As I have defined it, $\gamma$ may ``wrap around cycles in $\ms{R}(M)$,"  and may permeate $\sigma$ multiple times.  It is possible to choose $\gamma$ with tamer properties, but all that is necessary for the argument is to isolate {\it one} relation in $\gamma$ passing from the past of $\sigma$ to the future of $\sigma$.} Since $r$ is in the past of $\sigma$, there exists a chain $r\prec r_0\prec...\prec r_n$ in $\ms{R}(M)$, with  $r_n\in\sigma$.  Similarly, since $s$ is in the future of $\sigma$, there exists a chain $s_0\prec...\prec s_m\prec s$ in $\ms{R}(M)$, with $s_0\in\sigma$.   Now 
\[t(s_m)=i(s)=t(r)=i(r_0)\hspace*{.3cm} \tn{ in } \hspace*{.1cm} M,\]
since $s_m\prec s$, $r\prec s$, and $r\prec r_0$ in $\ms{R}(M)$.  This sequence of equalities is illustrated in figure \hyperref[figimpermeability]{\ref{figimpermeability}}b above, which shows the {\it element space viewpoint.}   Hence, $s_m\prec r_0$ in $\ms{R}(M)$, so the two chains $s_0\prec...\prec s_m$ and $r_0\prec...\prec r_n$ fit together to define a composite chain $s_0\prec...\prec s_m\prec r_0\prec...\prec r_n$, which is a chain between two elements $s_0$ and $r_n$ of $\sigma$.  This contradicts the hypothesis that $\sigma$ is an antichain.   
\end{proof}

\refstepcounter{textlabels}\label{preferrelspace}

Impermeability of maximal antichains in relation space is of great importance in the {\it dynamics} of directed sets, especially in the context of quantum theory.  In particular, the theory of path summation over a multidirected set, which is the causal analogue of {\it path integration} over continuum spacetime, relies on impermeability in a critical way, as explained in section  \hyperref[subsectionquantumpathsummation]{\ref{subsectionquantumpathsummation}} below.  Viewing spacetime structure in terms of relations rather than elements is much more than a philosophical preference, since some of the most important associated methods {\it do not apply in element space}.  Besides impermeability, there are other reasons to prefer relation space over element space for formulating {\it spatial notions} such as Cauchy surfaces.  One such reason is {\it adjacency.}  In the acyclic case, any pair of relations with the same initial or terminal element form a minimal nontrivial antichain in relation space, which may be viewed as a ``building block" of spatial structure.  Such relations are adjacent,  in the sense that they share a common element; i.e., they form a connected subgraph.  While it is {\it not} true that a maximal antichain in relation space is always connected as a subgraph, a ``high degree of connectivity" is common.  For example, the antichain of relations illustrated in in figure \hyperref[causalpathsum]{\ref{causalpathsum}}b of section \hyperref[subsectionquantumpathsummation]{\ref{subsectionquantumpathsummation}} below is connected as a subgraph.  Similarly, the maximal antichain of relations consisting of the ``first generation" of relations in the causal set illustrated in figure \hyperref[semiordinalspatial]{\ref{semiordinalspatial}} above has just three connected components, each corresponding to a copy of $\ZZ$ in an obvious way.  This supports the vague statement made in section \hyperref[relativeacyclicdirected]{\ref{relativeacyclicdirected}} that this causal set has ``spatial size greater than $\ZZ$."


\refstepcounter{textlabels}\label{analogymorphism}

{\bf Analogy between $\ms{R}(M)$ and Morphism Categories.} A {\it morphism category} is a category whose objects are the morphisms of another category, and whose morphisms are {\it morphisms between morphisms}, called {\it $2$-morphisms}.  An {\it augmented category} possessing objects, morphisms, and $2$-morphisms, is called a {\it $2$-category.}  Let $W,X,Y,$ and $Z$ be objects of some category.  Then a $2$-morphism between morphisms $f:X\rightarrow Y$ and $g:W\rightarrow Z$ is defined to be a {\it commutative square} of the form illustrated in figure \hyperref[2cat]{\ref{2cat}}a below:

\begin{figure}[H]
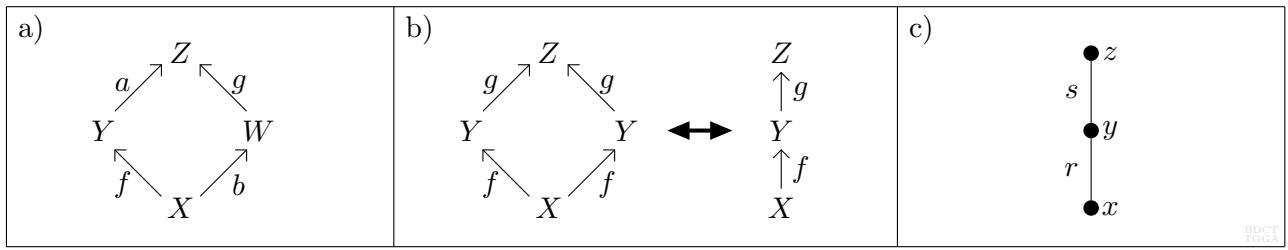


\caption{a) A $2$-morphism; b) ``commutativity" without composition; c) $2$-chain analogue. }
\label{2cat}
\end{figure}
\vspace*{-.5cm}


A basic requirement for commutativity to make sense is the {\it composition axiom} of category theory, which guarantees that the ordered pairs of morphisms $(f,a)$ and $(b,g)$ compose to define morphisms $a\circ f:X\rightarrow Z$ and $g\circ b:X\rightarrow Z$.  As discussed in section \hyperref[subsectiontransitivitydeficient]{\ref{subsectiontransitivitydeficient}} above, the composition axiom is roughly analogous to the causal set axiom of transitivity.  However, there remains at least one circumstance under which the square may reasonably be regarded as ``commutative" even without the composition axiom: namely, when $Y=W$, $b=f$, and $a=g$.  In this case, the square is equivalent to the sequence of morphisms $X\overset{f}{\rightarrow}Y\overset{g}{\rightarrow}Z$, as illustrated in figure  \hyperref[2cat]{\ref{2cat}}b above.  This sequence, in turn, is analogous to a $2$-chain in a multidirected set $M=(M,R,i,t)$, consisting of relations $r$ and $s$, with initial and terminal elements $i(r)=x$, $t(r)=i(s)=y$, and $t(s)=z$, as illustrated in figure  \hyperref[2cat]{\ref{2cat}}c.  By the definition of relation space, such a $2$-chain corresponds to a relation $r\prec s$ in $\ms{R}(M)$. 

One way to interpret this analogy is to view the relation space $\ms{R}(M)$ over a multidirected set $M$, together with its induced relation $\prec$, as a {\it generalized morphism category.}  More holistically, the original multidirected set $M$ may be viewed as a {\it generalized $2$-category,} whose ``objects" are its elements, whose ``morphisms" are its relations, and whose ``$2$-morphisms" are its $2$-chains.   This provides an example of a phenomenon mentioned in section \hyperref[settheoretic]{\ref{settheoretic}}, in the context of Isham's topos theoretic approach to physics: causal theory naturally presents structures that are {\it not} categories but that play similar roles.  Unmodified appropriation of category-theoretic notions can do more harm than good in this context.   For example, the na\"{\i}vest causal analogue of a $2$-morphism is a {\it causal diamond,} but this analogy leads in the wrong direction in the context of relation space.  Many other such ``category-like" structures appear in succeeding sections.  


\refstepcounter{textlabels}\label{starrevisited}

{\bf The Star Model Revisited.}  The star model $M_\star$ of a multidirected set $M=(M,R,i,t)$, used in section \hyperref[sectioninterval]{\ref{sectioninterval}} above to facilitate the study of local finiteness, may be elevated to a multidirected set $(M_\star,R_\star,i_\star, t_\star)$ in its own right, by means of the induced relation $\prec$ on the relation space $\ms{R}(M)$ over $M$.  As a set, the star model $M_\star$ is just the union $M\cup R$ of (the underlying set of) $M$ and its relation set $R$.  This set may be endowed with multidirected structure by taking its relation set $R_\star$ to be the union $R\cup\prec$, where elements of $R$ relate elements of $M$, and elements of $\prec$ relate elements of $R$.  Initial and terminal element maps $i_\star$ and $t_\star$ may then be defined in the obvious way, by combining the maps $i$ and $t$ with corresponding maps from $\prec$ to $R$, sending each induced relation $r\prec s$ to its initial and terminal elements $r$ and $s$ in $R$, respectively.  This multidirected version of the star model combines the ``lowest-order graded pieces" of the {\it causal concatenation spaces} over $M$ and $\ms{R}(M)$, described in more detail in section \hyperref[subsectionpathspaces]{\ref{subsectionpathspaces}} below.   The star model of a directed set is directed, rather than multidirected, since both $R$ and $\prec$ define binary relations in this case. 


\refstepcounter{textlabels}\label{inducedtwoelt}

{\bf An Induced Binary Relation on Two-Element Subsets.} Alternatively, the induced relation $\prec$ on the relation space $\ms{R}(M)$ over a multidirected set $M=(M,R,i,t)$ may be used to construct a binary relation on the set $M_2$ of {\it two-element subsets} of $M$.   This relation is defined by taking subsets $\{x,y\}$ and $\{u,v\}$ to be related if and only if there is a relation $x\prec y$ or $y\prec x$, a relation $u\prec v$ or $v\prec u$, and the terminal element of the first relation is the initial element of the second.  This construction generally does not preserve information, even in the interior of $M$, since merely specifying a pair of elements of $M$ neither determines which element is initial and which is terminal, nor distinguishes among different relations having the same initial and terminal elements.   Further, information about reflexive relations is lost entirely.  Less information loss occurs in the special case of an acyclic directed set $A=(A,\prec)$, since each two-element subset $\{x,y\}$ of $A$ corresponds to at most one relation in $A$ involving $x$ and $y$, and since no reflexive relations exist in this case.   For these reasons, the induced relation on two-element subsets may be viewed as a trivial extension of the induced relation on the relation space $\ms{R}(A)$.

Most of the multidirected sets used as examples in this paper induce very {\it sparse} relations on their sets of two-element subsets, since most of these subsets do not consist of the initial and terminal elements of a relation.  Figure \hyperref[twoelement]{\ref{twoelement}} below illustrates this sparseness, using the familiar $35$-element multidirected set $M$ appearing repeatedly in previous examples.  Figure \hyperref[twoelement]{\ref{twoelement}}a shows that a ``typical" two-element subset of $M$ does not correspond to relation; indeed, $M$ has $\binom{35}{2}=595$ two-element subsets, but only $75$ relations.  Hence, most of the elements of $M_2$ are isolated in the induced relation on $M_2$.   Figure \hyperref[twoelement]{\ref{twoelement}}b shows $M_2$ with its induced relation; isolated elements of $M_2$ are represented by small square nodes, while non-isolated elements are represented by large square nodes.  Since $M$ is in fact an acyclic directed set in this example, the induced relation on $M_2$ may be viewed as a trivial extension of the induced relation on $\ms{R}(M)$.  This may be seen by comparing figure \hyperref[twoelement]{\ref{twoelement}}b to figure \hyperref[relationspace]{\ref{relationspace}}b above.  

\begin{figure}[H]
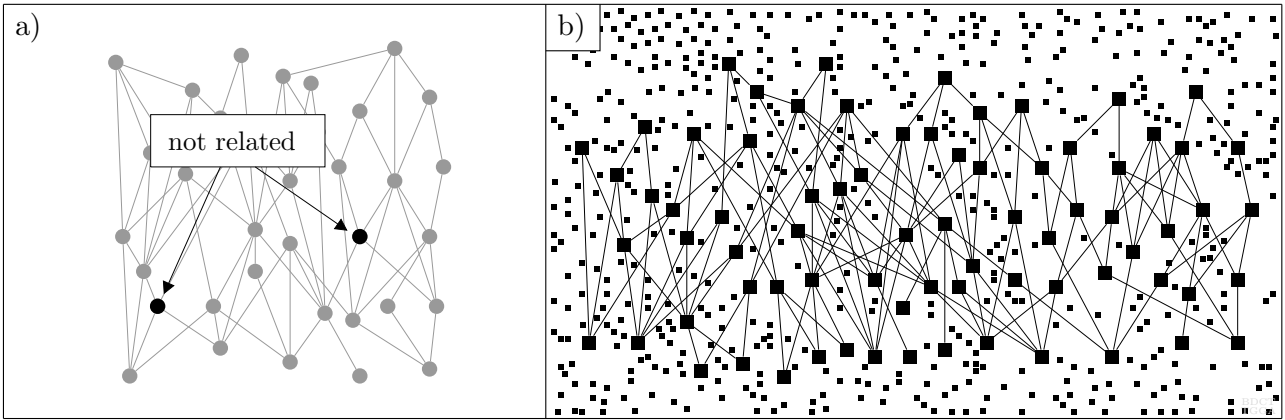


\caption{a) Most two-element subsets of $M$ do not consist of the initial and terminal elements of a relation; b) induced relation on $M_2$.}
\label{twoelement}
\end{figure}

\subsection{Power Spaces}\label{subsectionpowerset}

\refstepcounter{textlabels}\label{powerspaces}

{\bf Power Spaces; Induced and Holistic Power Spaces.}  The notion of relations between pairs of two-element subsets of a multidirected set $M=(M,R,i,t)$, discussed at the end of section \hyperref[subsectionrelation]{\ref{subsectionrelation}} above, naturally motivates consideration of relations between pairs of subsets of $M$ of arbitrary cardinality.  Since subsets of $M$ correspond to {\it elements} of the power set $\ms{P}(M)$ over $M$, families of relations between pairs of subsets of $M$ correspond to multidirected structures on $\ms{P}(M)$, called {\it power spaces over} $M$.  The following general set-theoretic definition makes this idea precise:

\vspace*{.2cm}
\begin{defi}\label{defpowerspaces} Let $M$ be an arbitrary set.  A {\bf power space} over $M$ is a multidirected set whose underlying set is the power set $\ms{P}(M)$ of $M$. 
\end{defi}

\refstepcounter{textlabels}\label{inducedpowerspaces}

The reason for taking $M$ to be an arbitrary set in definition \hyperref[defpowerspaces]{\ref{defpowerspaces}} is that the ``general power space viewpoint" takes multidirected structure on $M$ to be {\it residual structure} left over after ignoring relations on $\ms{P}(M)$ involving nonsingleton subsets.  As discussed below, it is generally not possible to recover a power space from this residual structure.   Power spaces that {\it do} arise naturally from the structure of a ``preexisting" multidirected set $M=(M,R,i,t)$ are called {\bf induced power spaces}.  The majority of power spaces considered in this paper are induced, but the reader should be aware that these are exceptional cases.

A ``trivial" example of an induced power space over a multidirected set $M=(M,R,i,t)$ is given by taking the relation set over $\ms{P}(M)$ to be $R$, viewed as a set of relations between pairs of singleton subsets of $M$, with no relations involving subsets of higher cardinality.  A slightly more interesting example may be constructed by combining the relation set $R$ of $M$ with the induced relation on two-element subsets of $M$, introduced at the end of section \hyperref[subsectionrelation]{\ref{subsectionrelation}} above.  This defines a power space over $M$ with relations between distinguished pairs of one and two-element subsets.  If $A$ is an acyclic directed set, then the star model $A_\star$ of $A$, viewed as a directed set as described in section \hyperref[subsectionrelation]{\ref{subsectionrelation}} above, is the ``nontrivial part" of this power space, from a generalized order-theoretic perspective.  However, the star model of a general multidirected set $M$ does not embed naturally into a power space over $M$, since relations in $M$ generally cannot be identified with two-element subsets of $M$.  A third induced power space over $M$ is the space with one relation between each pair of subsets $M'$ and $M''$ of $M$ for each relation in $R$ between an element of $M'$ and an element of $M''$.   

\refstepcounter{textlabels}\label{holisticpowerspaces}

In contrast to these examples, most power spaces over a set $M$ are {\bf holistic,} meaning that their information content consists at least partly of {\it irreducibly complex relationships} between pairs of nonsingleton subsets of $M$.  Such power spaces are {\it not} induced in any natural way by any preexisting multidrected structure on $M$.\footnotemark\footnotetext{The caveat {\it naturally} is included because one {\it can} define bijections between the set of {\it finite} multidirected structures on a {\it finite} set and the set of {\it finite} multidirected structures on its power set, since both sets are countably infinite.  Under such a bijection, a typical power space could be encoded in an unnatural and arbitrary way by invoking large numbers of relations between typical elements of $M$.} It has recently become popular in certain scientific fields, not limited to physics, to take such holistic relationships seriously, even at the classical level.  I briefly discuss a few instances of this notion below, including the general concept of {\it top-down causation,} and some specific holistic approaches to fundamental physics. 


\refstepcounter{textlabels}\label{higherinduced}

{\bf Higher Induced Relations; Splice Relations.}  Since relations in a multidirected set $M=(M,R,i,t)$ may be identified with $1$-chains in $M$, an obvious way to define induced relations between pairs of distinguished nonsingleton subsets of $M$ is to modify the induced binary relation $\prec$ on the relation space $\ms{R}(M)$ over $M$ to apply to longer chains.  Complications immediately arise in the general multidirected context.  For example, an $n$-chain in a  multidirected set $M$ does not always correspond to an $n$-element subset of $M$, since chains may intersect themselves.  Further, multiple chains may share the same sequence of elements.   Hence, this approach generally involves a choice between accepting significant loss of information, as in the case of induced relations between pairs of two-element subsets of $M$, discussed at the end of section \hyperref[subsectionrelation]{\ref{subsectionrelation}} above, and leaving the realm of power spaces, as in the theory of {\it causal path spaces,} introduced in section \hyperref[subsectionpathspaces]{\ref{subsectionpathspaces}} below.  However, in the special case where $M$ is an acyclic directed set, each $n$-chain in $M$ uniquely identifies an $n$-element subset of $M$, permitting relations between pairs of chains to be viewed as relations between elements of the power set $\ms{P}(M)$ of $M$. 

A straightforward way to define such relations between pairs of chains is to {\it iterate the process of forming relation spaces.}  Let $A$ be an acyclic directed set, with relation space $\ms{R}(A)$, and set of $2$-chains $\tn{Ch}_2(A)$.   Here, it is convenient to denote the binary relation on $A$ by $\prec_0$, and the induced binary relation on $\ms{R}(A)$ by $\prec_1$.  Now let $w\prec_0 x\prec_0 y\prec_0 z$ be a $3$-chain in $A$, and denote the relations $w\prec_0 x$, $x\prec_0 y$, and $y\prec_0 z$, by $q$, $r$, and $s$, respectively.   Denote the $2$-chains $w\prec_0 x\prec_0 y$ and $x\prec_0 y\prec_0 z$ by $\gamma$ and $\delta$, respectively. Then there exists an {\it induced $2$-chain} $q\prec_1 r\prec_1 s$ in $\ms{R}(A)$.   The {\it second relation space} $\ms{R}^2(A):=\ms{R}(\ms{R}(A))$ of $A$ corresponds to $\tn{Ch}_2(A)$ in an obvious way, and the {\it second induced relation} $\prec_2$ on $\ms{R}^2(A)$, induced by $\prec_1$, may be naturally identified with the binary relation on $\tn{Ch}_2(A)$ defined by taking one $2$-chain to precede another if and only if these chains ``overlap" in a single relation in $A$.  For example, in the case described above, $\gamma\prec_2 \delta$, since $\gamma$ and $\delta$ overlap in the relation $x \prec_0 y$.  One may proceed iteratively to define {\bf higher induced relations} $\prec_n$ on the chain sets $\tn{Ch}_n(A)$ of $A$.   As described so far, this construction applies to a general multidirected set.  However, since $A$ is an acyclic directed set, the higher induced relations $\prec_n$ descend to binary relations on the corresponding sets of $n$-element subsets of $A$.   Any subfamily of these relations may be used to define a power space over $A$. 

Whenever two overlapping $n$-chains $\gamma=\{x_0\prec x_1\prec...\prec x_n\}$ and $\delta=\{x_1\prec...\prec x_n\prec x_{n+1}\}$ in $A$ each extend to $n$-simplices $\overline{\gamma}$ and $\overline{\delta}$ in $A$; i.e., transitive linear suborders of $A$, the relation $\gamma\prec_n \delta$ may be viewed as {\it joining the maximal face of the ``initial" simplex $\overline{\gamma}$ to the minimal face of the ``terminal" simplex $\overline{\delta}$.}  This viewpoint motivates the definition of alternative, ``sparser" power spaces over $A$ than those defined in terms of the higher induced relations described above. These {\bf simplicial power spaces} are defined by recognizing only relations between pairs of $n$-element subsets of $A$ that define simplices in $A$.   The ``nontrivial parts" of simplicial power spaces are generalizations of the star model $A_\star$ of $A$, introduced in section \hyperref[subsectiontopology]{\ref{subsectiontopology}} above.  Despite obvious complications, these ideas may be extended in interesting ways to apply to general multidirected sets.\footnotemark\footnotetext{This is one possible way of viewing causal dynamical triangulations.}  Power spaces defined in terms of higher induced relations, and simplicial power spaces, are both examples of {\bf binary power spaces,} which involve only {\it directed,} rather than multidirected, structures on power sets.

\refstepcounter{textlabels}\label{splicerel}

Higher induced relations, and the binary relations defining simplicial power spaces, are special cases of {\bf splice relations}, in which pairs of chains, or more complicated ``generalized elements" associated with a multidirected set $M$, are taken to be related if they may be ``spliced together" in some specified way to form a ``generalized element of greater temporal size."   These ``generalized elements" may often be formalized in terms of {\it spaces of morphisms into} $M$, such as the {\it causal path spaces} introduced in section \hyperref[subsectionpathspaces]{\ref{subsectionpathspaces}} below.   These spaces are generally {\it not} power spaces, since morphisms into $M$ generally involve information that cannot be recovered from the images of their element maps.  Moreover, since there is generally more than one way to ``splice together" a given pair of ``generalized elements," splice relations generally define multidirected, rather than merely directed, structures.   The splice relations of greatest interest in this paper are {\it not} higher induced relations, nor relations defining simplicial power spaces, which relate pairs of chains or simplices overlapping in increasingly long subchains or subsimplices.   More useful are splice relations relating pairs of ``generalized elements of small or empty intersection,"  such as the {\it concatenation relation} and the {\it directed product relation} for pairs of paths, introduced in section \hyperref[subsectionpathspaces]{\ref{subsectionpathspaces}}.   These relations define causal path spaces called the {\it causal concatenation space} and the {\it causal directed product space,} respectively.

\newpage

\refstepcounter{textlabels}\label{causalatoms}

{\bf Causal Atoms; Causal Atomic Resolution.}  A useful application of induced power spaces is to {\it organize information in a multidirected set} into families of subobjects, for the purposes of approximation and computation.  Here I present a method of using {\it nested families of subobjects} of a multidirected set to approximate structure at different scales.  Such approximation methods are important for practical purposes in the study of fundamental spacetime structure, since the scale of individual elements may be many orders of magnitude too small to be directly accessible to experimental observation in the short term. The most interesting such approximation methods are those that {\it organize information in a manner compatible with multidirected structure.}   Here, I briefly discuss one such method, in which elements of a multidirected set are grouped into subsets, called {\it causal atoms,} or more generally, {\it local causal atoms,} which may be viewed as elements of a ``coarser" multidirected set.   The reader should compare this to the {\it decimation} approach to {\it coarse-graining} of causal sets, appearing, for example, in \cite{SorkinEvidenceofContLimVer208}.  The computational utility of causal atoms has already been illustrated in the technique of {\it atomic accretion} used in the proof of theorem \hyperref[theoremcausalsetrelint]{\ref{theoremcausalsetrelint}} above. 

A {\bf causal atom} $\alpha$ in a multidirected set $M=(M,R,i,t)$ is a {\it convex subset} of $M$ in a generalized order-theoretic sense, satisfying the property that every element lying on a chain in $M$ between a pair of elements in $\alpha$ also belongs to $\alpha$.  This means that every element in the complement $M-\alpha$ is either exclusively in the past of $\alpha$, exclusively in the future of $\alpha$, or unrelated to any element of $\alpha$.  In particular, $\alpha$ {\it absorbs} any cycle it intersects.  Hence, the possible relationships between a causal atom and an element in its complement are analogous to the possible relationships between a pair of elements in an acyclic directed set.  Causal atoms may therefore be viewed as ``generalized elements," which explains the choice of terminology.  To temper the absorption property in multidirected sets involving large cycles, one may define {\bf local causal atoms,} which include all elements lying on {\it sufficiently short} chains between pairs of their elements.  I do not explore these details here, however.   Causal atoms may be endowed with internal multidirected structure by elevating them to full subobjects of $M$.  

\refstepcounter{textlabels}\label{atomictop}

Relationships between pairs of causal atoms in a multidirected set $M=(M,R,i,t)$ are generally more complicated than relationships between pairs of elements of $M$, or between a causal atom in $M$ and an element of $M$.  If $M$ is acyclic, its individual elements are causal atoms, but elements lying on cycles are ``smaller than any causal atom,"  due to the absorption property.  The open interval $\llangle x,z\rrangle$ between two elements $x$ and $z$ in $M$ is a causal atom, since every chain between two elements in $\llangle x,z\rrangle$ extends to a chain from $x$ to $z$.  Antichains are causal atoms, since they admit no chains between pairs of their elements.  Maximal cycles are causal atoms, since any element belonging to both the past and future of a cycle lies on an intersecting cycle. The family of causal atoms in $M$ is closed under intersection, so appropriate subfamilies can serve as {\it bases,} rather than merely {\it subbases,} for topologies on the underlying set of $M$, called {\bf atomic topologies}.  Intersections of open intervals are causal atoms, so the interval topology is an atomic topology.  The discrete topology is atomic only in the acyclic case.  Many other examples of atomic topologies exist.   The star topology on the star model $A_\star$ of an acyclic directed set $A$ is an atomic topology for the directed structure on $A_\star$ described in section \hyperref[subsectionrelation]{\ref{subsectionrelation}} above.   In the general case, however, the star $\tn{St}(x)$ at an element $x$ in $M$ may be ``smaller than any causal atom containing $x$."  {\bf Local atomic topologies} may be defined in terms of local causal atoms.   

\refstepcounter{textlabels}
\label{causalatomicdec}

A {\bf causal atomic decomposition} of a multidirected set $M=(M,R,i,t)$ is a partitioning of the underlying set of $M$ into causal atoms.  In this context, it is sometimes useful to relabel $M$ as $M^0$, while the causal atoms in the decomposition are viewed as elements of a ``coarser" set $M^1$.  This set may be endowed with multidirected structure in a variety of different ways.  In particular, $M^1$ may have ``higher-level cycles," even though every cycle in $M$ is absorbed into a unique causal atom.  For example, if a pair of causal atoms $\alpha$ and $\alpha'$ in $M$ each have elements in the past of the other, relations may be defined in both directions between $\alpha$ and $\alpha'$ in $M^1$.   {\bf Local causal atomic decompositions} may also be defined.   Multidirected structures arising from either type of decomposition define power spaces over $M$, since causal atoms in $M$ correspond uniquely to subsets of $M$.  

\begin{figure}[H]
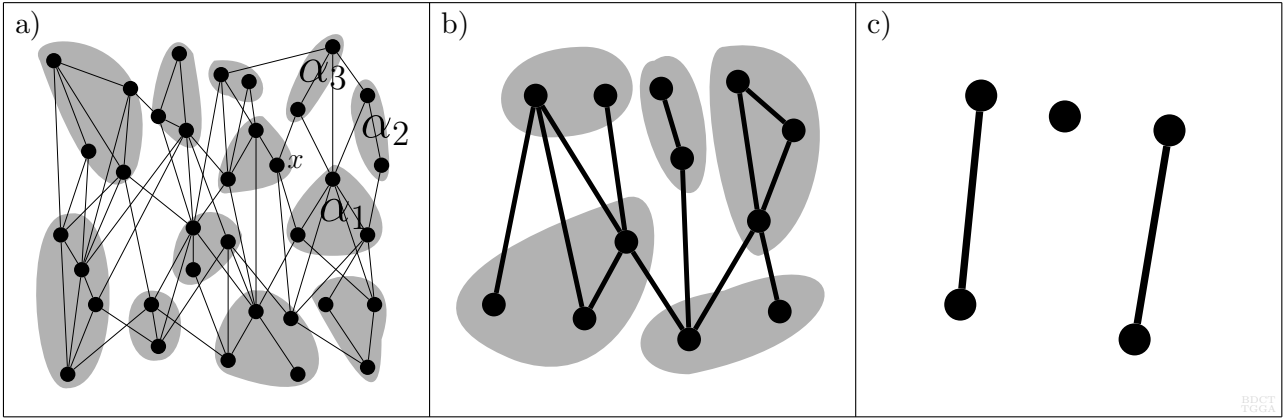


\caption{a) Grouping elements of a multidirected set $M=M^0$ into causal atoms; b) causal atomic decomposition $M^1$, with elements grouped for further decomposition; c) second decomposition $M^2$.}
\label{atomic}
\end{figure}

A causal atomic decomposition of a multidirected set $M=M^0$ is illustrated in figure \hyperref[atomic]{\ref{atomic}}a below, with the elements represented by the nodes in each grey region grouped together as causal atoms.  Figure \hyperref[atomic]{\ref{atomic}}b illustrates one possible method of assigning a multidirected structure to the set $M^1$ of causal atoms in this decomposition.  In this case, the chosen structure is a binary relation, defined as follows: $\alpha\prec\alpha'$ in $M^1$ if and only if 1) every element of $\alpha$ precedes some element of $\alpha'$ in $M$; 2) no element of $\alpha$ succeeds any element of $\alpha'$ in $M$; and 3) some element of $\alpha$ {\it directly} precedes some element of $\alpha'$ in $M$.  This is a very ``sparse" choice of structure for $M^1$.  The further grouping of elements of $M^1$ indicated by the grey regions in figure \hyperref[atomic]{\ref{atomic}}b represents a second causal atomic decomposition, as discussed below. 

\vspace*{-.5cm}

An important classical case of causal atomic decomposition is the {\it foliation of a spacetime region by Cauchy surfaces,} as mentioned in section \hyperref[subsectionrelation]{\ref{subsectionrelation}} above. The analogue of such foliation in the discrete causal context may be called {\bf generational dynamics}, in contrast to the sequential growth dynamics introduced by Sorkin and Rideout \cite{SorkinSequentialGrowthDynamics99}.  I briefly return to the topic of generational dynamics in the context of kinematic schemes in section \hyperref[subsectionkinematicschemes]{\ref{subsectionkinematicschemes}} below.  A simple example of {\it local} causal atomic decomposition in the cyclic interpolative case is the partitioning of a directed circle into a pair of half-open arcs.  

\refstepcounter{textlabels}\label{causalatomicres}

A {\bf causal atomic resolution} of a multidirected set $M=(M,R,i,t)$ is a sequence 
\[\mbf{M}:=\{M=M^0,M^1,M^2,...\}\]
of multidirected sets, where $M^{n+1}$ is a causal atomic decomposition of $M^n$ for all $n\ge0$.  The multidirected set $M^n$ is called the $n$th {\it level} of the causal atomic resolution $\mbf{M}$.  Note that causal atomic decomposition is generally {\it not} a transitive operation, meaning that $M^{n+2}$ is generally not a causal atomic decomposition of $M^n$.  More precisely, if $\alpha_1,...,\alpha_k$ are causal atoms in $M^n$, viewed as elements of $M^{n+1}$, such that $\{\alpha_1,...,\alpha_k\}$ is a causal atom in $M^{n+1}$, the union of the elements of $\alpha_1,...,\alpha_k$ is generally not a causal atom in $M^n$.  For example, the element $x$ in figure \hyperref[atomic]{\ref{atomic}}a above is in the intersection of the past and future of the set $\alpha_1\cup \alpha_2\cup\alpha_3$, which is therefore not a causal atom in $M=M^0$.  However, $\{\alpha_1,\alpha_2,\alpha_3\}$ {\it is} a causal atom in $M^1$.   This is yet another example in which a notion analogous to category-theoretic composition fails to apply in discrete causal theory.\footnotemark\footnotetext{More generally, failure of ``composition" is not unusual in "information-filtering constructions."  A familiar example is given by {\it composition series} of finite groups in the context of the Jordan-H\"{o}lder theorem, since formation of normal subgroups is nontransitive.}   

There are a number of different ways to characterize a causal atomic resolution $\mbf{M}$ in terms of the properties of its levels $M^n$, and the relationships among them.  For example, the $n$th {\it growth factor} $\lambda_n$ of $\mbf{M}$ is the average number of elements of $M^{n-1}$ making up an element of $M^{n}$.  If the sequence $\Lambda(\mbf{M}):=\{\lambda_1,\lambda_2,...\}$ of growth factors of $\mbf{M}$ is constant over a certain range of indices, then $\mbf{M}$ is called {\it exponential} over this range, since the sizes of typical atoms at different levels within this range are related by a power of the common growth factor.  Similarly, if the sequence $\Lambda(\mbf{M})$ is decreasing or increasing over a certain range of indices, then $\mbf{M}$ is called {\it subexponential} or {\it superexponential} over this range, respectively.  The growth factors of the causal atomic resolution illustrated in figure \hyperref[atomic]{\ref{atomic}} above are $\lambda_1=35/12\approx 2.92$ and $\lambda_2=12/5=2.4$, so this particular resolution is subexponential.  One may also characterize the {\it uniformity} of a given level $M^n$ in a number of obvious ways.  The method of causal atomic resolution is similar in spirit to {\it multiresolution analysis} in the study of {\it wavelets} in the field of {\it harmonic analysis.}  The general aim of such methods is to provide a unified way of dealing with aspects of structure existing at different scales.  This viewpoint is of particularly interest in physical theories where emergent phenomena are prominent, since emergence is often scale-dependent.  


\refstepcounter{textlabels}\label{classicalhol}
\refstepcounter{textlabels}\label{ellis}

{\bf Top-Down Causation; Classical Holism.} A number of physicists, computer scientists, chemists, biologists, neuroscientists, and philosophers have recently devoted attention to the concept of classical {\bf top-down causation}, which challenges the philosophy of {\bf classical reductionism}.  Classical reductionism assumes that classical causal structure may be completely described in terms of relations between pairs of individual events, as implied by the binary axiom (\hyperref[b]{B}) of causal set theory.  Representative samplings of the top-down viewpoint appear in the recent collections {\it Top-Down Causation} \cite{EllisNobleOConnor12}, and {\it Downward Causation and the Neurobiology of Free Will} \cite{MurphyEllisOConnor09}.  As George Ellis \cite{EllisNature05} puts it, {\it ``the higher levels in the hierarchy of complexity have autonomous causal powers that are functionally independent of lower-level processes."} Depending on its specific version, top-down causation either approaches or crosses the threshold of {\bf classical holism}, the antithesis of the reductionist philosophy.  To evaluate the legitimacy and implications of top-down causation, it is necessary to formulate the idea in precise mathematical terms.  Since top-down causation involves influences exerted by entire families of events, the theory of power spaces is a natural tool to use.\footnotemark\footnotetext{Other alternatives exist, even in the context of discrete causal theory.  In particular, even the most general power spaces preserve a certain {\it binary} character, by focusing on relationships {\it between pairs} of families of elements, rather than relationships {\it among families} of families of elements.  There are excellent physical motivations for this focus, but one may choose to abstain from it.  I briefly mention some alternatives in section \hyperref[subsectionomitted]{\ref{subsectionomitted}} below.} Different implementations are possible, some conservative, others more radical.  

\refstepcounter{textlabels}\label{degreeshol}

A conservative example of the top-down viewpoint involving power spaces appears in the method of causal atomic resolution, where influences are ascribed to families of events purely for convenience.  The power spaces arising in this context merely represent ways of organizing and filtering information; they do not involve any profound hypotheses about the nature of the universe.  In particular, the ``approximate" multidirected structures given by causal atomic decompositions may {\it innocuously} exhibit properties that would arouse controversy in more fundamental contexts.  For example, two causal atoms may be ``related to each other," in the sense that each contains events in the other's past, yet this is irrelevant to the controversial issue of  classical causal cycles.   Similarly, one causal atom may be related to another, even though some of the corresponding events are {\it unrelated,} yet this is irrelevant to the controversial issue of classical nonlocal interaction.   

More interesting are cases in which relationships involving complex families of events are considered ``exact," rather than mere approximations.  The corresponding power spaces may still be induced by relations between pairs of individual elements, but there may be alternative families of relations that better represent the information contained in these power spaces.   Choosing such a family is analogous to choosing a ``natural basis" from among a linearly dependent set of vectors.  For example, the relation space $\ms{R}(M)$ over a multidirected set $M$, together with its induced relation $\prec$, gives an alternative representation of the information contained in $M$, except at the boundary $\partial M$ of $M$.  The resolution of the impermeability problem for maximal antichains upon passage to $\ms{R}(M)$, proven in theorem \hyperref[theoremrelimpermeable]{\ref{theoremrelimpermeable}} above, illustrates that advantages may sometimes be gained by working in terms of one representation rather than another.  This example demonstrates that the reductionist viewpoint is not always the most useful one, even when reductionism is information-theoretically adequate.  The argument that ``emergence" and reductionism are not mutually exclusive has been recognized for some time; see, for example, the classic paper by Anderson \cite{AndersonMoreisDifferent72}, or the recent paper by Butterfield \cite{ButterfieldReductionism11}.

\refstepcounter{textlabels}\label{twistor}

A classical alternative representation of ``spacetime information" is given by the {\it twistor space} $T$ over Minkowski spacetime $X$ in Roger Penrose's {\it twistor theory}, introduced in his 1967 paper {\it Twistor Algebra} \cite{PenroseTwistor67}.  A useful later reference for twistor theory is Ward and Wells' standard text {\it Twistor Geometry and Field Theory} \cite{WardWellsTwistor90}.  Twistor space $T$ may be identified with a five-dimensional real hypersurface in the complex projective space $\CC\PP^3$, which is six-dimensional as a real manifold.   Elements of $T$ correspond to {\it null lines} in $X$, physically interpreted as the spacetime paths of rays of light.  Interesting analogies exist between twistor space and the relation space $\ms{R}(M)$ over a multidirected set $M=(M,R,i,t)$.  For example, at a given element $x$ in $M$, the relations incident at $x$ are may be viewed as {\it minimal null segments}\footnotemark\footnotetext{In causal set theory, {\it irreducible} relations; i.e., ``links," are treated as ``null segments;" see, for example, Sorkin's paper \cite{SorkinLightLinksCausalSets09}.  In the present case, this structural analogy is not intended to fix a rigid ``geometric" interpretation of either irreducible or reducible relations.} at $x$.  One of the most attractive features of twistor theory is that it permits the conversion of certain important but intractable equations over classical spacetime into more natural and manageable forms.  In an analogous way, passage to relation space allows a natural and straightforward derivation of {\it causal Schr\"{o}dinger-type equations} in section \hyperref[subsectionquantumcausal]{\ref{subsectionquantumcausal}} of this paper.

\refstepcounter{textlabels}\label{shapedynamics}

The most radical versions of top-down causation involve holistic power spaces, which contain more information than can be naturally encoded in any multidirected structure on their underlying sets.  Such {\it irreducible holism} at the classical level is broadly agreed to be at least {\it strongly constrained} by well-supported principles such as classical locality, but this view is not universal.   A modern theory incorporating a radical degree of irreducible holism is Julian Barbour's {\it shape dynamics} \cite{BarbourShape12}.  Shape dynamics does not fall under the umbrella of causal theory, taking a nearly opposite, ``space-first" approach to fundamental spacetime structure, in which {\it ``one \tn{[spatial]} state of the universe succeeds another along a unique evolution curve"} in {\it shape space,} the {\it conformal superspace of all conformal $3$-geometries.}\footnotemark\footnotetext{This quotation is taken from Barbour, Koslowsi, and Mercati \cite{BarbourShape13}. Conformal superspace is only one possible choice of a shape space; for example, one may define {\it discrete shape spaces.}} Underlying shape dynamics is the philosophy of Ernst Mach (1838-1916), that {\it global structure determines local effects.}\footnotemark\footnotetext{Mach's ideas influenced Einstein in the development of general relativity.  ``Mach's principle," under various interpretations, has since become a ``strange attractor" for dubious theorizing.  Carlo Rovelli demonstrates the disunity of modern Machian ideas in an amusing way by listing {\it eight} different versions of ``Mach's principle" on page 76 of his book {\it Quantum Gravity} \cite{Rovelli04}.}  Shape dynamics represents the utmost extreme in classical holism: every aspect of each ``state of the universe" is relevant to all subsequent behavior.  Among other ``eyebrow-raising" implications, this entails a {\it universal time parameter} and {\it absolute simultaneity.}   Shape dynamics also favors a {\it spatially closed universe,} disfavored by current cosmological observations.  Despite this, the theory succeeds in reproducing certain results of the {\it ADM formulation} of general relativity. 


\subsection{Causal Path Spaces}\label{subsectionpathspaces}


\refstepcounter{textlabels}\label{causalpaths}

{\bf Causal Paths.}  A {\it directed path} in a multidirected set $M=(M,R,i,t)$ is a {\it morphism from a linear directed set into} $M$.  A relation in $M$ may be regarded as a directed path of length one, and an $n$-chain in $M$ may be regarded as a directed path of length $n$, whose source is irreducible.   In the special case where $M$ is a directed set, viewed as a model of classical spacetime structure, directed paths are {\it causal paths,} in the sense that an event represented by an element $x$ in $M$ influences an event represented by an element $y$ in $M$ if and only if there is a directed path from $x$ to $y$ in $M$.   Such causal paths differ from causal curves in general relativity in two important ways.  First, following the causal metric hypothesis (\hyperref[cmh]{CMH}), causal paths represent {\it actual influence,} not merely {\it potential influence.}   In the language of section \hyperref[naturalphilosophy]{\ref{naturalphilosophy}} above, causal paths are {\it descriptive,} rather than {\it prescriptive.}  Second, whereas causal curves in general relativity are equivalence classes of paths with the same continuum source, there exist an immense variety of distinct types of linear directed sets that can serve as sources for causal paths in discrete causal theory.  In this paper, I focus almost entirely on the case of chains; i.e., causal paths with irreducible sources.  The physical intuition associated with causal paths in directed sets is useful for multidirected sets in general.   For this reason, I refer to a number of general constructions involving directed paths in multidirected sets as {\it causal.}  For directed paths themselves, I usually abbreviate and simply use the term {\it paths,} since ``undirected paths" play no role in this paper.

{\it Causal path sets} associated with a multidirected set $M=(M,R,i,t)$ are natural generalizations of the relation set $R$ of $M$.  The recognition of natural directed structures on causal path sets, supplied by {\it splice relations,} such as the {\it concatenation relation} and the {\it directed product relation,} elevates these sets to {\it causal path spaces,} analogous to the relation space $\ms{R}(M)$ over $M$.   This directed structure may be defined algebraically, in terms of {\it splice products,} and the corresponding causal path spaces may be viewed algebraically, as {\it causal path semicategories.}  These, in turn, supply the multiplicative structure of {\it causal path algebras,} which are crucial for converting the physical information represented by a multidirected set into a mathematically tractable form.  Multidirected, rather than merely directed, structures, may also be defined on causal path sets, but these are not explored in this paper.  More generally, it is not feasible here to replicate, for the various spaces and algebraic objects introduced in this section, the careful examination of functorial, information-theoretic, and structural properties carried out for the relation space $\ms{R}(M)$ in section \hyperref[subsectionrelation]{\ref{subsectionrelation}} above.


\refstepcounter{textlabels}\label{pathspacephys}

{\bf General Contextual Remarks.} The use of path spaces in mathematics and theoretical physics goes back too far, and originates from too many independent sources, for a historical synopsis to be of any use to the reader.  I therefore limit explicit reference to modern sources of more direct relevance to discrete causal theory.   However, it is worthwhile, for contextual purposes, to briefly mention some of the conventional uses of path spaces.  In mathematics, path spaces form the basis for {\it homotopy theory}\footnotemark\footnotetext{That is, {\it ordinary} homotopy theory.  Other variants of the same idea are possible, such as Voevodsky's theory of $\AA^1$-{\it homotopy} in algebraic geometry.  {\it Causal homotopy} is an interesting topic, but I do not pursue it here.} in algebraic topology, via the {\it fundamental groupoid} and its higher-dimensional analogues.   Path spaces also feature prominently in the theory of {\it topological groups,} and particularly {\it Lie groups,} due to the importance of {\it one-parameter subgroups.}  In virtually every area of theoretical physics based on continuum geometry, path spaces are central.  At the simplest level, they represent trajectories of point particles.  For more complex systems, they appear in the theory of {\it phase space,} with particularly iconic status in the field of {\it nonlinear dynamics.}  Perhaps most importantly, they underlie the {\it Lagrangian} and {\it Hamiltonian formulations} of classical mechanics, extend to analogous methods in quantum theory, and culminate in the {\it path integral approach} to quantum mechanics and quantum field theory.  Path spaces find additional applications in string theory, due to the one-dimensional nature of strings, and in loop quantum gravity, due to the use of {\it holonomy} in the construction of the {\it loop representation.}  However, the {\it directedness} inherent in causal structure makes path spaces especially important in the context of causal theory. 


\refstepcounter{textlabels}\label{pathmorph}
\refstepcounter{textlabels}\label{linsource}

{\bf Causal Path Sets.}  Before giving formal definitions of paths and causal path sets, I briefly discuss the source objects of paths in causal theory, which are {\it linear directed sets.}  Linear directed sets are generally nontransitive analogues of the linearly ordered sets discussed in section \hyperref[settheoretic]{\ref{settheoretic}} above.   In this section, I often suppress binary relations, relation sets, etc., to avoid notational clutter.  A directed set $L$ is {\bf linear} if it is acyclic, and if every pair of elements of $L$ is connected by a chain.  The {\bf length} of a finite linear directed set $L$ is defined to be the chain length of its skeleton $\tn{sk}(L)$ in the sense of part 2 of definition \hyperref[definitionchains]{\ref{definitionchains}} above; i.e., its maximal number of consecutive {\it relations}.  Infinite linear directed sets may be ``longer" and ``more complicated" than any acyclic chain.  In particular, every acyclic chain may be represented as a relative directed set over $\ZZ$, but a linear directed set may a ``too large" to be represented in this way.  For example, the directed set I refer to as {\it Jacob's ladder,} illustrated in figure \hyperref[jacobs]{\ref{jacobs}}c above, is linear, but does not admit a bijective morphism into any linear suborder of $\ZZ$.   The analogue of length for an infinite linear directed set $L$ is the {\it isomorphism class of its transitive closure} $\tn{tr}(L)$ as a linearly ordered set. \\

\refstepcounter{textlabels}\label{pathsets}

\begin{defi}\label{defipaths} Let $M=(M,R,i,t)$ be a multidirected set. 
\begin{enumerate}
\item A {\bf (directed) path} in $M$ is a morphism $\gamma:L\rightarrow M$ for some linear directed set $L$, called the {\bf source} of $\gamma$, considered up to isomorphism.  A  path may also be called a {\bf causal path}.
\item If the source $L$ of a path $\gamma$ is finite, the {\bf length} of $\gamma$ is defined to be the length of $L$.  If $L$ is infinite, the isomorphism class of $\tn{tr}(L)$ supersedes the usual notion of length.
\item The {\bf causal path set} $\Gamma(M)$ of $M$ is the set whose elements are paths in $M$. 
\item A path $\gamma:L\rightarrow M$ is called {\bf irreducible} if its source $L$ is irreducible.  The {\bf irreducible path set} $\Gamma_{\tn{\fsz{irr}}}(M)$ of $M$ is the subset of $\Gamma(M)$ whose elements are irreducible paths.
\end{enumerate}
\end{defi}

The image of an irreducible path $\gamma:L\rightarrow M$ is a chain in $M$, {\it not} necessarily irreducible, since ``pairs of elements may acquire new relations between them in the image."   Every chain in $M$ may be identified with an appropriate irreducible path in $M$, so the chain set $\tn{Ch}(M)$ of $M$ may be identified with the irreducible path set $\Gamma_{\tn{\fsz{irr}}}(M)$ of $M$.   Supplying appropriate directed structure to $\Gamma(M)$ and $\tn{Ch}(M)$, respectively, as described below, elevates these sets to {\it causal path spaces} and {\it chain spaces over} $M$, respectively.   It is convenient in many cases to define this structure {\it algebraically,} in terms of {\it products of paths.}   One of many ways to generalize the above definitions is to consider sets whose elements are paths {\it together with} distinguished subpaths of length zero, physically representing choices of {\it present} in linear sequences of events.  Such sets are structurally analogous to {\it flag varieties} in algebraic geometry.   Extending the analogy between causal theory and twistor theory mentioned in section \hyperref[subsectionpowerset]{\ref{subsectionpowerset}} above, such ``enriched causal path sets" may be used to define {\it correspondences,} analogous to those described in \cite{WardWellsTwistor90}.   However, besides a footnote in section \hyperref[subsectionkinematicschemes]{\ref{subsectionkinematicschemes}} mentioning an analogous ``higher-level" construction in the context of kinematic schemes, I do not explore enriched causal path sets further in this paper.  



\refstepcounter{textlabels}\label{pathprod}

\begin{figure}[H]
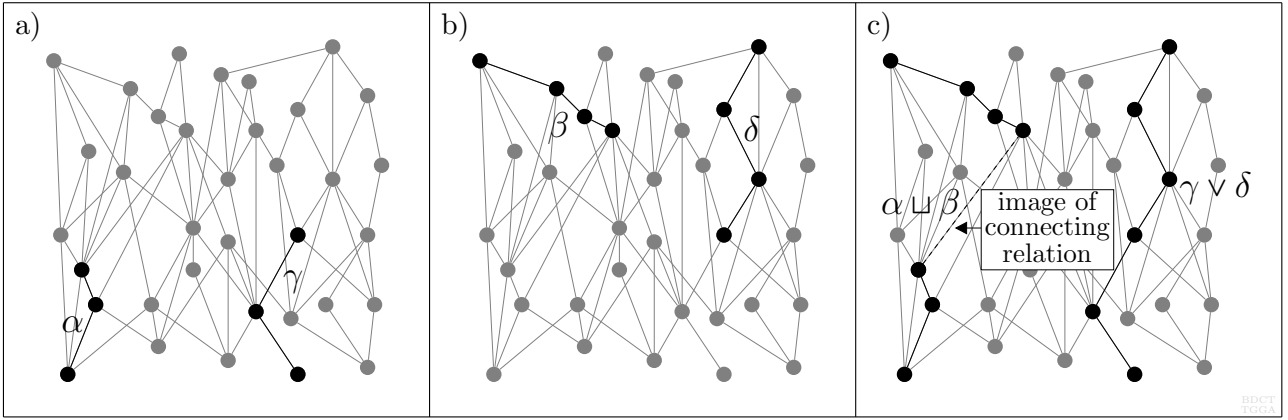


\caption{a) Paths $\alpha$ and $\gamma$ in a directed or multidirected set; b) paths $\beta$ and $\delta$; c) concatenation product $\alpha\sqcup\beta$ and directed product $\gamma\vee\delta$.}
\label{pathalgebras}
\end{figure}

{\bf Products of Paths; Causal Path Spaces.}  A wide variety of algebraic operations may be defined on causal path sets over directed sets and multidirected sets.  Two of the most important are special cases of {\it splice products,} which are {\it partially defined binary operations} formalizing different notions of ``joining morphisms together end-to-end in a temporal sense."   Russling \cite{RusslingConcatenation95} discusses essentially equivalent notions in the special case of chains, and applies these constructions to problems in algorithmic graph theory.  These operations define natural directed structures on causal path sets, elevating them to {\it causal path spaces.}  The first such operation, called the {\it concatenation product,} applies to pairs of paths in a directed set $(D,\prec)$, for which the terminal element in the image of the first path {\it directly precedes} the initial element in the image of the second path.   The images of such a pair of paths $\alpha$ and $\beta$ are illustrated in figures \hyperref[pathalgebras]{\ref{pathalgebras}}a and \hyperref[pathalgebras]{\ref{pathalgebras}}b below. 

\refstepcounter{textlabels}\label{concatprod}
\vspace*{.2cm}
\begin{defi}\label{deficoncatprod} Let $\alpha:L\rightarrow D$ and $\beta:L'\rightarrow D$ be paths in a directed set $D=(D,\prec)$.  Suppose that $L$ has a maximal element $\ell_+$, and $L'$ has a minimal element $\ell_-'$, such that $\alpha(\ell_+)\prec\beta(\ell_-')$ in $D$.
\begin{enumerate}
\item Denote by $L\sqcup L'$ the linear directed set whose underlying set is the disjoint union $L\coprod L'$, and whose binary relation is given by augmenting the union of the binary relations on $L$ and $L'$ with the single relation $\ell_+\prec\ell_-'$, called the {\bf connecting relation}.  Then the {\bf concatenation product} $\alpha\sqcup\beta: L\sqcup L' \rightarrow D$ of $\alpha$ and $\beta$ is defined by setting
\[\alpha\sqcup\beta(\ell)=\begin{cases} \alpha(\ell) & \tn{if } \ell\in L\\ \beta(\ell) &\tn{if } \ell\in L' \end{cases}.\]
\item The {\bf concatenation relation} $\prec_\sqcup$ on the causal path set $\Gamma(D)$ of $D$ is the binary relation given by setting $\alpha\prec_\sqcup\beta$ if and only if $\alpha\sqcup\beta$ exists.  The {\bf causal concatenation space} over $(D,\prec)$ is the directed set $\big(\Gamma(D),\prec_\sqcup\big)$.
\end{enumerate}
\end{defi}

\vspace*{-.5cm}

The concatenation product is uniquely defined only for directed sets.  In a multidirected set, there may be multiple connecting relations between $\alpha(\ell_+)$ and $\beta(\ell_-')$.  Unlike the concatenation product, the concatenation {\it relation} $\prec_\sqcup$ may be generalized in an obvious way to yield a multidirected structure on the causal path set over a multidirected set, but I do not pursue this generalization here.  The notation $\sqcup$ for the concatenation product is deliberately chosen to evoke the disjoint union $\coprod$ of sets.  The concatenation relation $\prec_\sqcup$ on $\Gamma(D)$ extends the binary relation $\prec$ on $D$.  Hence, $(D,\prec)$ embeds as a full subobject into its causal concatenation space.   

The second operation on causal path sets, called {\it directed product,} applies to pairs of paths in a multidirected set $M=(M,R,i,t)$, for which the terminal element in the image of the first path {\it coincides with} the initial element in the image of the second path. The images of such a pair of paths $\gamma$ and $\delta$ are illustrated in figures \hyperref[pathalgebras]{\ref{pathalgebras}}a and \hyperref[pathalgebras]{\ref{pathalgebras}}b above.\\

\refstepcounter{textlabels}\label{dirprod}

\begin{defi}\label{defidirprod} Let $\alpha:L\rightarrow M$ and $\beta:L'\rightarrow M$ be paths in a multidirected set $M=(M,R,i,t)$.  Suppose that $L$ has a terminal element $\ell_+$, and $L'$ has an initial element $\ell_-'$, such that $\alpha(\ell_+)=\beta(\ell_-')$ in $M$. 
\begin{enumerate}
\item Denote by $L\vee L'$ the linear directed set whose underlying set is the quotient space of the disjoint union $L\coprod L'$, given by identifying $\ell_+$ and $\ell_-'$, and whose binary relation is the union of the binary relations on $L$ and $L'$.  Then the {\bf directed product} $\alpha\vee\beta: L\vee L' \rightarrow M$ of $\alpha$ and $\beta$ is defined by setting
\[\alpha\vee\beta(\ell)=\begin{cases} \alpha(\ell) & \tn{if } \ell\in L \\ \beta(\ell) &\tn{if } \ell\in L' \end{cases}.\]
\item The {\bf directed product relation} $\prec_\vee$ on the causal path set $\Gamma(M)$ of $M$ is the binary relation given by setting $\alpha\prec_\vee\beta$ if and only if $\alpha\vee\beta$ exists.  The {\bf causal directed product space} over $M$ is the directed set $\big(\Gamma(M),\prec_\vee\big)$.
\end{enumerate}
\end{defi}

The notation $\vee$ for the directed product is deliberately chosen to evoke the {\it wedge sum} $\bigvee$, familiar from topology, which joins topological spaces at distinguished basepoints.  The directed product relation $\prec_\vee$ on $\Gamma(M)$ extends the induced relation on the relation space $\ms{R}(M)$ over $M$.  Hence, $\ms{R}(M)$ embeds as a full subobject into the causal directed product space over $M$.  

\refstepcounter{textlabels}\label{spliceprod}

The causal concatenation space and the causal directed product space are prototypical examples of {\bf causal path spaces}, which are causal path sets over a directed or multidirected set equipped with natural directed or multidirected structures.  The concatenation product $\sqcup$ and the directed product $\vee$ are the simplest examples of {\bf splice products}, defined by ``splicing together morphisms."  The corresponding relations $\prec_\sqcup$ and $\prec_\vee$ are the simplest examples of {\it splice relations,} already introduced in section \hyperref[subsectionpowerset]{\ref{subsectionpowerset}} above.  


\refstepcounter{textlabels}\label{catsemicat}

{\bf Categories and Semicategories.}  I now introduce another analogy between causal theory and category theory, similar to those already cited in the fifth argument against transitivity in section \hyperref[subsectiontransitivitydeficient]{\ref{subsectiontransitivitydeficient}}, and the discussion of morphism categories near the end of section \hyperref[subsectionrelation]{\ref{subsectionrelation}}. These analogies arise essentially because both causal theory and category theory involve notions of {\it multidirectedness}, {\it preservation of structure,} and {\it hierarchy.}   Category theory was first developed in the 1940's by Eilenberg and MacLane, to study relationships among different {\it cohomology theories}, which are special families of functors, often from a category of topological spaces to a category of groups.  In this context, additional notions such as {\it composition,} {\it associativity,} and {\it existence of identities} are natural, but these notions are perhaps less relevant to applications of directed and multidirected structure in theoretical physics.\footnotemark\footnotetext{Note the contrast with Chris Isham's viewpoint; Isham {\it adds additional properties} to categories to obtain {\it topoi,} whereas I emphasize certain category-theoretic properties, and de-emphasize others.  Of course, Isham works with topoi of many different kinds, not only those whose objects are intended to model causal structure.}  For example, elements and relations in a multidirected set may be compared to objects and morphisms of a category, respectively, but there are important qualitative differences in their formal properties: the ``composition" of relations is generally not defined, at least, not as a relation, and associativity makes no sense without composition.  Identity morphisms, and more generally, endomorphisms, correspond to reflexive relations, which are not guaranteed to exist in the general case, and are forbidden in the acyclic case.    

{\it Semicategories,} or {\it non-unital categories,} are similar to categories, but without the guaranteed existence of identity morphisms.  Causal path spaces over directed or multidirected sets, such as the causal concatenation space and the causal directed product space, may be viewed as {\it generalized semicategories,} admitting ``morphisms without sources and/or targets."\footnotemark\footnotetext{Heuristically, one may think of a field in which field lines either terminate on charges or extend to infinity.}   More precisely, a {\bf semicategory} consists of a class of objects $\ms{S}$, and a class of morphisms $\Gamma(\ms{S})$, such that every morphism $\gamma$ has an initial object $\mbf{i}(\gamma)$ and a terminal object $\mbf{t}(\gamma)$.   In the context of causal theory, $\ms{S}$ may be viewed as an abstraction of the underlying set of elements of a multidirected set $M=(M,R,i,t)$, while $\Gamma(\ms{S})$ may be viewed as an abstraction of the class of {\it finite chains} in $M$.  However, generalizations of these analogies, involving the entire causal path set $\Gamma(M)$ of $M$, are needed below. The maps $\mbf{i}:\Gamma(\ms{S})\rightarrow \ms{S}$ and $\mbf{t}:\Gamma(\ms{S})\rightarrow \ms{S}$ may be viewed as abstractions of the initial and terminal element maps $i$ and $t$ of $M$, extended to apply to finite chains of arbitrary length, rather than merely relations.  A semicategory $\ms{S}$ also admits {\it associative compositions} of morphisms, defined whenever the terminal object of the first morphism coincides with the initial object of the second.  This composition may be viewed as an abstraction of the directed product of chains.  Note that although the ``composition" of relations is not a {\it relation} in this context, it {\it is} a chain, more specifically, a $2$-chain.  

The analogy between complex chains and morphisms should not be carried too far.  In section \hyperref[subsectionrelation]{\ref{subsectionrelation}} above, I compared $2$-chains to $2$-morphisms, rather than morphisms, and this is the {\it preferred structural analogy for causal theory,} at least in my opinion.  The semicategory viewpoint regarding causal path spaces is closer to the category-theoretic approach to {\it groups}, mentioned in sections  \hyperref[settheoretic]{\ref{settheoretic}} and  \hyperref[subsectiontransitivitydeficient]{\ref{subsectiontransitivitydeficient}} above, in which the elements of a group are viewed as morphisms in a category with a single object.  In this sense, a semicategory is a {\it grouplike structure} whose morphisms are its {\it elements,} and composition is viewed as a {\it partially defined algebraic operation} on these elements.   From this perspective, a semicategory is simply a class together with an associative partially defined operation, generally noncommutative, and without identity or inverses.  Familiar examples of such structures are more common than one might expect.  These include the space of continuum paths in a topological space under the operation of directed product, the class of {\it all} matrices over a given ring under the operation of matrix multiplication, and the class of algebraic cycles on an algebraic variety under the operation of {\it proper} intersection.\footnotemark\footnotetext{In all three of these examples, information may be discarded to obtain a structure that is better-behaved mathematically, but less complete and natural; namely, fundamental groupoids or groups, matrix groups, and Chow rings, respectively.}


\refstepcounter{textlabels}\label{causalpathsemicat}

{\bf Causal Path Semicategories.}  The causal path set $\Gamma(D)$ of a directed set $D=(D\prec)$, equipped with the partially defined operation of concatenation product $\sqcup$, is referred to in this paper as the {\bf concatenation semicategory} over $D$.   Similarly, the causal path set $\Gamma(M)$ of a multidirected set $M=(M,R,i,t)$, equipped with the partially defined operation of directed product $\vee$, is referred to as the {\bf directed product semicategory} over $M$.   Both of these structures, viewed as ``group-like objects," are examples of what I refer to as {\bf causal path semicategories}, which represent the {\it algebraic viewpoint} regarding directed structure on causal path sets, in contrast to the {\it generalized order-theoretic viewpoint} represented by causal path spaces.  The ``if and only if" statements relating $\sqcup$ and $\prec_\sqcup$ in part 2 of definition \hyperref[deficoncatprod]{\ref{deficoncatprod}}, and $\vee$ and $\prec_\vee$ in part 2 of definition \hyperref[defidirprod]{\ref{defidirprod}} above, ensure that the two viewpoints are equivalent in the present context.\footnotemark\footnotetext{Of course, the concatenation relation $\prec_\sqcup$ generalizes to yield a multidirected structure on causal path sets over multidirected sets, while the concatenation product $\sqcup$ is uniquely defined only for directed sets, as explained above.}  Causal path semicategories are {\it generalized semicategories,} whose morphisms, in this case paths, do not necessarily possess initial or terminal objects, and whose composition law, for morphisms $\gamma$ and $\delta$, is not necessarily defined in terms of the condition $\mbf{t}(\gamma)=\mbf{i}(\delta)$.  In particular, different splice products on the same underlying causal path set define different causal path semicategories, with different composition laws.  

Interesting relationships exist among semicategory operations on causal path sets, and functors such as the relation space functor $\ms{R}$ and abstract element space functor $\ms{E}$.  For example, the concatenation semicategory over the relation space $\ms{R}(M)$ over a multidirected set $M=(M,R,i,t)$ corresponds to a subsemicategory of the directed product semicategory over $M$ itself, as illustrated by the following example: if $r$ and $s$ are elements of $\ms{R}(M)$, with initial and terminal elements $i(r)=x$, $t(r)=i(s)=y,$ and $t(s)=z$, in $M$, then the directed product in $\Gamma(M)$ of $r$ and $s$, viewed as paths of length $1$, corresponds to the concatenation product in $\Gamma(\ms{R}(M))$ of $r$ and $s$, viewed as paths of length zero.  This illustrates once again how the relation space functor $\ms{R}$ {\it reduces multidirected structure to directed structure,} since concatenation products are uniquely defined only in the directed case.  The physical importance of {\it both} element space and relation space is the reason for introducing both types of semicategories. 

Since the operations $\sqcup$ and $\vee$ preserve irreducibility of paths, useful {\it subsemicategories} of $(\Gamma(D),\sqcup)$ and $(\Gamma(M),\vee)$ may be defined by restricting attention to the irreducible path sets of $M$ and $D$, which may be identified with the chain sets $\tn{Ch}(D)$ and $\tn{Ch}(M)$, respectively.  I refer to these subsemicategories as the {\bf chain concatenation semicategory} over $D$, and the {\bf chain directed product semicategory} over $M$, respectively.   These {\bf chain semicategories} $\big(\tn{Ch}(D),\sqcup\big)$ and $\big(\tn{Ch}(M),\vee\big)$ belong to a special class of causal path semicategories which may be studied without explicit reference to morphisms.\footnotemark\footnotetext{Another case in which explicit reference to morphisms can be avoided is the case in which all paths under consideration are {\it isomorphisms onto their images,} viewed as subobjects $M$.  However, this case does not behave well algebraically.  In particular, splice products of such paths are generally {\it not} isomorphisms onto their images.}   They sometimes appear in the literature as implicit structural substrata for multiplicative operations in special types of {\it path algebras,} discussed in detail below.  Though I focus mostly on the special case of chains in this paper, there are good reasons to develop the theory in terms of general paths: it makes the relative viewpoint (\hyperref[rv]{RV}) explicit, enables more refined treatment of target objects, and includes paths of transfinite length.    

Special {\it quotient semicategories} also play a significant role in the theory of causal path semicategories.  In this context, quotient semicategories may be viewed concretely as families of {\it conguence classes of paths.}  A {\it congruence relation} $\sim$ on an arbitrary semicategory $(\Gamma,\sqcup)$ is a reflexive, symmetric, transitive binary relation on $\Gamma$, preserving the semicategory operation $\sqcup$, in the sense that for any four elements $\alpha,\beta,\gamma,\delta$ of $\Gamma$, if $\alpha\sim\beta$ and $\gamma\sim\delta$, then $\alpha\sqcup\gamma\sim\beta\sqcup\delta$.   Important congruence relations on $(\Gamma(D),\sqcup)$ and $(\Gamma(M),\vee)$ are given by taking two paths to be congruent if and only if their images share the same initial and terminal elements.  Such congruence relations appear in the theory of {\it co-relative histories,} developed in section \hyperref[subsectionquantumcausal]{\ref{subsectionquantumcausal}} below, where they identify ``congruent families of kinematic accounts of evolutionary processes."


\refstepcounter{textlabels}\label{causalpathalg}

{\bf Causal Path Algebras.}  Causal path semicategories are ``primitive" algebraic objects, in the sense that they involve partially defined operations that fail to satisfy a number of familiar algebraic properties.  Richer objects, called {\it causal path algebras,} may be constructed from causal path semicategories, by incorporating the structure of a ring.  An analogy from elementary algebra is useful to motivate this construction.  Multiplication of monomials with real coefficients involves {\it two} distinct operations: multiplication of coefficients in $\RR$, and addition of exponents in $\NN$.  Thus, the additive structure of $\NN$ serves as a {\it substratum for multiplication} in the polynomial algebra $\RR[x]$.  More generally, {\it group} and {\it semigroup algebras} combine the structure of a ring, analogous to $\RR$, with the structure of a group or semigroup, analogous to $\NN$, which participates only in the multiplicative operation of the resulting algebra.  Still more general are {\it semicategory algebras}, which derive a portion of their multiplicative structure from a semicategory operation.  

Two important types of (generalized) semicategory algebras arising naturally in discrete causal theory are {\it concatenation algebras} and {\it directed product algebras}, constructed by combining the structure of a ring $T$ with appropriate semicategory operations on causal path sets.   I refer to both types of algebras as {\bf causal path algebras}, since they are generated by causal path semicategories.  The notation $T$ in this context is intended to evoke the word {\it ``target,"} since $T$ serves as a {\it target object for path functionals,} as discussed in section \hyperref[subsectionpathsummation]{\ref{subsectionpathsummation}}. $T$ may be replaced with an object having less structure than a ring, such as an abelian group, or by a structure at a different level of algebraic hierarchy, such as a category.  Correspondingly, causal path algebras may be replaced by {\it causal path modules,} or by higher-level structures such as {\it generalized enriched functor categories.}   Generalizing beyond the context of paths, any splice product for a class of morphisms into a multidirected set $M$ may be used to define a {\it causal splice module} or {\it causal splice algebra} over $M$.  Alternatively, one may simultaneously consider many distinct splice products on the same class of morphisms.\footnotemark\footnotetext{{\it Vertex operator algebras,} used extensively in string theory, provide familiar examples of algebraic objects involving many distinct operations.}

Concatenation algebras are defined in the context of {\it directed sets,} since their multiplicative structures arise from concatenation products:\\

\begin{defi}\label{deficoncatalg} Let $D=(D,\prec)$ be a directed set, with causal path set $\Gamma(D)$.  Let $T$ be a ring, usually assumed to be commutative with unit. 
\begin{enumerate}
\item The {\bf concatenation algebra} $T^\sqcup(D)$ over $D$, with coefficients in $T$, is the algebra whose elements are formal sums $\sum_\gamma t_\gamma\gamma$, where $\gamma\in\Gamma(D)$ and $t_\gamma\in T$, and whose multiplication is defined by multiplying coefficients in $T$, and forming concatenation products of paths in $\Gamma(D)$, whenever these products exist: 
\[\Big(\sum_\gamma t_\gamma\gamma\Big)\sqcup\Big(\sum_{\gamma'} t_{\gamma'}\gamma'\Big):=\sum_{\gamma\prec_\sqcup\gamma'}t_\gamma t_{\gamma'}\gamma\sqcup\gamma'.\]
\item The {\bf chain concatenation algebra} $T^\sqcup[D]$ over $D$, with coefficients in $T$, is the subalgebra of $T^\sqcup(D)$ given by restricting $\gamma$ to $\Gamma_{\tn{\fsz irr}}(D)=\tn{Ch}(D)$.  
\end{enumerate}
The empty sum is identified with the additive identity $0$ in both algebras. 
\end{defi}

Directed product algebras are defined in the context of multidirected sets, since their multiplicative structures arise from directed products:\\

\begin{defi}\label{defidirprodalg} Let $M=(M,R,i,t)$ be a multidirected set, with path set $\Gamma(M)$.  Let $T$ be a ring, usually assumed to be commutative with unit.
\begin{enumerate}
\item The {\bf directed product algebra} $T^\vee(M)$ over $M$, with coefficients in $T$, is the algebra whose elements are formal sums $\sum_\gamma t_\gamma\gamma$, where $\gamma\in\Gamma(M)$ and $t_\gamma\in T$, and whose multiplication is defined by multiplying coefficients in $T$, and forming directed products of paths in $\Gamma(M)$, whenever these products exist: 
\[\Big(\sum_\gamma t_\gamma\gamma\Big)\vee\Big(\sum_{\gamma' }t_{\gamma'}\gamma'\Big):=\sum_{\gamma\prec_\vee\gamma'}t_\gamma t_{\gamma'}\gamma\vee\gamma'.\]
\item The {\bf chain directed product algebra} $T^\vee[M]$ over $M$, with coefficients in $T$, is the subalgebra of $T^\vee(M)$ given by restricting $\gamma$ to $\Gamma_{\tn{\fsz irr}}(M)=\tn{Ch}(M)$.  
\end{enumerate}
The empty sum is identified with the additive identity $0$ in both algebras.  
\end{defi}

The causal path algebras introduced in definitions \hyperref[deficoncatalg]{\ref{deficoncatalg}} and \hyperref[defidirprodalg]{\ref{defidirprodalg}} above are noncommutative {\it generalized graded algebras,} since the ``length functionals" encoding the lengths of paths in $D$ and $M$ are ``additive"\footnotemark\footnotetext{This vague description is due to obvious caveats, such as the fact that the length of $\gamma\sqcup\delta$, when this product exists, is not the sum of the lengths of $\gamma$ and $\delta$, but the sum of these lengths {\it plus one.}  Further, additivity has a generalized meaning, in terms of isomorphism classes, for paths of infinite length.} in a generalized sense.  In particular, for each nonnegative integer $n$, the additive subgroups of these algebras generated by paths or chains of length $n$ are their ``$n$th graded pieces."  From a purely algebraic viewpoint, this generalized graded structure is ultimately responsible for many of the most useful properties of causal path algebras, but I do not emphasize this perspective in this paper.  It is sometimes useful to consider subalgebras of {\it finite} formal sums in causal path algebras.  In  this paper, I usually avoid this subtlety by working with {\it finite} subsets of the causal path sets $\Gamma(D)$ and $\Gamma(M)$.   This restriction also avoids the use of cumbersome notation, such as $T^\sqcup\llbracket D\rrbracket$, to denote algebras of formal sums without finiteness assumptions.  


\refstepcounter{textlabels}\label{conventionalpath}

{\bf Conventional Applications of Path Algebras.} Path algebras similar or equivalent to causal path algebras arise naturally in graph theory, information theory, computer science, network theory, category theory, noncommutative geometry, algebraic $K$-theory, and of course causal set theory.  For example, in the special case where the coefficient ring $T$ is a field, the subalgebra of finite formal sums in the chain directed product algebra $T^\vee[M]$ is the {\it standard path algebra} appearing in \cite{AbramsPathAlgebra05}.  The chain concatenation algebra $T^\sqcup[D]$ also appears in the literature as a special type of {\it quiver algebra}.  Similar quiver algebras play important roles in recent and ongoing research in noncommutative algebra and noncommutative geometry, particularly in the study of so-called {\it quantum groups}, which are {\it deformations of Hopf algebras.}  

Path algebras possess a number of attractive mathematical properties, besides the generalized graded structure already noted above.  The chain directed product algebra $T^\vee[A]$ over a  finite acyclic directed set $A=(A,\prec)$ will serve for illustrative purposes.  $T^\vee[A]$  has a multiplicative identity given by the formal sum of the elements of $A$.  Each element $x$ in $A$ defines an {\it idempotent operator} on $T^\vee[A]$, whose image is an {\it indecomposable projective} $T^\vee[A]$-{\it module} $P(x)$, called the {\it punctual module} of $x$.  Every other chain in $A$ has square zero.  $T^\vee[A]$ is a {\it hereditary algebra}, which means that any submodule of a projective module over $T^\vee[A]$  is itself projective; this follows from the fact that every projective module over $T^\vee[A]$  is built up from the punctual modules $P(x)$.  These properties play an important role in the study of related {\it module categories,} {\it derived categories,} and {\it cluster categories.}  For further details, I suggest Bernhard Keller's recent paper {\it Cluster Algebras, Quiver Representations and Triangulated Categories} \cite{KellerClusterAlgebras10}.   Some of these properties, or suitable analogues thereof, still apply for general multidirected sets, but others break down.  For example, the formal sum of elements of a multidirected set $M$ is not a multiplicative identity for $T^\vee[M]$ if $M$ has unbounded chains, and a reflexive relation in $M$ does not have square zero, but generates an infinite cyclic multiplicative subgroup of $T^\vee[M]$.  


\refstepcounter{textlabels}\label{raptisincidence}

{\bf Raptis' Incidence Algebra.}  In his 2000 paper {\it Algebraic Quantization of Causal Sets} \cite{RaptisAlgebraicQuantization00}, Ioannis Raptis employs a noncommutative algebra called an {\it incidence algebra,} which is a close transitive analogue of the directed product algebra, with coefficients in $\mbb{C}$, over a locally finite acyclic directed set.  As described by Raptis, a suggestive `ket-bra' notation may be used to represent elements of this algebra.  In particular, each relation $x\prec y$ in a causal set $C$ may be viewed as an {\it operator} $|x\rangle\langle y|$, and relations may be ``multiplied" according to the formula
\begin{equation}
\label{shortcircuit}|x\rangle\langle y|u\rangle\langle v|=\langle y|u\rangle|x\rangle\langle v|, \hspace*{.5cm}\tn{where}\hspace*{.5cm} \langle y|u\rangle:=\delta_{yu}=\begin{cases} 1& \tn{if } y=u \\ 0&\tn{if } y\ne u,\end{cases}
\end{equation}
where the expression $\delta_{yu}$ denotes the {\it Kronecker delta function}, defined by the cases on the far right of equation \hyperref[shortcircuit]{\ref{shortcircuit}}.  This definition requires the causal set axiom of transitivity (\hyperref[tr]{TR}), since the relation $x\prec v$ may not exist in the nontransitive case.  Equation \hyperref[shortcircuit]{\ref{shortcircuit}} defines a semicategory operation on the relation space over a causal set, with Raptis' incidence algebra arising as the corresponding semicategory algebra with complex coefficients.  The operation described by equation \hyperref[shortcircuit]{\ref{shortcircuit}} involves {\it loss of information} compared with the product operation in the corresponding directed product algebra, since there may be many distinct $2$-chains between a given pair of elements $x$ and $v$ in a causal set.  Hence, Raptis' incidence algebra may be viewed as a special type of ``short-circuit algebra," a descriptive term for a  ``degenerate splice algebra," which fails to preserve information.  Raptis describes addition in the incidence algebra over a causal set $C$ as representing {\it quantum-theoretic superposition.}  While this is a reasonable physical interpretation, it can only apply in the background-dependent context, since it involves combining algebraic data associated with substructures of a {\it particular} causal set.  More relevant to the approach developed in this paper are the {\it monoids of arrow fields} used by Chris Isham in his {\it quantization on a category} \cite{IshamQuantisingI05}, described in section \hyperref[subsectionquantumprelim]{\ref{subsectionquantumprelim}} below, which may be viewed as ``higher-level multiplicative analogues" of Raptis' incidence algebras. 


\subsection{Path Summation over a Multidirected Set}\label{subsectionpathsummation}

\refstepcounter{textlabels}\label{pathfunctcont}
\refstepcounter{textlabels}\label{lagrangehamilton}

{\bf Path Functionals; Motivation from Continuum Theory.} I begin this section by briefly reviewing some standard notions from continuum mechanics, in preparation for analogous constructions in the context of multidirected sets.  The reader should be aware that some of the notation appearing in this section, such as the standard notation for the {\it classical action} and the {\it Lagrangian,} is similar to notation I have used for completely different purposes elsewhere in the paper.\footnotemark\footnotetext{Actually, the mathscript symbol $\ms{L}$, which I use for the Lagrangian, is often used to denote a Lagrangian {\it density} in the context of field theory.  In this paper, however, mathscript symbols generally appear at the ``category-functor level of hierarchy," which is appropriate for the Lagrangian. In any case, densities do not arise in the locally finite context.} {\bf Path functionals} are maps from a space of paths into a target object, such as an abelian group, a ring, or a field.  In the study of particle motion in classical mechanics, the {\bf classical action} $\ms{S}$ is a real-valued path functional, given by integrating the {\bf Lagrangian} $\ms{L}$, which describes the difference between a particle's kinetic and potential energies, with respect to time, over a spacetime path $\gamma$, viewed as a possible particle trajectory:
\begin{equation}
\ms{S}(\gamma)=\int_\gamma \ms{L}dt.
\end{equation}  

{\bf Hamilton's principle} states that {\it the classical trajectory of the particle renders the classical action stationary.}  Heuristically, this means that the Lagrangian $\ms{L}$ ``chooses" a trajectory, determined by how $\ms{S}$ varies with $\gamma$. This is a more {\it holistic} interpretation of classical mechanics than the interpretation provided by Newtonian theory, since the value of the action $\ms{S}(\gamma)$ depends on the entire trajectory $\gamma$.  Thus, Hamilton's principle provides a classical example of the general rule, mentioned in section \hyperref[subsectionpowerset]{\ref{subsectionpowerset}} above, that a holistic viewpoint may be {\it useful}, even when the reductionist approach is information-theoretically adequate.   The Lagrangian $\ms{L}$ is an {\bf infinitesimal path functional}, meaning that it depends only on the position and motion of the particle at a given time, not on the details of its trajectory.  In fact, the Lagrangian is usually taken to depend only on {\it information up to first order}; i.e., on position, time, and velocity.  As such, it may be viewed as a function on the {\it real tangent bundle} over continuum spacetime.\footnotemark\footnotetext{Similarly, important causal analogues of Lagrangians may be viewed as {\it relation functions,} as described later in this section. The reader may recall here the comparison between the relation set $R$ of a multidirected set $M=(M,R,i,t)$ and the {\it tangent sheaf} in geometry, mentioned in the last footnote in section \hyperref[subsectionacyclicdirected]{\ref{subsectionacyclicdirected}} above.}  This restriction makes certain computations easier; for example, it simplifies the calculation of Feynman's path integral.  Generalizations of  $\ms{L}$ and $\ms{S}$ are many and varied, but the primitive versions cited here will serve for the purposes of illustration and analogy.  

In Feynman's original version of the histories approach to quantum theory, described in more detail in section \hyperref[subsectionquantumpathsummation]{\ref{subsectionquantumpathsummation}} below, Feynman's {\bf phase map} $\Theta$ is a path functional assigning to each particle trajectory $\gamma$ the {\it complex exponential of its classical action, in units of Planck's reduced constant:}
\begin{equation}\label{feynmanphase}
\Theta(\gamma)=e^{\frac{i}{\hbar}\ms{S}(\gamma)}.
\end{equation}
The classical action $\ms{S}$ is additive for directed products of paths\footnotemark\footnotetext{This means essentially the same thing in the continuum context as in the context of multidirected sets: a pair of continuum paths has a directed product ``joining the two paths together" if and only if the terminal element in the image of the first path coincides with the initial element in the image of the second.  This product is particularly familiar in homotopy theory, where it descends to multiplication in the fundamental groupoid.} by the definition of integration.  Hence, Feynman's phase map is multiplicative, since exponentiation converts addition to multiplication.   In more sophisticated terms, $\ms{S}$ is a ``semicategory morphism"  from an appropriate causal path semicategory over spacetime to the underlying additive group of $\RR$, and $\Theta$ is a ``semicategory morphism" from the same causal path semicategory into the unit circle, viewed as a multiplicative subgroup of the complex numbers $\CC$.


\refstepcounter{textlabels}\label{pathfunctac}

{\bf Path Functionals for Multidirected Sets. } The analogue of an infinitesimal path functional, such as a Lagrangian, on the causal path set $\Gamma(M)$ over a multidirected set $M=(M,R,i,t)$, is a {\bf relation function}; i.e., a map $\theta:R\rightarrow T$, for an appropriate target object $T$.  The analogue of an ordinary path functional is a map $\Theta:\Gamma(M)\rightarrow T$, which I refer to as a {\bf causal path functional}, or simply a {\bf phase map.}  Using the latter term, the individual value $\Theta(\gamma)$ of $\Theta$ on a path $\gamma$ may be referred to as the {\bf phase} of $\gamma$ with respect to $\Theta$, which is both concise and physically suggestive.  In the cases of principal interest in this paper, $\Gamma(M)$ is equipped with a semicategory operation, such as the directed product, and $\Theta$ is a ``semicategory morphism," analogous to Feynman's phase map.  Depending on the context, the phases $\Theta(\gamma)$ of paths $\gamma$ with respect to a phase map $\Theta$ may reside in a group, a ring, a field, or perhaps an algebraic object physically suggestive in some specific way, such as an operator algebra.\footnotemark\footnotetext{For example, Isham uses functionals whose values are Hilbert space operators in his {\it quantization on a category} \cite{IshamQuantisingI05}.} At higher levels of algebraic hierarchy, such as in the theory of kinematic schemes, path functionals may be viewed as {\it assignments,} analogous to functors.  For example, in the language of section \hyperref[subsectionkinematicschemes]{\ref{subsectionkinematicschemes}} below, there exist ``higher-level path functionals" assigning to each co-relative history in a kinematic scheme its {\it causal Galois groups.}\footnotemark\footnotetext{It is actually preferable in many ways to think of phase maps as generalized functors even when working over an individual directed set, rather than a kinematic scheme.  However, it is not feasible to undertake a formal exploration of hierarchical considerations in this paper.} 

A phase map $\Theta$ on the causal path space $\Gamma(M)$ over a multidirected set $M$ may be regarded as a {\it formal sum} $\sum_{\gamma\in\Gamma(M)}\Theta(\gamma)\gamma$, since this expression uniquely pairs each path with its corresponding phase.  This viewpoint immediately evokes the causal path algebras described in section \hyperref[subsectionpathspaces]{\ref{subsectionpathspaces}} above.  Indeed, in the case where $T$ is a ring, a phase map $\Theta$ may be regarded as {\it supplying the coefficients of a causal path algebra element.}  Conversely, each element of a causal path algebra specifies a unique phase map.  Hence, the causal path algebras $T^\sqcup(D)$ and $T^\vee(M)$ supply multiplicative structures on the {\bf mapping spaces} $T^{\Gamma(D)}$ and $T^{\Gamma(M)}$ of maps $\Gamma(D)\rightarrow T$ and $\Gamma(M)\rightarrow T$, respectively.  For example, if $\Theta\leftrightarrow\sum\Theta(\gamma)\gamma$ and $\Psi\leftrightarrow\sum\Psi(\gamma)\gamma$ are elements of $T^{\Gamma(D)}$ or $T^{\Gamma(M)}$; i.e., phase maps, then the formulae
\begin{equation}\label{convolutions}
\Phi\sqcup\Psi=\sum_{\gamma\prec_\sqcup\gamma'}\Phi(\gamma)\Psi(\gamma')\gamma\sqcup\gamma'\hspace*{.7cm}\tn{and}\hspace*{.7cm}\Phi\vee\Psi=\sum_{\gamma\prec_\vee\gamma'}\Phi(\gamma)\Psi(\gamma')\gamma\vee\gamma',
\end{equation}
given by multiplication in $T^\vee(D)$ and $T^\sqcup(M)$, respectively, define analogues of {\it convolution}\footnotemark\footnotetext{For similar reasons, certain types of path algebras are sometimes called {\it convolution algebras} in the literature.} on $T^{\Gamma(D)}$ and $T^{\Gamma(M)}$, respectively.  Using the concatenation algebra as an example, the similarity of such a product to ordinary convolution of real valued functions, given by the formula 
\[f*g(t)=\int_\tau f(\tau)g(t-\tau)d\tau,\] 
may be seen by expressing the value $\Phi\sqcup\Psi(\beta)$, for a particular path $\beta$, as
\[\Phi\sqcup\Psi(\beta)=\sum_{\alpha}\Phi(\alpha)\Psi(\beta-\alpha),\]
where $\alpha$ ranges over all paths that ``may be extended in an appropriate way\footnotemark\footnotetext{This may be formalized via the theory of {\it transitions,} introduced in section \hyperref[subsectionquantumprelim]{\ref{subsectionquantumprelim}} below, since the image of $\alpha$ embeds as a {\it full originary subobject} of the image of $\beta$.  } to produce $\beta$," and where the expression $\beta-\alpha$ means ``the path such that $\alpha\sqcup(\beta-\alpha)=\beta$."  Similarly, the chain algebras $T^\sqcup[D]$ and $T^\vee[M]$  supply multiplicative structures on the mapping spaces $T^{\tn{Ch}(D)}$ and $T^{\tn{Ch}(M)}$, while the corresponding causal path algebras of {\it finite} formal sums supply multiplicative structures on the spaces of {\it finitely supported} maps $\Gamma(D)\rightarrow T$ and $\Gamma(M)\rightarrow T$, where the {\it support} of a phase map $\Theta$ is the set of all paths $\gamma$ in $\Gamma(D)$ or $\Gamma(M)$ {\it not} mapping to $0_T$ under $\Theta$.  


\refstepcounter{textlabels}\label{pathsumac}

{\bf Path Summation over a Multidirected Set.}  Fortunately, there is no need to discuss the general theory of path integration in the continuum context before introducing the analogous theory of path summation over a multidirected set.   Indeed, many of the difficulties of continuum path integration are irrelevant to the latter theory, at least in the locally finite case.   In section \hyperref[subsectionquantumpathsummation]{\ref{subsectionquantumpathsummation}} below, I discuss in some detail the specific methods of continuum path integration used by Richard Feynman to formulate his original version of the histories approach to quantum theory, since these constructions are directly relevant to the corresponding quantum causal theory.   Here, I merely describe the rudiments of path summation over a multidirected set, in the case where the target object for phase maps is a commutative ring with unit.  I treat phase maps and path algebra elements interchangeably, as needed.  

Let $M=(M,R,i,t)$ be an a multidirected set, with causal path set $\Gamma(M)$, and let $T$ be a commutative ring with unit.  The {\bf global evaluation map} $e_M:T^{\Gamma(M)}\dashedrightarrow T$ is the {\it partially defined} map given by ``collapsing" the formal sum associated with a phase map $\Theta$, to yield the corresponding {\it concrete sum} of elements of $T$, whenever it exists:
\begin{equation}\label{globalevaluation}
e_M(\Theta)=e_M\Bigg(\sum_{\gamma\in\Gamma(M)}\Theta(\gamma)\gamma\Bigg)= \sum_{\gamma\in\Gamma(M)}\Theta(\gamma).
\end{equation}
The global evaluation map $e_M$ may be expressed in terms of the {\bf indicator functions} $e_\gamma:\Gamma(M)\rightarrow T$, defined by the formulae
\[e_\gamma(\gamma')=\delta_{\gamma\gamma'}=\begin{cases} 1_T & \tn{if } \gamma'=\gamma \\ 0_T &\tn{if } \gamma'\ne\gamma.\end{cases}\]
The indicator functions $e_\gamma$ extend by $T$-linearity to yield totally defined functions $T^{\Gamma(M)}\rightarrow T$, given by the formulae $e_\gamma(\Theta)=e_\gamma\big(\sum_{\gamma'\in\Gamma(M)}\Theta(\gamma')\gamma'\big)=\Theta(\gamma)$. This particular application of $e_\gamma$ is roughly analogous to integrating a continuum function against a Dirac delta function, to obtain a function value at a particular point.   Restricting to the case of finitely-supported phase maps, the formal sum of indicator functions gives a totally defined evaluation map.   In the general case, however, the right-hand side of equation \hyperref[globalevaluation]{\ref{globalevaluation}} may not converge to an element of $T$.  

Let $\Theta:\Gamma(M)\rightarrow T$ a phase map.  The $T$-valued {\bf path sum} $\Theta(M)$ over $M$ with respect to $\Theta$, {\it if it exists,} is the image of the global evaluation map:
\begin{equation}\label{pathsum}
\Theta(M):=e_{M}(\Theta)=\sum_{\gamma\in\Gamma(M)}\Theta(\gamma).
\end{equation}
If $M'$ is a subset of $M$, viewed as a full subobject, then $\Theta$ also defines an $T$-valued path sum $\Theta(M')$ over $M'$, given by restricting $\gamma$ to belong to $\Gamma(M')$.  This sum may also be expressed in terms of indicator functions in the obvious way.  The assignment $M'\mapsto \Theta(M')$ is a partially-defined {\it power set function} $\ms{P}(M)\dashedrightarrow T$, which may be viewed as a $T$-valued {\it generalized measure} on $M$.  More generally, one may restrict summation to other subsets of the causal path set $\Gamma(M)$, not necessarily those consisting of {\it all} paths into a particular subobject of $M$.  In particular, sets of {\it maximal chains} in finite subsets of directed sets, particularly acyclic directed sets, are important index sets for summation in quantum causal theory. 

\newpage

\section{Quantum Causal Theory}\label{subsectionquantumcausal}

In this section, I introduce a new version of quantum causal theory, using locally finite directed sets as underlying models of classical spacetime.   Passage to the quantum regime is accomplished via an appropriate adaptation of the {\it histories approach to quantum theory,} expressed in terms of path summation over multidirected sets. The presentation in this section is not intended to be comprehensive, and involves a number of simplifying assumptions.   Despite this, the approach introduced here is very general at the conceptual level.  In particular, it may be used to construct {\it either} background-dependent theories of fields and particles existing {\it on} directed sets, {\it or} background-independent theories involving {\it configuration spaces of directed sets.} The former theories are analogous to quantum field theories in curved spacetime, while the latter theories provide suitable tools for studying the problems of fundamental spacetime structure, quantum gravity, and unification in the discrete quantum causal context.  Applicability of the same abstract approach to both types of theories is due to the principle of {\it iteration of structure} (\hyperref[is]{IS}), a quantum-theoretic expression of the fact that configuration spaces of directed sets possess {\it higher-level abstract multidirected structures,} collectively induced by their member sets, and by relationships between pairs of these sets. 

Section \hyperref[subsectionquantumprelim]{\ref{subsectionquantumprelim}} below consists of preliminary material on quantum theory, beginning with a brief explanation of how path integration in conventional quantum physics relates to the general quantum-theoretic principle of {\it superposition.}   Paths in distinguished continuum manifolds represent {\it histories,} whose superposition may be quantified, via path integration, to define {\it quantum amplitudes,} conventionally interpreted in terms of probabilities.  Background independence elevates each history to a complete {\it universe.}  In particular, under the classical causal metric hypothesis (\hyperref[ccmh]{CCMH}), histories are represented by directed sets.  The na\"{i}ve analogues of path integrals in this context are therefore {\it sums over configuration spaces of directed sets.}   Important examples of such sums appear in existing approaches to background-independent quantum theory, such as Isham's {\it quantization on a category} \cite{IshamQuantisingI05}, and Sorkin's {\it quantum measure theory} \cite{SorkinQuantalMeasure12}, \cite{SorkinQuantumMeasure94}.  These approaches provide motivation for the principle of iteration of structure (\hyperref[is]{IS}).  For both conceptual and technical reasons, however, it is advantageous to adopt a modified perspective, emphasizing {\it relationships between pairs of histories,} again utilizing Grothendieck's relative viewpoint (\hyperref[rv]{RV}).  I refer to such relationships as {\it co-relative histories.}   In the discrete causal context, they are represented by special morphisms of directed sets, called {\it transitions.}

In section \hyperref[subsectionquantumpathsummation]{\ref{subsectionquantumpathsummation}}, I introduce an abstract causal analogue of Feynman's theory of path integration, expressed in terms of path summation over multidirected sets.  As explained above, the same general theory applies in both the background-dependent and background-independent contexts.   I review the construction of Feynman's path integral over Euclidean spacetime, and define an analogous path sum on the relation space over a multidirected set.  Impermeability of maximal antichains in relation space, proven in theorem \hyperref[theoremrelimpermeable]{\ref{theoremrelimpermeable}} above, plays a crucial role here.  Feynman's path integral yields quantum amplitudes associated with the motion of particles through Euclidean spacetime, while the analogous causal construction yields generalized quantum amplitudes, associated either with ``particles moving through relation space," or with  ``evolution of families of initial directed sets into families of terminal directed sets."   In section \hyperref[subsectionschrodinger]{\ref{subsectionschrodinger}}, I review Feynman's re-derivation, via path integrals, of Schr\"{o}dinger's equation, which serves as a dynamical law in ordinary quantum theory.  I then proceed to derive analogous {\it causal Schr\"{o}dinger-type equations,} which play the same role in the discrete quantum causal context.  In section \hyperref[subsectionkinematicschemes]{\ref{subsectionkinematicschemes}}, I introduce the theory of {\it kinematic schemes,} which are configuration spaces of directed sets, endowed with ``higher-level multidirected structures," supplied by families of co-relative histories, according to the principle of iteration of structure (\hyperref[is]{IS}).   I discuss in some detail the {\it positive sequential kinematic scheme} $(\ms{A}_{\tn{\fsz{fin}}},\mc{H}_1)$, which is the analogue, in the nontransitive case, of Sorkin and Rideout's kinematic scheme describing the sequential growth of causal sets.   I briefly mention analogies between kinematic schemes and number systems, such as $\QQ$ and $\RR$.  I also introduce {\it kinematic space}, which represents ``all possible frames of reference for all possible histories."   


\subsection{Quantum Preliminaries; Iteration of Structure; Co-Relative Histories}\label{subsectionquantumprelim}

\refstepcounter{textlabels}\label{superposition}
\refstepcounter{textlabels}\label{pathintegration}
\refstepcounter{textlabels}\label{histories}

{\bf  Superposition; Path Integration; Histories Approach to Quantum Theory.} A generic feature of quantum theory is the notion of {\it superposition}, in which multiple independent physical scenarios coexist.   In ordinary quantum mechanics, each physical scenario is a solution of Schr\"{o}dinger's equation, and superposition is a consequence of its {\it linearity.}   In Feynman's theory of path integration, each physical scenario may be understood as a possible {\it history} of a physical system, traced out as a {\it path} on a background spacetime manifold or configuration space.  Predictions are made in this context by summing over these histories to obtain {\it quantum amplitudes,} which may be translated into probabilities.  All ``possible" histories, even those that seem classically absurd, contribute to these amplitudes.  Incorporating fields and special relativity, path integration becomes one of the principal tools, and one of the principal technical difficulties, of quantum field theory.  The {\it histories approach to quantum theory} is a very general approach that grew out of path integration.  The histories approach weighs contributions from members of a configuration space of histories.   In the background-dependent context, these histories may be represented by particle trajectories, or field configurations, or some other auxiliary structure {\it inhabiting} spacetime.   In the background-independent context, each history is an entire {\it classical universe,} which may be represented by a geometry, or a spinfoam, or a dynamical triangulation, or a causal set, or some other model.  The flexibility and generality of the histories approach allows it to be applied in contexts where most other versions of quantum theory do not even make sense. 


\refstepcounter{textlabels}\label{historiesquantumcausal}
\refstepcounter{textlabels}\label{backgrounddependentqct}
\refstepcounter{textlabels}\label{backgroundindependentqct}

{\bf Histories Approach to Quantum Causal Theory.}  In the discrete causal context, it is straightforward, at least in a formal sense, to construct a {\bf background-dependent quantum causal theory}, by simply replacing the spacetime background in Feynman's theory of path integration with a locally finite directed set, and applying the theory of path summation introduced in section \hyperref[subsectionpathsummation]{\ref{subsectionpathsummation}} above.  In this context, a {\it history} is represented by a ``particle trajectory," or some more complex auxiliary structure, inhabiting a {\it fixed} locally finite directed set.  This approach supplies flexible discrete causal alternatives to quantum field theories in curved spacetime and lattice field theories.  It is conceptually similar to existing treatments of particles and fields on causal set backgrounds, but may be considerably strengthened by the systematic use of relation space, and by the other improvements to the causal set viewpoint described in sections \hyperref[sectiontransitivity]{\ref{sectiontransitivity}}, \hyperref[sectioninterval]{\ref{sectioninterval}}, and \hyperref[sectionbinary]{\ref{sectionbinary}} of this paper.\footnotemark\footnotetext{Obvious complications arise in the cyclic case, where one must decide how to handle ``trajectories" wrapping around cycles arbitrary numbers of times.   In this section, however, I work mostly with the acyclic case.}  Much deeper, more difficult, and more interesting is {\bf background-independent quantum causal theory}, in which each history is represented by a {\it different} directed set.  This requires consideration of entire configuration spaces of directed sets, in a manner analogous to Isham's {\it quantization on a category} and Sorkin's {\it quantum measure theory.}   Of particular interest are {\it kinematic schemes,} which are special configuration spaces adapted to an {\it evolutionary} viewpoint.  The abstract structure of kinematic schemes may be understood in terms of {\it relationships between pairs of histories,} which I refer to as {\it co-relative histories.} Along with the theory of relation space, the theory of co-relative histories is the most important physical application of Grothendieck's now-familiar relative viewpoint (\hyperref[rv]{RV}) appearing in this paper.     


\refstepcounter{textlabels}\label{ishamquantcat}

{\bf Isham's Quantization on a Category.}  Chris Isham has explored a number of ideas relevant to quantum causal theory in his groundbreaking work on physical applications of category theory.  For example, in his 2005 paper {\it Quantising on a Category} \cite{IshamQuantisingI05}, Isham introduces a category-theoretic approach to ``quantizing"  a broad variety of physical models, including causal sets. This represents only a small facet of Isham's ongoing research program to develop a general {\it topos-theoretic} framework for physics.  A {\it topos} is an ``enriched" category with properties rendering it suitable to serve as a ``mathematical universe."  For example, there is a {\it topos of sets,} which serves as Isham's topos for classical physics.  Topos theory is yet another crucial structural idea arising from Alexander Grothendieck's approach to commutative algebraic geometry, where it was conceived as a way of organizing information about {\it presheaves} and {\it sheaves} on topological spaces and generalized topological spaces called {\it sites}.  For the purposes of this paper, it suffices to highlight only a few specific aspects of Isham's approach. 

Isham \cite{IshamQuantisingI05} writes that his 

\begin{quotation}\noindent{\it ``...main topic of interest \tn{[is]} to construct a quantum framework for a system whose ``history configuration space"... ...is a collection of possible space-times, such as... ...causal sets... ...or... topological spaces."} (page 277)
\end{quotation}
and later,
\begin{quotation}\noindent{\it ``...when quantising a system whose configuration space... ... is a collection... ...of sets with structure, the starting point will be an association to each pair \tn{[of such sets]} a set... ...of transformations \tn{[between them]} that, in some suitable sense, `respect,' or `reflect,' \tn{[their]} internal structures..."} (page 278)
\end{quotation}
Isham then proceeds to explain that such a family of ``structured sets" may be organized into a category $\ms{Q}$, whose morphisms are the aforementioned transformations.  Abstracting this, he proposes to quantize general categories by 
\begin{quotation}\noindent{\it ``...representing \tn{[morphisms]} with operators on a Hilbert space in some way."} (page 280)
\end{quotation}

Without going into details, Isham's method involves a {\it presheaf of Hilbert spaces} and a {\it monoid of arrow fields} over $\ms{Q}$.  An {\it arrow field} assigns to each object $Q$ of $\ms{Q}$ a morphism $Q\rightarrow Q'$ to some other object $Q'$.  A natural monoid structure on the class of arrow fields over $\ms{Q}$ is given by composition of morphisms.  As mentioned at the end of section \hyperref[subsectionpathspaces]{\ref{subsectionpathspaces}} above, Isham's monoid of arrow fields may be viewed as a higher-level, purely multiplicative analogue of Raptis' incidence algebras \cite{RaptisAlgebraicQuantization00}.  All the information in either object is contained in the corresponding directed product algebra.\footnotemark\footnotetext{Isham mentions Raptis' incidence algebras in the preprint version of \cite{IshamQuantisingI05}.}   Isham gives a simple example of this construction, using a category consisting of two causal sets, on pages 293-296 of \cite{IshamQuantisingI05}.  However, for the purposes of this paper, it is Isham's {\it general reasoning,} represented by the above quotations, that is important.  The technical details of the approach presented here are much different.  In particular, I do not {\it a priori} assign Hilbert spaces, or any other ``ready-made" quantum-theoretic devices, to directed sets.  I also prefer to work with path algebras than with incidence algebras, arrow fields, or similar constructions, which ``lose information" under composition. 


\refstepcounter{textlabels}\label{sorkinquantal}

{\bf Sorkin's Quantum Measure Theory.} Rafael Sorkin has done very general work on the histories approach to background-independent quantum theory, and has applied these ideas in specific ways to causal set theory.  In particular, in his 2012 paper {\it Toward a Fundamental Theorem of Quantal Measure Theory} \cite{SorkinQuantalMeasure12}, Sorkin describes quantum-theoretic measures in the context of the Sorkin-Rideout kinematic scheme $\ms{K}$ describing the sequential growth of causal sets, illustrated in figure \hyperref[sequential]{\ref{sequential}} of section \hyperref[subsectionapproach]{\ref{subsectionapproach}} above.  Sorkin works mostly in terms of {\it labeled} causal sets in this paper, which leads to the replacement of $\ms{K}$ with an appropriate {\it tree.}  Sorkin writes,

\begin{quotation}\noindent{\it ``...the collection of naturally labeled finite \tn{[causal sets]}... ...has itself the structure of a \tn{[tree]}, because its elements are labeled... ...Clearly, a particular realization of the growth process... ...can be conceived of as an upward path through this tree." } (page 7)
\end{quotation}

Here, a ``realization of the growth process" means a history, represented by a causal set $C$, together with a generalized frame of reference for $C$, represented by a specific labeling of $C$.   In section \hyperref[subsectionkinematicschemes]{\ref{subsectionkinematicschemes}} below, I revisit such ``upward paths" in the more general context of {\it co-relative kinematics,} and {\it completions of kinematic schemes.}


\refstepcounter{textlabels}\label{iteration}

{\bf Iteration of Structure.} As mentioned in the introduction to section \hyperref[subsectionquantumcausal]{\ref{subsectionquantumcausal}}, the same theory of path summation over directed sets and multidirected sets, outlined in section \hyperref[subsectionpathsummation]{\ref{subsectionpathsummation}} above, applies to both the background-dependent quantum theory of fields and particles on a directed set, and the background-independent quantum theory of co-relative histories and kinematic schemes.  This fact is by no means obvious, since the path sums in section \hyperref[subsectionpathsummation]{\ref{subsectionpathsummation}} involve the causal path set over a {\it particular} directed or multidirected set, while background-independent quantum causal theory requires simultaneous consideration of entire configuration spaces of directed sets. The link between the two approaches is the principle of {\it iteration of structure}, which is based on the observation that configuration spaces of directed sets possess {\it abstract multidirected structures,} collectively induced by their member sets, and by relationships between pairs of these sets.  The following statement of this principle is sufficiently general for the purposes of this paper, although other examples of the same basic idea may be recognized at other levels of hierarchy in causal theory.\footnotemark\footnotetext{For example, the relationship between kinematic schemes and {\it kinematic spaces,} briefly outlined at the end of section \hyperref[subsectionkinematicschemes]{\ref{subsectionkinematicschemes}}, is an example of iteration of structure beyond the quantum level.}  

\refstepcounter{textlabels}\label{is}

\hspace*{.3cm} IS. {\bf Iteration of Structure:} {\it Background independent quantum causal theory may be described \\ \hspace*{.9cm}  in terms of ``multidirected sets whose elements are directed sets."}

Iteration of structure is perhaps the most important generalization of the binary axiom (\hyperref[b]{B}).  A prototypical example of iteration of structure is given by Sorkin and Rideout's kinematic scheme describing the sequential growth of causal sets
\cite{SorkinSequentialGrowthDynamics99}, illustrated in figure \hyperref[sequential]{\ref{sequential}} of section \hyperref[subsectionapproach]{\ref{subsectionapproach}} above.  A closely related example, more relevant to the approach developed in this paper, is given by the {\it positive sequential kinematic scheme,} illustrated in figure \hyperref[figsequentialscheme]{\ref{figsequentialscheme}} of section \hyperref[subsectionkinematicschemes]{\ref{subsectionkinematicschemes}} below.  Both of these kinematic schemes possess the abstract structures of countably infinite, locally finite, acyclic multidirected sets.  Iteration of structure is a striking property, conceptually unifying classical and quantum theory as {\it different levels of algebraic hierarchy} in a single universal construction.  This is a distinguishing characteristic of discrete causal theory, compared with other approaches to the study of fundamental spacetime structure, quantum gravity, and unification.  For comparison, the abstract structures of configuration spaces of relativistic spacetime geometries, up to suitable equivalence, are generally nothing like individual spacetime geometries.\footnotemark\footnotetext{For example, the moduli space of {\it Einstein metrics} on the single diffeomorphism class of K3 surfaces admits a local diffeomorphism into a $57$-dimensional orbifold \cite{AndersonEinsteinMetrics09}.}  


\refstepcounter{textlabels}\label{transitions}

{\bf Transitions.}  Iteration of structure arises naturally in the search for suitable notions of {\it initial} and {\it terminal conditions} in causal theory.  It is useful to express  this idea in terms of {\it relationships between histories,} which I refer to as {\it co-relative histories.}  In this paper, I focus on the special case of co-relative histories that formalize the ``evolution of one directed set into another."\footnotemark\footnotetext{As usual, only {\it isomorphism classes} of directed sets are significant in this context, but the necessary information may be described in terms of representatives.}   Such co-relative histories are represented by {\it transitions,} which are special morphisms of directed sets. The term ``co-relative" is explained below.\\

\begin{defi}\label{defitransition} A {\bf transition} in the category $\ms{D}$ of directed sets is a monomorphism $\tau:D_i\rightarrow D_t$, embedding its {\bf initial set}, or {\bf source}, $D_i$, into its {\bf terminal set}, or {\bf target}, $D_t$, as a proper, full, originary subobject.  This means that $\tau(D_i)$ has nontrivial complement in $D_t$ (proper), that $\tau(x)\prec\tau(y)$ in $D_t$ if and only if $x\prec y$ in $D_i$ (full), and that the isomorphic image $\tau(D_i)$ of $D_i$ in $D_t$ contains its own past (originary). 
\end{defi}

Heuristically, a transition $\tau$ {\it augments} its source $D_i$, by adding directed structure in a manner compatible with its existing structure, thereby ``producing" its target $D_t$.   There is no difficulty in extending the definition of a transition to multidirected sets, but such a generalization is not needed in this paper.  Figure \hyperref[transitionfig]{\ref{transitionfig}} below illustrates a transition. 

\begin{figure}[H]
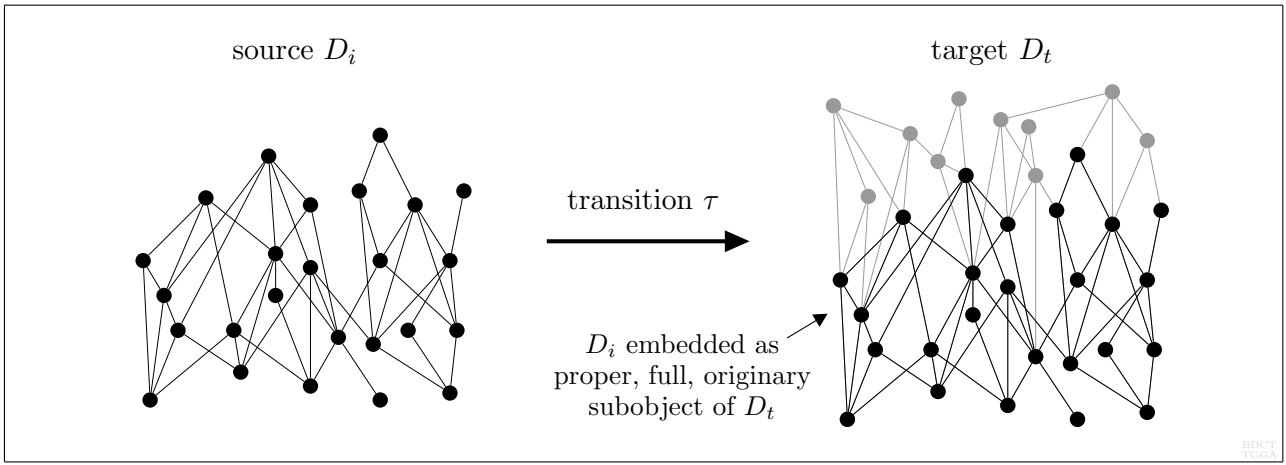


\caption{A transition $\tau:D_i\rightarrow D_t$.}
\label{transitionfig}
\end{figure}
\vspace*{-.5cm}

\refstepcounter{textlabels}\label{benderrobinson} 

Transitions are {\it a priori} too specific to be regarded as physically fundamental, since they generally contain extra, nonphysical information, due to symmetries of their sources and targets.  For example, the composition of a transition $\tau$ with an automorphism of its source defines a ``physically equivalent" transition. The automorphism group of the target of $\tau$  also plays a role, whenever this group ``mixes the image of $\tau$ with its complement."  Given the ``irregularity" of an arbitrary directed set, one might expect most such sets to lack ``large-scale symmetries;" i.e., automorphisms mapping large subsets to their complements.   On the other hand, one might expect ``small-scale symmetries;" i.e., automorphisms permuting a few elements in a small symmetric subset, to be fairly common.  An interesting result, in the acyclic case, is that {\it a typical finite acyclic directed set is rigid;} i.e., it has no nontrivial automorphisms at all. This result was proven, in an asymptotic sense, by Edward Bender and Robert Robinson, in their 1988 paper {\it The Asymptotic Number of Acyclic Digraphs II} \cite{BenderRobinsonAuto88}.   

The physical implications of Bender and Robinson's rigidity result in the context of discrete causal theory depend on a variety of factors.  For example, a {\it typical} finite directed set has a ``large number of relations," due to elementary counting considerations.  This tends to favor rigidity, since any given relation can break a symmetry.  However, dynamical considerations may ultimately prove to favor ``sparser" directed sets, allowing nontrivial automorphism groups to play a significant role.  This is an issue of considerable physical importance, since automorphisms of directed sets are analogous to ``external" or ``spacetime" symmetries in conventional physics, such as those encoded by the Poincar\'{e} symmetry group of Minkowski spacetime.  Such external symmetries impose basic constraints on the properties of elementary particles in quantum field theory, via Lie representation theory.  


\refstepcounter{textlabels}\label{symmetry} 

{\bf Symmetry Preservation, Extension, Breaking, Generation.} Assuming, despite Bender and Robinson's rigidity result, that (``external") symmetries are relevant in discrete causal theory, it is worthwhile to briefly consider how such symmetries interact with transitions.   In this context, a {\it symmetry} means simply an automorphism of a particular directed set.  A transition may {\it preserve,} {\it extend,} or {\it break} symmetries of its source, or it may {\it generate} symmetries of its target, or some combination of the four.  Care is required in the physical interpretation of these notions, particularly because the too-specific nature of individual transitions adds arbitrary information to the picture.  For example, {\it different but equivalent} transitions mapping a common source to different subobjects of a common target generally preserve, extend, break, and/or generate {\it different but equivalent} families of symmetries.   

Let $\tau:D_i\rightarrow D_t$ be a transition of directed sets, and let $\tn{Aut}(D_i)$ and $\tn{Aut}(D_t)$ be the automorphism groups of the source $D_i$ and target $D_t$ of $\tau$, respectively.  Let $F_\tau:=\tn{Aut}(D_t;\tau(D_i))$ be the subgroup of $\tn{Aut}(D_t)$ whose elements {\it permute $\tau(D_i)$ and its complement separately.}  Each element of $F_\tau$ defines an automorphism of $D_i$, via restriction and conjugation by $\tau$.  Hence, I refer to $F_\tau$ as the {\bf group of extensions} of symmetries of $D_i$.  More precisely, there is a group homomorphism 
\[\rho_\tau:F_\tau\longrightarrow\tn{Aut}(D_i),\] 
\[\hspace*{.8cm}\beta\hspace*{.2cm}\mapsto\hspace*{.2cm}\tau^{-1}\circ\beta\circ\tau,\] 
which I refer to as the {\bf restriction homomorphism}, where the factors $\tau^{-1}$ and $\beta$ in the composition are understood to be restricted to the image $\tau(D_i)$ of $\tau$.  If $\rho_\tau$ is surjective, then $\tau$ {\bf preserves the symmetries} of the source $D_i$ of $\tau$.  The kernel $K_\tau$ of $\rho_\tau$ is the normal subgroup of $F_\tau$ whose elements consist of automorphisms of $D_t$ fixing $\tau(D_i)$.  Hence, if $\tau$ preserves the symmetries of $D_i$, there is a short exact sequence
\[1\rightarrow K_\tau\overset{\iota}{\rightarrow} F_\tau\overset{\rho_\tau}{\rightarrow} \tn{Aut}(D_i)\rightarrow 1,\]
where $\iota$ is the inclusion of the kernel $K_\tau$ into $\tn{Aut}(D_i)$.   Hence, $F_\tau$ is a {\it group extension} of $\tn{Aut}(D_i)$ by $K_\tau$ if and only if $\tau$ preserves the symmetries of $D_i$.   The cokernel $F_\tau/K_\tau$ of $\rho_\tau$ may be identified with the normal subgroup of $F_\tau$ fixing the {\it complement} of $\tau(D_i)$ in $D_t$.  The group $F_\tau$ is isomorphic to the direct product of $K_\tau$ and $G_\tau$:
\[F_\tau\cong K_\tau\times G_\tau.\]

\refstepcounter{textlabels}\label{galois} 

Whether or not $\rho_\tau$ is surjective, it induces a quotient monomorphism $\overline{\rho_\tau}:G_\tau\rightarrow  \tn{Aut}(D_i)$, identifying $G_\tau$ with the {\it subgroup of automorphisms of $D_i$ preserved by} $\tau$.   In the finite case, the subgroup index $[F_\tau:G_\tau]$ of $G_\tau$ in $F_\tau$ measures the number of different ways automorphisms of $D_i$ preserved by $\tau$ are extended by $\tau$.  If $\rho_\tau$ is surjective, then $\overline{\rho_\tau}$ is an isomorphism.  If $\rho_\tau$ is not surjective, then every element of the complement $\tn{Aut}(D_i)-\overline{\rho_\tau}(G_\tau)$ represents a {\bf symmetry broken by} $\tau$.   In the finite case, the index $[\tn{Aut}(D_i):G_\tau]$ measures the extent of this symmetry breaking.   The groups $K_\tau$ and $G_\tau$ may be called {\bf causal Galois groups}, since they are groups of automorphisms fixing distinguished subobjects of directed sets in the context of discrete causal theory.\footnotemark\footnotetext{Many interesting analogies exist between causal theory and Galois theory, which cannot be explored in this paper.  For example, I have not mentioned the obviously relevant theory of {\it Galois connections.}}  

Figure \hyperref[transitiongalois]{\ref{transitiongalois}} below illustrates simple examples of symmetry preservation, extension, and breaking.  Symmetry generation, also depicted in the figure, is discussed later in this section.  Let $\tau_0:D_0\rightarrow D_1$ be the transition sending the elements $t_0$, $u_0,$ and $v_0$ of $D_0$ to the elements $t_1$, $u_1,$ and $v_1$ of $D_1$, respectively.  The automorphism group $\tn{Aut}(D_0)$ of the source $D_0$ of $\tau_0$ is isomorphic to $\ZZ_2$, generated by the unique automorphism $\alpha$ of $D_0$ interchanging $u_0$ and $v_0$.  The automorphism group $\tn{Aut}(D_1)$ of the target $D_1$ of $\tau_0$ is trivial.  Hence, the subgroups $F_{\tau_0}$, $K_{\tau_0},$ and $G_{\tau_0}$ of $\tn{Aut}(D_1)$ are trivial.   The index $[F_{\tau_0}:G_{\tau_0}]$ is therefore equal to $1$, which means that $\tau_0$ extends the trivial automorphism of $D_0$ to the trivial automorphism of $D_1$, with no other extensions.  The index $[\tn{Aut}(D_0):G_{\tau_0}]$ is equal to $2$, which means that $\tau_0$ breaks the unique nontrivial symmetry $\alpha$ of $D_0$.    

\begin{figure}[H]
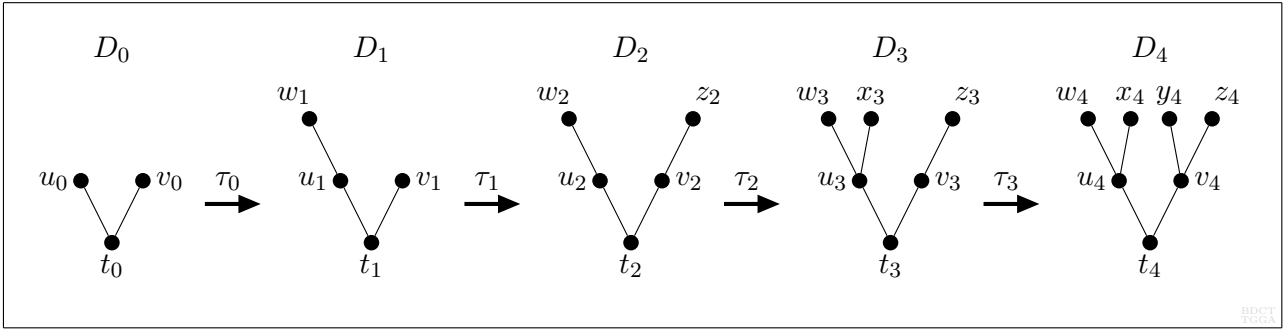


\caption{Symmetry preservation, extension, breaking, and generation, via transitions.}
\label{transitiongalois}
\end{figure}
\vspace*{-.5cm}

It is instructive to examine the significance of this broken $\ZZ_2$ symmetry in an informal sense, first ``from the perspective of the source $D_0$," then ``from the perspective of the target $D_1$."   There are obvious physical motivations for considering such viewpoints.   In particular, one might begin with $D_0$, and examine the possible ways in which it could ``evolve into $D_1$;" conversely, one might begin with $D_1$, and investigate how it could have ```evolved from $D_0$."  From the perspective of the source $D_0$ of $\tau_0$, the broken $\ZZ_2$ symmetry reflects the existence of two ``equivalent" ways to obtain a directed set isomorphic to $D_1$ by adding a single element to $D_0$.  This ``new element" may be added either in the future of $u_0$, or in the future of $v_0$.   These two constructions yield transitions from $D_0$ into two {\it different, but isomorphic,} targets.  It is important to note, however, that the existence of multiple isomorphic targets is not always associated with a broken symmetry of the source.   In particular, there may be multiple ``inequivalent" ways to construct the isomorphism class of a specified target from a specified source {\it even if both source and target have trivial automorphism groups.}  This curious phenomenon, explained in more detail below, can occur even in cases where only a single new element is added.   From the perspective of the target $D_1$ of $\tau_0$, there is a unique subobject of $D_1$ that may be identified with the source $D_0$.  From this perspective, the broken $\ZZ_2$ symmetry reflects the fact that the unique nontrivial symmetry of this subobject does not extend to a symmetry of $D_1$. 

Returning for the moment to the case of an arbitrary transition $\tau:D_i\rightarrow D_t$, the complement of the subgroup $F_\tau$ in $\tn{Aut}(D_t)$ consists of all automorphisms of $D_t$ that {\it mix} the image $\tau(D_i)$ of $\tau$ with its complement in $D_t$.  These automorphisms obviously do not form a subgroup of $\tn{Aut}(D_t)$, since they do not include the identity.  Further, $F_\tau$ is generally not a normal subgroup of $\tn{Aut}(D_t)$,\footnotemark\footnotetext{For example, consider the automorphism $\beta$ of $D_4$ in figure \hyperref[transitiongalois]{\ref{transitiongalois}} interchanging $u_4,w_4$, and $x_4$ with $v_4, y_4,$ and $z_4,$ respectively.  It is easy to see that $\beta^{-1}F_{\tau_3}\beta\ne F_{\tau_3}$.} so there is generally no quotient group to study.   These automorphisms are new ``from the perspective of the transition $\tau$," since they do not restrict to automorphisms of the image $\tau(D_i)$.  Hence, I refer to them as symmetries {\bf generated by} $\tau$.   However, these automorphisms are not necessarily new ``from the perspective of the source $D_i$ of $\tau$."  The meaning of this distinction is that there may exist other transitions $\tau':D_i\rightarrow D_t$, corresponding to different, but isomorphic, subgroups $F_{\tau'}$ of $\tn{Aut}(D_t)$, whose elements extend symmetries of $D_i$ broken by $\tau$, and which are {\it equivalent to} $\tau$, in the sense that their images are mapped isomorphically onto the image of $\tau$ by automorphisms of $D_t$.   This notion of equivalence is made precise in definition \hyperref[deficorelative]{\ref{deficorelative}} below.  From a physical perspective, this distinction reflects the fact that transitions are too specific to be considered physically fundamental. 

Referring again to figure \hyperref[transitiongalois]{\ref{transitiongalois}} above, let $\tau_1:D_1\rightarrow D_2$ be the transition sending the elements $t_1$, $u_1$, $v_1$, and $w_1$ of $D_1$ to the elements $t_2$, $u_2$, $v_2$, and $w_2$ of $D_2$, respectively.  The automorphism group $\tn{Aut}(D_1)$ of the source $D_1$ of $\tau_1$ is trivial, while the automorphism group $\tn{Aut}(D_2)$ of the target $D_2$ of $\tau_1$ is isomorphic to $\ZZ_2$, generated by the unique automorphism $\beta$ of $D_2$ interchanging $u_2$ and $v_2$, and $w_2$ and $z_2$, respectively.   The extension group $F_{\tau_1}$ is trivial, since $\beta$ mixes $\tau_1(D_1)$ with its complement.   Hence, the causal Galois groups $K_{\tau_1}$ and $G_{\tau_1}$ are also trivial.   The complement of $F_{\tau_1}$ in $\tn{Aut}(D_2)$ is the singleton $\{\beta\}$; therefore, $\beta$ is a symmetry generated by $\tau_1$.  From the perspective of the source $D_1$ of $\tau_1$, there is only one way to add an extra element to $D_1$ to obtain a directed set isomorphic to the target $D_2$.   However, from the perspective of the target $D_2$ of $\tau_1$, there are {\it two} different subobjects that may be identified with the source $D_1$.  This reflects the fact that there is an equivalent transition ${\tau_1}':D_1\rightarrow D_2$, sending the elements $t_1$, $u_1$, $v_1,$ and $w_1$ of $D_1$ to the elements $t_2$, $v_2$, $u_2,$ and $z_2$ of $D_2$, respectively.   The transition ${\tau_1}'$ {\it also} generates the unique nontrivial symmetry $\beta$ of $D_2$.   

The remaining transitions $\tau_2:D_2\rightarrow D_3$ and $\tau_3:D_3\rightarrow D_4$ appearing in figure  \hyperref[transitiongalois]{\ref{transitiongalois}} involve more complex phenomena.  The transition $\tau_2$ breaks the $\ZZ_2$-symmetry of $D_2$, but generates a {\it new} $\ZZ_2$ symmetry in $D_3$.  Hence, the symmetry groups of $D_2$ and $D_3$ are ``accidentally" isomorphic.   The transition $\tau_3$, meanwhile, generates multiple new symmetries.  One may also consider compositions of the transitions in the figure; for example, the composition $\tau_1\circ\tau_0:D_0\rightarrow D_2$ preserves the symmetries of $D_0$, without generating any new symmetries.   In this case, the fact that the symmetry groups of the source and target are isomorphic is {\it not} ``accidental." 


\refstepcounter{textlabels}\label{labelcorelative}

{\bf Co-Relative Histories.}  In defining co-relative histories, the objective is to precisely capture the {\it physical essence of particular relationships between pairs causal structures,} represented by directed sets, following Grothendieck's relative viewpoint (\hyperref[rv]{RV}).  In this context, a na\"{\i}ve attempt to represent such relationships by individual morphisms in the category $\ms{D}$ of directed sets leads in the wrong direction, at least in the case in which the directed sets under consideration possess nontrivial automorphism groups.  A better choice is to represent such relationships by {\it equivalence classes of morphisms whose images are related by automorphisms of their target sets.}   Here, I focus on equivalence classes of {\it transitions,} since the goal is to formalize the evolution of causal structure.   Such equivalence classes define {\it proper, full, originary co-relative histories} relating their common sources and targets.  One may imagine more general relationships between pairs of histories, but maximal generality is not the goal in the present context.   

\refstepcounter{textlabels}\label{corelmulti}

\begin{figure}[H]
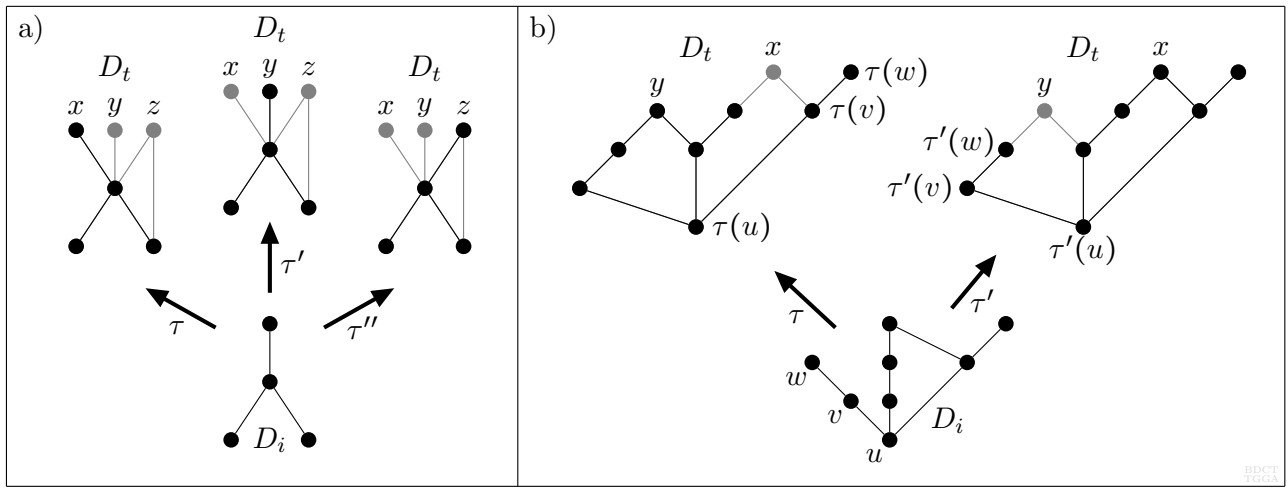


\caption{a) Equivalent and inequivalent transitions; b) McKay's example of inequivalent single-element transitions between the same pair of directed sets.}
\label{corelativeexamples}
\end{figure}

Examples of equivalent and inequivalent transitions between pairs of directed sets are illustrated in figure \hyperref[corelativeexamples]{\ref{corelativeexamples}}a below, with the labels on some of the elements suppressed to avoid clutter.   The transitions $\tau$ and $\tau'$ between $D_i$ and $D_t$ are equivalent, since their images are related by the unique automorphism $\alpha$ of $D_t$ interchanging the elements $x$ and $y$.   The transition $\tau''$ is {\it not} equivalent to $\tau$ or $\tau'$, since $\alpha$ is the only nontrivial automorphism of $D_t$, and $\alpha$ fixes the element $z$.  

\vspace*{-.5cm}

An interesting fact, already mentioned above in the context of symmetry breaking, is that there may exist multiple ``inequivalent" ways to construct the isomorphism class of a specified target $D_t$ from a specified source $D_i$, even if both $\tn{Aut}(D_i)$ and  $\tn{Aut}(D_t)$ are trivial, and even if $D_i$ and $D_t$ differ by only a single element.   Equivalently, viewing $D_i$ and $D_t$ as ``given," rather than approaching the problem constructively, there may exist multiple inequivalent transitions between $D_i$ and $D_t$.  Figure \hyperref[corelativeexamples]{\ref{corelativeexamples}}b above illustrates an example of this phenomenon, supplied by Brendan McKay \cite{McKayPrivate13}.  The element labels $u$, $v$, $w$, etc., are included to clarify the existence of the transitions $\tau$ and $\tau'$.  The elements $x$ and $y$ in the complements of $\tau$ and $\tau'$, respectively, are called {\it pseudosimilar vertices,} in the language of graph theory.\footnotemark\footnotetext{The theory of pseudosimilarity is of interest in the {\it graph reconstruction problem.}  McKay tells me \cite{McKayPrivate13} that little work has been done on pseudosimilarity in the case of {\it directed} graphs.}  Viewing $D_i$ and $D_t$ as histories, it is reasonably obvious that the relationships between them represented by $\tau$ and $\tau'$  differ in a physically significant way.  The presence of this type of behavior has significant structural consequences in discrete causal theory.  For example, as discussed below, it implies that the abstract structures of kinematic schemes are generally multidirected, rather than merely directed.  In particular, Sorkin and Rideout's kinematic scheme describing the sequential growth of causal sets, illustrated in figure \hyperref[sequential]{\ref{sequential}} of section \hyperref[subsectionapproach]{\ref{subsectionapproach}} above, has the abstract structure of a multidirected set, {\it not} the skeleton of a partial order. 

The foregoing discussion motivates the following formal definition of co-relative histories, suitably general for the purposes of this paper. The terminology is chosen to suggest a {\it fixed source,} reflecting the intended evolutionary viewpoint.  However, there are other equally valid perspectives.  

\vspace*{.2cm}
\begin{defi}\label{deficorelative}  A {\bf proper, full, originary co-relative history} $h:D_i\Rightarrow D_t$ between directed sets $D_i$ and $D_t$ is an equivalence class of transitions from $D_i$ to $D_t$, where two transitions $\tau$ and $\tau'$ are equivalent if and only if there exists an automorphism $\beta$ of $D_t$ mapping $\tau(D_i)$ onto $\tau'(D_i)$.
The common initial set $D_i$ of the transitions making up $h$ is called the {\bf cobase} of $h$, and the common terminal set $D_t$ of these transitions is called the {\bf target} of $h$.
\end{defi}

The {\it double-arrow notation} $\Rightarrow$ in definition \hyperref[deficorelative]{\ref{deficorelative}} is intended to convey the idea that a co-relative history consists of a {\it family} of morphisms, ``bundled together" to encode a {\it single physical relationship.} More generally, one may specify a {\it cobase family} $\mbf{D}_i$ of directed sets, and consider {\it histories co-relative to} $\mbf{D}_i$.  The reason for using the terms {\it co-relative} and {\it cobase,} rather than {\it relative} and {\it base,} is that $D_i$ is taken to represent ``initial," or ``known" information, while $D_t$ is allowed to vary over ``possible evolutionary outcomes."  This viewpoint contrasts with the viewpoint accompanying the theory of relative multidirected sets over a fixed base, introduced in section \hyperref[relativeacyclicdirected]{\ref{relativeacyclicdirected}} above, in which the base is a fixed target set, and the source is allowed to vary.\footnotemark\footnotetext{The terminology in both cases originates from the category-theoretic notions of {\it covariant} and {\it contravariant} constructions, related to one another by ``reversing the arrows."  Unfortunately, these terms carry many different meanings in both mathematics and physics.} The hyphen in {\it co-relative} is added to avoid confusion with the too-similar word {\it correlative}.  The physical sense of definition \hyperref[deficorelative]{\ref{deficorelative}} is reasonably clear when the cobase $D_i$ and target $D_t$ are both finite.  In the infinite case, {\it cycles of co-relative histories} are possible under definition \hyperref[deficorelative]{\ref{deficorelative}}, even when the directed sets involved are acyclic.  This raises interesting interpretive challenges. 

\begin{figure}[H]
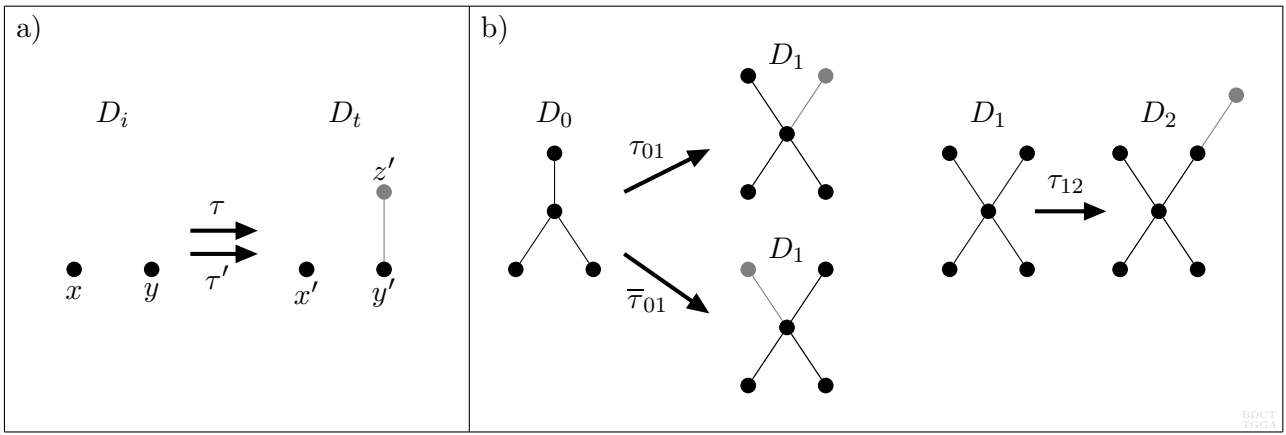


\caption{a) Equivalent transitions may be unrelated as {\it morphisms} due to symmetry breaking; b) equivalence of transitions is not a congruence relation under composition in $\ms{D}$.}
\label{transitionnotcongruence}
\end{figure}

Note that definition \hyperref[deficorelative]{\ref{deficorelative}} does {\it not} require that $\tau'=\beta\circ\tau$ as {\it morphisms;} it is only the {\it images} of $\tau$ and $\tau'$ that must be related by an automorphism of the target.  For example, if $\tau$ breaks a symmetry $\alpha$ of its cobase $D_i$, then the transition $\tau'=\tau\circ\alpha$ cannot be recovered by applying an automorphism to its target $D_t$ {\it after} application of $\tau$.   The simplest nontrivial example of this is illustrated in figure \hyperref[transitionnotcongruence]{\ref{transitionnotcongruence}}a below, in which $D_i$ consists of a pair of unrelated elements $x$ and $y$, and $D_t$ consists of three elements $x'$, $y'$, and $z'$, with one relation $y'\prec z'$. In this case, $\tn{Aut}(D_i)$ is isomorphic to $\ZZ_2$, generated by the automorphism interchanging $x$ and $y$, while $\tn{Aut}(D_t)$ is trivial. There is a unique co-relative history $h:D_i\Rightarrow D_t$, represented by two equivalent transitions $\tau$ and $\tau'$, where $\tau$ sends the elements $x$ and $y$ of $D_i$ to the elements $x'$ and $y'$ of $D_t$, respectively, and where $\tau'$ reverses these images.   Both transitions break the unique nontrivial symmetry of $D_i$.  These transitions are equivalent, via the identity automorphism of $D_t$, since they have the same image, but there is no automorphism of $D_t$ relating them as morphisms. 

\refstepcounter{textlabels}\label{corelativesubtle}

Co-relative histories are subtle from a category-theoretic perspective.  For example, equivalence of transitions is {\it not} a congruence relation under composition in the category $\ms{D}$ of directed sets.  This means that, given equivalent transitions from a common source $D_0$ to a common target $D_1$, and from $D_1$ to a third directed set $D_2$, it is generally {\it not} true that the compositions of these transitions are equivalent.  This is illustrated in figure \hyperref[transitionnotcongruence]{\ref{transitionnotcongruence}}b.  The transitions $\tau_{01}$ and $\overline{\tau}_{01}$ from $D_0$ to $D_1$ are related by any automorphism of $D_1$ interchanging its two terminal elements, but the compositions $\tau_{12}\circ\tau_{01}$ and $\tau_{12}\circ\overline{\tau}_{01}$ represent different co-relative histories between $D_0$ and $D_2$.  This example demonstrates that equivalence of transitions is not even {\it cancellative,} since the ``left-hand factor" $\tau_{12}$ is the same in both cases.  More generally, pairs of ``consecutive" co-relative histories generally do not compose to give unique ``products."   Instead, the ``composition" of a pair of ``consecutive" co-relative histories is a {\it family of co-relative histories.}  

\vspace*{-.5cm}

Congruence relations are so convenient that it is tempting to try to work with congruence classes of transitions, rather than the equivalence classes specified in definition \hyperref[deficorelative]{\ref{deficorelative}} above.  One way to do this is to define a stricter equivalence relation, in which two transitions $\tau$ and $\tau'$ from $D_i$ to $D_t$ are taken to be equivalent if and only if they {\it share the same image.}  This equivalence is obviously a congruence; the corresponding congruence classes may be called {\bf image-fixed co-relative histories}.  A co-relative history may then be viewed as an equivalence class of image-fixed co-relative histories, with members indexed by an appropriate subgroup of the target $\tn{Aut}(D_t)$, moving the image of a representative transition around its orbit in $D_t$.  Heuristically, transitions {\it freeze the actions} of both the automorphism groups $\tn{Aut}(D_i)$ and $\tn{Aut}(D_t)$, while image-fixed co-relative histories allow $\tn{Aut}(D_i)$ to act, and co-relative histories allow both groups to act.  

Co-relative histories embody the principle of iteration of structure (\hyperref[is]{IS}), by supplying ``higher-level relations" between pairs of directed sets in a configuration space of directed sets.   Specifying an appropriate class of co-relative histories, ``suitable for the evolutionary viewpoint," elevates such a configuration space to a {\it kinematic scheme,} which is a ``category-like structure encoding evolutionary pathways for histories" in discrete causal theory, as described in section  \hyperref[subsectionkinematicschemes]{\ref{subsectionkinematicschemes}} below.  Figure \hyperref[corelative]{\ref{corelative}} below illustrates how a particular family of four co-relative histories sharing a common cobase fits into a kinematic scheme called the {\it positive sequential kinematic scheme,} which is a close analogue of Sorkin and Rideout's kinematic scheme describing the sequential growth of causal sets.   

\begin{figure}[H]
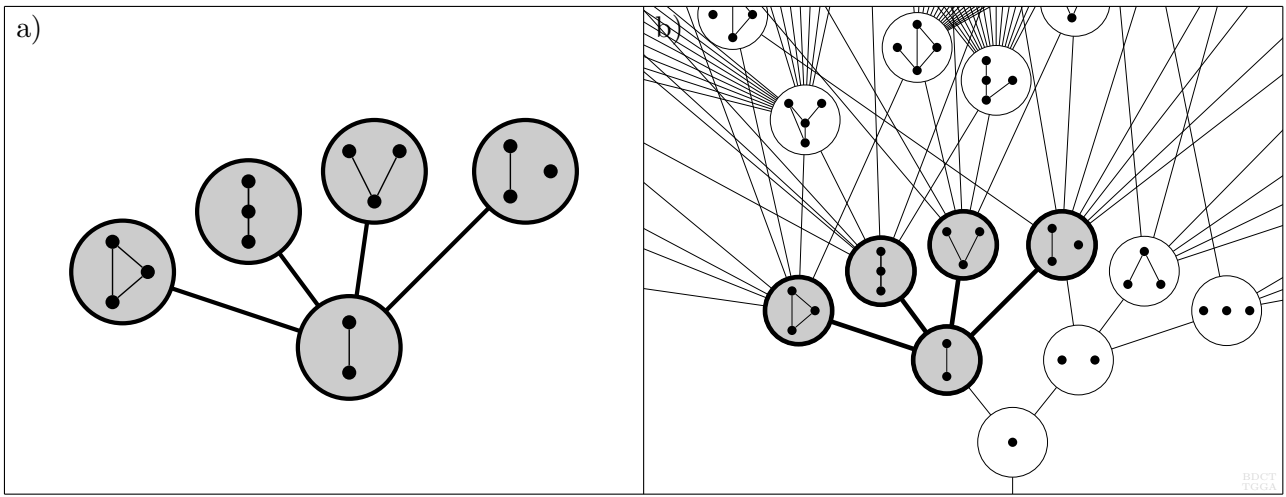


\caption{a) a) Four irreducible co-relative histories sharing a common cobase; b) co-relative histories embedded in a kinematic scheme.}
\label{corelative}
\end{figure}
\vspace*{-.5cm}

Due to the nonuniqueness of co-relative histories between a given cobase and target, illustrated by McKay's example in figure \hyperref[corelativeexamples]{\ref{corelativeexamples}}b above, the abstract structure of a kinematic scheme is generally multidirected, rather than merely directed.  ``Higher-level analogues" of irreducibility and independence are of interest in this context.  For example, the four co-relative histories illustrated in figure \hyperref[corelative]{\ref{corelative}}a above are {\it irreducible,} since each target differs from the common cobase by a ``minimal amount;" i.e., by a single element.  McKay's example implies that {\it irreducible co-relative histories are generally nonunique,} which is just another way of restating the fact that multidirected sets are necessary to encode the abstract structure of kinematic schemes.  Turning to the subject of independence, a basic requirement built into the definition of a kinematic scheme is that it must include ``enough independent co-relative histories" to admit ``evolutionary pathways for every possible history." This requirement is formalized by the condition of {\it weak accessibility} (\hyperref[wa]{WA}), defined in section \hyperref[subsectionkinematicschemes]{\ref{subsectionkinematicschemes}} below.

\subsection{Abstract Quantum Causal Theory via Path Summation}\label{subsectionquantumpathsummation}

\refstepcounter{textlabels}\label{adaptinghistories}

{\bf Adapting the Histories Approach.}  The histories approach to quantum theory, first implemented in the continuum setting via Feynman's path integral, may be adapted without {\it formal} difficulty to the discrete causal context, by means of the theory of path summation over a multidirected set, outlined in section \hyperref[subsectionpathsummation]{\ref{subsectionpathsummation}} above.   In this section, I carry out this adaptation in an important special case, involving a {\it finite region of a locally finite acyclic multidirected set.}   Following such an adaptation, the problem ceases to be a purely formal one, and the actual suitability of the resulting theory becomes the primary concern.   Conceptually, one must consider the range of physical interpretations that may be assigned to the mathematical structures appearing in the theory, and must choose those interpretations most appropriate and useful for the intended applications.  Technically, one must examine whether or not the chosen approach treats the information involved in the theory in an adequate fashion, and produces intelligible results. 

\refstepcounter{textlabels}\label{backgrounddepadapt}

The most na\"{i}ve adaptation of the histories approach to discrete causal theory involves simply replacing continuum spacetime with a directed set $D$, and performing path summation over $D$.  Examples of this approach, in the special case where $D$ is a causal set, already appear in the literature.  These efforts suffer from multiple problems, including the structural deficiencies associated with transitivity and interval finiteness, the shortcomings of element space, the inadequacy of algebraic devices such as incidence algebras, and the problem of background dependence.   With the exception of this latter problem, the results of sections  \hyperref[sectiontransitivity]{\ref{sectiontransitivity}}, \hyperref[sectioninterval]{\ref{sectioninterval}}, and \hyperref[sectionbinary]{\ref{sectionbinary}} of this paper offer the prospect of significant improvements, without requiring any radical conceptual changes.   These results point the way toward flexible new alternatives to quantum field theories in curved spacetime and lattice field theories, involving {\it path summation on relation spaces over fixed locally finite directed sets.} 

\refstepcounter{textlabels}\label{backgroundindepadapt}

The desire for a {\it background-independent} theory is the final catalyst for the radical change in viewpoint represented by the principle of iteration of structure (\hyperref[is]{IS}), and the theory of co-relative histories, introduced in section \hyperref[subsectionquantumprelim]{\ref{subsectionquantumprelim}} above, and further explored in the context of kinematic schemes in section \hyperref[subsectionkinematicschemes]{\ref{subsectionkinematicschemes}} below.   While there is existing precedent for these ideas in the work of Sorkin, Isham, and others, it seems that a truly clear structural picture emerges only by combining the causal metric hypothesis (\hyperref[cmh]{CMH}), discreteness, the independence convention (\hyperref[ic]{IC}), local finiteness (\hyperref[lf]{LF}), hidden structure (\hyperref[hs]{HS}), the relative viewpoint (\hyperref[rv]{RV}), the histories approach, and background independence.   This leads to the following conclusion, which I call the {\it path summation principle:}

\refstepcounter{textlabels}\label{appliestoboth}
\refstepcounter{textlabels}\label{ps}

\hspace*{.3cm} PS. {\bf Path Summation:} {\it A background-independent version of the histories approach to discrete
\\\hspace*{1cm} quantum causal theory may be constructed using the same abstract methods that apply in the
\\\hspace*{1cm} background-dependent context, expressed in terms of path summation over multidirected sets. 
\\\hspace*{1cm} }

The proper object over which to perform path summation in the background-independent context is the {\it relation space over the underlying multidirected set of a kinematic scheme of directed sets.}  This relation space represents the abstract directed structure of a distinguished class of co-relative histories.   The physical interpretation of this conclusion is that the quantum theory of fundamental spacetime structure ultimately involves not a space of elements, nor a space of relations, nor even a space of classical universes, but a space of {\it relationships between pairs of classical universes.}   This viewpoint is further developed in section \hyperref[subsectionkinematicschemes]{\ref{subsectionkinematicschemes}} below. 



\refstepcounter{textlabels}\label{contpathintegral}

{\bf Feynman's Continuum Path Integral.}  For the convenience of the reader, I briefly review Feynman's construction before discussing its discrete causal analogues.  Following Feynman's 1948 paper \cite{FeynmanSOH48}, path integrals over continuum manifolds have found broad use, particularly in quantum field theory, in determining complex-valued {\it quantum amplitudes,} interpreted as encoding relative probabilities for the values of physical measurements.  Although the credit for this approach rightly belongs to Feynman, it has important antecedents.  Feynman explicitly credits the 1935 edition of Dirac's classic book {\it The Principles of Quantum Mechanics} \cite{Dirac35}, while Norbert Wiener's earlier work on Brownian motion, published in the early 1920's just before the introduction of ``modern quantum theory" by Schr\"{o}dinger and Heisenberg, seems to represent the first appearance of path integrals in the literature.  Here, I give a sketch of continuum path integration using the case originally considered by Feynman: the {\it quantum-theoretic behavior of a single particle in a finite region of Euclidean spacetime}.  This purpose of this sketch is to motivate the corresponding abstract quantum causal theory to follow.  Perhaps surprisingly, there is little conceptual advantage to be gained, for this specific purpose, by considering more ``realistic" or elaborate scenarios involving, for instance, special relativity or field theory. In particular, most of the familiar technical machinery for evaluating path integrals in quantum field theory is wholly irrelevant in this context.  

Let $X$ be a finite subset of Euclidean spacetime, bounded between two spatial sections $t=t^\pm$, where $t$ is the universal Newtonian time parameter, and $t^\pm$ are specific time values.  The {\it lower boundary} $\sigma^-$ of $X$ is the set of all lower endpoints of temporal intervals in $X$ at fixed spatial points, and the {\it upper boundary} $\sigma^+$ of $X$ is the set of all upper endpoints of such intervals.\footnotemark\footnotetext{Mathematically, one may imagine all manner of perverse spacetime regions.  For the present illustrative purpose, it does no harm to assume that $X$ is bounded by a smooth hypersurface, or even that $X$ is convex.}  Let $\Gamma=\Gamma_{\tn{\fsz max}}(X)$ be the set of all {\it maximal directed paths} in $X$; i.e., continuous future-directed paths in $X$ beginning somewhere on $\sigma^-$ and terminating somewhere on $\sigma^+$.  Consider the motion of a single particle, with Lagrangian $\ms{L}$ and classical action $\ms{S}=\int_\gamma \ms{L}dt$, along a path $\gamma$ in $\Gamma$.  Feynman's approach to studying the quantum behavior of such a particle is based on the two following ``postulates:"

\refstepcounter{textlabels}\label{feynpost}

\refstepcounter{textlabels}\label{f1}

\hspace*{.3cm} F1. {\bf Quantum Amplitude}: {\it If an ``ideal measurement" is performed, to determine whether 
\\\hspace*{1.05cm} the trajectory of the particle belongs to $\Gamma=\Gamma_{\tn{\fsz max}}(X)$, then the probability of an affirmative \\\hspace*{1.05cm} result is the square modulus of a complex-valued {\bf quantum amplitude} $\psi(X;\ms{L})$, given 
\\\hspace*{1.05cm} by summing complex contributions from each path $\gamma\in\Gamma$.}(Adapted from \cite{FeynmanSOH48}, page 8.)

\refstepcounter{textlabels}\label{f2}

\hspace*{.3cm} F2. {\bf Phase Map}: {\it Each path $\gamma\in\Gamma$ contributes equally in magnitude to the amplitude $\psi(X;\ms{L})$, \\\hspace*{1.1cm} with complex {\bf phase} $\Theta(\gamma)=e^{\frac{i}{\hbar}\ms{S}(\gamma)}$, where $\ms{S}$ is the classical action corresponding to $\ms{L}$, \\\hspace*{1.1cm} and where $\hbar=h/2\pi$ is Planck's reduced constant.}(Adapted from \cite{FeynmanSOH48}, page 9.)

\refstepcounter{textlabels}\label{feynamp}

What these ``postulates" really describe is the {\it heuristic intent of a limiting procedure,} in which the interval $[t^-,t^+]$ is subjected to a sequence of increasingly fine {\it partitions} $\Delta=\{t^-=t_0,t_1,...,t_{N}=t^+\}$, and a ``path" $\gamma$ is specified by a sequence $\mbf{x}:=\{x_0,x_1,...,x_{N}\}$ of position values at each time in $\Delta$.  A suitable interpolating convention must be chosen for the ``path increments" $\gamma_i$ of $\gamma$ between each pair of spacetime points $(x_i,t_i)$ and $(x_{i+1},t_{i+1})$.   For example, $\gamma_i$ might be taken to be the {\it classical trajectory} between its endpoints, as determined by the Lagrangian $\ms{L}$, according to Hamilton's principle.  Denote by $\sigma_i=\sigma_i(\Delta)$ the spatial section of $X$ at time $t=t_i$ in the partition $\Delta$, and let $|\Delta|$ be the norm of $\Delta$; i.e., the supremum of the values $t_{i+1}-t_i$.  Then the amplitude $\psi(X;\ms{L})$ is {\it defined} to be the limit
\[\psi(X;\ms{L}):=\lim_{|\Delta|\to0}\int_{\mathbf{x}}\psi(\Delta;\mathbf{x};\ms{L})d\mathbf{x}\]
\begin{equation}\label{continuumpathintegral}
:=\lim_{|\Delta|\to0}\int_{\sigma_N}... \int_{\sigma_1}\int_{\sigma_0}    C\cdot\tn{exp}\Bigg(\sum_{i=1}^{N}\frac{i}{\hbar}\ms{S}(\gamma_i)\Bigg)dx_0dx_1...dx_{N},
\end{equation}
if this limit exists.  The first (i.e., top) expression in equation \hyperref[continuumpathintegral]{\ref{continuumpathintegral}} is purely symbolic, meaning ``integrate $N+1$ times, over all values of $x_0,...,x_N$, a complex quantity $\psi(\Delta;\mbf{x};\ms{L})$, depending on a particular choice of $x_0,...,x_N$ for the partition $\Delta$, and on the Lagrangian $\ms{L}$, then take the limit as $|\Delta|\rightarrow 0$."   The second expression makes this more explicit at a particular level of approximation.   The sum in the exponential comes from {\it multiplying} the phases of each ``increment" $\gamma_i$ of a typical ``path" $\gamma$, determined by $\Delta$ and $\mbf{x}$.  The number $C$ appearing in the integrand is a proportionality factor which may be ignored in the present context. 

\begin{figure}[H]
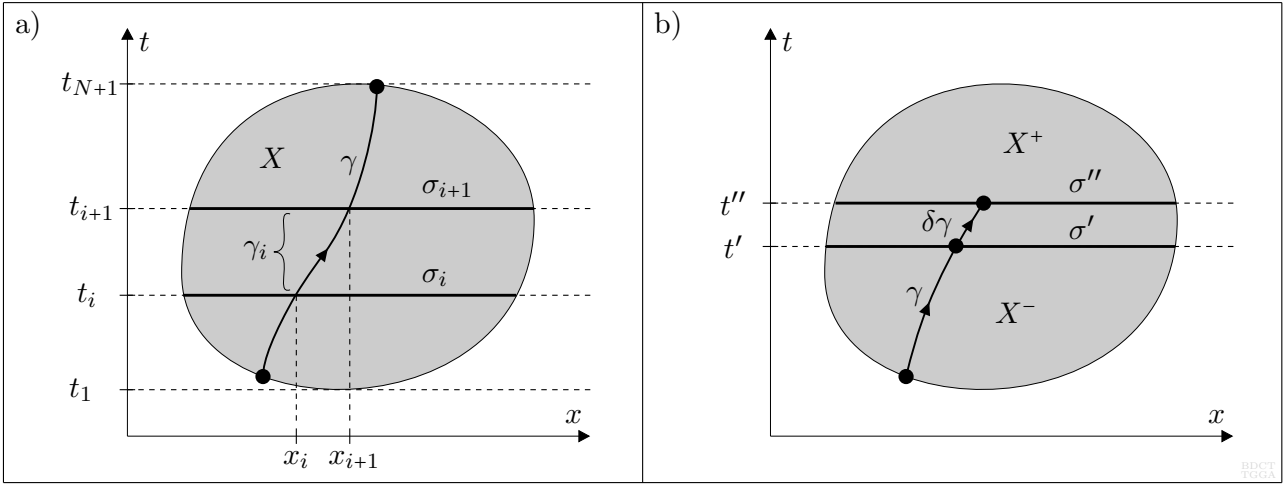


\caption{a) Setting for Feynman's path integral; b) deriving Schr\"{o}dinger's equation via Feynman's path integral.}
\label{feynman}
\end{figure}

Figure \hyperref[feynman]{\ref{feynman}}a below illustrates the setting for Feynman's path integral.  Comparing this with figure \hyperref[permeability]{\ref{permeability}}a of section \hyperref[subsectionrelation]{\ref{subsectionrelation}} above, one may observe that the spatial sections $\sigma_i$ serve as {\it Cauchy surfaces} for $X$, since every future-directed path $\gamma$ in $X$ from the past of $\sigma_i$ to the future of $\sigma_i$ intersects $\sigma_i$ uniquely.  The physical significance of these Cauchy surfaces is that Feynman's path integral is {\it information-theoretically adequate;} i.e., it captures all information flowing from the lower boundary $\sigma^-$ of $X$ to the upper boundary $\sigma^+$ of $X$.   In particular, there is no {\it permeability problem} of the type studied by Major, Rideout, and Surya \cite{RideoutSpatialHypersurface06} in the context of causal set theory.   Figure  \hyperref[feynman]{\ref{feynman}}b illustrates the derivation of Schr\"{o}dinger's equation via Feynman's path integral, which is described in section \hyperref[subsectionschrodinger]{\ref{subsectionschrodinger}} below. 

\vspace*{-.5cm}


\refstepcounter{textlabels}\label{causalpathint}
\refstepcounter{textlabels}\label{dependsimperm}

{\bf Abstract Analogues of Feynman's Path Integral in the Discrete Causal Context.} Using the theory of path summation over a multidirected set, introduced in section \hyperref[subsectionpathsummation]{\ref{subsectionpathsummation}} above, it is straightforward, at least in a formal sense, to define abstract analogues of Feynman's path integral, with maximal antichains playing the role of spatial sections of spacetime.  However, new issues, as well as new choices, arise in the discrete causal context.  First, one must select a suitable class of directed sets to serve as models of classical spacetime.  The results of sections \hyperref[sectiontransitivity]{\ref{sectiontransitivity}} and \hyperref[sectioninterval]{\ref{sectioninterval}} of this paper suggest that certain classes of nontransitive locally finite directed sets are potentially better choices than the class of causal sets for this purpose.  Second, the generic problem of permeability of maximal antichains in directed sets, discussed in section \hyperref[subsectionrelation]{\ref{subsectionrelation}} above, renders na\"{i}ve path summation over an arbitrary directed set information-theoretically inadequate. The obvious solution is {\it to consider path sums over relation space}, where the permeability problem disappears, by theorem \hyperref[theoremrelimpermeable]{\ref{theoremrelimpermeable}}.   Third, one must decide what kinds of paths to consider.  For example, is it necessary to include {\it all} maximal paths in path sums, or does it suffice to work in terms of maximal chains?  Assuming that chains do suffice, do they provide the most natural and insightful point of view?  For elementary information-theoretic reasons, I work exclusively with chains in this section, but do not attempt to argue that more complicated paths are always irrelevant.

A remaining issue, absent both in Feynman's construction and in the context of causal sets, is the question of whether or not to allow directed sets including cycles, and if so, how to treat path summation when cycles are involved.  Even a small finite region of a directed set containing a cycle admits an infinite number of distinct chains wrapping around this cycle different numbers of times, potentially dominating the corresponding path sum.  Moreover, as already mentioned in section \hyperref[subsectionquantumprelim]{\ref{subsectionquantumprelim}} above, ``higher-level cycles" arise in the theory of co-relative histories and kinematic schemes, even when the individual directed sets involved are acyclic, unless one also restricts to the finite case.   Hence, the issue of cycles remains relevant in the {\it background-independent context} even for relatively conservative models of classical spacetime.  A number of possible approaches to this problem may be devised.  First, one may simply apply the same formal methods as in the acyclic case, and try to show that the resulting path sums converge in some suitable sense.  Second, one might attempt to treat cycles ``holistically" in some way.  A simple method would be to define an equivalence relation $\sim$ on paths, in which $\gamma\sim\gamma'$ if and only if $\gamma$ coincides with $\gamma'$ ``outside of cycles," then assign a single phase to each equivalence class.   A more elaborate version of the same idea is to perform a causal atomic decomposition, in which the causal atoms are maximal cycles, and the resulting directed set is a ``maximal acyclic approximation" of the original directed set.  Paths in this acyclic approximation then correspond to the above equivalence classes, and the phases assigned to its paths are partially determined by the internal cyclic structures of the ``elements" in their images.    A mathematically attractive holistic approach to cycles is to adopt the relative viewpoint (\hyperref[rv]{RV}) and {\it generalize cycles to morphisms whose sources are cyclic directed sets,}\footnotemark\footnotetext{That is, {\it totally} cyclic directed sets; i.e., sets in which there are chains in both directions between any pair of elements.} just as paths generalize chains to morphisms whose sources are linear directed sets.  Finally, if all cycles in a directed set happen to be ``large," like many of the closed timelike curves appearing in solutions to general relativity, one could investigate local problems without considering cycles.  In this section, I sidestep these issues by focusing on the abstract acyclic case.  

\begin{figure}[H]
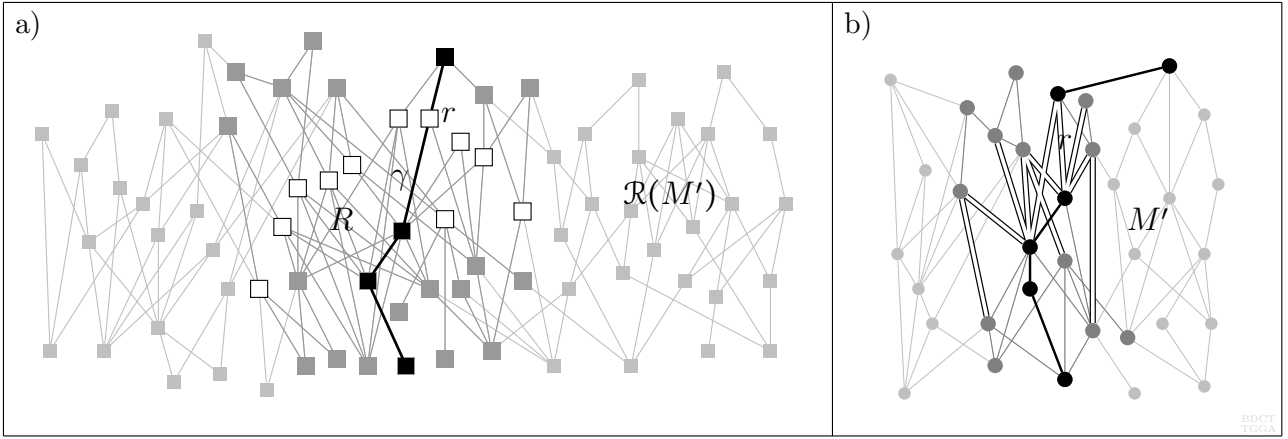


\caption{a) Path summation over a finite subset $R$ of relation space, represented by large nodes; maximal antichain $\sigma\subset R$ represented by large open nodes; maximal chain $\gamma$ in $R$ represented by large dark nodes and edges; b) corresponding view in element space; thick white lines represent $\sigma$.}
\label{causalpathsum}
\end{figure}

Applying these simplifying assumptions, let $R$ be a finite subspace of the relation space $\ms{R}(M')$ over a locally finite acyclic multidirected set $M'=(M',R',i',t')$.\footnotemark\footnotetext{Primes are used to denote the underlying multidirected set $M'=(M',R',i',t')$, in order to leave the ``unprimed" symbol $R$ free to denote the object of principal interest; namely, a finite subspace of $R'$.}  Let $\big(\tn{Ch}(R),\sqcup\big)$ be the chain concatenation semicategory over $R$, and let $\Gamma=\tn{Ch}_{\tn{\fsz max}}(R)$ be the subset\footnotemark\footnotetext{From the algebraic viewpoint, the concatenation product of any pair of elements of $\Gamma$ is zero, since these elements are maximal chains.  Similarly, from the order-theoretic viewpoint, no element of $\Gamma$ directly precedes any other element of $\Gamma$ under the causal concatenation relation.  Hence, it makes no difference whether $\Gamma$ is called a ``subset," a ``subsemicategory," or a ``subspace" in this context.  I choose the first option.} of $\big(\tn{Ch}(R),\sqcup\big)$ consisting of all maximal chains in $R$.   Let $\sigma$ be a maximal antichain in $R$, separating $R$ into the disjoint union $R^-\coprod\sigma\coprod R^+$, and let $\gamma\in \Gamma$ be a typical maximal chain in $R$, intersecting $\sigma$ in a unique element $r$.  Comparing these designations to Feynman's continuum construction, $\ms{R}(M')$ corresponds to Euclidean spacetime, $R$ to the finite spacetime region $X$, $\Gamma$ to the set of maximal directed paths in $X$, $\sigma$ to a spatial section of $X$, and $\gamma$ to a maximal directed path in $X$. This setup is illustrated in figure \hyperref[causalpathsum]{\ref{causalpathsum}}a below, with the corresponding picture in the original element space $M'$ shown in figure \hyperref[causalpathsum]{\ref{causalpathsum}}b. 

\vspace*{-.5cm}

Referring to Feynman's path integral formula for the quantum amplitude $\psi(X;\ms{L})$ in the continuum context, appearing in equation \hyperref[continuumpathintegral]{\ref{continuumpathintegral}} above, note that the ``path increments" $\gamma_i$, appearing in the second (i.e., bottom) expression, are defined in terms of {\it pairs} of consecutive spatial sections $\sigma_i$ and $\sigma_{i+1}$, since a single section of continuum spacetime has ``no thickness."  In the discrete causal context, however, the {\it single} maximal antichain of relations $\sigma$ accomplishes an analogous purpose.  Each relation $r$ belonging to $\sigma$ constitutes an ``increment" of any chain $...\prec r^-\prec r\prec r^+\prec...$ passing through $r$.\footnotemark\footnotetext{For general paths, the situation is more complicated, since the portion of such a chain containing $r$ may be ``short-circuited" by a reducible relation.} 

Having specified the necessary data for path summation {\it intrinsic to} $R$, I now consider the target object $T$ in which the values of path sums over $R$ are taken to lie, and the phase map $\Theta:\tn{Ch}(R)\rightarrow T$ that supplies the terms of the particular path sum analogous to Feynman's path integral.  Both $T$ and $\Theta$ are {\it a priori} independent of $R$, but physical and philosophical considerations narrow the field of possibilities considerably.  For example, the target object $T$ is strongly constrained if one wishes to interpret path summation in terms of probabilities.\footnotemark\footnotetext{As noted in section \hyperref[settheoretic]{\ref{settheoretic}} above, Chris Isham advocates replacing probabilities with ``generalized truth values."  See, for example, \cite{Isham11}, page 6.}  The phase map $\Theta$ specifies the dynamics in discrete causal theory, just as Feynman's phase map specifies the dynamics in continuum theory, via the classical action.  I will not attempt to conceal that I do not presently know precisely what the corresponding ``action" in the discrete causal context should be.  As mentioned in section \hyperref[subsectionapproach]{\ref{subsectionapproach}} near the beginning of this paper, in the discussion of causal set phenomenology, this problem has been studied in the case of causal sets by Fay Dowker and a few others \cite{DowkerScalarCurvature10}.   Under the strong form of the causal metric hypothesis (\hyperref[cmh]{CMH}), the phase map $\Theta$ ought to be somehow determined by ``underlying directed structure," perhaps via natural ``mediating structures," such as the valence fields introduced at the end of section \hyperref[subsectiontopology]{\ref{subsectiontopology}} above.  In particular, in the background-independent context, the elements of $R$ represent co-relative histories, whose entire internal structures as equivalence classes of transitions are relevant to the choice of phase map under the {\it quantum causal metric hypothesis} (\hyperref[qcmh]{QCMH}), introduced in section \hyperref[subsectionkinematicschemes]{\ref{subsectionkinematicschemes}} below.  It is reasonable to surmise that {\it entropy} plays a significant role in the action in the background independent setting. 

In the present abstract context, I make simplifying assumptions regarding $T$ and $\Theta$, while retaining enough generality to yield reasonably interesting and relevant {\it causal Schr\"{o}dinger-type equations,} derived in section \hyperref[subsectionschrodinger]{\ref{subsectionschrodinger}} below.  In particular, I assume that the target object $T$ is a commutative ring with unit, and that the phase map $\Theta$ is generated under the concatenation product on $\tn{Ch}(R)$ by a relation function $\theta:R\rightarrow T$.  The latter assumption means that the phase $\Theta(\gamma)$ of any finite chain $\gamma=r_1\sqcup...\sqcup r_N$ in $R$ is the product
\begin{equation}\label{chainfactorization}
\Theta(\gamma)=\prod_{i=1}^N\theta(r_i).
\end{equation}
Since $R$ is a finite directed set, any chain $\gamma$ in $R$ has a unique factorization of this form.\footnotemark\footnotetext{Again, this simplification does not apply to general paths in $R$.}  At a more technical level, $\Theta$ may be regarded as a ``semicategory morphism," from the chain concatenation semicategory $\big(\tn{Ch}(R),\sqcup\big)$ over $R$, to the target object $T$.  Factorization of $\gamma$ is analogous to the partitioning of a continuum ``path" $\gamma$ into ``path increments" $\gamma_i$, via a partition $\Delta$, in Feynman's construction.  The product formula for $\Theta(\gamma)$ in equation \hyperref[chainfactorization]{\ref{chainfactorization}} is analogous to Feynman's formula
\[\Theta(\gamma)=\prod_{i=1}^{N}\Theta(\gamma_i)=\prod_{i=1}^{N}e^{\frac{i}{\hbar}\ms{S}(\gamma_i)}=\tn{exp}\Bigg(\sum_{i=1}^N\frac{i}{\hbar}\ms{S}(\gamma_i)\Bigg).\]
As mentioned above, much more general treatments are possible.  For example, the target object $T$ might be replaced by some higher-level algebraic structure, such as a {\it monoidal category.}  

\refstepcounter{textlabels}\label{abstractamp}

Under these assumptions, the analogue of Feynman's quantum amplitude $\psi(X;\ms{L})$ is an element $\psi(R;\theta)$ of $T$, given by summing phases over the set $\Gamma$ of maximal chains in $R$.   In the present simplified setting, the existence of $\psi(R;\theta)$ is automatic.  The choice of notation for this {\it generalized quantum amplitude} reflects the fact that the relation function $\theta$ is {\it information-theoretically analogous to a Lagrangian.}  It is useful to define an element $\Psi(R;\theta)$ of the chain concatenation algebra $T^\sqcup[R]$ to serve as an {\it algebraic precursor} to $\psi(R;\theta)$.\footnotemark\footnotetext{There is nothing to prevent consideration of such precursor functionals in continuum theory, but they are {\it a priori} cumbersome as formal sums due to their large numbers of terms.}  As a causal path functional, $\Psi(R;\theta)$ is merely the {\it restriction to} $\Gamma$ of the phase map $\Theta$ generated by the relation function $\theta$.  The corresponding amplitude $\psi(R;\theta)$ is then given by applying the global evaluation map $e_R$ to $\Psi(R;\theta)$. The following definition makes these notions precise:\\

\begin{defi}\label{maximalfunctional} Let $M'=(M',R',i',t')$ be a locally finite acyclic multidirected set with relation space $\ms{R}(M')$, and let $R$ be a finite subset of $\ms{R}(M')$, viewed as a full subobject.  Let $\Gamma=\tn{Ch}_{\tn{\fsz max}}(R)$ be the set of maximal chains in $R$, viewed as a subset of the chain concatenation semicategory $\big(\tn{Ch}(R),\sqcup\big)$.  Let $T$ be a commutative ring with unit, and let $\Theta:\tn{Ch}(R)\rightarrow T$ be a multiplicative phase map generated by a relation function $\theta:R\rightarrow T$.   
\begin{enumerate}
\item The {\bf maximal chain functional} $\Psi(R;\theta)$ of $R$ with respect to $\theta$ is the causal path functional $\Gamma\rightarrow T$ given by restricting the phase map $\Theta$ to $\Gamma$:
\begin{equation}\label{precursoramplitude}
\Psi(R;\theta)=\Theta|_\Gamma=\sum_{\gamma\in \Gamma}\Theta(\gamma)\gamma.
\end{equation}
\item The {\bf (generalized quantum) amplitude} $\psi(R;\theta)$ of $R$ with respect to $\theta$ is the element of $T$ given by applying the global evaluation map $e_R$ to $\Psi(R;\theta)$:
\begin{equation}\label{amplitude}
\psi(R;\theta)=e_R\big(\Psi(R;\theta)\big)=\sum_{\gamma\in\Gamma}\Theta(\gamma)=\sum_{\gamma \in \Gamma}\Bigg(\prod_{i=1}^{N_\gamma}\theta(r_{\gamma,i})\Bigg),
\end{equation}
where $\gamma=r_{\gamma,1}\sqcup...\sqcup r_{\gamma,N_\gamma}$ is the unique factorization of $\gamma$ in $R\subset \big(\tn{Ch}(R),\sqcup\big)$.
\end{enumerate}
\end{defi}
The sum $\sum_{\gamma \in \Gamma}()$ appearing in equation \hyperref[amplitude]{\ref{amplitude}} is analogous to the integral $\int_\mbf{x}()d\mbf{x}$ in Feynman's continuum construction, appearing in equation \hyperref[continuumpathintegral]{\ref{continuumpathintegral}} above.  However, in the discrete causal context, this sum is exact; no limiting process is necessary.   


\subsection{Schr\"{o}dinger-Type Equations in Quantum Causal Theory}\label{subsectionschrodinger}

\refstepcounter{textlabels}\label{schrodhistory}

{\bf Schr\"{o}dinger-type Equations via the Histories Approach.}  Schr\"{o}dinger's equation describes the dynamics of nonrelativistic quantum systems.  It may be derived in a conceptually satisfying manner via path integral methods, as demonstrated by Feynman.   Since these methods represent a special case of the general histories approach to quantum theory, Schr\"{o}dinger's equation is much more than merely a ``low-energy approximation of quantum field theory."  So far as the histories approach itself is valid, any shortcomings in Schr\"{o}dinger's equation arise from {\it shortcomings in the choice of history configuration space} used in this approach.  In nonrelativistic quantum theory, histories are represented in a manifestly unrealistic fashion, by paths traced out on a fixed copy of Euclidean spacetime.  Supplying better underlying models in a similar conceptual context may be expected to produce superior dynamical laws of the same general type.  This expectation is already borne out to some extent in more sophisticated modern implementations of path integration in quantum field theory and various approaches to quantum gravity.  In this section, I examine this problem in the context of discrete causal theory.  I begin by reviewing Feynman's derivation of Schr\"{o}dinger's equation, then derive discrete causal analogues.  Equation \hyperref[causalschrodinger]{\ref{causalschrodinger}} below is an example of a Schr\"{o}dinger-type equation derived via this method. 


\refstepcounter{textlabels}\label{conschrodequ}

{\bf Continuum Version of Schr\"{o}dinger's Equation via Path Integrals.}  One of the major achievements of Feynman's paper \cite{FeynmanSOH48} is the recovery, via path integral methods, of the {\bf nonrelativistic Schr\"{o}dinger equation}:
\begin{equation}\label{hamiltonschrod}
i\hbar\frac{\partial\psi^-}{\partial t}=\mathbf{H}\psi^-,
\end{equation}
which describes the mathematical behavior of the {\it past wave function} $\psi^-$, associated with the motion of a particle through a distinguished region $X$ of Euclidean spacetime, as described in section \hyperref[subsectionschrodinger]{\ref{subsectionschrodinger}} above.  This is the ``usual wave function" encountered in elementary quantum theory.  The adjective ``past" indicates that it depends only on events {\it preceding} a specified choice of ``present," given by an appropriate spatial section of $X$.  Here $\mathbf{H}$ is the {\bf Hamiltonian operator}, defined in the single-particle case by the formula
\begin{equation}\label{classham}
\mathbf{H}\psi^-=\frac{-\hbar^2}{2m}\frac{\partial^2\psi^-}{\partial x^2}+V\psi^-,
\end{equation}
where $m$ is the mass of the particle and $V$ is its potential energy.   Schr\"{o}dinger's equation is a {\it dynamical law} for nonrelativistic quantum theory, permitting quantitative predictions of observable physical phenomena.

To derive Schr\"{o}dinger's equation, Feynman first expressed his quantum amplitude $\psi(X;\ms{L})$ as the {\it complex inner product} of the past wave function $\psi^-$ and a corresponding {\it future wave function} $\psi^+$:
\begin{equation}\label{innerprod}
\psi(X;\ms{L})=\big\langle \psi^+(x',t'),\psi^-(x',t')\big\rangle:=\int_{\sigma'} \psi^+(x',t')^*\psi^-(x',t')dx'.
\end{equation}
The domain of integration of the variable $x'$ in equation \hyperref[innerprod]{\ref{innerprod}} is a {\it single} spatial section $\sigma'$ of $X$ at time $t=t'$.  This spatial section partitions the spacetime region $X$ into a disjoint union $X=X^-\coprod\sigma'\coprod X^+$, where $X^{\pm}$ are the {\bf past} and {\bf future regions} with respect to the choice of ``present" represented by $\sigma'$.  The past and future wave functions $\psi^{\pm}$ depend on {\it both position and time}; hence, different choices of $\sigma'$ produce different values of $\psi^{\pm}$.   However, the quantum amplitude $\psi(X;\ms{L})$ is a complex number depending {\it only on the spacetime region $X$ and the Lagrangian $\ms{L}$,} so different choices of $\sigma'$ produce the same inner product in equation \hyperref[innerprod]{\ref{innerprod}}. 

\refstepcounter{textlabels}\label{contwavefunc}

Without belaboring the details, the value $\psi^-(x',t')$ of the past wave function at a point $(x',t')$ in $X$ is determined by ``integrating over all maximal paths" in the past region $X^-$ of $X$ {\it terminating} at $(x',t')$.  Similarly, the value of the future wave function at $(x',t')$ is determined by ``integrating over all maximal paths" in the future region $X^+$ of $X$ {\it beginning} at $(x',t')$.  Like Feynman's description of how to determine his quantum amplitude $\psi(X;\ms{L})$, given in his first ``postulate" (\hyperref[f1]{F1}) above, these are {\it heuristic descriptions of a limiting process.}  Every path contributing to the quantum amplitude $\psi(X;\ms{L})$ is given by ``joining paths in $X^-$ and $X^+$ at a common point $(x',t')$ in the spatial section $\sigma'$," since $\sigma'$ is a Cauchy surface.  Hence, combining the wave functions $\psi^{\pm}$, via the inner product formula in equation \hyperref[innerprod]{\ref{innerprod}}, suffices to recover $\psi(X;\ms{L})$, with the $x'$-independence ``integrated out."  

The {\bf past wave function} $\psi^-$ may be {\it defined} as a limit:
\begin{equation}\label{psiminuslim}
\psi^-(x',t'):=\lim_{|\Delta^-|\to0}\int_{\mbf{x}^-}\psi^-(\Delta^-;\mathbf{x}^-;\ms{L})d\mbf{x}^-,
\end{equation}
where $\{\Delta^-\}$ is a sequence of partitions of the time interval corresponding to the past region $X^-$ of $X$, bounded above by the spatial section $\sigma'$.  The integration is performed over all sequences $\mbf{x}^-$ of values of spatial coordinates corresponding to such partitions, subject to the additional condition that these sequences must terminate at the specific spatial value $x'$.  The integrand $\psi^-(\Delta^-;\mathbf{x}^-;\ms{L})$ is given, up to a constant, by products of phases of ``path increments" determined by $\Delta^-$ and $\mbf{x}^-$. 

\refstepcounter{textlabels}\label{contschrodequ}

Schr\"{o}dinger's equation \hyperref[hamiltonschrod]{\ref{hamiltonschrod}} is derived by comparing the values of the past wave function $\psi^-$ for ``nearby" time values $t'$ and $t''$, where $t'<t''$, then taking the limit as the difference $t''-t'$ approaches zero.   For simplicity, I replace the limit notation with the ``approximation symbol" $\approx$ below, since no limiting process is necessary in the discrete causal case.  Let $t'$ and $t''$ be ``nearby" time values, and let $\sigma'$ and $\sigma''$ be the corresponding spatial sections.  Let $\Delta^-$ be a partition of the time interval up to and including $t'$.  This setup is illustrated in figure \hyperref[feynman]{\ref{feynman}}b of section \hyperref[subsectionquantumpathsummation]{\ref{subsectionquantumpathsummation}} above.  

The approximation 
\[\psi^-(x',t')\approx\int_{\mbf{x}^-}\psi^-(\Delta^-;\mathbf{x}^-;\ms{L})d\mbf{x}^-,\]
computed at time $t=t'$, given by dropping the limit in equation \hyperref[psiminuslim]{\ref{psiminuslim}} above, leads to the ``subsequent" approximation
\[\psi^-(x'',t'')\approx\int_{\sigma'}\int_{\mbf{x}^-}\psi^-(\Delta^-;\mathbf{x}^-;\ms{L})d\mbf{x}^-\tn{exp}\Big(\frac{i}{\hbar}\ms{S}(\delta\gamma)\Big)dx',\]
computed at time $t=t''$, where $\delta\gamma$ ranges over all ``path increments" beginning somewhere on the spatial section $\sigma'$, and terminating at the specific point $(x'',t'')$ of the spatial section $\sigma''$.  This expression, in turn, leads to the ``approximate recursion"
\begin{equation}\label{schrodinger}
\psi^-(x'',t'')\approx\int_{\sigma'}\psi^-(x',t')\tn{exp}\Big(\frac{i}{\hbar}\ms{S}(\delta\gamma)\Big)dx',
\end{equation}
which yields Schr\"{o}dinger's equation \hyperref[hamiltonschrod]{\ref{hamiltonschrod}} in the limit.  

\begin{figure}[H]
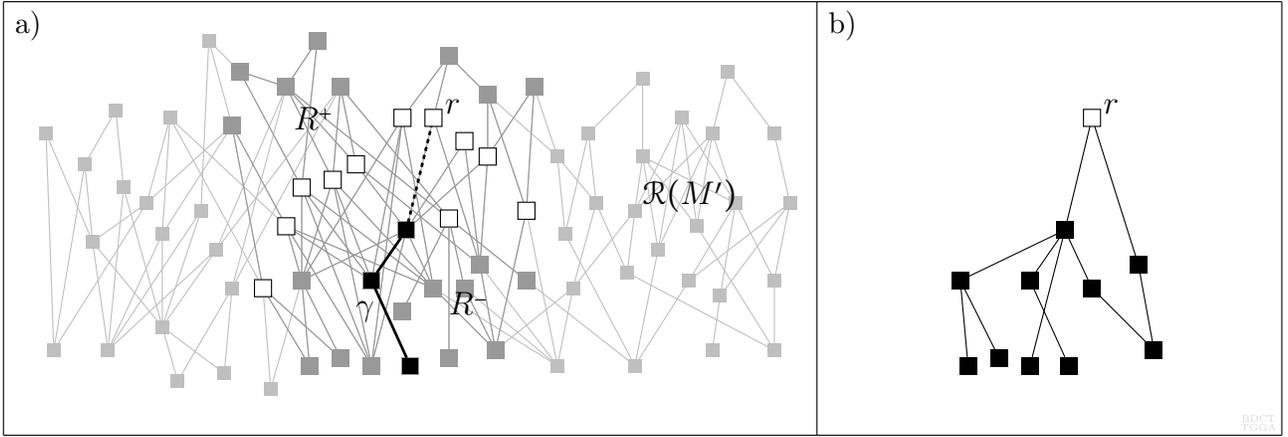


\caption{a) Setup for deriving Schr\"{o}dinger-type equation in the discrete causal context: $R$ represented by large nodes; $\sigma$ represented by large open nodes; b) multiplication on the right by $r$ selects chains admitting extension by $r$.}
\label{figcausalschrodinger}
\end{figure}

{\bf Discrete Causal Analogues of Schr\"{o}dinger's Equation.}  Returning to the discrete causal context, let $M'$, $R$, $\sigma$, $R^\pm$, $T$, and $\Theta$ be as designated in section \hyperref[subsectionquantumpathsummation]{\ref{subsectionquantumpathsummation}} above, and let $r$ be an element of $\sigma$.  Let $\gamma$ be a maximal chain in $R^-$ admitting {\it extension by} $r$; i.e., a maximal chain in $R^-$ such that $\gamma'=\gamma\sqcup r$ is a maximal chain in $R^-\coprod\sigma$.  In what follows, $r$, $\gamma$, and $\gamma'$ are allowed to vary as needed. For example, $r$ is allowed to vary over $\sigma$, and later, over all of $R$.  This setup is illustrated in figure \hyperref[figcausalschrodinger]{\ref{figcausalschrodinger}}a below. 

\vspace*{-.5cm}

Now consider the restriction $\Theta|_\sigma$ of the phase map $\Theta:\tn{Ch}(R)\rightarrow T$ to the maximal antichain $\sigma$, viewed as an element of the chain concatenation algebra $T^\sqcup[R]$ over $R$:
\[\Theta|_\sigma=\sum_{r\in\sigma}\theta(r)r,\]
where I have replaced $\Theta$ in the sum by its generating relation function $\theta$, since the two coincide on individual relations; i.e., individual elements of $R$.  Consider also the restriction $\Theta|_{\Gamma^-}$ of $\Theta$ to the set $\Gamma^-:=\tn{Ch}_{\tn{\fsz max}}(R^-)$ of maximal chains in the past region $R^-$: 
\[\Theta|_{\Gamma^-}=\sum_{\gamma\in \Gamma^-}\Theta(\gamma)\gamma,\]
where I have chosen not to factor $\Theta(\gamma)$ in terms of $\theta$ in the summand.  I claim that the product in $T^\sqcup[R]$ of these two elements, suitably ordered, is the element representing the restriction of $\Theta$ to the set $\Gamma_\sigma^-:=\tn{Ch}_{\tn{\fsz max}}(R^-\coprod\sigma)$ of maximal chains in $R^-\coprod\sigma$:
\begin{equation}\label{restrictions}
\Theta|_{\Gamma^-}\sqcup\Theta|_\sigma=\sum_{\gamma'\in \Gamma_\sigma^-}\Theta(\gamma')\gamma'.
\end{equation}
To see that this is true, observe that for any chain $\gamma$ in $\Gamma^-$, and for any relation $r$ belonging to $\sigma$, the concatenation product $\gamma'=\gamma\sqcup r$ is nonzero in $T^\sqcup[R]$ if and only if $\gamma$ admits extension by $r$.  An equivalent condition is that the terminal relation $r^-$ of $\gamma$ directly precedes $r$ in $R$.  Hence, the summand $\theta(r)r$ in $\Theta|_\sigma$, multiplying the element $\Theta|_{\Gamma^-}$ on the right, annihilates all chains $\gamma$ in $\Gamma^-$ {\it except those admitting extension by $r$,} as illustrated in figure \hyperref[figcausalschrodinger]{\ref{figcausalschrodinger}}b above.  The surviving terms in the product $\Theta|_{\Gamma^-}\sqcup\theta(r)r$ are therefore chains $\gamma'$ in  $\Gamma_\sigma^-$ terminating at $r$, weighted by their appropriate phases, since $\Theta$ is generated by $\theta$.\footnotemark\footnotetext{The qualifier ``weighted by their appropriate phases" makes this statement true even when $\theta(r)=0_T$.}   Letting $r$ vary over $\sigma$, it remains to show that {\it every} such chain $\gamma'$ is included in the product $\Theta|_{\Gamma^-}\sqcup\Theta|_\sigma$, appearing on the left-hand side of equation \hyperref[restrictions]{\ref{restrictions}}.   This follows from the impermeability of the maximal antichain $\sigma$, guaranteed by theorem \hyperref[theoremrelimpermeable]{\ref{theoremrelimpermeable}} of section \hyperref[subsectionrelation]{\ref{subsectionrelation}} above, which implies that every maximal chain $\gamma$ in $R^-$ terminates in a relation $r^-$ directly preceding some relation $r$ belonging to $\sigma$.  

\refstepcounter{textlabels}\label{abstractpathfunct}
\refstepcounter{textlabels}\label{abstractwavefunct}

Now let $r$ vary over all of $R$.  Denote by $\Gamma_r^-:=\tn{Ch}_{\tn{\fsz max}}^-(R;r)$ the set of all maximal chains in $R$ terminating at $r$, and by $\Gamma_r^+:=\tn{Ch}_{\tn{\fsz max}}^+(R;r)$ the set of all maximal chains in $R$ beginning at $r$.  Proceeding as in the construction of the maximal chain functional $\Psi(R;\theta)$ and the abstract quantum amplitude $\psi(R;\theta)$ in section \hyperref[subsectionquantumpathsummation]{\ref{subsectionquantumpathsummation}} above, I define {\it past and future chain functionals} $\Psi_{R;\theta}^\pm(r)$ on the chain sets $\Gamma_r^\pm$, viewed as elements of the chain concatenation algebra $T^\sqcup[R]$, which then serve as algebraic precursors to {\it past and future wave functions} $\psi_{R;\theta}^\pm(r)$, obtained from the functionals $\Psi_{R;\theta}^\pm(r)$ by application of the global evaluation map $e_R$.\\

\begin{defi}\label{pastfuturepathfunctionals} Let $M'=(M',R',i',t')$ be a locally finite acyclic multidirected set, with relation space $\ms{R}(M')$, and let $R$ be a finite subset of $\ms{R}(M')$, viewed as a full subobject.   Let $r$ be an element of $R$, and let $\Gamma_r^-=\tn{Ch}_{\tn{\fsz max}}^-(R;r)$ and $\Gamma_r^+=\tn{Ch}_{\tn{\fsz max}}^+(R;r)$ be the sets of maximal chains in $R$ terminating at $r$ and beginning at $r$, respectively, viewed as subsets of the chain concatenation semicategory $\big(\tn{Ch}(R),\sqcup\big)$.  Let $T$ be a commutative ring with unit, and let $\Theta:\tn{Ch}(R)\rightarrow T$ be a multiplicative phase map generated by a relation function $\theta:R\rightarrow T$. 
\begin{enumerate}
\item The elements 
\[\Psi_{R;\theta}^-(r):=\sum_{\gamma\in \Gamma_r^-}\Theta(\gamma)\gamma\hspace*{.5cm}\tn{and}\hspace*{.5cm}\Psi_{R;\theta}^+(r):=\sum_{\gamma\in \Gamma_r^+}\Theta(\gamma)\gamma,\]
of the chain concatenation algebra $T^\sqcup[R]$, are called the {\bf past} and {\bf future chain functionals} of $r$ in $R$ with respect to $\theta$, respectively.
\item The relation functions 
\[\psi_{R;\theta}^-(r):=\sum_{\gamma\in \Gamma_r^-}\Theta(\gamma)\hspace*{.5cm}\tn{and}\hspace*{.5cm}\psi_{R;\theta}^+(r):=\sum_{\gamma\in \Gamma_r^+}\Theta(\gamma),\]
given by applying the global evaluation map $e_R$ to the past and future chain functionals $\Psi_{R;\theta}^-(r)$ and $\Psi_{R;\theta}^+(r)$, respectively, are called the {\bf past} and {\bf future wave functions} over $R$ with respect to $\theta$.  
\end{enumerate}
\end{defi}

Given two different elements $r$ and $\overline{r}$ of $R$, $\Psi_{R;\theta}^-(r)$ and $\Psi_{R;\theta}^-(\overline{r})$ are  {\it two different functionals,} mapping two different chain sets $\Gamma_r^-$ and $\Gamma_{\overline{r}}^-$ to the common target object $T$, respectively, while $\psi_{R;\theta}^-(r)$ and $\psi_{R;\theta}^-(\overline{r})$ are merely {\it two different values of the same past wave function} $\psi_{R;\theta}^-$, mapping $R$ to $T$.   Similar statements apply to the future chain functionals $\Psi_{R;\theta}^+(r)$ and $\Psi_{R;\theta}^+(\overline{r})$, and the corresponding values $\psi_{R;\theta}^+(r)$ and $\psi_{R;\theta}^+(\overline{r})$ of the future wave function $\psi_{R;\theta}^+$.  Heuristically, passage from past and future chain functionals to values of past and future wave functions is a ``diagonal process," similar to taking a {\it category-theoretic limit,} reducing a ``function-like entity" at one level of algebraic hierarchy to a ``value-like entity" at the next lower level of hierarchy. 


\refstepcounter{textlabels}\label{causalschrodequ}

Now consider in more detail the past chain functional $\Psi_{R;\theta}^-(r)$ of $r$ in $R$ with respect to $\theta$.  If $r$ is a minimal element of $R$, then $\Psi_{R;\theta}^-(r)$ consists of the single summand $\theta(r)r$.   Otherwise, $\Psi_{R;\theta}^-(r)$ may be {\it decomposed} in terms of the set $D_0^-(R;r)$ of direct predecessors of $r$ in $R$; i.e., those elements $r^-$ in $R$ such that $r^-\prec r$.\footnotemark\footnotetext{The reason for using the notation $D_0^-(R;r)$, rather than merely $D_0^-(r)$, is to make it clear that only direct predecessors of $r$ belonging to $R$ are  considered here.  Other direct predecessors of $r$ in the ``ambient relation space" $\ms{R}(M')$ are excluded.}   Every maximal chain $\gamma'$ in $R$ terminating at $r$ may be expressed as a concatenation product $\gamma\sqcup r$ for an appropriate maximal chain $\gamma$ terminating at some $r^-\in D_0^-(R;r)$.  Fixing $r^-$, and summing in $T^\sqcup[R]$ over all phase-weighted chains $\gamma$ of this form, yields the past chain functional $\Psi_{R;\theta}^-(r^-)$ of $r^-$ in $R$ with respect to $\theta$.  Now letting $r^-$ vary over  {\it all} direct predecessors of $r$ in $R$, the past chain functional $\Psi_{R;\theta}^-(r)$ of $r$ in $R$ with respect to $\theta$ may be written as a product:
\begin{equation}\label{causschrodprecursor}
\Psi_{R;\theta}^-(r)=\Big(\sum_{r^-\prec r}\Psi_{R;\theta}^-(r^-)\Big)\theta(r)r.
\end{equation}
Equation \hyperref[causschrodprecursor]{\ref{causschrodprecursor}} serves as an algebraic precursor to the causal Schr\"{o}dinger-type equation introduced in definition \hyperref[causschrod]{\ref{causschrod}} below, just as the functionals $\Psi(R;\theta)$ and $\Psi_{R;\theta}^{\pm}$ are algebraic precursors to the generalized quantum amplitude $\psi(R;\theta)$ and the past and future wave functions $\psi_{R;\theta}^{\pm}$, respectively.\\
\begin{defi}\label{causschrod} Under the hypotheses of definition \hyperref[pastfuturepathfunctionals]{\ref{pastfuturepathfunctionals}}, the equation 
\begin{equation}\label{causalschrodinger}
\psi_{R;\theta}^-(r)=\theta(r)\sum_{r^-\prec r}\psi_{R;\theta}^-(r^-),
\end{equation}
given by applying the evaluation map $e_R$ to equation \hyperref[causschrodprecursor]{\ref{causschrodprecursor}}, is called the {\bf causal Schr\"{o}dinger equation} over $R$ with respect to $\theta$.  
\end{defi}


The method used to derive equation \hyperref[causschrodprecursor]{\ref{causschrodprecursor}}, in which the past chain functional of a relation is decomposed in terms of its direct predecessors, may be iterated, in a process I refer to as {\bf past fractal decomposition.}  The adjective {\it fractal} refers to the similarity of the nested chain sets $\Gamma_r^-$ across the various levels of the decomposition. Past fractal decomposition leads to {\it higher-order analogues} of Schr\"{o}dinger's equation. 

\refstepcounter{textlabels}\label{causalfeynamp}  

To conclude this section, I briefly discuss two analogues of Feynman's complex inner product formula, given in equation \hyperref[innerprod]{\ref{innerprod}} above, which expresses his quantum amplitude $\psi(X;\ms{L})$ in terms of the past and future wave functions $\psi^\pm$.  To construct the first analogue, begin by expressing the maximal chain functional $\Psi(R;\theta)$ as a path algebra element in the usual way; i.e., as the sum of all chains in $\Gamma=\tn{Ch}_{\tn{\fsz max}}(R)$, weighted by their appropriate phases.  Let $\sigma$ be a maximal antichain in $R$.  By the impermeability of $\sigma$, every term in the sum for $\Psi(R;\theta)$ is a concatenation product of the form $\Theta(\gamma^-)\gamma^-\sqcup\theta(r)r\sqcup\Theta(\gamma^+)\gamma^+$ for appropriate chains $\gamma^\pm$ in $\Gamma^\pm=\tn{Ch}_{\tn{\fsz max}}^\pm(R)$, and an appropriate relation $r$ in $\sigma$.  The sum of all such terms is therefore 
\begin{equation}
\Psi(R;\theta)=\Psi(R^-;\theta)\sqcup\Theta|_\sigma\sqcup\Psi(R^+;\theta).
\end{equation}
To construct the second analogue, observe that for any relation $r$ in $R$, the directed product $\Psi_{R;\theta}^-(r)\vee\Psi_{R;\theta}^+(r)$ {\it almost} coincides with the sum of phase-weighted maximal chains in the total domain of influence $D(r)=D^-(r)\cup\{r\}\cup D^+(r)$ of $r$ in $R$.  The only difference between the product and the sum is that the phase of $r$ is counted twice in the product.  Given a maximal antichain $\sigma$ in $R$, every maximal path in $R$ passes through a unique relation $r$ in $\sigma$.  Therefore, $\Psi(R;\theta)$ is given by the sum:
\begin{equation}\label{directedinnerprod}
\Psi(R;\theta)=\sum_{r\in\sigma}\frac{1}{\theta(r)}\Psi_{R;\theta}^-(r)\vee\Psi_{R;\theta}^+(r),
\end{equation}
where the factors of $\theta(r)$ in the denominators correct for the corresponding extra factors in the products.  Equation \hyperref[directedinnerprod]{\ref{directedinnerprod}} easily reduces to an expression for the corresponding amplitude $\psi(R;\theta)$. To obtain this expression, note that since every path contributing to $\Psi_{R;\theta}^-(r)$ precedes every path contributing to $\Psi_{R;\theta}^+(r)$ in the directed product relation, the global evaluation map ``commutes with the directed product" in this case.  Hence, 
\begin{equation}
\psi(R;\theta)=\sum_{r\in\sigma}\frac{1}{\theta(r)}\psi_{R;\theta}^-(r)\psi_{R;\theta}^+(r),
\end{equation}
which is directly analogous to equation \hyperref[innerprod]{\ref{innerprod}}.


\subsection{Kinematic Schemes}\label{subsectionkinematicschemes}

\refstepcounter{textlabels}\label{pathsumbackground}

{\bf Path Summation in the Background-Independent Context.} The principle of path summation (\hyperref[ps]{PS}), appearing at the beginning of section \hyperref[subsectionquantumpathsummation]{\ref{subsectionquantumpathsummation}} above, states that the abstract quantum causal theory outlined in sections \hyperref[subsectionquantumpathsummation]{\ref{subsectionquantumpathsummation}} and \hyperref[subsectionschrodinger]{\ref{subsectionschrodinger}} may be used to construct a {\it background independent} version of the histories approach to quantum theory in the discrete causal context.  In this section, I describe the rudiments of how this construction may be carried out.  The proper objects over which to perform path summation in this context are {\it relation spaces over ``higher-level multidirected sets," whose ``elements" are directed sets.}  This approach combines the theory of relation space, developed in section \hyperref[subsectionrelation]{\ref{subsectionrelation}}, with the principle of iteration of structure (\hyperref[is]{IS}), introduced in section \hyperref[subsectionquantumprelim]{\ref{subsectionquantumprelim}} above.  The ``higher-level multidirected sets" involved in this approach are called {\it kinematic schemes}.  Their ``relation spaces" are {\it spaces of co-relative histories,}  viewed as relationships between pairs of classical universes.  The theory of kinematic schemes formalizes the {\it evolutionary viewpoint} in discrete causal theory, generalizing Sorkin and Rideout's theory of sequential growth dynamics for causal sets \cite{SorkinSequentialGrowthDynamics99}.  Kinematic schemes play a structural role similar to that of categories, and their ``relation spaces" play a role similar to that of morphism categories.  The abstract structure of a kinematic scheme is generally more complicated than the structures of its individual members.  For example, kinematic schemes of finite acyclic directed sets generally have multidirected, rather than merely directed, structures,  as illustrated by McKay's example appearing in figure \hyperref[corelativeexamples]{\ref{corelativeexamples}} of section \hyperref[subsectionquantumprelim]{\ref{subsectionquantumprelim}} above.  Kinematic schemes of countable acyclic directed sets generally have ``higher-level cycles."    


\refstepcounter{textlabels}\label{kinversusdyn}

\refstepcounter{textlabels}\label{kinpreschemes}

{\bf Kinematic Preschemes; Classes of Co-Relative Histories.}  In classical mechanics, {\bf kinematics} is the study of the general types of behavior {\it permissible} within a theory, while {\bf dynamics} is the study of which behaviors are {\it determined} or {\it favored} under specified conditions.   Under the strong form of the causal metric hypothesis (\hyperref[cmh]{CMH}), all physical behavior is represented by the structure of directed sets.  {\it Permissible behavior} is therefore defined by specifying a distinguished class $\ms{K}$ of directed sets.\footnotemark\footnotetext{As usual, only the isomorphism classes of directed sets are significant.   For simplicity, I work with representatives, with the implicit understanding that at most one representative of each isomorphism class appears in $\ms{K}$.}  This class may be endowed with a ``higher-level multidirected structure," by specifying a class $\mc{H}$ of distinguished co-relative histories between pairs of members of $\ms{K}$, which represent possible instances of evolution from one directed set into another.  An arbitrary choice of $\mc{H}$ and $\ms{K}$ yields a structure called a {\it kinematic prescheme.}   Imposing additional conditions, described below, yields a {\it kinematic scheme.}  Dynamics is then supplied by specifying which co-relative histories are ``favored," in a suitable sense.   This may be accomplished by identifying a distinguished phase map, or a suitable higher-level analogue.\\

\begin{defi}\label{defikinprescheme} A {\bf kinematic prescheme} is a pair $(\ms{K},\mc{H})$, where $\ms{K}$ is a distinguished class of directed sets, and $\mc{H}$ is a distinguished class of co-relative histories between pairs of members of $\ms{K}$. 
\end{defi}

The class $\ms{K}$ is called the {\bf object class} of the kinematic prescheme $(\ms{K},\mc{H})$, and the class $\mc{H}$ is called the {\bf class of co-relative histories} of $(\ms{K},\mc{H})$.  This definition may be immediately generalized to define kinematic preschemes of {\it multidirected} sets, but the physical relevance of such kinematic preschemes is unclear, at least in the context of causal theory.   Following the precedent of section  \hyperref[subsectionquantumprelim]{\ref{subsectionquantumprelim}} above, I focus attention on {\it proper, full, originary} co-relative histories, which are represented by equivalence classes of transitions, as expressed in definition  \hyperref[deficorelative]{\ref{deficorelative}}.   A kinematic prescheme whose class of co-relative histories $\ms{H}$ includes only proper, full, originary co-relative histories is called a {\bf proper, full, originary kinematic prescheme.}  All kinematic preschemes in this section are assumed to be proper, full, and originary, unless stated otherwise.  

The class $\mc{H}$ of co-relative histories of a kinematic prescheme $(\ms{K},\mc{H})$ ``bundles equivalent transitions together into single morphism-like entities," for the purpose of distinguishing essential physical information from ``coordinate-like" or ``gauge-like" information.  Despite this, it is often useful to {\it remember the internal structures} of co-relative histories in a kinematic prescheme; i.e., to preserve explicit knowledge of their constituent transitions, rather than simply treating them as ``abstract arrows."   One motivation for maintaining this viewpoint is that co-relative histories exhibit ``unfamiliar behavior"  from a category-theoretic perspective, as discussed in section \hyperref[subsectionquantumprelim]{\ref{subsectionquantumprelim}} above.  For example, the ``composition" of two co-relative histories $h:D\Rightarrow D'$ and $h':D'\Rightarrow D''$ is generally {\it not} a co-relative history $D\Rightarrow D''$, but rather a {\it family of such co-relative histories.}   The reason for this is that different representatives of $h$ and $h'$ may compose to yield inequivalent transitions $D\rightarrow D''$; i.e., equivalence of transitions is not a congruence relation under composition in the category $\ms{D}$ of directed sets.  An example of this phenomenon appears in figure \hyperref[transitionnotcongruence]{\ref{transitionnotcongruence}} of section \hyperref[subsectionquantumprelim]{\ref{subsectionquantumprelim}} above.   Transitions, on the other hand, are morphisms in $\ms{D}$, so they compose to yield unique transitions.   For such reasons, it is best to regard a kinematic prescheme $(\ms{K},\mc{H})$ as a ``category-like entity" whose ``objects" and ``morphisms" both encode nontrivial ``internal" structure.  Whenever any aspect of this structure is not needed, it may be systematically discarded via a quotient operation, or a {\it decategorification-like procedure,} as explained below. 


\refstepcounter{textlabels}\label{underlyingdirclass}

{\bf Underlying Directed Sets and Multidirected Sets of Kinematic Preschemes.}   I have repeatedly referred to kinematic schemes, and more generally, kinematic preschemes, as ``higher-level multidirected sets."  In fact, a variety of different ``higher-level directed or multidirected structures" may be ascribed to a kinematic prescheme $(\ms{K},\mc{H}$), each defined by taking ``relations" between members of its object class $\ms{K}$ to correspond to one of several types of ``morphism-like entities" associated with its class $\mc{H}$ of co-relative histories.  The ``preferred" notion of a ``morphism-like entity" in this context is just a member of $\mc{H}$, but variations of this notion are sometimes useful.  Here, I describe three such structures, in increasing order of complexity. 
\vspace*{.2cm}
\begin{defi}\label{defiunderlying} Let $(\ms{K},\mc{H})$ be a kinematic prescheme.  
\begin{enumerate}
\item The {\bf underlying directed set} $\ms{U}(\ms{K},\mc{H})$ of $(\ms{K},\mc{H})$ is the directed set possessing one element $x(D)$ for each directed set $D$ in $\ms{K}$, and one relation $x(D_i)\prec x(D_t)$ for each pair of directed sets $D_i$ and $D_t$ in $\ms{K}$ with a co-relative history between them in $\mc{H}$. 
\item The {\bf underlying multidirected set} $\ms{V}(\ms{K},\mc{H})$ of $(\ms{K},\mc{H})$ is the multidirected set possessing one element $x(D)$ for each directed set $D$ in $\ms{K}$, one relation $r(h)$ for each co-relative history $h:D_i\Rightarrow D_t$ in $\mc{H}$, and initial and terminal element maps $i$ and $t$ sending $r(h)$ to $x(D_i)$ and $x(D_t)$, respectively. 
\item The {\bf underlying transition structure} $\ms{W}(\ms{K},\mc{H})$ of $(\ms{K},\mc{H})$ is the multidirected set possessing one element $x(D)$ for each directed set $D$ in $\ms{K}$, one relation $r(\tau)$ for each transition $\tau$ representing a co-relative history $h:D_i\Rightarrow D_t$ in $\mc{H}$, and initial and terminal element maps $i$ and $t$ sending $\tau$ to $x(D_i)$ and $x(D_t)$, respectively. 
\end{enumerate}
\end{defi}
Additional ``higher-level" directed or multidirected sets may be associated with a kinematic prescheme $(\ms{K},\mc{H})$, each defined by preserving or ignoring different amounts and/or types of structure.  For example, relations may be chosen to represent {\it image-fixed co-relative histories,} mentioned near the end of section \hyperref[subsectionquantumprelim]{\ref{subsectionquantumprelim}} above.  However, the three alternatives appearing in definition \hyperref[defiunderlying]{\ref{defiunderlying}} suffice for the purposes of this paper.  The underlying multidirected set $\ms{V}(\ms{K},\mc{H})$ of $(\ms{K},\mc{H})$ is a quotient of its underlying transition structure $\ms{W}(\ms{K},\mc{H})$, given by identifying relations corresponding to transitions representing the same co-relative history in $\mc{H}$.  The underlying directed set $\ms{U}(\ms{K},\mc{H})$ of $(\ms{K},\mc{H})$, in turn, is a quotient of $\ms{V}(\ms{K},\mc{H})$, given by identifying all relations corresponding to co-relative histories in $(\ms{K},\mc{H})$ sharing the same cobase and target.  All three sets possess the same abstract element set $\{x(D)|D\in\ms{K}\}$.  $\ms{U}, \ms{V}$ and $\ms{W}$ may be viewed more formally in a functorial sense, but I do not elaborate on this viewpoint in this paper.

The directed set $\ms{U}(\ms{K},\mc{H})$, and the multidirected sets $\ms{V}(\ms{K},\mc{H})$ and $\ms{W}(\ms{K},\mc{H})$, each {\it forget internal structure} associated with members of $\ms{K}$ and $\mc{H}$.   Passage to $\ms{U}, \ms{V}$ or $\ms{W}$ may be viewed as a {\it decategorification-like procedure,} systematically reducing the ``category-like entity" $(\ms{K},\mc{H})$ to an ``object-like entity."  It is important to emphasize that $\ms{U}, \ms{V}$ and $\ms{W}$ all derive their relations from the class of co-relative histories $\mc{H}$ of $(\ms{K},\mc{H})$, {\it not} from the ambient category $\ms{D}$ of directed sets.  In particular, there may be pairs of directed sets $D$ and $D'$ in $\ms{K}$ related by morphisms in $\ms{D}$, or even by transitions in $\ms{D}$, whose corresponding co-relative histories are {\it not chosen} as members of $\mc{H}$, and whose underlying elements $x(D)$ and $x(D')$ are therefore not related in $\ms{U}, \ms{V}$ or $\ms{W}$. 

The abstract structural similarities between underlying directed or multidirected sets of kinematic preschemes, and the individual members of their object classes, is the essence of the principle of iteration of structure (\hyperref[is]{IS}) in discrete quantum causal theory.   Application of the relation space functor $\ms{R}$ to such underlying directed or multidirected sets renders this similarity ``perfect," in the sense that all structures under consideration are then directed sets ``at different levels of algebraic hierarchy," since $\ms{R}$ reduces multidirected structure to directed structure.  However, a more practical reason for applying $\ms{R}$ in this context is that it circumvents the problem of permeability of maximal antichains in multidirected sets, as proven in theorem \hyperref[theoremrelimpermeable]{\ref{theoremrelimpermeable}} of section \hyperref[subsectionrelation]{\ref{subsectionrelation}}.  This allows for effective implementation of the abstract quantum causal theory outlined in sections \hyperref[subsectionquantumpathsummation]{\ref{subsectionquantumpathsummation}} and \hyperref[subsectionschrodinger]{\ref{subsectionschrodinger}} above. 


\refstepcounter{textlabels}\label{kinschemes}

{\bf Kinematic Schemes.} A kinematic prescheme $(\ms{K},\mc{H})$ is called a {\bf kinematic scheme} if it satisfies the following two additional properties:

\refstepcounter{textlabels}\label{hereditary}
\refstepcounter{textlabels}\label{h}

\hspace*{.35cm}H. \hspace*{.3cm}{\bf Hereditary Property}: {\it $\ms{K}$ is closed under the formation of proper, full, originary  \\ \hspace*{1.1cm} subobjects.} 
 
 \refstepcounter{textlabels}\label{weakaccessibility}
 \refstepcounter{textlabels}\label{wa}

 \hspace*{.3cm}WA. {\bf Weak Accessibility}: {\it Suppose that $D$ and $D'$ are members of $\ms{K}$.  If there exists a 
 \\ \hspace*{1.1cm} transition $\tau:D\rightarrow D'$ in $\ms{D}$ with finite complement, then there exists a chain from $x(D)$ \\
 \hspace*{1.1cm} to $x(D')$ in $\ms{W}(\ms{K},\mc{H})$.}

The intuition underlying the hereditary property is that if a given directed set is a permissible model of causal structure, then so are its ``ancestors;" i.e., its ``earlier stages of development."  The intuition underlying weak accessibility is that any permissible directed set should be {\it realizable,} in the sense that it should have at least one ``evolutionary pathway" leading to it.   The qualifier {\it weak} is included because of the finite complement condition in the definition; without this caveat, certain pairs of directed sets differing by an infinite number of elements would perforce be ``connected by a finite number of evolutionary steps" in $(\ms{K},\mc{H})$.   While there are interesting kinematic schemes that {\it do} exhibit such behavior, it is nonetheless too restrictive to include as part of the definition.  Weak accessibility is closely related to the notion of {\it inaccessible cardinals} in order theory.  In particular, it is analogous to the fact that one may access any positive integer in a finite number of steps by taking minimal successors, but may {\it not} access Cantor's first transfinite ordinal in this manner.  Note that weak accessibility does not depend on the choice to use the underlying transition structure $\ms{W}(\ms{K},\mc{H})$ of $(\ms{K},\mc{H})$ in the definition, rather than one of its quotients $\ms{U}(\ms{K},\mc{H})$ or $\ms{V}(\ms{K},\mc{H})$.  

The two classes $\ms{K}$ and $\mc{H}$ making up a kinematic scheme $(\ms{K},\mc{H})$ stand on very different footings.  The object class $\ms{K}$ of $(\ms{K},\mc{H})$ embodies {\it a hypothesis about the structure of the physical universe,} while the class $\mc{H}$ of co-relative histories of $(\ms{K},\mc{H})$ is chosen for convenience among many equally valid alternatives.  This difference is analogous to the difference in general relativity between choosing a class of pseudo-Riemannian manifolds as models of classical spacetime, and choosing particular families of coordinate systems on these manifolds.    The arbitrary choice of the class of co-relative histories in a kinematic scheme requires an appropriate notion of {\it covariance,} to ensure that the physical predictions of the theory do not depend on this choice.  This topic is more subtle than one might expect, essentially because the ``frames of reference" encoded by kinematic schemes tend to be much more general and varied than relativistic coordinate systems.\footnotemark\footnotetext{Sorkin and Rideout discuss ``general discrete covariance" in the context of sequential growth dynamics.  This is closely related, but not identical, to the notion required here.} 


\refstepcounter{textlabels}\label{quantumcausalmetric}

{\bf Quantum Causal Metric Hypothesis.} Just as the classical causal metric hypothesis (\hyperref[ccmh]{CCMH}) takes the properties of classical spacetime to arise from directed structure, so the {\it quantum causal metric hypothesis} takes the properties of quantum spacetime to arise from higher-level multidirected structure, via the principle of iteration of structure (\hyperref[is]{IS}).  The theory of kinematic schemes enables a precise statement of this idea.  

\hspace*{.3cm} QCMH.\refstepcounter{textlabels}\label{qcmh} {\bf Quantum Causal Metric Hypothesis.} {\it The properties of quantum spacetime\\ \hspace*{1.8cm} arise from the structure of a kinematic scheme of directed sets.} 

As noted above, the class $\mc{H}$ of co-relative histories of a kinematic scheme $(\ms{K},\mc{H})$ involves a {\it choice,} similar to a choice of coordinate system.   Hence, the quantum causal metric hypothesis may be realized via {\it any suitable kinematic scheme.}  Uniqueness may be achieved by using a {\it universal kinematic scheme,} as described below, but this is not always convenient.  


\refstepcounter{textlabels}\label{pathsumkinscheme} 
\refstepcounter{textlabels}\label{corelkin} 

{\bf Path Summation over a Kinematic Scheme.}  Path summation over a kinematic scheme $(\ms{K},\mc{H})$, or more precisely, over the relation space of an underlying directed or multidirected set of $(\ms{K},\mc{H})$, provides a precise realization of the quantum causal metric hypothesis (\hyperref[qcmh]{QCMH}).  Although the formal aspects of this approach are automatic, due to the principle of path summation (\hyperref[ps]{PS}) stated in section \hyperref[subsectionquantumpathsummation]{\ref{subsectionquantumpathsummation}} above, faithful organization and interpretation of the corresponding physical information is a subtle matter, and I can only briefly sketch the rudiments here.  Let $(\ms{K},\mc{H})$ be a kinematic scheme, $D$ and $D'$ objects of $\ms{K}$, and $\tau:D\rightarrow D'$ a transition in $\ms{D}$.  Suppose, for simplicity, that $\tau$ has finite complement.   By weak accessibility (\hyperref[wa]{WA}), there exists at least one chain $\gamma$ in $\ms{W}(\ms{K},\mc{H})$ from $x(D)$ to $x(D')$.  Such a chain represents a {\it generalized order extension,} in which ``relativity of simultaneity" is broken by arbitrary choices of succession among causally unrelated subsets of ``histories lying between $D$ and $D'$ in $(\ms{K},\mc{H})$."   Hence, in addition to specifying the co-relative history $h$ represented by $\tau$, such a chain also provides a ``kinematic account," or {\it generalized frame of reference,} for $h$.  I refer to such a chain as a {\bf maximal co-relative kinematics for $h$ in} $(\ms{K},\mc{H})$.\footnotemark\footnotetext{There is an interesting analogy here arising from algebraic geometry: a co-relative kinematics may be compared to a point in a {\it flag variety,} since it involves a ``distinguished sequence of subobjects in a space."}

Analogous chains in the ``coarser" sets $\ms{U}(\ms{K},\mc{H})$ and $\ms{V}(\ms{K},\mc{H})$ correspond to equivalence classes of co-relative kinematics, representing less-detailed generalized order extensions.  Generalizing this, any directed path in an underlying directed or multidirected set associated with the kinematic scheme $(\ms{K},\mc{H})$, with or without initial and terminal elements, may be called a {\bf kinematics} in $(\ms{K},\mc{H})$.  Path summation over a kinematic scheme $(\ms{K},\mc{H})$ is therefore better characterized as summation over {\it kinematics} than as summation over {\it histories.}  The underlying reason for this is simply that any particular path in $(\ms{K},\mc{H})$ involves arbitrary information, analogous to the arbitrary information treated by covariance and/or gauge theory in ordinary physics.   Sorkin's quote about  {\it upward directed paths in a tree of causal sets,} in the context of quantum measure theory, cited in section \hyperref[subsectionquantumprelim]{\ref{subsectionquantumprelim}} above, foreshadows the general idea of co-relative kinematics. 

Path summation over a kinematic scheme $(\ms{K},\mc{H})$ requires specification of a suitable phase map $\Theta$.   In order to respect the quantum causal metric hypothesis, $\Theta$ must depend only on the class $\mc{H}$ of co-relative histories of $(\ms{K},\mc{H})$.  In this way, the information contained in $\Theta$ derives ultimately from the causal preorders of the individual members of $\ms{K}$.   The notion of a phase map must be accorded a broad interpretation in this context.  In particular, since the ``elements" in the ``domain" of $\Theta$ represent paths in a space of co-relative histories, whose individual relations represent relationships between pairs of classical universes, it may be asking too much to expect individual ``numbers;" i.e., elements of a ring, to preserve enough information as images in this context.  Hence, $\Theta$ may be a ``higher-level map," such as a generalized functor, and path sums may be ``higher-level evaluations," such as generalized category-theoretic limits. 


\refstepcounter{textlabels}\label{possequkinscheme}
\refstepcounter{textlabels}\label{relsorkinrideout}

{\bf Positive Sequential Kinematic Scheme.} The prototypical example of a kinematic scheme $(\ms{K},\mc{H})$ appears in Sorkin and Rideout's theory of sequential growth dynamics of causal sets \cite{SorkinSequentialGrowthDynamics99}.  In this case, $\ms{K}$ is the class $\ms{C}_{\tn{\fsz{fin}}}$ of finite causal sets, and $\mc{H}$ is the class $\mc{H}_1$ of co-relative histories $h:C_i\Rightarrow C_t$ such that $C_t$ differs from $C_i$ by the addition of a single element.   In section \hyperref[subsectionapproach]{\ref{subsectionapproach}} above, I briefly discussed this kinematic scheme in a simplified manner, denoting it by the single symbol $\ms{K}$, and describing its abstract structure in terms of transitions, rather than co-relative histories.  Due to the structural shortcomings of causal set theory,  I use a slightly different kinematic scheme for illustrative purposes in this section, replacing $\ms{C}_{\tn{\fsz{fin}}}$ with the class $\ms{A}_{\tn{\fsz{fin}}}$ of finite acyclic directed sets, and enlarging the class of co-relative histories $\mc{H}_1$ to accommodate the larger class of objects, without changing its abstract definition or its notation.   I refer to $(\ms{A}_{\tn{\fsz{fin}}},\mc{H}_1)$ as the {\bf positive sequential kinematic scheme}, since each originary chain\footnotemark\footnotetext{That is, each directed path beginning at $x(\oslash)$, the element of $\ms{W}(\ms{A}_{\tn{\fsz{fin}}},\mc{H}_1)$ corresponding to the empty set.} in its underlying transition structure $\ms{W}(\ms{A}_{\tn{\fsz{fin}}},\mc{H}_1)$ defines a total sequential labeling of the elements of its terminal acyclic directed set, or of its countably infinite {\it limit set,} in the case of an infinite chain.  A portion of the positive sequential kinematic scheme $(\ms{A}_{\tn{\fsz{fin}}},\mc{H}_1)$ is illustrated in figure \hyperref[figsequentialscheme]{\ref{figsequentialscheme}} below.   

\begin{figure}[H]
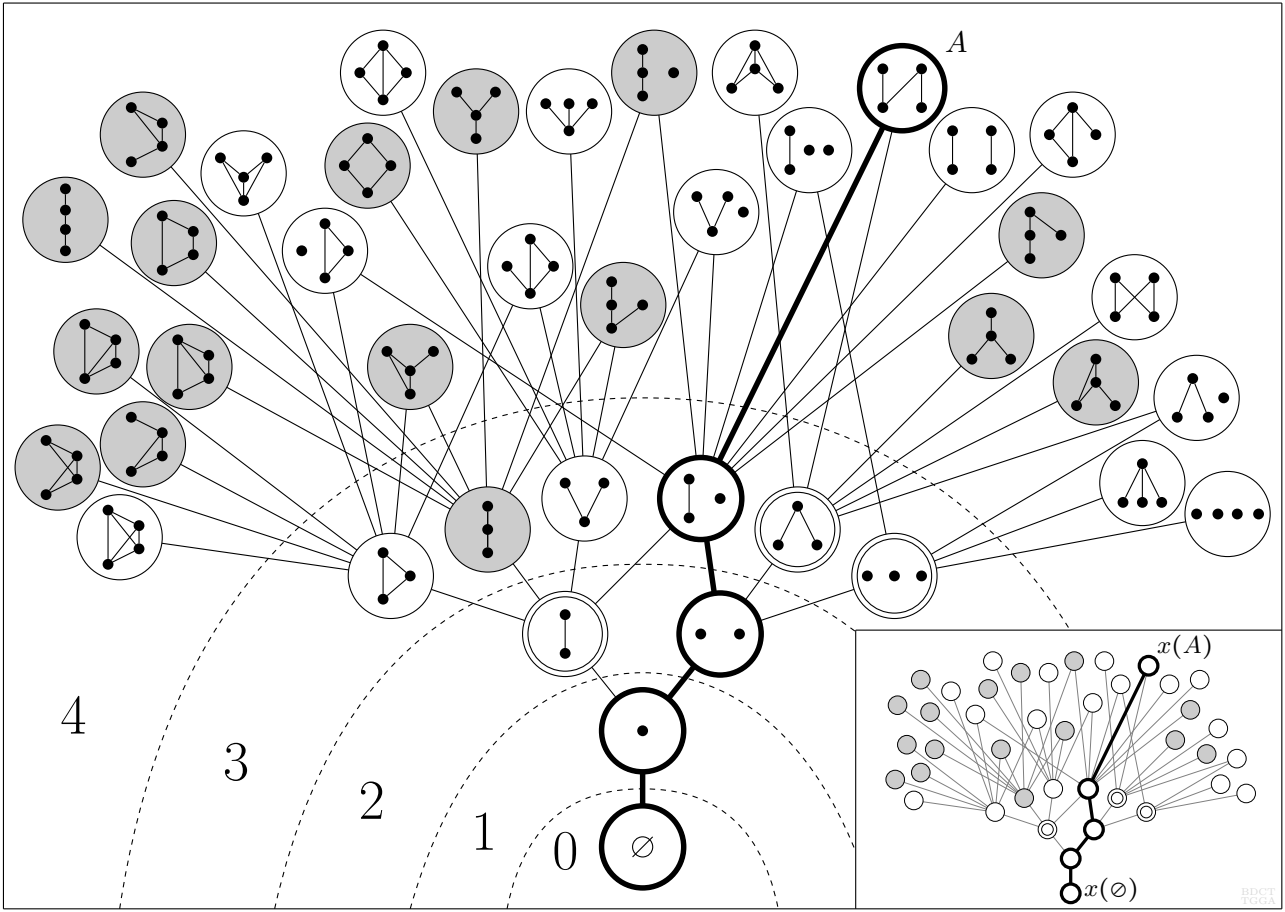


\caption{A portion of the positive sequential kinematic scheme $(\ms{A}_{\tn{\fsz{fin}}},\mc{H}_1)$; inset shows the underlying multidirected set $\ms{V}(\ms{A}_{\tn{\fsz{fin}}},\mc{H}_1)$; large-font numbers indicate generations.}
\label{figsequentialscheme}
\end{figure}
\vspace*{-.5cm}

The reader should compare figure \hyperref[figsequentialscheme]{\ref{figsequentialscheme}} to figure \hyperref[sequential]{\ref{sequential}} of section \hyperref[subsectionapproach]{\ref{subsectionapproach}} above, and also to figure $1$ of Sorkin and Rideout's paper {\it Classical sequential growth dynamics for causal sets} \cite{SorkinSequentialGrowthDynamics99}, which illustrate portions of Sorkin and Rideout's kinematic scheme $(\ms{C}_{\tn{\fsz{fin}}},\mc{H}_1)$.  The large numbers $0$ through $4$ in figure \hyperref[figsequentialscheme]{\ref{figsequentialscheme}} indicate how each member set $D$ of $(\ms{A}_{\tn{\fsz{fin}}},\mc{H}_1)$ may be assigned to a positive integer-valued {\it generation} by its cardinality, or equivalently, by the common length of all originary chains terminating at $x(D)$ in the underlying multidirected set $\ms{V}(\ms{A}_{\tn{\fsz{fin}}},\mc{H}_1)$.  The nodes shaded grey in the figure represent member sets which are {\it nontransitive}, and which therefore belong to $(\ms{A}_{\tn{\fsz{fin}}},\mc{H}_1)$ but {\it not} to Sorkin and Rideout's kinematic scheme $(\ms{C}_{\tn{\fsz{fin}}},\mc{H}_1)$.  In particular, generations $0$ through $3$ of $(\ms{A}_{\tn{\fsz{fin}}},\mc{H}_1)$ differ from the corresponding portion of $(\ms{C}_{\tn{\fsz{fin}}},\mc{H}_1)$ by a single member, the nontransitive chain of length three.  However, about half of the fourth generation of $(\ms{A}_{\tn{\fsz{fin}}},\mc{H}_1)$ consists of nontransitive acyclic directed sets, and subsequent generations are dominated by them.   The inset in figure \hyperref[figsequentialscheme]{\ref{figsequentialscheme}} illustrates the portion of the underlying multidirected set $\ms{V}(\ms{A}_{\tn{\fsz{fin}}},\mc{H}_1)$ of $(\ms{A}_{\tn{\fsz{fin}}},\mc{H}_1)$ corresponding to the portion of $(\ms{A}_{\tn{\fsz{fin}}},\mc{H}_1)$ appearing in the main figure. $\ms{V}(\ms{A}_{\tn{\fsz{fin}}},\mc{H}_1)$  is a {\it a countably infinite, locally finite, acyclic multidirected set.}   The portion shown here does not contain multiple edges between individual pairs of elements, but multiple edges begin to appear by the  seventh and eighth generations.  The fact that $\ms{A}_{\tn{\fsz{fin}}}$, and hence $\ms{V}(\ms{A}_{\tn{\fsz{fin}}},\mc{H}_1)$,  is countable, was proven in section \hyperref[settheoretic]{\ref{settheoretic}} above, while the local finiteness and acyclicity of $\ms{V}(\ms{A}_{\tn{\fsz{fin}}},\mc{H}_1)$ follow from the finite cardinality of the member sets of $\ms{A}_{\tn{\fsz{fin}}}$, and the definition of a co-relative history.  The remaining labels in figure \hyperref[figsequentialscheme]{\ref{figsequentialscheme}} are to aid in the discussion below. 

\refstepcounter{textlabels}\label{transindices}
\refstepcounter{textlabels}\label{causalgalois}

Underlying directed sets and multidirected sets of kinematic schemes generally suffer from the problem of permeability of maximal antichains in multidirected sets, discussed in section \hyperref[subsectionrelation]{\ref{subsectionrelation}} above.  The underlying multidirected set $\ms{V}(\ms{A}_{\tn{\fsz{fin}}},\mc{H}_1)$ of the positive sequential kinematic scheme $(\ms{A}_{\tn{\fsz{fin}}},\mc{H}_1)$ is no exception.  For example, the three nodes distinguished by double circles in figure \hyperref[figsequentialscheme]{\ref{figsequentialscheme}} represent a maximal antichain in $\ms{V}(\ms{A}_{\tn{\fsz{fin}}},\mc{H}_1)$, permeated by the chain from $x(\oslash)$ to $x(A)$ indicated by the thickened nodes and edges.  This indicates that passage to relation space remains a crucial tool in the context of kinematic schemes.  

A variety of natural relation functions may be associated with underlying directed or multidirected sets of kinematic schemes. Perhaps the most obvious choices for such relation functions involve the automorphism groups of the directed sets corresponding to the initial and terminal elements of each relation.  For example, the subgroup indices of the {\it causal Galois groups,} mentioned in section \hyperref[subsectionquantumprelim]{\ref{subsectionquantumprelim}} above, provide examples of such relation functions.  Analogous ``higher-level" relation functions are given by simply assigning the entire causal Galois groups to their corresponding relations. As mentioned in section \hyperref[subsectionquantumpathsummation]{\ref{subsectionquantumpathsummation}}, such relation functions may be viewed as partial precursors for the phase maps appearing in the path summation approach to abstract quantum causal theory.  


\refstepcounter{textlabels}\label{genkinematics}

{\bf Generational Kinematics for  $\ms{A}_{\tn{\fsz{fin}}}$.} The positive sequential kinematic scheme $(\ms{A}_{\tn{\fsz{fin}}},\mc{H}_1)$ is not the only interesting kinematic scheme whose object class is the set $\ms{A}_{\tn{\fsz{fin}}}$ of isomorphism classes of finite acyclic directed sets.  An alternative kinematic scheme over $\ms{A}_{\tn{\fsz{fin}}}$, which provides closer analogues of relativistic frames of reference than $(\ms{A}_{\tn{\fsz{fin}}},\mc{H}_1)$, is the {\bf generational kinematic scheme} $(\ms{A}_{\tn{\fsz{fin}}},\mc{H}_{\tn{\fsz{gen}}})$.  The class $\mc{H}_{\tn{\fsz{gen}}}$ of co-relative histories of the generational kinematic scheme includes every co-relative history adding a single {\it generation;} i.e., antichain, to its cobase, rather than merely a single element.  The generational kinematic scheme provides formal analogues of {\it spacelike foliations} of members of $\ms{A}_{\tn{\fsz{fin}}}$, and of countably infinite limit sets of $\ms{A}_{\tn{\fsz{fin}}}$, discussed below.  The underlying multidirected set of $(\ms{A}_{\tn{\fsz{fin}}},\mc{H}_{\tn{\fsz{gen}}})$ is acyclic, but is not locally finite, since each new generation may have any positive integer cardinality.  


\refstepcounter{textlabels}\label{completions}

{\bf Completions of Kinematic Schemes.}  In addition to organizing the set of isomorphism classes of finite acyclic directed sets into a higher-level multidirected structure, the positive sequential kinematic scheme $(\ms{A}_{\tn{\fsz{fin}}},\mc{H}_1)$ also encodes the structure of certain isomorphism classes of {\it countably infinite} acyclic directed sets, corresponding to {\it equivalence classes of maximal chains in the underlying transition structure} $\ms{W}(\ms{A}_{\tn{\fsz{fin}}},\mc{H}_1)$ of $(\ms{A}_{\tn{\fsz{fin}}},\mc{H}_1)$.  These countably infinite sets may be viewed as {\it limits,} in a generalized category-theoretic sense.  At a more elementary level, they are analogous to limits of Cauchy sequences of rational numbers.   More precisely, let
 \[x(\oslash)\prec x(A_1)\prec x(A_2)\prec...\prec x(A_n)\prec...\]
be a maximal chain in the underlying transition structure $\ms{W}(\ms{A}_{\tn{\fsz{fin}}},\mc{H}_1)$ of $(\ms{A}_{\tn{\fsz{fin}}},\mc{H}_1)$, where each precursor symbol $\prec$ represents a {\it specific} transition.  Each finite acyclic directed set $A_n$ in this chain may be regarded as a subset of the succeeding set $A_{n+1},$, and each relation in $A_n$ may be regarded as a relation in $A_{n+1}$.  Hence, the union $A_\infty:=\cup_{n=1}^\infty A_n$ is a well-defined, countably infinite, acyclic directed set.   The maximal chain in $\ms{W}(\ms{A}_{\tn{\fsz{fin}}},\mc{H}_1)$ corresponding to $A_\infty$ determines a total labeling of $A_\infty$, which corresponds to a bijective morphism $A_\infty\rightarrow \NN$.  Conversely, every bijective morphism from a directed set\footnotemark\footnotetext{The existence of such a morphism guarantees that the source is in fact a countably infinite acyclic directed set.} into $\NN$ determines a unique maximal chain in $\ms{W}(\ms{A}_{\tn{\fsz{fin}}},\mc{H}_1)$.   Hence, the class of maximal chains in $\ms{W}(\ms{A}_{\tn{\fsz{fin}}},\mc{H}_1)$ corresponds bijectively to the class of countably infinite relative directed sets over $\NN$, viewed as morphisms.\footnotemark\footnotetext{As usual, only isomorphism classes are significant.}  Defining an appropriate equivalence relation on maximal chains in $\ms{W}(\ms{A}_{\tn{\fsz{fin}}},\mc{H}_1)$, corresponding to the equivalence relation identifying relative directed sets over $\NN$ sharing common sources,  identifies a class of countably infinite acyclic directed sets called {\bf limits} of the kinematic scheme $(\ms{A}_{\tn{\fsz{fin}}},\mc{H}_1)$.    Augmenting $(\ms{A}_{\tn{\fsz{fin}}},\mc{H}_1)$ by its class of limits yields a new kinematic scheme $\overline{(\ms{A}_{\tn{\fsz{fin}}},\mc{H}_1)}$ called the {\bf chain completion} of $(\ms{A}_{\tn{\fsz{fin}}},\mc{H}_1)$.  The object class of $\overline{(\ms{A}_{\tn{\fsz{fin}}},\mc{H}_1)}$ is the familiar set of isomorphism classes of countable acyclic directed sets representable as relative directed sets over $\NN$, which has cardinality $2^{\aleph_0}$.  As suggested by its name, its definition in terms of maximal chains, and its cardinality, $\overline{(\ms{A}_{\tn{\fsz{fin}}},\mc{H}_1)}$ is related to $(\ms{A}_{\tn{\fsz{fin}}},\mc{H}_1)$ in a manner analogous to the relationship between the real numbers $\RR$ and the rational numbers $\QQ$.   
 
More generally, let $(\ms{K},\mc{H})$ be a kinematic scheme, and let $\widetilde{\tn{Ch}_{\tn{\fsz{max}}}}=\widetilde{\tn{Ch}_{\tn{\fsz{max}}}}(\ms{K},\mc{H})$ be the class of limits of $(\ms{K},\mc{H})$; i.e., the class of directed sets represented by equivalence classes of maximal chains in $\ms{W}(\ms{K},\mc{H})$.\footnotemark\footnotetext{Care is required when $\ms{W}(\ms{K},\mc{H})$ includes cycles.} Then the {\bf chain completion} $\overline{(\ms{K},\mc{H})}$ of $(\ms{K},\mc{H})$ is defined to be the kinematic prescheme $\big(\ms{K}\cup\widetilde{\tn{Ch}_{\tn{\fsz{max}}}},\mc{H}\big)$ given by enlarging the object class $\ms{K}$ to include equivalence classes of maximal chains, {\it without adding new co-relative histories.}  It is easy to see that the chain completion of a kinematic scheme is a kinematic scheme: the addition of directed sets corresponding to maximal chains preserves the hereditary property (\hyperref[h]{H}) by construction, while the finite complement condition in the definition of weak accessibility (\hyperref[wa]{WA}) exempts the added sets from the necessity of accessibility by a finite chain in $\ms{W}(\ms{K},\mc{H})$.  This, in fact, is one of the principal  reasons for imposing the finite complement condition.  
 

\refstepcounter{textlabels}\label{kinschemecountablyinf}

{\bf Kinematic Schemes of Countably Infinite Acyclic Directed Sets.} The chain completion $\overline{(\ms{A}_{\tn{\fsz{fin}}},\mc{H}_1)}$ of the positive sequential kinematic scheme $(\ms{A}_{\tn{\fsz{fin}}},\mc{H}_1)$ provides a simple example of a kinematic scheme including infinite directed sets among its objects.  However, these infinite sets are of a very restricted type, even in the locally finite context, being representable as relative directed sets over $\NN$.  Further, they are ``mere limits," in the sense that no co-relative histories involving these sets, either as cobases or targets, are admitted into the higher-level multidirected structure of $\overline{(\ms{A}_{\tn{\fsz{fin}}},\mc{H}_1)}$.   There are good physical reasons to consider more complex kinematic schemes including countably infinite member sets; for example, kinematic schemes whose object classes are the classes $\ms{C}$ and $\ms{A}_{\aleph_0}$ of {\it all} countably infinite causal sets and locally finite acyclic directed sets, respectively.  Such kinematic schemes have the same cardinality $2^{\aleph_0}$ as $\overline{(\ms{A}_{\tn{\fsz{fin}}},\mc{H}_1)}$, but are generally much more complicated.   For example, they usually contain cycles, such as the ``self-transition" of causal sets $-\NN\rightarrow-\NN$ sending $-n$ to $-n-1$, whose ``new" element is the maximal element $0$ in the second copy of $-\NN$. 


\refstepcounter{textlabels}\label{morphkinscheme}

{\bf ``Functors" of Kinematic Schemes.} The principle of iteration of structure (\hyperref[is]{IS}) may be extended to apply in more general contexts than the relationship between directed sets and kinematic schemes.  At least one higher level of hierarchy is also physically relevant.  This level of hierarchy involves {\it relationships between pairs of kinematic schemes,} following Grothendieck's relative viewpoint (\hyperref[rv]{RV}).  These relationships may be viewed as ``higher-level morphisms," or ``higher-level co-relative histories."  Drawing on the relativistic analogy, they represent a vast generalization of coordinate transformations.  From an algebraic perspective, in which kinematic schemes are viewed as generalized categories, relationships between kinematic schemes may be viewed as generalized functors.   In this paper, I do not attempt to undertake any systematic analysis of physically essential structure for such relationships, as I did for co-relative histories in section \hyperref[subsectionquantumprelim]{\ref{subsectionquantumprelim}} above, nor do I even give general definitions.  Instead, I  briefly discuss a particular class of examples, for purely illustrative purposes.  

Let $(\ms{K},\mc{H})$ and $(\ms{K}',\mc{H}')$ be kinematic schemes such that $\ms{K}\subset\ms{K}'$ and $\mc{H}\subset\mc{H}'$. In this case, there is a natural inclusion ``functor" $\mc{I}:(\ms{K},\mc{H})\rightarrow (\ms{K}',\mc{H}')$.   I focus on the case where the inclusion of objects is proper, and where the inclusion of co-relative histories is full, meaning that any ``new" co-relative histories involve ``new" objects.  Further, I assume that every object of $\ms{K}$ is finite, and every object of $\ms{K}'$ is countable.  In this case, the cardinalities of $\ms{K}$ and $\ms{K}'$ are ``typically" $\aleph_0$ and $2^{\aleph_0}$, respectively.   The ``inclusion functor" $\mc{I}$ is a higher-level analogue of a transition, since it embeds $(\ms{K},\mc{H})$ into $(\ms{K}',\mc{H}')$ as a ``proper, full, originary subscheme."  The prototypical example is the inclusion of the positive sequential kinematic scheme into its chain completion. 

\refstepcounter{textlabels}\label{numberanalogies}

\begin{figure}[H]
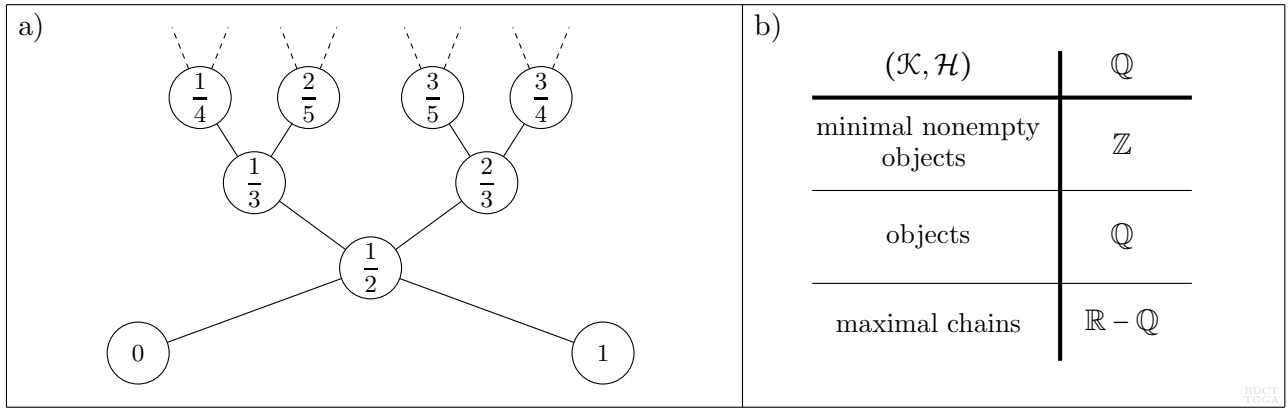


\caption{a) Farey ``tree" $(F,\prec)$; chain completion is $[0,1]\subset\RR$; b) analogies between a countable kinematic scheme and familiar number systems.}
\label{farey}
\end{figure}

Interesting analogies exist between such ``functors" and inclusions of number systems of cardinality $\aleph_0$, such as the integers or rational numbers, into number systems of cardinality $2^{\aleph_0}$, such as the real numbers, $p$-adic numbers, complex numbers, etc.   For example, consider the set of rational numbers in the unit interval $[0,1]$, represented by irreducible fractions, with binary relation $\prec$ induced by the usual linear order on $\QQ$ via the {\it Farey sequences,} in which each fraction is related to ``the next two irreducible fractions bracketing it in the usual order."  The resulting acyclic directed set $F=(F,\prec)$ is a {\it tree} from its ``second generation" onward, sometimes called the {\it Farey tree}.  A small portion of the Farey tree is illustrated in figure \hyperref[farey]{\ref{farey}}a below.  A maximal chain in $F$ represents a Cauchy sequence of rational numbers in $[0,1]$, and hence an irrational number in the interval $[0,1]$.  Moreover, every irrational number in $[0,1]$ is represented by a maximal chain in $(F,\prec)$ in this way.   In this particular case, the representation is unique, precisely because $F$ is a tree from its second generation onward.  In a similar way, a maximal chain in the underlying transition structure $\ms{W}(\ms{K},\mc{H})$ of a kinematic scheme $(\ms{K},\mc{H})$ represents an object in its chain completion.  In this case, the representation is generally {\it not} unique, since $\ms{W}(\ms{K},\mc{H})$ is generally not a tree.  

\vspace*{-.5cm}


\refstepcounter{textlabels}\label{universalkinschemes}

{\bf Universal Kinematic Schemes; Kinematic Spaces.} For a given class $\ms{K}$ of directed sets, considered up to isomorphism, there are generally many different classes $\mc{H}$ of co-relative between pairs of members of $\ms{K}$, and in particular, many different kinematic schemes $(\ms{K},\mc{H})$.  The class $\mc{H}(\ms{K})$ of {\it all} proper, full, originary co-relative histories in $\ms{D}$ between pairs of members of $\ms{K}$ defines a ``transitive" kinematic scheme over $\ms{K}$, which is {\it maximal} in the class of all proper, full, originary kinematic preschemes over $\ms{K}$.  Due to this property, I refer to $(\ms{K},\mc{H}(\ms{K}))$ as the {\bf universal kinematic scheme} over $\ms{K}$.   Despite its uniqueness, $(\ms{K},\mc{H}(\ms{K}))$ can be cumbersome to work with, since it encodes ``many redundant accounts of each evolutionary process."  

At the opposite extreme are ``irreducible" kinematic schemes, in which descriptions of evolutionary processes are ``as unique as possible."  In the locally finite case, a kinematic scheme may be reduced to an irreducible kinematic scheme by applying the skeleton operation to its underlying multidirected set, since this operation does not effect the hereditary property (\hyperref[h]{H}) or weak accessibility (\hyperref[wa]{WA}) in this context.  For example, the skeleton of the universal kinematic scheme $(\ms{A}_{\tn{\fsz{fin}}},\mc{H}(\ms{A}_{\tn{\fsz{fin}}}))$ over the set $\ms{A}_{\tn{\fsz fin}}$ of isomorphism classes of finite acyclic directed sets is the positive sequential kinematic scheme $(\ms{A}_{\tn{\fsz fin}},\mc{H}_1)$.  Since ``transitive" kinematic schemes over classes of {\it finite} directed sets are recoverable from their skeleta, $(\ms{A}_{\tn{\fsz{fin}}},\mc{H}_1)$ inherits a universal property of sorts from $(\ms{A}_{\tn{\fsz{fin}}},\mc{H}(\ms{A}_{\tn{\fsz{fin}}}))$.  This provides a purely structural reason for regarding the positive sequential kinematic scheme as important. 

Every proper, full, originary kinematic prescheme with object class $\ms{K}$ is a subobject of the universal kinematic scheme $(\ms{K},\mc{H}(\ms{K}))$ over $\ms{K}$.  In terms of underlying multidirected sets, these kinematic preschemes are given by deleting relations of $\ms{V}(\ms{K},\mc{H}(\ms{K}))$, without altering its class of elements. The class of proper, full, originary kinematic preschemes over $\ms{K}$ is then naturally ordered by proper inclusion of classes of co-relative histories, with $(\ms{K},\mc{H}(\ms{K}))$ being the unique maximal element.  These inclusions correspond to ``functors" of kinematic schemes, bijective on objects, but neither full nor originary.  They induce a natural partial order $\prec_{\ms{K}}$ on the class of proper, full, originary kinematic preschemes over $\ms{K}$, in which $(\ms{K},\mc{H})\prec_{\ms{K}}(\ms{K},\mc{H}')$ if and only if there is a proper inclusion $\mc{H}\subset\mc{H}'$.   Restricting to kinematic schemes over $\ms{K}$ yields a structure $(\tn{\bf Kin}(\ms{K}),\prec_{\ms{K}})$, called the {\bf kinematic space over} $\ms{K}$, whose ``elements" correspond to proper, full, originary kinematic schemes over $\ms{K}$.

\newpage

\section{Conclusions}\label{sectionconclusions}

This final section of the paper is devoted to two main tasks. The first, undertaken in section \hyperref[subsectionalternative]{\ref{subsectionalternative}}, is to summarize the principal conclusions of the preceding sections.  Included are a short recapitulation of the axiomatic analysis of sections \hyperref[sectiontransitivity]{\ref{sectiontransitivity}}, \hyperref[sectioninterval]{\ref{sectioninterval}}, and \hyperref[sectionbinary]{\ref{sectionbinary}}; a specific proposal for a new set of axioms for discrete causal theory; a brief discussion of possible alternatives to this proposal, both conservative and radical; and a concise review of the new perspectives and technical methods introduced here.  The second task, pursued in section \hyperref[subsectionomitted]{\ref{subsectionomitted}}, is more difficult and open-ended.  It consists of briefly outlining important omitted topics.  Some of these topics were included in earlier versions of the paper, but were excised due to length considerations, or due to my dissatisfaction with my own understanding of them.  Other omitted topics involve elaboration of the material presented here.   Included in this section are further remarks about the axioms of countability and irreflexivity/acyclicity, not examined in detail in the main body of the paper; an additional word or two about covariance; a glimpse of algebraic and structural topics only encountered obliquely in the paper; a speculative elaboration regarding phase maps; a short description of the mathematical field of {\it random graph dynamics,} likely to play a crucial role in mechanisms of emergence in discrete causal theory; and an outline of alternatives to the theory of power spaces for generalizing order theory in physics, particularly in the context of classical holism. The section concludes with a list of other physical and mathematical theories with potentially interesting connections to the ideas developed in this paper.  Finally, section \hyperref[subsectionacknowledgements]{\ref{subsectionacknowledgements}} consists of acknowledgements and a few miscellaneous notes. 


\subsection{New Axioms, Perspectives, and Technical Methods}\label{subsectionalternative}

\refstepcounter{textlabels}\label{axiomaticanalysis}

{\bf Summary of Axiomatic Analysis.} Sections \hyperref[sectiontransitivity]{\ref{sectiontransitivity}} and \hyperref[sectioninterval]{\ref{sectioninterval}} of this paper present the case that the causal set axioms of {\it transitivity} (\hyperref[tr]{TR}) and {\it interval finiteness} (\hyperref[if]{IF}) are unsuitable for discrete causal theory under Sorkin's version of the {\it causal metric hypothesis} (\hyperref[cmh]{CMH}).  Section \hyperref[sectiontransitivity]{\ref{sectiontransitivity}} addresses the shortcomings of transitivity, focusing on the case of directed sets, and particularly acyclic directed sets.  The principal conclusion is that transitive causal orders, or generalizations of such orders containing cycles, should be viewed as derivative constructs of generally nontransitive binary relations called {\it causal preorders,} physically interpreted via the {\it independence convention} (\hyperref[ic]{IC}), introduced in section \hyperref[subsectionchains]{\ref{subsectionchains}}.  Section \hyperref[subsectiontransitivitydeficient]{\ref{subsectiontransitivitydeficient}} presents six arguments that transitive relations are information-theoretically deficient.   Section \hyperref[sectioninterval]{\ref{sectioninterval}} details the many shortcomings of interval finiteness as a local finiteness condition for directed sets and multidirected sets.  The proposed solution is to replace interval finiteness with the alternative axiom of {\it local finiteness} (\hyperref[lf]{LF}).   Sections \hyperref[subsectionintervalfinitenessdeficient]{\ref{subsectionintervalfinitenessdeficient}} and \hyperref[subsectionarglocfin]{\ref{subsectionarglocfin}} summarize a few of the principal arguments against interval finiteness and in favor of local finiteness. 

The remaining axioms of causal set theory are the {\it binary axiom} (\hyperref[b]{B}), the {\it measure axiom} ({\hyperref[m]{M}), {\it countability} (\hyperref[c]{C}), and {\it irreflexivity} (\hyperref[ir]{IR}).   None of these axioms seem to present problems necessitating their outright abandonment, but more subtle changes are worth considering.  Most obviously, irreflexivity no longer rules out causal cycles in the absence of transitivity, leading to a choice to either admit cycles, or elevate acyclicity (\hyperref[ac]{AC}) to an axiom.  Most of the results of this paper do {\it not} depend on acyclicity, but some of the most important results of section \hyperref[subsectionquantumcausal]{\ref{subsectionquantumcausal}} are developed and presented only in the acyclic case.  Turning to the measure axiom, sections \hyperref[subsectioncausalmetric]{\ref{subsectioncausalmetric}} and \hyperref[subsectionintervalfiniteness]{\ref{subsectionintervalfiniteness}} include brief remarks regarding the possibility of weakening this axiom to admit consideration of {\it local causal structure} in the emergent definition of spacetime volume.  In the locally finite case, this change is not so radical as to annul the basic insight that counting can serve as a proxy for the missing conformal factor in Malament's metric recovery theorem.  Regarding the binary axiom, section \hyperref[sectionbinary]{\ref{sectionbinary}} provides ample illustrations of how this axiom {\it could} be altered; for example, by invoking holistic power spaces.  However, as elaborated below, the primary attitude toward the binary axiom adopted in this paper is to {\it expand} the catalogue of structures that may be viewed in essentially the same way, culminating in the principle of iteration of structure (\hyperref[is]{IS}) in section \hyperref[subsectionquantumcausal]{\ref{subsectionquantumcausal}}.   Finally, there are excellent reasons to retain the axiom of countability (\hyperref[c]{C}), at least in the locally finite case.


\refstepcounter{textlabels}\label{suggestedalt}

{\bf Suggested Alternative Axioms.} The following set of axioms for classical discrete causal theory represents, in my judgment, a reasonable balance of relevance and generality, in light of the analysis presented in sections \hyperref[sectiontransitivity]{\ref{sectiontransitivity}}, \hyperref[sectioninterval]{\ref{sectioninterval}}, and \hyperref[sectionbinary]{\ref{sectionbinary}} of this paper.   The inclusion of acyclicity as an axiom is a conservative, provisional choice, based on the lack of clear physical evidence for the existence of causal cycles.   Due to this choice, the underlying set referenced in the following list of axioms is denoted by the letter $A$, for ``acyclic directed set."  For comparison, the letter $C$, for ``causal set," is used in some of the same axioms in section \hyperref[subsectionaxioms]{\ref{subsectionaxioms}}. 

\hspace*{.4cm}B. \hspace*{.2cm}{\bf Binary Axiom}: {\it Classical spacetime may be modeled as a set $A$, whose elements 
\\ \hspace*{1cm} represent spacetime events, together with a binary relation $\prec$ on $A$, whose elements 
\\ \hspace*{1cm} represent causal relations between pairs of spacetime events.}

\hspace*{.35cm}M$^*$. \hspace*{0cm}{\bf Measure Axiom}: {\it Elements of $A$ are assigned volume via a discrete measure 
\\ \hspace*{1.1cm} $\mu:\ms{P}(A)\rightarrow\RR^+$, whose value depends only on local causal structure.}

\hspace*{.35cm}C. \hspace*{.25cm}{\bf Countability}: {\it $A$ is countable.}

\hspace*{.3cm}LF. \hspace*{.1cm}{\bf Local Finiteness}: {\it Every element of $A$ has a finite number of direct predecessors\\ \hspace*{1.1cm}and successors.}

\hspace*{.3cm}AC. \hspace*{.05cm}{\bf Acyclicity}: {\it $(A,\prec)$ has no cycles.}

Note the change to the measure axiom to admit consideration of local causal structure, as compared to the original measure axiom (\hyperref[m]{M}).  Also observe that local finiteness is stated here in terms of {\it direct predecessors and successors,} rather than relation sets, since no multidirected structure is involved.   Since the binary axiom is included in this list of axioms, the resulting alternative formulation of classical discrete causal theory is {\it not irreducibly holistic.}  However, relation spaces, path spaces, and other important directed sets and multidirected sets whose elements do not correspond to spacetime events, remain crucially important in this context. 


\refstepcounter{textlabels}\label{conservativealt}

{\bf Conservative Alternative.} The alternative list of axioms presented above is conservative, excluding both causal cycles and classical holism.  However, it is worthwhile to consider an even more conservative theory, incorporating only those conclusions from sections \hyperref[sectiontransitivity]{\ref{sectiontransitivity}} and \hyperref[sectioninterval]{\ref{sectioninterval}} of this paper {\it absolutely necessary} to repair the most serious structural shortcomings of causal set theory.  This hyper-conservative approach permits only {\it irreducible} relations at the level of the causal preorder, thus obviating the need for the independence convention (\hyperref[ic]{IC}).  Interval finiteness is also included, so that only relative acyclic directed sets over $\ZZ$ are considered.  Local finiteness is {\it added,} to rule out pathologies such as the infinite bouquet.   Many existing methods and results from causal set theory may be carried over, or adjusted, with a minimal amount of effort, under these axioms.  The technical methods of relation space, path spaces, and kinematic schemes may then be brought to bear, without drastically altering the information content of causal set theory. 

\hspace*{.4cm}B. \hspace*{.25cm}{\bf Binary Axiom}: {\it Classical spacetime may be modeled as a set $A$, whose elements 
\\ \hspace*{1.1cm} represent spacetime events, together with a binary relation $\prec$ on $A$, whose elements 
\\ \hspace*{1.1cm} represent causal relations between pairs of spacetime events.}

 \hspace*{.4cm}M. \hspace*{.2cm}{\bf Measure Axiom}: {\it The volume of a spacetime region corresponding to a subset  
\\ \hspace*{1.1cm} $S$ of $A$ is equal to the cardinality of $S$ in fundamental units, up to Poisson-type 
\\ \hspace*{1.1cm} fluctuations.}

\hspace*{.4cm}C. \hspace*{.25cm}{\bf Countability}: {\it $A$ is countable.}
 
\hspace*{.25cm}IRR. \hspace*{0cm}{\bf Irreducibility}: {\it All relations between pairs of elements of $A$ are irreducible.}

\hspace*{.35cm}IF. \hspace*{.18cm}{{\bf Interval Finiteness}: {\it For every pair of elements $x$ and $z$ in $A$, the open interval \\
\hspace*{1.1cm}$\llangle x,z\rrangle:=\{y\in A\hspace*{.1cm}|\hspace*{.1cm} x\prec y\prec z\}$ has finite cardinality.}  

\hspace*{.35cm}LF. \hspace*{.1cm}{\bf Local Finiteness}: {\it Every element of $A$ has a finite number of direct predecessors\\ \hspace*{1.1cm}and successors.}

\hspace*{.35cm}AC. {\bf Acyclicity}: {\it $(A,\prec)$ has no cycles.}

The three axioms of countability, acyclicity, and interval finiteness may be replaced by the single axiom that $(A,\prec)$ admits a morphism of finite index into the integers. 


\refstepcounter{textlabels}\label{radicalalt}

{\bf Radical Alternatives.} Many, but not all, of the methods and results of this paper, also apply to more speculative approaches to classical discrete causal theory, involving either causal cycles, modification of the binary axiom to admit irreducible classical holism, or a combination of the two.  Consideration of causal cycles has strong supporting precedent from general relativity.   Perhaps the most reasonable alternative along these lines is to simply drop the axiom of acyclicity from the list of suggested alternative axioms above.  Regarding classical holism, probably the simplest versions, at least from a conceptual standpoint, involve ``families of events influencing other families of events" in an irreducible fashion, a notion admitting straightforward formalization in terms of power spaces, as described in section \hyperref[subsectionpowerset]{\ref{subsectionpowerset}}.   Modified versions of the measure axiom, countability, local finiteness, and acyclicity still make sense in this context.  Alternatively, one might prefer to regard ``spacetime events" as merely a way of referencing some aspect of structure defined in other terms.   Penrose's twistor theory provides classical precedent for this perspective.   Grothendieck's relative viewpoint (\hyperref[rv]{RV}), Isham's topos theory, and other structural notions from modern mathematics provide many alternatives, besides those arising from causal theory itself. 


\refstepcounter{textlabels}\label{summaryperspectives}

{\bf Summary of New Perspectives and Technical Methods.} The axiomatic changes suggested above serve to correct what I view as actual {\it deficiencies} in the current formulation of causal set theory.  However, the potential exists to further improve discrete causal theory by adopting a broader structural viewpoint and introducing new technical methods.   This paper includes a variety of suggestions along these lines.  The principles of natural philosophy listed in section \hyperref[naturalphilosophy]{\ref{naturalphilosophy}} provide a general preamble to this effort.  The guiding physical idea, throughout the paper, is the {\it causal metric hypothesis} (\hyperref[cmh]{CMH}), presented in section \hyperref[subsectioncausalmetric]{\ref{subsectioncausalmetric}}, which states that the properties of the physical universe are manifestations of causal structure.  This idea itself is not new; in particular, Sorkin's phrase, {\it``order plus number equals geometry,"} represents a version of it.   However, the choice to explicitly isolate and study the causal metric hypothesis in its own right, independently of any specific choice of mathematical model, enables a much clearer, more general, and more flexible approach, incorporating many new features.   In particular, it facilitates the recognition of distinct classical (\hyperref[ccmh]{CCMH}) and quantum (\hyperref[qcmh]{QCMH}) versions of this hypothesis, representing different hierarchical aspects of discrete causal theory.   A number of other basic structural concepts, transcending any specific choice of model, appear in the latter part of section \hyperref[sectionaxioms]{\ref{sectionaxioms}}.  On an individual basis, these concepts are standard, or elementary, or both, but they may be combined and utilized in novel ways in the context of discrete causal theory.  These include the {\it independence convention} (\hyperref[ic]{IC}), the {\it order extension principle} (\hyperref[oep]{OEP}), the principle of {\it hidden structure} (\hyperref[hs]{HS}), and Grothendieck's {\it relative viewpoint} (\hyperref[rv]{RV}). 

``New physics" is introduced beginning in section \hyperref[sectiontransitivity]{\ref{sectiontransitivity}} of this paper.   The first step is to reassign the fundamental role of the transitive causal order to a generally nontransitive binary relation called the {\it causal preorder,} which generates the causal order under the operation of transitive closure.  The terms {\it order} and {\it preorder} are accorded nonstandard generalized meanings in this context, to allow for the possibility of causal cycles.  The causal preorder provides a more nuanced view of fundamental causal structure than the causal order in the discrete setting.  In particular, it accommodates the possibilities of {\it irreducibility} and {\it independence} of relations between pairs of events, features which are absent in the interpolative continuum setting where the transitive paradigm originated.  Previous work of Finkelstein and Raptis involves structures formally equivalent to acyclic causal preorders, but with different physical interpretations.  Among other improvements, recognition of the causal preorder rectifies the {\it Kleitman-Rothschild pathology,} in which configuration spaces of causal sets are dominated by members of negligible ``temporal size." 

The second major change of perspective introduced in this paper is a new viewpoint regarding {\it local structure,} presented in section \hyperref[sectioninterval]{\ref{sectioninterval}}.  This viewpoint is much different than the interval-theoretic viewpoint prevailing in the current formulation of causal set theory, expressed, for example, in the causal set axiom of interval finiteness (\hyperref[if]{IF}). It is based on the concept of {\it causal locality,} under which a condition is taken to be local if and only if it may be checked by examining the {\it independent causes and effects associated with each event.}   Again, there are formal precursors in the work of Finkelstein and Raptis.   The foremost application of causal locality in section \hyperref[sectioninterval]{\ref{sectioninterval}} is the replacement of interval finiteness with the alternative axiom of {\it local finiteness} (\hyperref[lf]{LF}), as discussed in the summary of axiomatic analysis above.  The {\it star topology,} introduced in section \hyperref[subsectiontopology]{\ref{subsectiontopology}}, provides a technical means to implement this new local viewpoint.  The theory of {\it relative multidirected sets over a fixed base,} appearing in section \hyperref[relativeacyclicdirected]{\ref{relativeacyclicdirected}}, is the first major application of Grothendieck's relative viewpoint (\hyperref[rv]{RV}) appearing in this paper.  The interplay of relative multidirected sets and the concept of causal locality provides an exceptionally clear picture of the structural limitations of causal sets, revealing how drastically the supposedly ``local" condition of interval finiteness restricts {\it global} structure.  

Section \hyperref[sectionbinary]{\ref{sectionbinary}} greatly amplifies the new perspectives developed in sections \hyperref[sectiontransitivity]{\ref{sectiontransitivity}} and \hyperref[sectioninterval]{\ref{sectioninterval}}.  The structural theme of this section is to expand the viewpoint embodied in the binary axiom (\hyperref[b]{B}) of causal set theory, to encompass ``causal relationships" between pairs of structures more complex than individual events, following Grothendieck's relative viewpoint (\hyperref[rv]{RV}).  The simplest example of this theme is the theory of {\it relation space}, explored in section \hyperref[subsectionrelation]{\ref{subsectionrelation}}, which studies ``causal relationships" between pairs of {\it relations} in a multidirected set.   The global structure arising from all such relationships elevates this set of relations to the status of a directed set in its own right, called the {\it relation space} of the original multidirected set.  This viewpoint leads to the solution of important technical problems in discrete causal theory, such as the {\it permeability of maximal antichains} in multidirected sets.  Relation spaces have appeared previously in purely mathematical contexts as {\it line digraphs,} but do not seem to have been invoked in the study of fundamental spacetime structure.   Comparison of the {\it relation space functor} and the {\it abstract element space functor} reveals the striking fact that ``typical" multidirected sets are {\it nearly interchangeable} with their relation spaces.   In particular, the relation space functor {\it reduces multidirected structure to directed structure,} while {\it preserving information, except at the boundary.}   Section \hyperref[subsectionpowerset]{\ref{subsectionpowerset}} introduces {\it power spaces,} which provide many more examples and applications of natural relationships between pairs of nontrivial structures in discrete causal theory.  An example is the theory of {\it causal atoms,} already used in section \hyperref[relativeacyclicdirected]{\ref{relativeacyclicdirected}} to give a conceptually satisfying proof that causal sets are relative directed sets over the integers.   Power spaces also provide a natural framework for the study of {\it classical holism} in the discrete causal context.  {\it Causal path spaces,} introduced in section \hyperref[subsectionpathspaces]{\ref{subsectionpathspaces}}, are important generalizations of relation spaces, formalizing the flow of information or causal influence in a wide variety of contexts.  Modern algebraic methods provide powerful techniques for studying causal path spaces.  In particular, algebraic objects called {\it causal path semicategories} and {\it causal path algebras} play central roles in the general theory {\it path summation over a multidirected set,} introduced in section \hyperref[subsectionpathsummation]{\ref{subsectionpathsummation}}.  

Section \hyperref[subsectionquantumcausal]{\ref{subsectionquantumcausal}} brings these new perspectives and methods to bear on the quantum theory of spacetime.  The crucial link between the classical and quantum realms in discrete causal theory is the principle of {\it iteration of structure} (\hyperref[is]{IS}), introduced in section \hyperref[subsectionquantumprelim]{\ref{subsectionquantumprelim}}, which permits effective implementation of the {\it histories approach to quantum theory} in the {\it background-independent context,} via path summation over multidirected sets. Iteration of structure formalizes the observation that configuration spaces of directed sets possess natural {\it higher-level multidirected structures,} collectively induced by their member sets, and by relationships between pairs of these sets.   I refer to such relationships as {\it co-relative histories.}   Of particular interest in discrete causal theory are {\it proper, full, originary co-relative histories,} which may be viewed as equivalence classes of special morphisms called {\it transitions,} representing the ``evolution" of one directed set into another.   In section \hyperref[subsectionquantumpathsummation]{\ref{subsectionquantumpathsummation}}, I carry out a discrete causal adaptation of the histories approach to quantum theory, motivated by Feynman's continuum path integral.   This adaptation is {\it abstract,} in the sense that the multidirected sets involved are not {\it a priori} assigned any specific physical interpretation.  The {\it path summation principle} (\hyperref[ps]{PS}), motivated by iteration of structure, states that this abstract approach applies to {\it both} the background-dependent theory of particles and fields on directed sets, and the background-independent theory of co-relative histories.   In section \hyperref[subsectionschrodinger]{\ref{subsectionschrodinger}}, I apply this theory to derive {\it causal Schr\"{o}dinger-type equations,} which serve as dynamical laws for discrete quantum causal theory.  The impermeability of maximal antichains in relation space implies that these equations adequately account for the information encoded in the underlying causal structure.  Section \hyperref[subsectionkinematicschemes]{\ref{subsectionkinematicschemes}} introduces the theory of {\it kinematic schemes,} which are special configuration spaces of directed sets encoding ``evolutionary pathways" for all ``permissible histories."  Kinematic schemes are endowed with higher-level multidirected structures supplied by classes of co-relative histories between pairs of their member sets, providing a prototypical example of iteration of structure (\hyperref[is]{IS}).  They provide ``global formalizations" of the {\it evolutionary viewpoint} in background independent discrete causal theory, generalizing Sorkin and Rideout's theory of sequential growth dynamics for causal sets. The abstract quantum causal theory developed in sections \hyperref[subsectionquantumpathsummation]{\ref{subsectionquantumpathsummation}} and  \hyperref[subsectionschrodinger]{\ref{subsectionschrodinger}} may be applied to kinematic schemes by isolating this higher-level multidirected structure and moving to relation space.   The resulting Schr\"{o}dinger-type equations provide dynamical laws for discrete quantum spacetime.  


\subsection{Omitted Topics and Future Research Directions}\label{subsectionomitted}

A number of topics of obvious interest in discrete causal theory are omitted from this paper, due both to length considerations and to my own inability to locate or develop certain results.  In this section, I make remarks about a few of these topics, and give some hints regarding interesting directions for future research.  


\refstepcounter{textlabels}\label{remainingax}

{\bf Further Remarks on Countability, Irreflexivity, and Acyclicity.}  As stated in section \hyperref[subsectionalternative]{\ref{subsectionalternative}} above, I know of no compelling reasons to abandon the causal set axioms of countability (\hyperref[c]{C}) and irreflexivity (\hyperref[ir]{IR}), though of course the latter axiom loses much of its strength in the nontransitive context, no longer implying acyclicity (\hyperref[ac]{AC}), for example.  Since this paper contains no significant analysis of these axioms, I give here a brief outline of what such analysis might entail.  Deeper examination of the countability axiom might include the following: 1) explanation of how Grothendieck's relative viewpoint (\hyperref[rv]{RV}) applies to the topic of cardinality; this explains why cardinality conditions are less objectionable as global constraints than order-theoretic conditions; 2) elaboration of the distinguished role of {\it finite} directed sets in the theory of experimentation and measurement; 3) further investigation of {\it entropy issues,} such as the Kleitman-Rothschild pathology; 4) discussion of the {\it continuum hypothesis} and related set-theoretic and order-theoretic topics; 5) fuller treatment of {\it qualitative} structural differences between classes of directed sets and multidirected sets satisfying different cardinality conditions. 

\refstepcounter{textlabels}\label{cycles}

Turning to irreflexivity and acyclicity, the choice to abstain from assumptions about causal cycles, even at the classical level, throughout much of sections \hyperref[sectiontransitivity]{\ref{sectiontransitivity}}, \hyperref[sectioninterval]{\ref{sectioninterval}}, and \hyperref[sectionbinary]{\ref{sectionbinary}} of this paper, reflects my conviction that their status is far from a settled issue.  General relativity admits many solutions involving continuum analogues of such cycles; i.e., closed timelike curves, and it seems premature to entirely dismiss these solutions as pathologies that may be expected to disappear in a suitable quantum theory of gravity.  After all, general relativity is the only well-supported theory available which {\it could} shed any light on the status of classical causal cycles, since it is the only such theory that leaves their existence to be determined dynamically.   The choice to rule out such cycles, by including acyclicity in the suggested list of alternative axioms in section \hyperref[subsectionalternative]{\ref{subsectionalternative}} above, is therefore provisional, and somewhat reluctant.   The argument tipping the balance in favor of this choice is simply that it is best to explore simple alternatives first.  However, it does go against my descriptive instincts. 

General {\it types} of objections to causal cycles include: 1) lack of observational evidence for their existence; 2) logical ``problems" such as the grandfather paradox; 3) the fact that certain existing theories ``conspire in unforeseen ways to avoid cycles," an example being the {\it no-cloning theorem} in quantum information theory; and 4) technical difficulties arising in theories permitting cycles, such as those discussed in the context of path summation in section \hyperref[subsectionquantumpathsummation]{\ref{subsectionquantumpathsummation}} of this paper. The strong form of the causal metric hypothesis (\hyperref[cmh]{CMH}) completely resolves the second type of objection, since it leads to theories with perfect background independence.  The other types of objections, however, remain relevant in causal theory.   Arguments in favor of considering cycles in discrete causal theory include the following: 1$^*$) general relativity permits closed timelike curves; 2$^*$) regardless of their classical status, cycles may be unavoidable at higher levels of hierarchy, appearing, for example, in the abstract structure of kinematic schemes of countable acyclic directed sets; 3$^*$) abstract element spaces over acyclic multidirected sets generally include cycles; 4$^*$) cycles are generally nonlocal and therefore do not generally present any local pathology; 5$^*$) cycles are reasonable candidates for {\it subclassical structure;} i.e., ``internal structure of classical elements."


\refstepcounter{textlabels}\label{covariance}

{\bf Covariance.}  Whenever a physical theory involves arbitrary choices, it must provide mechanisms to ensure that its substance remains unchanged if these choices are made otherwise.  This is the essence of the principle of {\it covariance} in physical law.   In the classical continuum context, covariance is closely related to the arbitrary choice of a coordinate system; for example, dynamical laws expressible as tensor equations are often called ``covariant," because their abstract forms are not coordinate-dependent.  In causal theory, this type of covariance generalizes in a straightforward manner in terms of generalized frames of reference for directed sets, represented by generalized order extensions.   The prototypical example of this idea appears in Sorkin and Rideout's theory of sequential growth dynamics for causal sets \cite{SorkinSequentialGrowthDynamics99}, as the principle of {\it discrete general covariance.}  The purpose of this principle is to capture the notion that ``all evolutionary accounts of a causal set are equally valid."   

At a formal level, the arbitrary choice associated with discrete general covariance may be viewed as a {\it particular choice of kinematics in a particular kinematic scheme;} in this case, Sorkin and Rideout's kinematic scheme $(\ms{C}_{\tn{\fsz{fin}}},\mc{H}_1)$.   However, a {\it prior} arbitrary choice, in the context of background-independent quantum causal theory, is the choice of an appropriate class of co-relative histories, necessary to {\it define} a kinematic scheme.   Indeed, as mentioned in section \hyperref[subsectionkinematicschemes]{\ref{subsectionkinematicschemes}} of this paper, the object class $\ms{K}$ of a kinematic scheme $(\ms{K},\mc{H})$ represents a hypothesis about the structure of the physical universe, and should therefore be uniquely determined, but the class of co-relative histories $\mc{H}$ is selected for convenience; i.e., arbitrarily, from among many equally valid choices.  The meaning of covariance in this context is that the essential physical properties of the theory should not depend on the choice of $\mc{H}$.  In practical terms, this means that {\it if one changes kinematic schemes, one must change phase maps to compensate.}  


\refstepcounter{textlabels}\label{algstructure}

{\bf Algebraic Structure and Hierarchy.} The analogy between kinematic schemes and categories prompts a closer examination of their similarities and differences.  This, in turn, leads to a broader consideration of algebraic structure in causal theory, and in theoretical physics more generally, with a particular emphasis on questions of {\it hierarchy}.   Isham's topos-theoretic framework for physics provides valuable and wide-ranging contributions in this direction.  The principle of iteration of structure (\hyperref[is]{IS}), introduced in section \hyperref[subsectionquantumprelim]{\ref{subsectionquantumprelim}} of this paper, raises the issue of hierarchy explicitly, but in a limited context, arising more or less directly from the ordinary superposition principle of quantum theory.  However, much more general notions of hierarchy are hinted at throughout sections \hyperref[sectionbinary]{\ref{sectionbinary}} and \hyperref[subsectionquantumcausal]{\ref{subsectionquantumcausal}}, particularly in the contexts of relation spaces, causal path spaces, and kinematic spaces.  One may discern at least four distinct ``vertical" levels of hierarchy, represented, for example, by elements, directed sets, kinematic schemes, and kinematic spaces.  An {\it infinite} number of distinct ``horizontal" levels of hierarchy are readily apparent, represented, for example, by spaces of paths of various lengths or isomorphism types in directed or multidirected sets.    

A deeper examination of algebraic structure and hierarchy in discrete causal theory might include the following: 1) more detailed comparison of causal theory and category theory, particularly topos theory; 2) more adequate identification and explanation of the relevance and utility of various category-theoretic properties in physical contexts; 3) general discussion of {\it local, global, vertical,} and {\it horizontal} hierarchy; 4) the use of {\it binary trees} to encode algebraic hierarchy; 5) detailed comparison of categories with operations, such as {\it monoidal categories,} and analogous structures in discrete causal theory; 6) systematic examination of causal analogues of the theory of {\it categorification} and {\it decategorification,} generalizing the relationship between a kinematic scheme and its underlying multidirected set;  7) discussion of possible physical interpretations of additional levels of vertical hierarchy, such as {\it subclassical} and {\it hyperquantum} theories; 8) analysis of the fate of Lie representation theory in theoretical physics under the causal metric hypothesis (\hyperref[cmh]{CMH}), including bifurcation into {\it group-theoretic} and {\it non-group-theoretic} aspects in the theory of ``external" symmetries. 


\refstepcounter{textlabels}\label{phasetheory}

{\bf Phase Theory.} When confronted with a causal Schr\"{o}dinger-type equation such as equation \hyperref[causalschrodinger]{\ref{causalschrodinger}}:
\[\psi_{R;\theta}^-(r)=\theta(r)\sum_{r^-\prec r}\psi_{R;\theta}^-(r^-),\]
 the most basic question is the following: 
 
 \hspace*{.3cm}{\bf Question:} {\it How is the relation function $\theta$ generating the phase map determined, and what is 
 \\ \hspace*{.3cm}its target object?}
 
More general versions of this question arise for arbitrary phase maps, not necessarily generated by relation functions.  To investigate these questions, it is natural to begin by considering Feynman's phase map $\Theta$ in the continuum context, appearing in equation \hyperref[feynmanphase]{\ref{feynmanphase}}:
\[\Theta(\gamma)=e^{\frac{i}{\hbar}\ms{S}(\gamma)}.\]
Feynman's phase map depends on the classical action $\ms{S}$, which in turn depends on the Lagrangian.  Hence, conversion of equation \hyperref[causalschrodinger]{\ref{causalschrodinger}}, or a more general analogue of this equation, into a form that can actually be used to make predictions, involves identification of a suitable ``action" or ``Lagrangian" for a suitable class of directed sets or kinematic schemes of directed sets.  Interesting work has already been done in this direction in the special case of causal sets, of which I have mentioned just one example: the recent paper {\it Scalar Curvature of a Causal Set} \cite{DowkerScalarCurvature10}, by Dionigi Benincasa and Fay Dowker.   I have not attempted to adapt or generalize these efforts here. 

However, it is possible to make a few general remarks on elementary mathematical and information-theoretic grounds.  Feynman's phase map takes values in the unit circle $S^1$, viewed as a subset of  the complex numbers $\CC$.  These values are summed in $\CC$ to yield quantum amplitudes, modulo appropriate constants of proportionality, which are then translated into real-valued probabilities.  $S^1$ and $\CC$ are natural objects in the continuum setting: both admit continuum manifold structures, and $\CC$ is the algebraic closure of $\RR$.  In discrete causal theory, however, one would expect objects such as $S^1$ and $\CC$ to somehow {\it emerge as continuum limits of more primitive target objects.}  For example, working with a kinematic scheme over the class $\ms{A}_{\tn{\fsz{fin}}}$ of finite acyclic directed sets, it would be reasonable to expect phases to belong to some {\it finite algebraic object.}  Speculating further, different phases might have different natural targets, perhaps organized into a {\it filtered system,} whose category-theoretic limit might resemble a countable subset of $\CC$.   {\it Counting} or {\it entropy} considerations may be expected to contribute in this context.  Covariance also plays a role; {\it a priori,} different kinematic schemes require families of targets.  Finally, as suggested in section \hyperref[subsectionkinematicschemes]{\ref{subsectionkinematicschemes}} of this paper, it may be preferable in the background-independent context to view phase maps as {\it generalized functors,} and path sums as {\it generalized limits.}   A basic constraint on all these considerations is that one must eventually {\it interpret} the resulting generalized quantum amplitudes physically, in terms of probabilities or some generalization thereof. 
 

\refstepcounter{textlabels}\label{randomgraph}

{\bf Random Graph Dynamics.} In the late 1950's, mathematicians Paul Erd\"{o}s and Alfr\'{e}d R\'{e}nyi introduced the theory of {\it random graphs} in a long series of papers, beginning with \cite{ErdosRenyi59}.  A relatively modern reference on the subject is {\it Random Graphs} \cite{Bollobas01} by B\'{e}la Bollob\'{a}s, one of the leaders in the field.  The original context of this work was pure graph theory, with information-theoretic and physical applications appearing only afterwards.  A simple example of an {\it undirected} random graph may be constructed via the following process: beginning with a finite vertex set $V$ of cardinality $N$, consider each pair of elements $(x,y)$ of $V$, in some order, adding an edge between $x$ and $y$ with some {\it fixed} probability $p$.  When every pair of vertices has been considered, the process stops, and the result is a {\bf random graph} of type $(N,p)$.  

The evolutionary viewpoint in discrete causal theory, epitomized by Sorkin and Rideout's theory of sequential growth dynamics for causal sets \cite{SorkinSequentialGrowthDynamics99}, is closely related to random graph dynamics.  Sorkin and Rideout explicitly mention this connection in their 2000 paper {\it Evidence of a Continuum Limit in Causal Set Dynamics} \cite{SorkinEvidenceofContLimVer208}.   Transitions, co-relative histories, and kinematic schemes, discussed in section \hyperref[subsectionquantumcausal]{\ref{subsectionquantumcausal}} of this paper, represent vastly generalized versions of random graph processes.  A number of complicating factors distinguish these physically-oriented constructions from the simple models studied by Erd\"{o}s and R\'{e}nyi.  For example, the graphs involved are directed, their vertex sets have no fixed cardinality, more complex structures than single edges are added at each step of the process, and the ``probabilities" at each step depend on the previous steps.   

\refstepcounter{textlabels}\label{phasetrans}

One reason why the analogy between random graph dynamics and the evolutionary viewpoint in causal theory is interesting is because {\it peculiar and physically suggestive phenomena} occur in random graph dynamics, phenomena that no one seems to have even guessed at prior to Erd\"{o}s and R\'{e}nyi's papers.  One class of phenomena of particular interest is graph-dynamical {\it phase transitions.}\footnotemark\footnotetext{Use of the term {\it phase} here is entirely distinct from its use in the context of {\it phase maps.}}   In very general terms, a {\bf phase transition} is an abrupt change in some qualitative property of a class of random graphs that occurs at some critical value of some associated parameter.  For example, consider the class of random graphs of type $(N,p)$, mentioned above, and let $N$ and $p$ vary in such a way that $p$ is a function of $N$.  One may then consider the limit of the product $Np$ as $N$ approaches infinity; in particular, one may consider varying $p$ in such a way that the limit varies from less than $1$ to greater than $\log N$.   A sequence of phase transitions is then observed: for sufficiently small values of $p$, the largest connected component of a typical random graph of type $(N,p)$ has approximately $\log N$ vertices or fewer.  As the limit of $Np$ reaches the critical value of $1$, a single {\it giant component} with roughly $N^{\frac{2}{3}}$ vertices suddenly emerges, and rapidly increases in size until it contains an appreciable fraction of all the vertices, while still leaving many small components and isolated vertices.  This qualitative picture remains unchanged until the limit passes another critical value at $\log N$, at which point the typical graph suddenly becomes completely connected.  

Such striking behavior raises the possibility that certain qualitative properties of spacetime might be attributed to graph-dynamical processes in the context of discrete causal theory.  For example, one of the principal arguments in favor of the hypothesis of {\it cosmic inflation} in the early universe is the uniformity of the {\it cosmic microwave background,} which, following a conventional interpretation, suggests the likelihood that regions of spacetime with large separation were causally connected in the past.  A possible alternative to the inflationary hypothesis is the idea that {\it causal structure abruptly became sparser} in some sense, possibly due to a phase transition. 


\refstepcounter{textlabels}\label{holalt}

{\bf Alternatives to Power Spaces.}  Power spaces are only one of many ways to generalize binary relations.  For example, the causal path spaces studied in section \hyperref[subsectionpathspaces]{\ref{subsectionpathspaces}} of this paper are not power spaces, since the morphism defining a path is generally not information-theoretically reducible to its image.  However, even if one restricts consideration to {\it subobjects of individual structured sets,} rather than incorporating information about relationships between structured sets, a variety of alternatives remain.  In the theory of power spaces, the basic ``binary" device of relations between {\it pairs} is retained; it is the {\it members} of these pairs that are allowed greater complexity.  An alternative viewpoint is to exchange pairs for {\it $N$-tuples.}  For example, a {\it ternary relation} on a set $S$ is a subset of the third Cartesian power $S\times S\times S$; its elements are {\it ordered triples}.  Similarly, an {\it $N$-ary relation} is a subset of the $N$th Cartesian power $S^N$.   A ``multidirected version" of a set equipped with an $N$-ary relation is an $(N+2)$-tuple of the form $S=(S,R,e_1,...,e_N)$, where $S$ and $R$ are sets, and where the maps $e_i:R\rightarrow S$ are ``generalized initial and terminal element maps."  In this context, an element of $R$ may be viewed as an $(N-1)$-dimensional {\it simplex,} and the total structure of $S$ may be viewed as an $(N-1)$-dimensional {\it simplicial complex}.  One may also allow $N$ to vary, thereby obtaining simplicial complexes of nonconstant dimension.  In either case, random graph theory is subsumed by {\it random simplicial theory,} and mathematical problems such as {\it Whitehead's conjecture} come into play.  The associated physical theory may be viewed as a generalization of simplicial power spaces, discussed in section \hyperref[subsectionpowerset]{\ref{subsectionpowerset}} in the context of higher induced relations.  

Another way to generalize binary relations is to consider {\it multiple independent} binary or $N$-ary relations on a single set.  This may be accomplished in the context of multidirected sets, or $N$-ary analogues thereof, by adding {\it colors} to elements of relation sets or generalized relation sets.  One obvious motivation for such a construction in the discrete context would be the belief that the causal metric hypothesis (\hyperref[cmh]{CMH}), or at least the strong form of this hypothesis, is {\it false.}  For example, one might regard {\it spatial} as well as causal structure as fundamental.  One might then invoke {\it two} relations: a {\it symmetric} relation encoding spatial structure, and an {\it antisymmetric} relation encoding causal structure.   Two relatively familiar theories involving combinations of spatial and temporal structure on discrete manifolds at the fundamental scale are {\it causal dynamical triangulations} and {\it loop quantum gravity.}   At a broad conceptual level, there is at least a partial {\it duality} involving spatial/temporal structure and symmetric/antisymmetric binary relations in physics, observed already in relativity.  The extent and significance of this duality remains unsettled.  In particular, one finds ``time-first" theories, such as causal set theory, ``space-first" theories, such as shape dynamics, and hybrid theories, such as causal dynamical triangulations, all competing for the same ground.  It is conceivable that the basic approaches represented by these theories are to some degree interchangeable.  

Another possible application of multiple binary or $N$-ary relations might be to encode {\it different types of spatial or temporal structure.}  For example, a pair of independent antisymmetric relations on a set might be interpreted as a {\it pair of timelike dimensions.}  More generally, power spaces, $N$-ary relations, and colors may be combined in various ways as a ``theme and variations."  If nothing else, this creates fertile ground for speculation.  For example, one might invoke a whole family of independent relations, with ``time" emerging as a ``residual drift."   If history is any judge, however, the simplest nontrivial versions of ideas are usually the most important.  


\refstepcounter{textlabels}\label{othertheories}

{\bf Connections with Other Physical Theories.} Interesting analogies and potential connections exist between the version of discrete causal theory I have developed in this paper and a variety of other approaches to physics at the fundamental scale.  Although some of these connections have been partially explored for existing versions of discrete causal theory, the new perspectives presented here might justify reconsideration.  These theories include: 

\begin{enumerate}
\item {\bf Loop Quantum Gravity}.  Analogies involving spin networks and spin foams might be examined in detail. 
\item {\bf Causal Dynamical Triangulations}.  This theory {\it can} be expressed in terms of special types of directed sets, and methods described in this paper {\it can} be applied, setting aside questions of actual suitability. Significant work has been done on the histories approach to quantum theory for causal dynamical triangulations, and it would be interesting to compare this work to the approach presented here. 
\newpage
\item {\bf Domain Theory.} An interesting general project with potential applications in domain theory would be to examine which of the ideas presented in this paper admit ``limiting versions" in the interpolative context.  
\item {\bf Physical Applications of Category Theory.}  Connections are cited throughout the paper. 
\item {\bf Quantum Information Theory}.  Quantum circuits may be modeled as directed sets equipped with matrix-valued phase maps.  This leads to an interesting approach to quantum information theory involving ideas of a {\it relativistic} flavor, such as frames of reference, relativity of simultaneity, and so on.   Na\"{i}vely, one might hope to turn this relationship around and use quantum computers as ``virtual Planck-scale laboratories." 
\item {\bf Quantum Automaton Theory.}  G. M. D'Ariano and Paolo Perinotti \cite{DArianoDirac13} have recently recovered the {\it Dirac equation,} the prototypical ``relativistic Schr\"{o}dinger-type equation," via this approach.  It would be interesting to compare their approach to the method of deriving causal Schr\"{o}dinger-type equations presented in this paper.   
\item {\bf Physical Applications of Noncommutative Geometry}.  Noncommutative geometry gives a completely different geometric perspective regarding causal path algebras than that provided by the causal metric hypothesis (\hyperref[cmh]{CMH}).  Ioannis Raptis and Roman Zapatrin discuss related ideas in their paper {\it Algebraic description of spacetime foam} \cite{RaptisSpacetimeFoam01}.  There are also some relevant physical notions underlying the theory {\it deformed special relativity} that involve noncommutative geometry, in spite of the general lack of success of this theory to date. 
\item {\bf Twistor Theory}.  Analogies mentioned in section \hyperref[subsectionpowerset]{\ref{subsectionpowerset}} of this paper might be further explored. 
\item {\bf Entropic Gravity}.  Causal theory is {\it at least} a theory of gravity, and entropic considerations are of interest in the context of phase theory.  
\item {\bf Shape Dynamics}.  As mentioned in section \hyperref[subsectionpowerset]{\ref{subsectionpowerset}} of this paper, the viewpoint of this ``space-first" theory is nearly opposite to the ``time-first" approach of causal theory.   On occasion, such ``opposite approaches" are related by duality theories, and it would be interesting to determine if such a relationship exists in this case. 
\end{enumerate}

More tenuous ``connections" with additional physical theories may also be exhibited.  For example, it is mathematically tempting  to ``promote" directed sets to {\it tubular diagrams,} to obtain a ``background-independent causal theory of interacting closed spatial loops."  Similar diagrams appear in string theory and topological quantum field theory.  In this context, the direct future $D_0^+(x)$ of an element $x$ might be replaced by a {\it cobordism of $S^1$ to a disjoint union of copies of $S^1$}.  Beyond this, however, the analogy seems to break down.  For example, the status of individual relations is unclear, and a natural analogue of iteration of structure (\hyperref[is]{IS}) is not evident.


\refstepcounter{textlabels}\label{othermath}

{\bf Connections with Mathematical Topics.} Although this paper touches on many different mathematical fields, there is much more to be said about many of these relationships.  In addition, there are other potentially fruitful connections deserving at least brief mention. 

\begin{enumerate}
\item {\bf Graph Theory.} Numerous connections are cited throughout the paper, but many more exist.
\item {\bf Category Theory.}  Many additional connections exist besides those involving existing physical applications.
\item {\bf Noncommutative Algebra; Noncommutative Geometry; ``Quantum Groups."}  Again, there are additional relevant mathematical ideas not specifically associated with any existing physical theory.   
\newpage
\item {\bf Matroid Theory.}  As mentioned in section \hyperref[subsectionchains]{\ref{subsectionchains}} of this paper, matroid theory involves a notion of {\it independence} similar to linear independence of vectors, and distinct from the notion of independence specified by the independence convention (\hyperref[ic]{IC}).  Its potential interest in the context of discrete causal theory stems from the existence of many interesting and physically suggestive matroids arising in graph theory.   For example, {\it gammoids} are obviously relevant.
\item {\bf Numerical Methods.}  Usually appearing in the context of {\it approximation} in the fields of mathematical analysis and partial differential equations, such methods acquire new algebraic and arithmetic connotations when they are considered {\it exact,} as in the derivation of causal Schr\"{o}dinger-type equations.   Particularly relevant are {\it finite element methods} involving {\it variational techniques.} 
\item {\bf Cohomology Theories.}  Algebra cohomology theories such as {\it Hochschild cohomology} and {\it cyclic cohomology} are relevant to the study of causal path algebras and phase theory.  There are results to proven concerning relationships between such cohomology theories and corresponding theories involving more primitive objects such as semicategories.\footnotemark\footnotetext{James Madden \cite{MaddenPrivate13} is ultimately responsible for making me aware of these connections.}  Relative cohomology plays a role via congruences.  Lie cohomology and algebraic $K$-theory also enter the picture. 
\item {\bf Partial Differential Equations, Integral Equations, Boundary Value Problems.} There is a huge and sophisticated existing literature on Schr\"{o}dinger-type partial differential equations.  The concepts involved in sections \hyperref[subsectionquantumpathsummation]{\ref{subsectionquantumpathsummation}} and \hyperref[subsectionschrodinger]{\ref{subsectionschrodinger}} of this paper, involving ``flux through a boundary," the relationship between local (i.e., ``differential") and global (i.e., ``integral") equations, generalized Lagrangians, the superposition principle, etc., are all central to this field. 
\item {\bf Random Topology.}  This may be viewed as a generalization of random graph dynamics. 
\item {\bf Representation Theory.} Particularly important in modern physical continuum theories, especially quantum field theories, is the representation theory of Lie groups.   If continuum geometry is shown to break down, and if the causal metric hypothesis (\hyperref[cmh]{CMH}) withstands scrutiny, generalized order-theoretic analogues of Lie representation theory may eventually take over part of this workload. Other connections to representation theory arise via the theory of semicategory algebras, by abstracting the role of groups algebras in group representation theory. 
\item {\bf Galois Theory.}  The brief remarks about causal Galois groups and Galois connections in section \hyperref[subsectionquantumprelim]{\ref{subsectionquantumprelim}} of this paper may be greatly elaborated. 
\end{enumerate}


\subsection{Acknowledgements; Personal Notes}\label{subsectionacknowledgements}

\refstepcounter{textlabels}\label{ack}

{\bf Acknowledgements.}  Rafael Sorkin was kind enough to provide detailed answers to several queries about foundational issues, and to allow me to quote personal correspondence.  David Finkelstein answered questions about his papers on causal nets, and sent me some of his unpublished work.   David Rideout supplied informative answers to multiple questions about causal sets.   Jorge Pullin provided background on general relativity and cosmology, and gave me some valuable opinions about various approaches to quantum gravity.   Jimmie Lawson provided useful advice, both general and specific, and contributed significantly to my understanding of related topics in quantum information theory.  George Ellis made encouraging remarks about my suggestion to combine causal theory with classical holism via power spaces.  Manfred Droste wrote me a very nice response to a question about universal causal sets.  Giacomo Mauro D'Ariano provided useful information from the related viewpoint of quantum automata, as well as general encouragement.   

James Madden supplied important underlying ideas about algebraic notions involving multiplicative structures more primitive than groups, as well as order-theoretic background.  Brendan McKay and James Oxley provided helpful advice on graph-theoretic issues.   Marcel Ern\'{e} was kind enough to scan and send me copies of his unpublished notes on order theory.  Joel David Hamkins provided some helpful answers via the MathOverflow forum. My junior colleague Thomas Naugle, and my sister Stephanie Dribus, endured much tedious speculation on my part and made helpful suggestions.   Steph also helped me check some of the proofs,\footnotemark\footnotetext{Remaining errors should be attributed to her.} and Tommy pointed out some very useful references.  Al Vitter, as my first algebraic geometry teacher, acquainted me with the work and philosophy of Grothendieck.  Jerome W. Hoffman, as my thesis advisor, provided much of my general mathematical background, via years of advice and hundreds of references on algebraic geometry, algebraic $K$-theory, and category theory; it may surprise him to know how many I actually read!  Finally, my sister Marian Dribus, and my mother Virginia Dribus, provided invaluable editing assistance.

\refstepcounter{textlabels}\label{personal}

{\bf Personal Notes.} I began to develop the ideas behind this paper in 2009, with little knowledge of existing causal theory beyond a superficial awareness of domains, acquired during my presence at Tulane University when Keye Martin was on the faculty.   While no serious researcher would deliberately abstain from reading existing results relevant to a project, it is perhaps not such a disadvantage to devote considerable thought to a subject in a state of accidental ignorance, provided that one is willing afterwards to read thousands of pages of literature with an open mind, despite the disappointment of repeatedly discovering that cherished ideas are either already known, or are wrong.  My first interaction with the ``outside world" on the subject of discrete causal theory came in June 2010, when I emailed some questions about the current state of causal set theory to a leading exponent of the field.  In January of 2011, I suggested to him the basic premises of sections 3 and 4 of this paper, but never learned his view of the matter.  In August 2012, I outlined many of the principal ideas presented in this paper in an essay \cite{DribusFoundations12} for the Foundational Questions Institute (FQXi), which was one of the winners in their essay contest ``Questioning the Foundations."  

\newpage

\begin{appendix}

\section{Index of Notation}\label{appendixindex}

The following index gives a section-by-section listing of the notation used in this paper.   Some duplication of symbols was unavoidable; for example, the mathscript symbol $\ms{C}$ means an arbitrary category in one context, but the specific category of causal sets in another.   The Greek letter $\Gamma$ stands for a number of different sets or spaces of paths or chains.   To make this index easier to use on a section-by-section basis, some definitions are repeated.  My hope is that the notation in the paper is at least {\it locally} unambiguous.

\vspace*{.3cm}

{\bf Section \hyperref[sectionaxioms]{\ref{sectionintroduction}}:}

\begin{multicols}{2}
\begin{itemize}
\item $C$ \hspace*{2.18cm} causal set
\item $\prec$ \hspace*{2.25cm} binary relation on $C$
\item $x,y$ \hspace*{1.9cm} spacetime events or\\ \hspace*{2.6cm} elements of a causal set
\item $\mu$ \hspace*{2.2cm} discrete measure on $C$
\item $S$ \hspace*{2.2cm} subset of $C$
\item $X,X'$ \hspace*{1.5cm} pseudo-Riemannian \\ \hspace*{2.6cm} manifolds
\item $g,g'$ \hspace*{1.8cm} pseudo-Riemannian \\ \hspace*{2.6cm} metrics
\item $\ms{K}$ \hspace*{2.1cm} Sorkin and Rideout's  \\ \hspace*{2.55cm} kinematic scheme \\ \hspace*{2.55cm} for sequential growth
\item $\tau:C_i\rightarrow C_t$ \hspace*{.65cm} transition
\item $\theta(\tau)$ \hspace*{1.7cm} transition weight or phase
\item $\{x_0,x_1,x_2,...\}$ \hspace*{.1cm} total labeling \\ \hspace*{2.6cm} of a causal set
\item $C,A,D,M$ \hspace*{.7cm} structured sets \\ \hspace*{2.6cm} (causal, acyclic directed, \\ \hspace*{2.6cm} directed, multidirected)
\item $\ms{C},\ms{A},\ms{D},\ms{M}$ \hspace*{.75cm} categories of \\ \hspace*{2.6cm} structured sets
\item $\NN,\ZZ,\QQ,\RR,\CC$ \hspace*{.45cm} nonnegative integers, \\ \hspace*{2.6cm} integers, rationals, \\ \hspace*{2.6cm} reals, complex numbers
\item $\ZZ_n$ \hspace*{1.98cm} integers modulo $n$
\end{itemize}

\end{multicols}
{\bf Section \hyperref[sectionaxioms]{\ref{sectionaxioms}}:}
\begin{multicols}{2}
\begin{itemize}
\item CMH \hspace*{1.55cm} causal metric hypothesis
\item CCMH \hspace*{1.25cm} classical causal metric \\ \hspace*{2.6cm} hypothesis
\item $C$ \hspace*{2.15cm} causal set
\item B \hspace*{2.2cm} binary axiom
\item M \hspace*{2.15cm} measure axiom
\item C \hspace*{2.2cm} (axiom of) countability
\item TR \hspace*{1.9cm} (axiom of) transitivity
\item $\llangle w,y\rrangle$ \hspace*{1.4cm} open interval in \\ \hspace*{2.6cm} a directed or \\ \hspace*{2.6cm} multidirected set
\item IF \hspace*{2.05cm} (axiom of) interval \\ \hspace*{2.6cm} finiteness
\item IR \hspace*{2.05cm} (axiom of) irreflexivity
\item $\mu$ \hspace*{2.2cm} discrete measure on $C$
\item $\ms{P}(C)$ \hspace*{1.6cm} power set of $C$
\item $\RR^+$ \hspace*{2cm} set of positive \\ \hspace*{2.6cm} real numbers
\item $(C,\prec)$ \hspace*{1.5cm} causal set with \\ \hspace*{2.6cm}  explicit binary relation
\item $S$ \hspace*{2.2cm} subset of a causal set 
\item $\ms{C}$ \hspace*{2.2cm} category of causal sets
\item $\phi$ \hspace*{2.2cm} morphism of causal sets, \\ \hspace*{2.6cm} acyclic directed sets, \\ \hspace*{2.6cm} or directed sets
\item $\phi^{-1}(x')$ \hspace*{1.2cm} fiber of $\phi$ over $x'$
\item $\preceq$ \hspace*{2.25cm} partial order 
\newpage
\item $(P,\preceq)$ \hspace*{1.5cm} partially ordered \\ \hspace*{2.6cm} set with explicit \\ \hspace*{2.6cm} binary relation
\item $A$ \hspace*{2.2cm} acyclic directed set
\item AC \hspace*{1.9cm} (axiom of) acyclicity
\item $(A,\prec)$ \hspace*{1.45cm} acyclic directed set with \\ \hspace*{2.6cm}  explicit binary relation
\item $\ms{A}$ \hspace*{2.15cm} category of acyclic\\ \hspace*{2.6cm} directed sets
\item $D$ \hspace*{2.2cm} directed set
\item $(D,\prec)$ \hspace*{1.45cm} directed set with \\ \hspace*{2.6cm} explicit binary relation
\item $\ms{D}$ \hspace*{2.15cm} category of \\ \hspace*{2.6cm} directed sets
\item $M$ \hspace*{2.05cm} multidirected set
\item $R$ \hspace*{2.15cm} relation set of a \\ \hspace*{2.6cm} multidirected set
\item $i,t$ \hspace*{2cm} initial and terminal \\ \hspace*{2.6cm} element maps of a\\\hspace*{2.6cm} multidirected set
\item $x\prec y$ \hspace*{1.6cm} means there exists \\ \hspace*{2.6cm} $r\in R$ such that \\ \hspace*{2.6cm} $i(r)=x$ and $t(r)=y$
\item $(M,R,i,t)$ \hspace*{.65cm} multidirected set \\ \hspace*{2.6cm} expressed as an \\ \hspace*{2.6cm} ordered quadruple
\item $\ms{M}$ \hspace*{2.1cm} category of \\ \hspace*{2.6cm} multidirected sets
\item $\phi=(\phi_{\tn{\fsz elt}},\phi_{\tn{\fsz rel}})$ \hspace*{.1cm} morphism of \\ \hspace*{2.6cm} multidirected sets
\item $\phi_{\tn{\fsz elt}}$ \hspace*{1.85cm} map of elements
\item $\phi_{\tn{\fsz rel}}$ \hspace*{1.85cm} map of relations
\item $\gamma$ \hspace*{2.2cm} a chain in $M$
\item $\tn{Ch}(M)$ \hspace*{1.2cm} chain set of $M$
\item $\tn{Ch}_n(M)$ \hspace*{1.05cm} set of chains of \\ \hspace*{2.6cm} length $n$ in $M$
\item $\sigma$ \hspace*{2.2cm} an antichain in $M$
\item $V$ \hspace*{2.15cm} a vector space
\item $S$ \hspace*{2.2cm} linearly independent \\ \hspace*{2.6cm} subset of $V$
\item $\Gamma, \Gamma'$ \hspace*{1.75cm} sets of chains in\\ \hspace*{2.6cm} definition \hyperref[definitionindependence]{\ref{definitionindependence}}
\item IC \hspace*{2.05cm} independence convention
\item $J(x)$  \hspace*{1.55cm} total domain of influence \\\hspace*{2.6cm} of a spacetime event $x$
\item $I^\pm(x)$ \hspace*{1.5cm} chronological past and\\ \hspace*{2.6cm} future of a spacetime   \\\hspace*{2.6cm} event $x$
\item $J^\pm(x)$ \hspace*{1.5cm} causal past and future\\ \hspace*{2.6cm} of a spacetime event $x$
\item $D(x)$  \hspace*{1.58cm} total domain of influence \\\hspace*{2.6cm} of an element $x$ in\\\hspace*{2.6cm} a multidirected set

\item $D^\pm(x)$  \hspace*{1.35cm} past and future of $x$ 
\item $D_0^\pm(x)$  \hspace*{1.37cm} direct past and future of $x$
\item $D_{\tn{\fsz{max}}}^-(x),$  \hspace*{.88cm} maximal past and minimal \\ $D_{\tn{\fsz{min}}}^+(x)$\hspace*{1.2cm} future of $x$
\item $\partial M$ \hspace*{1.8cm} boundary of $M$ 
\item $\tn{Int}(M)$ \hspace*{1.2cm} interior of $M$ 
\item $\ms{A}_{\tn{\fsz fin}}$ \hspace*{1.75cm} class of isomorphism\\ \hspace*{2.6cm} classes of finite \\ \hspace*{2.6cm} acyclic directed sets
\item $\aleph_0$ \hspace*{2.05cm} countably infinite \\ \hspace*{2.6cm} cardinal
\item $\ms{A}_{\aleph_0}$ \hspace*{1.8cm} class of isomorphism\\ \hspace*{2.6cm} classes of countable\\ \hspace*{2.6cm} acyclic directed sets
\item $\ms{C}_{\tn{LF}}$ \hspace*{1.83cm} class of isomorphism\\ \hspace*{2.6cm} classes of locally\\ \hspace*{2.6cm} finite causal sets
\item $\ms{A}_{\aleph_0,\tn{LF}}$ \hspace*{1.35cm} class of isomorphism\\ \hspace*{2.6cm} classes of countable locally \\ \hspace*{2.6cm} finite acyclic directed sets

\item $S_{\aleph_0}$ \hspace*{1.9cm} fixed countably infinite set
\item $2^{\aleph_0}$ \hspace*{1.95cm} cardinality of $\ms{P}(S_{\aleph_0})$ 
\item OEP \hspace*{1.65cm} order extension principle
\item $\ms{C}$ \hspace*{2.3cm} a category; e.g., causal sets
\item $\tn{Obj}(\ms{C})$ \hspace*{1.25cm} objects  of $\ms{C}$
\item $\tn{Mor}(\ms{C})$ \hspace*{1.2cm} morphisms  of $\ms{C}$
\item $\mbf{i},\mbf{t}$ \hspace*{2cm} initial and terminal \\ \hspace*{2.6cm} object maps of a \\ \hspace*{2.6cm} category
\item $\gamma_C$ \hspace*{2.05cm} identity morphism of $C$
\item RV \hspace*{1.95cm} Grothendieck's relative \\ \hspace*{2.6cm} viewpoint
\item HS \hspace*{1.9cm} (principle of) hidden \\ \hspace*{2.6cm} structure
\end{itemize}
\end{multicols}

{\bf Section \hyperref[sectiontransitivity]{\ref{sectiontransitivity}}:}
\begin{multicols}{2}
\begin{itemize}
\item $\mbf{C}$  \hspace*{2.15cm} Finkelstein's transitive \\ \hspace*{2.6cm} causal relation
\item $\mbf{c}$  \hspace*{2.25cm} Finkelstein's nontransitive \\ \hspace*{2.6cm} local causal relation
\item $\prec$ \hspace*{2.25cm} binary relation, viewed as  \\ \hspace*{2.6cm} a causal preorder
\item $\tn{tr}(D)$ \hspace*{1.5cm} transitive closure of $D$
\item $\prec_{\tn{\fsz tr}}$ \hspace*{2cm} binary relation on $\tn{tr}(D)$
\item $\tn{sk}(D)$ \hspace*{1.5cm} skeleton of $D$
\item $\prec_{\tn{\fsz sk}}$ \hspace*{1.9cm} binary relation on $\tn{sk}(D)$
\item $\ms{D}_{\tn{\fsz tr}}$ \hspace*{1.9cm} category of transitive\\ \hspace*{2.6cm} directed sets
\item $\tn{Mor}_{\ms{D}}\big(D,D'\big)$ \hspace*{.25cm} set of morphisms \\ \hspace*{2.6cm} from $D$ to $D'$ in $\ms{D}$
\end{itemize}
\end{multicols}

{\bf Section \hyperref[sectioninterval]{\ref{sectioninterval}}:}
\begin{multicols}{2}
\begin{itemize}
\item $\ms{T}, \ms{T}'$ \hspace*{1.7cm} topologies 
\item $\overline{S}$ \hspace*{2.2cm} enlarged set containing $S$
\item $\overline{\ms{T}}$ \hspace*{2.2cm} topology on $\overline{S}$
\item $\ms{T}_{\tn{\fsz dis}}$ \hspace*{1.8cm} discrete topology 
\item $\ms{T}_{\tn{\fsz int}}$ \hspace*{1.8cm} interval topology
\item $R(x)$ \hspace*{1.6cm} relation set at $x$
\item $E(x)$ \hspace*{1.6cm} unit intervals \\ \hspace*{2.6cm} glued at $x$
\item $M_\RR$ \hspace*{1.9cm} continuum model of $M$
\item $\ms{T}_\RR$ \hspace*{2cm} continuum topology  
\item $\tn{St}_\rho(x)$ \hspace*{1.4cm} star of radius $\rho$ at $x$
\item $U$ \hspace*{2.15cm} neighborhood of $x$\\ \hspace*{2.6cm} in continuum topology
\item $M_\star$ \hspace*{1,9cm} star model of $M$
\item $\tn{St}(x)$ \hspace*{1.5cm} star at $x$ in $M_\star$
\item $\ms{T}_\star$ \hspace*{2.05cm} star topology
\item $\ms{T}_{\star,\tn{\fsz irr}}$ \hspace*{1.6cm} irreducible star topology
\item $\tn{St}_{\tn{\fsz irr}}(x)$ \hspace*{1.15cm} irreducible star \\ \hspace*{2.6cm} at $x$ in $M_\star$
\item $v(x)$ \hspace*{1.65cm} valence of $x$
\item $v$ \hspace*{2.25cm} cardinal-valued \\ \hspace*{2.6cm} scalar field on $M$
\item $v_M$ \hspace*{2cm} valence field on $M$
\item $v_M^{\pm}$ \hspace*{1.95cm} past and future \\ \hspace*{2.6cm}  valence fields on $M$
\item LF \hspace*{2cm} local finiteness
\item LFT \hspace*{1.7cm} local finiteness in the \\ \hspace*{2.6cm} transitive closure
\item LFS \hspace*{1.75cm} local finiteness in the \\ \hspace*{2.6cm} skeleton
\item $\alpha$ \hspace*{2.2cm} causal atom \\ \hspace*{2.6cm} in theorem \hyperref[theoremcausalsetrelint]{\ref{theoremcausalsetrelint}}
\item $\llangle \alpha, y_{k+1}\rrangle$ \hspace*{.93cm} generalized open interval \\ \hspace*{2.6cm} in theorem \hyperref[theoremcausalsetrelint]{\ref{theoremcausalsetrelint}}
\item $\NN\coprod-\NN$ \hspace*{1.25cm} linearly ordered base \\ \hspace*{2.6cm} for Jacob's ladder
\item $[n]$  \hspace*{2cm} finite simplex $\{0,...,n\}$. 
\item EF \hspace*{1.95cm} element finiteness
\item LEF \hspace*{1.75cm} local element finiteness
\item RF \hspace*{1.95cm} relation finiteness
\item PRF \hspace*{1.7cm} pairwise relation finiteness
\item CF \hspace*{1.95cm} chain finiteness
\item AF \hspace*{1.95cm} antichain finiteness
\end{itemize}
\end{multicols}

\newpage
{\bf Section \hyperref[sectionbinary]{\ref{sectionbinary}}:}
\begin{multicols}{2}
\begin{itemize}
\item $\ms{R}(M)$ \hspace*{1.45cm} relation space over \\ \hspace*{2.6cm} $M=(M,R,i,t)$ 
\item $\prec$ \hspace*{2.25cm} induced relation on $\ms{R}(M)$ 
\item $r,s,\overline{r},\overline{s}$ \hspace*{1.25cm} elements of relation \\ \hspace*{2.6cm} spaces 
\item $\ms{R}:\ms{M}\rightarrow\ms{D}$ \hspace*{.7cm} relation space functor
\item $\phi=(\phi_{\tn {\fsz elt}},\phi_{\tn {\fsz rel}})$ \hspace*{.1cm} morphism of \\ \hspace*{2.6cm} multidirected sets
\item $\ms{R}(\phi)$ \hspace*{1.6cm} induced morphism of \\ \hspace*{2.6cm} relation spaces 
\item $R=(R,\prec)$  \hspace*{.8cm} arbitrary directed set, \\ \hspace*{2.6cm} viewed as a relation set
\item $\ms{E}(R)$ \hspace*{1.6cm} abstract element space \\ \hspace*{2.6cm} over $R$
\item $M^{\pm}$ \hspace*{1.9cm} abstract sets yielding \\ \hspace*{2.6cm} $\ms{E}(R)$ as a quotient
\item $\widetilde{x_r},\widetilde{y_r}$ \hspace*{1.6cm} abstract elements \\ \hspace*{2.6cm} in $\ms{E}(R)$ of $r\in R$ 
\item $\ms{M}_{\overline{\partial}},\ms{D}_{\overline{\partial}}$ \hspace*{1.25cm} subcategories of $\ms{M}$ \\ \hspace*{2.6cm} and $\ms{D}$ whose objects\\ \hspace*{2.6cm} have empty boundary
\item $\tn{Id}_{\ms{M}_{\overline{\partial}}},\tn{Id}_{\ms{D}_{\overline{\partial}}}$ \hspace*{.7cm} identity functors on \\ \hspace*{2.6cm} $\ms{M}_{\overline{\partial}}$ and $\ms{D}_{\overline{\partial}}$
\item $\sigma$ \hspace*{2.2cm} Cauchy surface in \\ \hspace*{2.6cm} spacetime, or maximal \\ \hspace*{2.6cm} antichain in directed or \\ \hspace*{2.6cm} multidirected set
\item $W,X,Y,Z$ \hspace*{.75cm} objects of a category
\item $a,b,f,g$ \hspace*{1.2cm} morphisms of a category
\item $(M_\star,R_\star,i_\star, t_\star)$ \hspace*{0cm}  star model of $M$ as \\ \hspace*{2.65cm} a multidirected set
\item $M_2$ \hspace*{1.9cm} set of $2$-element \\ \hspace*{2.6cm} subsets of $M$
\item $\ms{R}^2(A)$ \hspace*{1.35cm} second relation space \\ \hspace*{2.6cm} over $A$
\item $\prec_n$ \hspace*{2.05cm} $n$th induced relation 
\item $\overline{\gamma},\overline{\delta}$  \hspace*{1.9cm} simplices in $M$
\item $\alpha,\alpha'$ \hspace*{1.7cm} causal atoms
\item $M^n$ \hspace*{1.85cm} $n$th degree causal atomic\\ \hspace*{2.6cm} decomposition of $M$
\item $\mbf{M}$ \hspace*{2.05cm} causal atomic \\ \hspace*{2.6cm} resolution of $M$
\item $\lambda_n$ \hspace*{2.05cm} $n$th growth factor for $\mbf{M}$
\item $\Lambda(\mbf{M})$ \hspace*{1.45cm} growth sequence for $\mbf{M}$
\item $T$ \hspace*{2.15cm} twistor space over \\ \hspace*{2.6cm} Minkowski spacetime $X$
\item $\CC\PP^3$ \hspace*{1.75cm} complex projective space 
\item $L,L',L''$ \hspace*{1.05cm} linear directed sets
\item $\alpha, \beta, \gamma, \delta$ \hspace*{1.1cm} paths in a directed \\ \hspace*{2.6cm} or multidirected set
\item $\Gamma(M)$ \hspace*{1.45cm} path set of $M$
\item $\Gamma_{\tn{\fsz irr}}(M)$ \hspace*{1.1cm} irreducible path\\ \hspace*{2.6cm} set of $M$
\item $\ell_+,\ell_-'$ \hspace*{1.6cm} terminal element of $L$ and\\ \hspace*{2.6cm} initial element of $L'$ in\\ \hspace*{2.6cm}  definitions \hyperref[deficoncatprod]{\ref{deficoncatprod}} and \hyperref[defidirprod]{\ref{defidirprod}}. 
\item $L\sqcup L'$ \hspace*{1.4cm} linear directed set given\\ \hspace*{2.6cm} by joining $L,L'$ via \\ \hspace*{2.6cm} connecting relation $\ell_+\prec\ell_-'$
\item $\alpha\sqcup\beta$ \hspace*{1.6cm} concatenation product \\ \hspace*{2.6cm} of $\alpha$ and $\beta$
\item $\prec_\sqcup$ \hspace*{2.05cm} concatenation relation \\ \hspace*{2.6cm} on $\Gamma(D)$
\item $\big(\Gamma(D),\prec_\sqcup\big)$ \hspace*{.55cm} causal concatenation space \\ \hspace*{2.6cm} over $D=(D,\prec)$
\item $L\vee L'$ \hspace*{1.45cm} linear directed set given\\ \hspace*{2.6cm} by identifying $\ell_+$ with $\ell_-'$
\item $\alpha\vee\beta$ \hspace*{1.55cm} directed product \\ \hspace*{2.6cm} of $\alpha$ and $\beta$
\item $\prec_\vee$ \hspace*{2.05cm} directed product \\ \hspace*{2.6cm} relation on $\Gamma(M)$
\item $\big(\Gamma(M),\prec_\vee\big)$ \hspace*{.45cm} causal directed product \\ \hspace*{2.6cm} space over $M=(M,R,i,t)$
\item $\ms{S}$ \hspace*{2.2cm} object class of a \\ \hspace*{2.6cm} semicategory
\item $\Gamma(\ms{S})$ \hspace*{1.65cm} morphism class of a \\ \hspace*{2.6cm} semicategory
\item $\mbf{i},\mbf{t}$ \hspace*{2cm} initial and terminal \\ \hspace*{2.6cm} object maps of a \\ \hspace*{2.6cm} semicategory
\item $\big(\Gamma(D),\sqcup\big)$ \hspace*{.8cm} causal concatenation \\ \hspace*{2.6cm} semicategory over $D$
\item $\big(\Gamma(M),\vee\big)$ \hspace*{.7cm} causal directed product \\ \hspace*{2.6cm} semicategory over $M$ 
\item $\big(\tn{Ch}(D),\sqcup\big)$ \hspace*{.5cm} chain concatenation \\ \hspace*{2.6cm} semicategory over $D$
\item $\big(\tn{Ch}(M),\vee\big)$ \hspace*{.45cm} chain directed product \\ \hspace*{2.6cm} semicategory over $M$
\item $\sim$ \hspace*{2.25cm} congruence relation \\ \hspace*{2.6cm} on a semicategory
\item $\RR[x]$  \hspace*{1.7cm} algebra of polynomials \\ \hspace*{2.6cm} with real coefficients
\item $T$ \hspace*{2.2cm} a ring, usually \\ \hspace*{2.6cm} commutative with unit
\item $t_\gamma$ \hspace*{2.2cm} element of $T$, appearing \\ \hspace*{2.6cm} as coefficient of path $\gamma$ \\ \hspace*{2.6cm} in path algebra
\item $T^{\sqcup}(D)$ \hspace*{1.35cm} concatenation algebra over\\ \hspace*{2.6cm} $D$ with coefficients in $T$
\item $T^{\vee}(M)$ \hspace*{1.25cm} directed product  \\ \hspace*{2.6cm} algebra over $M$ with \\ \hspace*{2.6cm} coefficients in $T$
\item $T^{\sqcup}[D]$ \hspace*{1.35cm} chain concatenation \\ \hspace*{2.6cm} algebra over $D$ with \\ \hspace*{2.6cm} coefficients in $T$
\item $T^{\vee}[M]$ \hspace*{1.3cm} chain directed product \\ \hspace*{2.6cm} algebra over $M$ with \\ \hspace*{2.6cm} coefficients in $T$
\item $P(x)$\hspace*{1.75cm} punctual module of $x$
\item $|x\rangle\langle y|$ \hspace*{1.6cm} ket-bra notation for \\ \hspace*{2.6cm} Raptis' incidence algebra
\item $\delta_{uv}$  \hspace*{1.95cm} Kronecker delta function
\item $\ms{L}$\hspace*{2.3cm} Lagrangian
\item $\ms{S}$\hspace*{2.35cm} classical action
\item $\Theta$\hspace*{2.3cm} phase map
\item $\hbar$ \hspace*{2.25cm} Planck's reduced \\ \hspace*{2.6cm} constant $h/2\pi$
\item $\theta$\hspace*{2.4cm} relation function
\item $T$\hspace*{2.3cm} target object of relation \\ \hspace*{2.6cm} function or phase map
\item $\sum_{\gamma\in\Gamma(M)}\Theta(\gamma)\gamma$\hspace*{.05cm} phase map $\Theta$ as a path \\ \hspace*{2.6cm} algebra element
\item $T^{\Gamma(D)}, T^{\Gamma(M)}$\hspace*{.35cm} mapping spaces of maps \\ \hspace*{2.6cm} $\Gamma(D)\rightarrow T$ and $\Gamma(M)\rightarrow T$
\item $\Theta,\Psi$\hspace*{1.8cm} pair of elements of  \\ \hspace*{2.6cm} $T^{\Gamma(D)}$ or $T^{\Gamma(M)}$
\item $\Phi\vee\Psi, \Phi\sqcup\Psi$\hspace*{.6cm} convolutions; i.e., products \\ \hspace*{2.6cm} in path algebras \\ \hspace*{2.6cm} $T^{\sqcup}(D)$ and $T^{\sqcup}(M)$
\item $f\star g$\hspace*{1.85cm} convolution of \\ \hspace*{2.6cm} real-valued functions
\item $\alpha-\beta$\hspace*{1.75cm} path such that\\ \hspace*{2.6cm} $\alpha\sqcup(\alpha-\beta)=\beta$
\item $T^{\tn{Ch}(D)},T^{\tn{Ch}(M)}$\hspace*{.05cm} chain mapping spaces 
\item $0_T$\hspace*{2.2cm} additive identity of $T$ 
\item $e_M$\hspace*{2.1cm} global evaluation \\ \hspace*{2.6cm} map $T^{\Gamma(M)}\dashedrightarrow T$
\item $e_\gamma$\hspace*{2.2cm} indicator function 
\item $\Theta(M)$\hspace*{1.53cm} shorthand for path\\ \hspace*{2.6cm} sum $\sum_{\gamma\in\Gamma(M)}\Theta(\gamma)$
\end{itemize}
\end{multicols}

{\bf Section \hyperref[subsectionquantumcausal]{\ref{subsectionquantumcausal}}:}
\begin{multicols}{2}
\begin{itemize}
\item $\ms{Q}$ \hspace*{2.2cm} category in Isham's \\ \hspace*{2.6cm} quantization
\item $Q,Q'$  \hspace*{1.6cm} objects of $\ms{Q}$
\item IS \hspace*{2.05cm} (principle of) iteration \\ \hspace*{2.6cm} of structure
\item $\tau$ \hspace*{2.25cm} transition in $\ms{D}$
\item $D_i$  \hspace*{2cm} initial object, or source, \\ \hspace*{2.6cm} of a transition
\item $D_t$  \hspace*{2cm} terminal object, or target, \\ \hspace*{2.6cm} of a transition
\item $\tn{Aut}(D_i),$ \hspace*{.95cm} automorphism groups
\\ $\tn{Aut}(D_t)$
\item $\alpha,\beta$ \hspace*{1.8cm} automorphisms of \\ \hspace*{2.6cm} directed sets
\item $F_\tau$ \hspace*{2.05cm} group of extensions \\ \hspace*{2.6cm} of symmetries of $D_i$
\item $\rho_\tau$ \hspace*{2.05cm} restriction homomorphism \\ \hspace*{2.6cm} $F_\tau\rightarrow \tn{Aut}(D_i)$
\item $K_\tau,G_\tau$ \hspace*{1.3cm} kernel and cokernel \\ \hspace*{2.6cm} of $\rho_\tau$, also called \\ \hspace*{2.6cm} causal Galois groups
\item $\iota$ \hspace*{2.3cm} inclusion $K_\tau\rightarrow F_\tau$
\item $\overline{\rho_\tau}$  \hspace*{2.1cm} quotient homomorphism \\ \hspace*{2.6cm} $G_\tau\rightarrow  \tn{Aut}(D_i)$
\item $[F_\tau:G_\tau]$  \hspace*{.95cm} subgroup index of \\ \hspace*{2.6cm} $G_\tau$ in $F_\tau$
\item $h:D_i\Rightarrow D_t$ \hspace*{.55cm} co-relative history
\item $D_i,D_t$  \hspace*{1.4cm} cobase and target \\ \hspace*{2.6cm} of $h$
\item $\mbf{D}_i$ \hspace*{2.05cm} cobase family 
\item PS \hspace*{1.95cm} path summation principle
\item $X$ \hspace*{2.15cm} finite subset of\\ \hspace*{2.6cm} Euclidean spacetime
\item $t=t^{\pm}$ \hspace*{1.6cm} spatial sections\\ \hspace*{2.6cm} bounding $X$
\item $\sigma^{\pm}$ \hspace*{2.05cm} upper and lower\\ \hspace*{2.6cm} boundaries of $X$
\item $\Gamma=\Gamma_{\tn{\fsz max}}(X)$ \hspace*{.3cm} space of maximal\\ \hspace*{2.6cm} directed paths in $X$
\item $\gamma$ \hspace*{2.25cm} element of $\Gamma_{\tn{\fsz max}}(X)$
\item $\ms{L}$ \hspace*{2.15cm} Lagrangian for single\\ \hspace*{2.6cm} particle moving in $X$
\item $\ms{S}$ \hspace*{2.25cm} classical action  \\ \hspace*{2.6cm} corresponding to \\ \hspace*{2.6cm} Lagrangian $\ms{L}$
\item F1, F2 \hspace*{1.35cm} Feynman's ``postulates"
\item $\psi(X;\ms{L})$ \hspace*{1.05cm} Feynman's quantum \\ \hspace*{2.6cm} amplitude 
\item $\hbar$ \hspace*{2.25cm} Planck's reduced \\ \hspace*{2.6cm} constant $h/2\pi$
\item $\Theta(\gamma)=e^{\frac{i}{\hbar}\ms{S}(\gamma)}$ \hspace*{.2cm} Feynman's phase map
\item $\Delta$ \hspace*{2.2cm} partition of $[t^-,t^+]$ 
\item $\mbf{x}$ \hspace*{2.3cm} sequence of \\ \hspace*{2.6cm} position values for $\Delta$
\item $\gamma_i$ \hspace*{2.1cm} ``path increment" \\ \hspace*{2.6cm} from $(x_i,t_i)$ to $(x_{i+1},t_{i+1})$
\item $\sigma_i=\sigma_i(\Delta)$ \hspace*{.75cm} spatial section $t=t_i$ \\ \hspace*{2.6cm} for $t_i\in\Delta$
\item $|\Delta|$ \hspace*{2cm} norm of partition $\Delta$
\item $\psi(\Delta;\mbf{x};\ms{L})$ \hspace*{.72cm} amplitude for a\\ \hspace*{2.6cm} particular sequence $\mbf{x}$ \\ \hspace*{2.6cm} of position values
\item $C$ \hspace*{2.15cm} proportionality factor \\ \hspace*{2.6cm}  in Feynman's integral
\item $R$ \hspace*{2.15cm} finite subset of\\ \hspace*{2.6cm}  relation space $\ms{R}(M')$
\item $\big(\tn{Ch}(R),\sqcup\big)$ \hspace*{.55cm} chain concatenation \\ \hspace*{2.6cm} semicategory over $R$
\item $\Gamma=\tn{Ch}_{\tn{\fsz max}}(R)$ \hspace*{.1cm} subset of maximal \\ \hspace*{2.6cm} chains in $R$
\item $\gamma$ \hspace*{2.3cm} element of $\tn{Ch}_{\tn{\fsz max}}(R)$
\item $\sigma$ \hspace*{2.25cm} maximal \\ \hspace*{2.6cm} antichain in $R$
\item $R^-\coprod\sigma\coprod R^+$ \hspace*{.37cm} separation of $R$ \\ \hspace*{2.6cm} into past and future \\ \hspace*{2.6cm} regions $R^\pm$ by $\sigma$
\item $T$ \hspace*{2.2cm} target object \\ \hspace*{2.6cm} for path sums 
\item $\Theta$ \hspace*{2.15cm} phase map \\ \hspace*{2.6cm} $\tn{Ch}(R)\rightarrow T$
\item $\theta$ \hspace*{2.25cm} relation function \\ \hspace*{2.6cm} generating $\Theta$
\item $\psi(R;\theta)$ \hspace*{1.25cm} generalized quantum \\ \hspace*{2.6cm} amplitude
\item $T^\sqcup[R]$ \hspace*{1.4cm} chain concatenation \\ \hspace*{2.6cm} algebra over $R$
\item $\Psi(R;\theta)$ \hspace*{1.2cm} maximal chain  \\ \hspace*{2.6cm} functional of $R$  \\ \hspace*{2.6cm} with respect to $\theta$
\item $\Theta|_\Gamma$ \hspace*{1.9cm} restriction of $\Theta$  \\ \hspace*{2.6cm} to $\Gamma=\tn{Ch}_{\tn{\fsz max}}(R)$
\item $\psi^-$ \hspace*{2cm} past wave function in \\ \hspace*{2.6cm} Schr\"{o}dinger's equation
\item $\mbf{H}$ \hspace*{2.1cm} Hamiltonian operator
\item $V$ \hspace*{2.15cm} potential energy
\item $m$ \hspace*{2.15cm} particle mass
\item $\big\langle,\big\rangle$ \hspace*{2.05cm} complex inner product
\item $\sigma'$ \hspace*{2.2cm} spatial section $t=t'$ 
\item $X^-\coprod\sigma'\coprod X^+$ \hspace*{.15cm} separation of $X$ \\ \hspace*{2.6cm} into past and future \\ \hspace*{2.6cm} regions $X^\pm$ by $\sigma'$
\item $\Delta^-$ \hspace*{1.95cm} partition of time \\ \hspace*{2.6cm} interval for $X^-$
\item $\approx$ \hspace*{2.25cm} signifies approximation 
\item $t''$ \hspace*{2.1cm} a time shortly after $t'$
\item $\delta\gamma$ \hspace*{2.05cm} path increment
\item $r$ \hspace*{2.3cm} element of $\sigma$
\item $\Gamma^\pm$ \hspace*{2.05cm} sets of maximal chains  \\ \hspace*{2.6cm} in $R^\pm$, respectively
\item $\gamma$ \hspace*{2.25cm} element of $\Gamma^-$
\item $\Theta|_\sigma$ \hspace*{1.9cm} restriction of phase \\ \hspace*{2.6cm} map $\Theta$ to $\sigma$ 
\item $\Gamma_\sigma^{\pm}$ \hspace*{2cm} sets of maximal chains in \\ \hspace*{2.6cm} $R$ terminating on $\sigma$ $(-)$; \\ \hspace*{2.6cm} beginning on $\sigma$ $(+)$
\item $\Gamma_r^{\pm}$ \hspace*{2cm} sets of maximal chains in\\ \hspace*{2.6cm} $R$ terminating at $r$ $(-)$; \\ \hspace*{2.6cm} beginning at $r$ $(+)$
\item $\Psi^{\pm}_{R,\theta}$ \hspace*{1.7cm} past and future \\ \hspace*{2.6cm} chain functionals
\item $\psi^{\pm}_{R,\theta}$ \hspace*{1.7cm} past and future \\ \hspace*{2.6cm} wave functions
\item $r^-$ \hspace*{2.1cm} maximal predecessor of $r$ 
\item $(\ms{K},\mc{H})$ \hspace*{1.3cm} kinematic prescheme \\ \hspace*{2.6cm} or scheme
\item $\ms{K}$ \hspace*{2.1cm} object class of $(\ms{K},\mc{H})$ 
\item $\mc{H}$ \hspace*{2.15cm} class of co-relative \\ \hspace*{2.6cm} histories of $(\ms{K},\mc{H})$ 
\item $\ms{U}(\ms{K},\mc{H})$ \hspace*{1cm} underlying directed set \\ \hspace*{2.6cm} of $(\ms{K},\mc{H})$
\item $\ms{V}(\ms{K},\mc{H})$ \hspace*{1.05cm} underlying multidirected \\ \hspace*{2.6cm} set of $(\ms{K},\mc{H})$
\item $\ms{W}(\ms{K},\mc{H})$ \hspace*{.95cm} underlying transition \\ \hspace*{2.6cm} structure of $(\ms{K},\mc{H})$
\item $x(D_i),x(D_t)$ \hspace*{.3cm} elements in an \\ \hspace*{2.6cm} underlying directed or \\ \hspace*{2.6cm} multidirected set of $(\ms{K},\mc{H})$
\item $r(\tau),r(h)$ \hspace*{.8cm} relations in an \\ \hspace*{2.6cm} underlying directed or \\ \hspace*{2.6cm} multidirected set of $(\ms{K},\mc{H})$
\item H  \hspace*{2.2cm} hereditary property
\item WA  \hspace*{1.85cm} weak accessibility
\item QCMH \hspace*{1.25cm} quantum causal metric \\ \hspace*{2.6cm} hypothesis
\item $(\ms{A}_{\tn{\fsz fin}},\mc{H}_1)$ \hspace*{.8cm} positive sequential \\ \hspace*{2.6cm} kinematic scheme
\item $\mc{H}_1$ \hspace*{2cm} class of transitions\\ \hspace*{2.6cm} adding a single element \\ \hspace*{2.6cm} to a directed set
\item $(\ms{C}_{\tn{\fsz fin}},\mc{H}_1)$ \hspace*{.85cm} Sorkin and Rideout's \\ \hspace*{2.6cm} kinematic scheme
\item $(\ms{A}_{\tn{\fsz fin}},\mc{H}_{\tn{\fsz gen}})$ \hspace*{.45cm} generational \\ \hspace*{2.6cm} kinematic scheme
\item $\mc{H}_{\tn{\fsz gen}}$ \hspace*{1.65cm} class of transitions\\ \hspace*{2.6cm} adding a single generation \\ \hspace*{2.6cm} to a directed set
\item $\widetilde{\tn{Ch}_{\tn{\fsz{max}}}}(\ms{K},\mc{H})$ \hspace*{.15cm} class of limits of $(\ms{K},\mc{H})$
\item $\overline{(\ms{K},\mc{H})}$ \hspace*{1.3cm} chain completion \\ \hspace*{2.6cm} of $(\ms{K},\mc{H})$
\item $\mc{I}$ \hspace*{2.2cm} inclusion ``functor" of \\ \hspace*{2.6cm} kinematic schemes
\item $(F,\prec)$ \hspace*{1.45cm} Farey ``tree"
\item $\mc{H}(\ms{K})$ \hspace*{1.45cm} class of all transitions \\ \hspace*{2.6cm} between pairs of \\ \hspace*{2.6cm} objects of $\ms{K}$
\item $\big(\ms{K},\mc{H}(\ms{K})\big)$ \hspace*{.6cm} universal kinematic \\ \hspace*{2.6cm} scheme over $\ms{K}$
\item $\big(\mbf{Kin}(\ms{K}),\prec_{\ms{K}}\big)$ \hspace*{0cm} kinematic space over $\ms{K}$
\end{itemize}
\end{multicols}

{\bf Section \hyperref[sectionconclusions]{\ref{sectionconclusions}}:}
\begin{multicols}{2}
\begin{itemize}
\item M$^*$ \hspace*{1.95cm} modified measure axiom
\item $e_i$ \hspace*{2.17cm} generalized initial or \\ \hspace*{2.6cm} terminal element map
\item $V$ \hspace*{2.15cm} vertex set of a \\ \hspace*{2.6cm} random graph
\item $(N,p)$ \hspace*{1.45cm} type of random graph 
\item $N$ \hspace*{2.1cm} number of vertices \\ \hspace*{2.6cm} in a random graph
\item $p$ \hspace*{2.25cm} probability of an edge \\ \hspace*{2.6cm} between two vertices \\ \hspace*{2.6cm} in a random graph
\end{itemize}
\end{multicols}

\end{appendix}

\newpage

\setcounter{secnumdepth}{0}

\section{References}

In the information age it is easy, and therefore not very helpful, to exhibit hundreds of papers on any given scientific subject not wholly obscure.  I have chosen instead to restrict attention to a minimal number of reliable and well-written references, subject to the requirement of conveying the proper priority of authorship. Many of the authors cited here have produced numerous other equally relevant papers, but I have made an effort to distill the necessary concepts and quotations from a more manageable cross-section.  A few of these references fall short in quality and/or clarity, but demand mention due to their priority in treating certain crucial topics.  I have cited peer-reviewed and/or edited versions of references whenever possible, but have also included blue-colored links to arXiv preprints when available; the reader should always be aware of the possibility of differences between the two, and of the transient nature of links in general.  Only the names of the actual articles cited are in italics; journal names, books containing articles, etc., are in normal font.  Bold digits in typical journal citations indicate the {\it volume} of the journal cited; succeeding digits indicate the {\it number} or {\it issue,} the {\it page numbers} (preceded by ``pp.") or {\it article number,} and finally the {\it year published.} References are listed in the order in which they are {\it first cited} in the paper.  The section headings below indicate in which sections these initial citations appear.  I beg indulgence for any unintended slights due to my own ignorance. 


\vspace*{-1.5cm}
\renewcommand{\refname}{}

\end{document}